\documentclass{article}

\usepackage[preprint]{neurips_2026}

\usepackage[utf8]{inputenc} 
\usepackage[T1]{fontenc}    
\usepackage{hyperref}       
\usepackage{url}            
\usepackage{booktabs}       
\usepackage{amsfonts}       
\usepackage{nicefrac}       
\usepackage{microtype}      
\usepackage{xcolor}         

\usepackage{parskip}

\usepackage{dsfont}

\usepackage[caption=false,font=footnotesize]{subfig}
\usepackage{wrapfig}
\usepackage[leftcaption]{sidecap}

\usepackage{array, makecell}
\usepackage{threeparttable}

\usepackage{csquotes}

\usepackage[normalem]{ulem}
\usepackage{soul}
\usepackage{nicefrac}

\usepackage{chngcntr}
\usepackage{apptools}
\usepackage[]{color-edits}

\usepackage{amsmath,amsfonts,bm}

\def\eqref#1{equation~\ref{#1}}

\def\1{\bm{1}}

\DeclareMathAlphabet{\mathsfit}{\encodingdefault}{\sfdefault}{m}{sl}
\SetMathAlphabet{\mathsfit}{bold}{\encodingdefault}{\sfdefault}{bx}{n}

\usepackage{mathrsfs}
\definecolor{tcbcolback}{RGB}{245,243,253}
\definecolor{tcbcolframe}{RGB}{245,243,253}

\usepackage[most]{tcolorbox}

\definecolor{custompurple}{HTML}{6A468D}
\definecolor{lightgray}{RGB}{230,230,230}
\definecolor{lightergray}{RGB}{250,250,250}

\usepackage{amsmath}
\usepackage{amssymb}
\usepackage{mathtools}
\usepackage{amsthm}

\usepackage{graphicx}
\usepackage{subcaption}

\usepackage{algorithm}
\usepackage{algorithmic}
\usepackage{nicefrac}

\usepackage[capitalize,noabbrev]{cleveref}

\usepackage[percent]{overpic}

\newcommand{\figpanel}[2]{%
\begin{minipage}[b]{0.32\linewidth}
    \centering
    \includegraphics[width=\linewidth]{#2}\\[-0.1em]
    {\scriptsize\bfseries #1}
\end{minipage}
}

\usepackage{hyperref}       
\hypersetup{
	colorlinks   = true, 
	urlcolor     = blue, 
	linkcolor    = blue, 
	citecolor   = blue 
}

\theoremstyle{plain}
\newtheorem{theorem}{Theorem}[section]
\newtheorem{proposition}[theorem]{Proposition}
\newtheorem{lemma}[theorem]{Lemma}
\newtheorem{corollary}[theorem]{Corollary}
\theoremstyle{definition}

\theoremstyle{remark}
\newtheorem{remark}[theorem]{Remark}

\usepackage[textsize=tiny]{todonotes}

\title{Evaluating Feedback-Driven Black-Box Processes via Stochastic Modeling and Simulation}

\author{%
  Mengfan Xu, Zirui Liu, Juan Zhai, and Shiqing Ma \\
  University of Massachusetts Amherst \\
  \texttt{\{mengfanxu,zliu,juanzhai,shiqingma\}@umass.edu}
  \AND
  Barry Nelson \\
  Northwestern University \\
  \texttt{nelsonb@northwestern.edu}
}

\begin{document}

\maketitle

\begin{abstract}
We study feedback-driven black-box processes across software engineering and
machine learning. These processes are inherently sequential: policies generate
inputs, observe outputs/feedback from systems, and use the feedback to guide
the design of subsequent inputs. While many policies are developed under
structural assumptions on the underlying systems, many real-world systems
operate without such knowledge, necessitating black-box policies that lack a
unified evaluation framework. We introduce a stochastic framework that models
and evaluates such policies using Markov chains and simulation, thereby
connecting empirical practice with theory. We instantiate the framework in
mutation-based fuzzing (widely used for discovering software bugs
via sequential input mutation), by a novel Markov chain model of fuzzing whose
transitions are governed by the mutation strategy and whose feedback is given
by program execution. We develop a rigorous evaluation paradigm with new
performance measures, sound estimators, and evaluation procedures with
provable guarantees. Guided by the theory, we evaluate 9
fuzzers on 18 widely used fuzzing benchmarks and conduct case studies in 4
non-fuzzing domains. These results show that the framework provides a
principled basis for evaluation while connecting to practice, with implications
beyond fuzzing.
\end{abstract}

\section{Introduction}



A broad class of problems in software engineering \citep{xia2024fuzz4all,
you2019profuzzer, klees2018evaluating} and machine learning
\citep{ziegler2016synthesis, bendkowski2022tuning, belete2022grid} can be
formulated as \emph{feedback-driven black-box processes}, in which inputs are
generated sequentially by policies guided by observed feedback from an unknown
system. This paradigm appears in many practical applications, including
evolutionary and genetic algorithms such as CMA-ES \citep{auger2005restart},
heuristic and rule-based parameter tuning in combinatorial optimization
\citep{bendkowski2022tuning, belete2022grid}, testing and search procedures
such as fuzzing \citep{arcuri2010black, braiek2020testing,
zolfagharian2023search, xia2024fuzz4all, you2019profuzzer,
klees2018evaluating}, and decision-making systems in dynamic pricing
\citep{luo2024distribution}, advertisement bidding \citep{feldman2007budget},
online learning \citep{provodin2024bandits, chen2021black}, and language models
\citep{zhang2026instruction, openai_clip_2021}. Across these settings, the system is treated as a black box, feedback is often
indirect but observable for each input, and the policy
selects the next input based on this feedback.

A defining theme of feedback-driven black-box processes is that policies
depend on system complexity and on how much system information can be learned
from aligned input-feedback pairs. Many learning and optimization methods
\citep{chen2021black, mcmahan2017survey, shalev2025online} have been
developed for systems with explicit state representations, feedback models, or
strong structural assumptions, which make learning tractable and have led to
provable policy success. Yet, many practical systems involve complex dynamics
and hidden states that violate these assumptions \citep{arcuri2010black, auger2005restart, openai_clip_2021}, leading to heuristic policies whose behavior is difficult for users to characterize
and is effectively \emph{black-box}. Hence, performance is
assessed empirically in an ad hoc manner, often without a unified objective or
principled guarantees that account for intrinsic complexity, leading 
to ongoing debates. This, in turn, reinforces the development of black-box policies, forming a
closed loop. Breaking this loop is fundamental and not confined to a single
domain, thereby motivating a new modeling and evaluation paradigm.



A representative, practically important, and challenging instance of such
processes is software fuzzing. Fuzzing is a core software testing methodology
that sequentially generates inputs (bit strings) using fuzzers, feeds them to a
program (e.g., software developed for machine learning
\citep{bendkowski2022tuning}, by machine learning
\citep{esteva2017dermatologist}, or by GenAI \citep{he2025llm}), and observes
execution outcomes to uncover hidden vulnerabilities or
bugs~\cite{godefroid2020fuzzing, godefroid2008grammar, wang2019superion,
jero2019leveraging, petsios2017nezha, petsios2017slowfuzz, she2024fox}. It has
long been central to software safety and is increasingly essential as
AI-assisted software development grows. Among existing approaches,
mutation-based fuzzing is one of the most widely deployed
techniques~\cite{cha2015program, lyu2019mopt, lemieux2018fairfuzz}: starting
from an initial input, a fuzzer iteratively mutates (modifies) it and uses
program feedback to guide future mutations.

Despite its extensive empirical success, fuzzing clearly illustrates the
broader difficulty of feedback-driven black-box processes. Its objective is not
rigorously defined, and commonly used evaluation metrics, such as code coverage
\citep{klees2018evaluating} or raw/unique bug counts \citep{li2021unifuzz}, are
ad hoc, sensitive to randomness \citep{klees2018evaluating, li2021unifuzz,
schloegel2024sok}, and often yield inconsistent conclusions across runs and
benchmarks. More fundamentally, the stochastic dynamics induced by policies
through sequential input generation remain poorly understood, limiting
principled evaluation and leading to ongoing debates. It further amplifies the trend of designing black-box policies rather
than stylized, theoretically grounded ones.

\emph{Stochastic modeling}, a branch of applied probability, represents system
complexity and variability via probabilistic models, avoiding deterministic or
exhaustive enumeration of all possible scenarios, as often encountered in
fuzzing. For example, some work~\cite{bohme2016coverage, bohme2017directed} has employed
stochastic modeling to guide mutation and develop new fuzzers. However, these
models 1) are tailored to specific fuzzers rather than general fuzzing or broader
black-box processes, and 2) focus on purely empirical performance rather than
resolving the evaluation debate. \emph{Simulation, not a simulator}, is a fundamental technique for executing
and analyzing \emph{real systems}: it repeatedly feeds inputs into systems and
analyzes the resulting outputs using stochastic analyses~\cite{nelson2010stochastic,
nelson2021foundations, rubino2009rare, wilson1984variance}. It avoids requiring
explicit knowledge of the underlying stochastic model and naturally provides a
framework for evaluating fuzzing processes. For instance, it has recently been applied in fuzzing to develop fuzzers empirically~\cite{vaisnor2024approximate, bohme2017directed}. Yet, these studies lack explicit modeling and evaluation of the fuzzing process, leaving a critical gap. The same applies to broader feedback-driven processes; a thorough literature review is in Appendix~\ref{app:related_work}.

These challenges motivate a meaningful question: \emph{Can we develop a general stochastic framework for modeling and evaluating feedback-driven black-box processes and apply it to the key area of fuzzing for software security?} We address this question by developing such a framework and instantiating it in fuzzing as a central case study, while confirming its applicability to related processes in machine learning domains beyond fuzzing.
Our \textbf{contributions} are as follows.  






We develop a general stochastic framework for \emph{feedback-driven black-box
processes} that formalizes the sequential interaction between an
input-generation policy and an unknown feedback-providing system, together
with the possibly hidden policy state, as a joint stochastic process.
\emph{In the context of fuzzing, this corresponds to the interaction between a
fuzzer and a target program.} 

Within this framework, the process is modeled via 1) a discrete Markov
chain on the joint space of inputs and hidden states, and 2) system feedback
as an output process generated by the chain. A feedback-driven policy is
characterized by an unknown \emph{transition matrix} over this joint space.
Policy-induced input transitions propagate through the black-box system to
generate stochastic outputs, while the underlying dynamics remain unobservable.
We define unified performance objectives that capture both long-run and
finite-time behavior under intrinsic randomness. \emph{For mutation-based fuzzing, each mutation corresponds to a Markov transition, the transition matrix encodes the fuzzer's mutation strategy, and the objectives are the bug detection rate and the adequacy of the duration.}

We then specialize the framework to \emph{mutation-based fuzzing} by introducing
a novel Markov chain model over the input space for \emph{stateless} fuzzers,
and an extension over the joint input-state space that encodes internal states
for \emph{stateful} fuzzers. We develop a new evaluation paradigm that simulates fuzzing processes via
actual repeated executions (\emph{not via a surrogate simulator}), construct estimators, and derive and analyze evaluation metrics
from their deviation from the bug rate. This provides a principled alternative to ad hoc metrics, such as code
coverage and raw/unique bug counts. Within this setting, we establish a comprehensive
evaluation paradigm with provable guarantees under different degrees of prior
knowledge, requiring only observable outputs as the minimal condition. These
results yield fundamental insights into evaluation and design: the key
performance drivers are governed by \emph{spectral parameters} of the
policy-induced transition matrix, which can be known or estimated to inform
evaluation. When such estimation is infeasible, the evaluation paradigm can in turn inform these parameters,
thereby guiding fuzzer design. It further informs the choice of evaluation period.

We further conduct an extensive empirical study of 9 representative fuzzers on 18 widely used real-world software benchmarks. The empirical results validate the proposed estimators and evaluation metrics, align closely with the theoretical analysis, and provide new insights into evaluation, robustness and generalization across benchmarks.  While our analysis and evaluation focus on fuzzing, the proposed paradigm
extends to other feedback-driven black-box processes by instantiating their
input, policy, and feedback in machine learning. We support this claim through 4 sets of experiments in non-fuzzing
domains, which validate the theory and reinforce the conclusions drawn from
fuzzing.

\section{Problem Formulation}\label{sec:problem_form}
In this section, we formally formulate the problem under its inherent randomness using stochastic modeling and simulation. This formulation involves two main components: the feedback-driven process and its stochastic representation, which we present in turn. 



\subsection{Feedback-Driven Black-Box Processes}

At each round \(1 \leq t \leq T\), a policy \(\Gamma\) generates an input \(x_t \in \mathcal{X}\), the black-box system $\mathcal{O}$ produces an output \(y_t = g(x_t)\), and the policy \(\Gamma\) leverages it to generate the next input \(x_{t+1}\). The policy \(\Gamma\) does not have access to \(\mathcal{O}\) or \(g\) and can observe only the outputs \(y_t\), so it may contain heuristics and hidden states \(G_t\) (may or may not be observable to users), i.e., the internal information carried from one step to the next and recursively updated. This abstraction captures a broad class of systems, e.g., evolutionary and genetic algorithms, heuristic optimization and parameter-tuning systems, and adaptive testing and decision making. A defining feature of such processes is their \emph{sequential dependence} and \emph{black-box} nature to users. Evaluation therefore lacks knowledge of what occurs inside the system or the policy; instead, it can rely only on input-feedback pairs from the sequential execution of the policy on the system. The challenge is to formulate these processes and principally evaluate the policies
\(\Gamma\) w.r.t. the quality of \(x_t\).



\subsubsection{Instantiation: Mutation-Based Fuzzing}\label{sec:design_fuzzing}

We consider a general fuzzing process for software programs. Specifically, let us denote the target program as \(f\). The input to the program is any input string \(x\) from a discrete input domain \(D\). Let the length of \(x\), i.e., the number of bits in \(x\), be \(l_x\).  The program \(f\) executes \(x\) and outputs program behavior \(g_f(x)\). 
Let \(B_f\) denote the set of abnormal program behaviors that indicate bugs in \(f\), and let \(B\) be the corresponding set of buggy inputs in \(D\), i.e., \(x \in B\) iff \(g_f(x) \in B_f\). 

A mutation-based fuzzer \(R\), takes a string \(x_t \in D\) as input and outputs a string \(x_{t+1} \in D\) by mutating \(x_t\) according to a mutation rule (mapping) \(\mathcal{R}\) written in unknown scripts (i.e., a policy \(\Gamma\)). A mutation is a bit-level operation on a string \(x\): it selects a position \(1 \leq l \leq l_x\) and modifies the bit at the \(l\)-th position, \(x[l]\). Examples of such mutations are: flipping the bit, i.e., \(x[l] = 1 - x[l]\); deleting \(x[l]\) from \(x\); inserting a random value \(v\) before \(x[l]\); setting \(x[l]\) to a randomly chosen value \(v\); and performing arithmetic operations on \(x[l]\). Fuzzers can be broadly categorized into \emph{stateless} and \emph{stateful} fuzzers w.r.t. the inputs fed into the rule \(\mathcal{R}\): for the former, mutations depend only on \(x_t\), i.e., \(x_{t+1} = \mathcal{R}(x_t)\); for the latter, mutations depend not only on the current input \(x_t\), but also on a recursively updated internal state \(G_t\) that may encode the corpus, scheduling state, priority scores, coverage information, or randomization state, i.e., \(x_{t+1} = \mathcal{R}(x_t, G_t)\). They can be black-box or grey-box w.r.t. design.

\begin{wrapfigure}{r}{0.5\textwidth}
    \centering
    \vspace{-22pt}
\includegraphics[width=\linewidth]{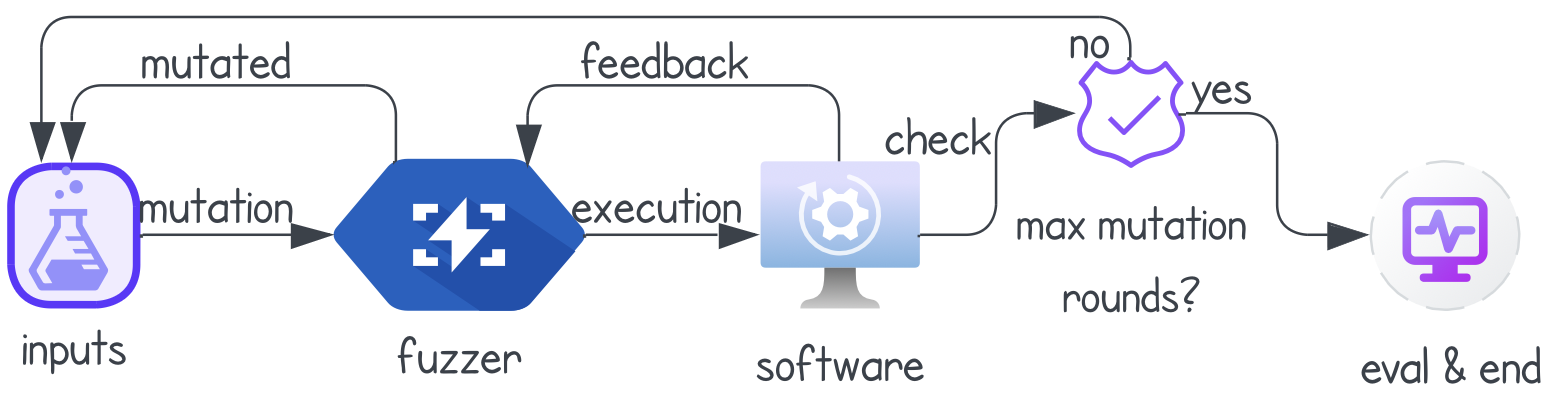}
    \caption{The workflow of a fuzzing process.}\vspace{-11pt}
    \label{fig:fuzzing_process}
\end{wrapfigure}

A fuzzing process proceeds as follows (shown in Figure~\ref{fig:fuzzing_process}), given a fuzzer \(R\) and a total number of mutation rounds (executions) \(T\). Starting with an initial input \(x_0\), we first execute \(x_0\) and observe the outputs \(g_f(x_0)\). If \(g_f(x_0) \in B_f\), we classify \(x_0\) as a bug. We then mutate \(x_0\) using \(R\) to obtain a new input string \(x_{1} = \mathcal{R}(x_0)\) ($\mathcal{R}(x_0, G_0)$), which is used as the next input. This process is repeated until \(T\) rounds. We denote the multisets of detected bugs and executed inputs as \(H_R^T\) and \(I_R^T\), respectively.

\subsection{Stochastic Modeling and Simulation}\label{sec:modeling_simulation}
Next, we introduce a stochastic modeling and simulation framework for \emph{feedback-driven black-box processes}. Transitions in the joint space of input and state are modeled as a Markov chain (MC), while transitions in the output space are modeled as stochastic processes induced by the input dynamics. Formally, let the sample space be \(\Omega\), with \(\sigma\)-algebra \(\mathcal{F}\) and probability measure \(\Pr\). On the probability space \((\Omega,\mathcal{F},\Pr)\), we define stochastic processes \((X_t, G_t, Y_t)\), where \(X_t\) denotes the input generated by policy $\Gamma$ at time \(t\) from the initial input/seed \(x_0\) (we suppress \(x_0\) and $\Gamma$ in the notation for convenience), $G_t$ is the state, and \(Y_t\) denotes the corresponding observable feedback. The MC evolves via an adaptive policy \(X_t=\Gamma(X_{t-1}, G_t)\) and $G_{t+1} = \Phi(G_t, X_{t+1})$ where $\Phi$ is the update rule of the state, and the output process is given by \(Y_t\doteq g(X_t)\). 

\textbf{Performance measure.} We next discuss performance objectives for feedback-driven black-box processes, e.g., fuzzing, adaptive testing, heuristic optimization, and online decision-making systems. In such systems, a natural objective is the long-run event rate of a performance signal \(Y_t\), which indicates a target event such as success, discovery, positive reward, or constraint violation. Finite-time proxies of this rate provide a principled way to reflect performance at a given execution horizon. Moreover, the choice of \(T\) determines the discrepancy between these proxies and the objective, so its effectiveness is evaluated accordingly. 


\subsubsection{Stateless Fuzzers}

We instantiate this formulation for \emph{mutation-based fuzzing}. We first consider stateless fuzzers: \(D\) is the sample space, \(\mathcal{R}\) is the mutation rule of a fuzzer $R$, \(X_t^R = \mathcal{R}(X_{t-1}^R)\) is the input after the \(t\)-th mutation starting from an initial seed, and \(Y_t^{R}=\mathbf{1}_{g_f(X_t^R)\in B_f} = g(X_t^R)\) is whether a bug is triggered at time \(t\). Then $X_t^R$ is a Markov chain on state space $D$ and its $\sigma$-algebra. Let \(P_R^{X}\) denote the \(|D|\times |D|\) transition matrix of \(X_t^R\), with entries \(P_R^{X}(x,x')=\Pr(X_{t+1}^R=x'\mid X_t^R=x)\), i.e., the probability of mutating \(x\) to \(x'\). We identify a \emph{novel Markov structure} in \(X_t^R\), whereas existing work places
it on \(Y_t^R\), which is generally \emph{not Markovian}.







\paragraph{Performance measure.} We specialize the general performance measure framework to fuzzing, where the target event corresponds to bug detection. Specifically, we define the \emph{bug detection rate} as
\(P_R=\lim_{T\to\infty}\nicefrac{1}{T}\sum_{t=1}^{T}Y_t^R\),
which represents the asymptotic fraction of bug-triggering executions among all executed inputs, i.e., \(|H_R^T|/|I_R^T|\). In practice, fuzzing is executed for a finite \(T\). We then define the finite-time bug detection rate as \(P_R(T)=\mathbb{E}[\nicefrac{1}{T}\sum_{t=1}^{T}Y_t^R]\). To assess the adequacy of \(T\), we measure the gap between \(P_R(T)\) and \(P_R\); smaller gaps indicate that \(T\) more accurately captures long-run behavior. Formally, we define the effectiveness of \(T\) as \(E(T)=\nicefrac{1}{\max_R |P_R(T)-P_R|}\).

Appendix~\ref{app:discussion} further justifies two modeling choices in fuzzing:
\emph{the single-seed formulation} and \emph{the proposed performance measure},
which connects to existing metrics, e.g., raw and unique bug counts.



\subsubsection{Extensions to stateful fuzzers}
Many modern fuzzers are adaptive and thus stateful, leveraging internal state \(G_t\). We model such fuzzers by \(X^R_{t+1}=\mathcal{R}(X_t^R,G_t)\) and identify a
\emph{novel Markov structure} accordingly.  Importantly, the output depends only on the input component \(X_t\), and hence can be modeled in the same way.

\textbf{Case 1.} If the state information $G_t$ is a deterministic function of the current input $X_t^R$ or its execution outcome (e.g., coverage delta), i.e., $G_t = G(X_t^R)$, then modeling the input space and the transition of $X_t$ yields a sufficient Markov chain, and these rules can be encoded in the transition matrix $P_R^X$. More precisely, $X_{t+1}^R = \mathcal{R}(X_t^R, G_t) = \mathcal{R}(X_t^R, G(X_t^R)) = \mathcal{R}^G(X_t^R)$. We use $\mathcal{R}$ to denote $\mathcal{R}^G$, the final mutation rule of fuzzer $R$. The stateless framework then directly applies without modification.

\textbf{Case 2.} If the internal state depends on the history, i.e., $G_t$ represents a sufficient summary of $X_1^R, X_2^R, \ldots, X_t^R$ and is updated recursively, then $G_{t+1} = \Phi(G_t, X_{t+1}^R)$. We then augment the stochastic process to $S_t^R = (X_t^R, G_t)$, so that $X_{t+1}^R = \mathcal{R}(X_t^R, G_t) = \mathcal{R}(S_t^R)$. Notably, $S_t^R$ is Markovian (Proposition~\ref{prop:S_t markovian}; proof is in Appendix~\ref{app:theorems}).

\begin{proposition}\label{prop:S_t markovian}
Let $S_t^R = (X_t^R, G_t)$. Then $\{S_t^R\}_{t \ge 0}$ is a Markov chain.
\end{proposition}

\textbf{Performance measure.} Importantly, the performance measure is defined with respect to $Y_t^R$, which is a function of $X_t^R$, i.e., $Y_t^R = \mathbf{1}_{g_f(X_t^R) \in B_f} = g(X_t^R) = g'(X_t^R, G_t)$. Hence, the definitions of the performance measures $P_R$, $P_R(T)$ and $E(T)$ follow the stateless case.  

\subsubsection{Properties of the induced transition matrix.} We use $P_R^Z$ to denote the transition matrix of $X_t^R$ in the stateless case and of $S_t^R$ in the stateful case, respectively. Notably, $P_R^Z$ is unknown and time-homogeneous for most existing fuzzers, whose mutation rules are fixed ($\mathcal{R}$ is determined by the code in scripts, thereby uniquely defining the transition matrix). It takes dynamic information such as $X_t^R$ or both $X_t^R$ and $G_t$ as inputs.

Stateless fuzzers trivially satisfy this property. In the stateful setting, Case 1, e.g., some coverage-guided fuzzers, also satisfies it. The driver, e.g., the change in coverage, \(G_t = G(X_t^R)\), is a deterministic function of \(X_t^R\). Hence, \(G\) is encoded in the fuzzer's script \(\mathcal{R}\) as part of the mutation rules taking \(X_t^R\) as input. The transition matrix depends on \(\mathcal{R}\) and is fixed. In Case 2, \(S_t^R\) is fed into the fixed script \(\mathcal{R}\), so the transition matrix over \(S_t^R\) remains fixed. When non-homogeneity arises, e.g., due to fuzzer-script upgrades, we can partition the horizon into multiple periods. Within each period, e.g., with a fixed script, \(P_R^Z\) is time-homogeneous, fitting the proposed framework. We then analyze outputs by aggregating across periods. 






\section{Evaluation Framework}\label{sec:evaluation}

This section presents a rigorous evaluation framework for fuzzers, which naturally extends to feedback-driven black-box processes by substituting the mutation and feedback mechanisms. Notably, as part of our unique contributions, the framework is designed for \textbf{\emph{unknown \(P_R^Z\)}}, consistent with the definition of black-box processes. Our main goal is to evaluate performance through stochastic analyses, using only the observed input-output pairs of the processes, both \emph{with and even without access to hidden-state information} \(G_t\) when it is used to decide the mutation (policies).

We first use simulation to collect data, where \textbf{\emph{simulation refers to executing the system under given policies and observing the input/output, not using a simulator}}. We then construct estimators for the performance measures using the data and finally define evaluation metrics accordingly.

\textbf{Monte Carlo Simulation.} We adopt Monte Carlo (MC) simulation to gather observations from real-world fuzzing systems (Algorithm~\ref{alg:mc} in Appendix~\ref{app:alg}). The idea is intuitive: for any \(M>0\), we execute the fuzzing process \(M\) times, i.e., \(M\) replications, thereby faithfully reflecting the fuzzing process with no fidelity gap. In replication \(m\), an input \(X_t^R(m)\) is executed on a program to produce output \(Y_t^R(m)\), then mutated by fuzzer \(R\) and re-executed; this cycle is repeated for \(T\) rounds. 


\textbf{Performance estimator.} We construct an estimator for the performance measure \(P_R\) based on the output. Formally, the estimator for a fuzzer \(R\) is defined as \(\Tilde{P}_R(T)=\nicefrac{1}{M}\sum_{m=1}^{M}\nicefrac{1}{T}\sum_{t=1}^T Y_t^R(m))\), which represents the average ratio of detected bugs to the total number of executions (mutations). 



\textbf{Evaluation measure.}
Given the probability measure $\Pr$, we denote \(\pi(0)=\Pr(x\notin B)\) and \(\pi(1)=\Pr(x\in B)\), where \(\pi(1)\) is called the bug rate. Based on the estimator \(\Tilde{P}_R(T)\), we evaluate a fuzzer by its deviation from the bug rate $\pi(1)$, defined as \(D_R(T)=|\Tilde{P}_R(T)-\pi(1)|\).  Importantly, we assess the effectiveness of \(T\), a novel and often overlooked dimension, with \(E(T)=\nicefrac{1}{\max_R|\Tilde{P}_R(T)-P_R|}\). This metric accounts for \emph{effort} and provides a rigorous criterion for selecting \(T\) in fuzzer evaluation, beyond fixed-duration heuristics, e.g., 5-6 or 24 hours.






\section{Analyses and Evaluations}\label{sec:analyses}

%

%



We present a rigorous assessment of both the fuzzer and the choice of \(T\) via spectral theory, which also helps practitioners design better fuzzers. We evaluate fuzzers using \(D_R(T)\) w.r.t. the performance measure \(\pi(1)\). When \(\pi(1)\) is known, \(D_R(T)\) can be computed directly, with smaller values indicating better performance, and \(T\) can be chosen so that \(\max_R D_R(T)\le \epsilon\) for a given tolerance \(\epsilon\). We therefore focus on the more practical case where \(P_R\) is unknown. The evaluation herein is built on the stochastic analyses of \(\Tilde{P}_R(T)\) and \(D_R(T)\); the results and proofs are provided in Appendix~\ref{app:theorems}. 

We first consider settings in which the spectral parameter \(\delta_R\) of
\(P_R^Z(\cdot,\cdot)\) is known (Section~\ref{4.2.1}) or can be estimated from
data (Section~\ref{4.2.2})~\cite{sherlaw1995estimating, hsu2019mixing,
wolfer2020mixing, wolfer2024improved}. Here, \(\delta_R \geq 1\) is the inverse
spectral gap; a larger \(\delta_R\) indicates slower convergence and is \emph{less desirable}. Crucially, it requires estimating only the
scalar parameter \(\delta_R\), rather than the full transition kernel
\(P_R^Z(\cdot,\cdot)\). Section~\ref{4.2.3} then handles cases where such estimation is infeasible, e.g., unobservable \(G_t\). We establish a general evaluation framework that applies regardless of the observability of \(G_t\) and without knowledge or estimation of \(\delta_R\); surprisingly, the evaluation can in turn inform \(\delta_R\). It is summarized in the following table. 

\vspace{-11pt}
\begin{table}[h]
\centering
\setlength{\tabcolsep}{2pt}
\renewcommand{\arraystretch}{0.8}
\label{tab:evaluation_regimes}
\begin{tabular}{p{0.35\linewidth}p{0.26\linewidth}p{0.16\linewidth}p{0.12\linewidth}}
\toprule
\textbf{$\delta_R$} & \textbf{Availability} & \textbf{Not Required} & \textbf{Evaluation} \\
\midrule
Known \& $<\infty$
& \(\delta_R\) is known 
&  \(P_R^Z(\cdot,\cdot)\)
& Sec. ~\ref{4.2.1}. \\
\midrule
Unknown but estimable \& $<\infty$
& \(G_t\) observable if present 
& \(P_R^Z(\cdot,\cdot)\) 
& Sec. ~\ref{4.2.2}. \\
\midrule
Infeasible estimation or $\infty$
& Observable $Y_t^{R}$ 
& \(\delta_R\) or \(P_R^Z(\cdot,\cdot)\) 
& Sec. ~\ref{4.2.3} \\
\bottomrule
\end{tabular}
\end{table}
\vspace{-6pt}

\begin{remark}
W.l.o.g., when discussing \(\delta_R\) in Sections~\ref{4.2.1} and~\ref{4.2.2}, we focus on non-degenerate fuzzers, i.e., ergodic fuzzers with \(\delta_R < \infty\). This means that the fuzzer has the possibility of exploring the input space when run indefinitely. This assumption helps exclude less desirable fuzzers whose mutations may become \emph{permanently trapped} or stuck in specific regions \((\delta_R=\infty)\). In practice, random or periodic restarts, or occasional injected noise, can help the fuzzer escape the current input region and avoid becoming stuck, thereby making the fuzzer ergodic. In Section~\ref{4.2.3}, we establish a general evaluation framework regardless of degeneracy/non-degeneracy. 
\end{remark}




\subsection{Evaluation with Known \(\delta_R\)}\label{4.2.1}

With known spectral parameters \(\delta_R\), we evaluate fuzzers using the upper bound on \(D_R(T)\) from Theorem~\ref{finite-time:matrix-spectral-nonrev}. Smaller bounds on \(D_R(T)\), i.e., smaller \(\delta_R\), indicate faster convergence to the bug rate, meaning that the fuzzers explore the input domain and discover bugs in \(B\) more quickly. 


\begin{theorem}[Evaluation of \(D_R(T)\)]\label{thm7} The fuzzer with the smallest \(\delta_R\) results in the smallest deviation \(D_R(T)\) w.p. \(1- \nicefrac{1}{T^2}\), and thus the best-performing.\end{theorem}


The key observation is that the upper bound on \(D_R(T)\) is dominated by the
finite-time convergence error. Let \(P_R=\pi^R(1)\) be the stationary bug rate
of \(X_t^R\). By the triangle inequality,
\(D_R(T)=|\tilde P_R(T)-\pi(1)|
\leq |\tilde P_R(T)-P_R|+|P_R-\pi(1)|\). The first term is bounded by
\(O(\sqrt{\nicefrac{\delta_R}{T}})\)
(Theorem~\ref{finite-time:matrix-spectral-nonrev}), while the second term is \(o(1/T)\)
(Lemma~\ref{lemma:0}). Thus, \(D_R(T)\) is 
controlled by $\delta_R$.


\begin{theorem}[Evaluation of \(E(T)\)]\label{thm: E_T_known}
For any \(T \geq argmax_{R}argmin_{T \in \mathbb{N}}\{\sqrt{\nicefrac{\log{T}}{T}} + \sqrt{\nicefrac{4\delta_{R}}{T}} \leq \epsilon\} \vee \nicefrac{1}{\sqrt{\alpha}}\), the effectiveness of \(T\) meets \(E(T) \geq \nicefrac{1}{\epsilon}\) with confidence level \(1 - \alpha\).
\end{theorem}

\subsection{Estimation-based Evaluation with Unknown \(\delta_R\)}\label{4.2.2}

We next consider evaluation with unknown \(\delta_R\), where the above evaluation is infeasible. We propose a data-driven method using trajectory data, where \(G_t\) is observable if present for stateful fuzzers.

\paragraph{Evaluation by estimating $\delta_R$.} We approximate \(\delta_R\) efficiently using the interaction history between the fuzzer and the program. The estimation depends on whether the underlying Markov chain is reversible. For reversible chains, we adopt the single-trajectory estimator of~\cite{hsu2019mixing}, which is well suited to fuzzing. For non-reversible chains, we use \cite{wolfer2024improved}, which provides sample-complexity guarantees. Algorithm~\ref{alg:data-driven} summarizes the procedure. In the first phase (\(t\le t_1\)), we estimate \(\delta_R=1/\gamma\) using~\cite{hsu2019mixing} for reversible chains or~\cite{wolfer2024improved} otherwise, and denote the estimate by \(\delta_{dps}(R)\). In the second phase (\(t_1\le t\le T\)), we evaluate fuzzers using the estimated metric \(\hat{U}_R(T)=\sqrt{\nicefrac{\log T}{T}}+\sqrt{\nicefrac{4\delta_{dps}(R)}{T}}\).



%

\begin{theorem}[\(D_R(T)\)]\label{thm:unknown_delta_U} The fuzzer with the smallest \(\delta_{dps}(R)\) ($\hat{U}_R(T)$) is optimal w.h.p.. 
\end{theorem}

\begin{theorem}[\(E(T)\)]\label{thm:unknown_delta_U_T}
    For \(T \geq argmax_{R}argmin_{T}\{\hat{U}_R(T) \leq \epsilon\} \vee \nicefrac{1}{\sqrt{\alpha}}\) , \(E(T) \geq \nicefrac{1}{\epsilon}\) w.h.p..
\end{theorem}

 \paragraph{Sample complexity and implementation.} For this data-driven procedure, the sample complexity/overhead
\(t_1\) is \(O(\nicefrac{1}{(1-\lambda_2)\epsilon^2})\), where \(\lambda_2\)
is the second-largest eigenvalue and \(\epsilon\) is the estimation error of
\(\delta_R\). Notably, \(t_1\) is independent of the size of the input
domain \(D\). The value of \(t_1\) can be selected using a doubling
scheme. In practice, the procedure can be implemented directly using the Python
package \texttt{pydtmc}~\citep{belluzzo_pydtmc}, and our numerical study in
Appendix~\ref{app:non-fuzzing} verifies its practical usability. 

\subsection{Estimation-free Evaluation with Unknown/degenerate \(\delta_R\)}\label{4.2.3}

When estimating \(\delta_R\) is infeasible, e.g., unobservable \(G_t\) or degeneracy \((\delta_R=\infty)\), we propose an estimation-free evaluation approach that may still inform \(\delta_R\). This also avoids estimation complexity. 

\paragraph{Evaluation by the proxy of \(P_R\).}

We represent \(P_R\) via a novel proxy \(\tilde{P}=\nicefrac{\tilde{N}}{T}\), where \(\tilde{N}\) is the union of all detected bugs across fuzzers by time \(T\), and approximate \(D_R(T)\) using \(|\tilde{P}_R(T)-\tilde{P}|\). 

By using the ideal curve \(\tilde{P}\), the estimator \(|\tilde{P}_R(T)-\tilde{P}|\) is \emph{biased with respect to \(D_R(T)\), but the bias is shared across all fuzzers and thus supports accurate evaluation}. By definition, the bug detection ratio of a fuzzer satisfies \(|\tilde{P}_R(T)-\pi(1)| \leq |\tilde{P}_R(T)-\tilde{P}|+|\tilde{P}-\pi(1)|\). The first term is fuzzer-dependent, while the second term is common to all fuzzers. Thus, although the estimator is biased, the shared bias does not affect comparative evaluation. We therefore abstract this bias out from the evaluation with the closest available approximation \(\tilde{P}\). This leads to the following corollaries. 


\begin{corollary}[\(D_R(T)\)]\label{coro:1} The fuzzer with the smallest $|\tilde{P}_R(T) - \tilde{P}|$ results in the smallest deviation \(D_R(T)\) w.p. \(1- \nicefrac{1}{T^2}\), and thus the best-performing.
\end{corollary}

\begin{corollary}[\(E(T)\)]\label{coro:2}
\(T \geq argmax_{R}argmin_{T \in \mathbb{N}}\{|\tilde{P}_R(T) - \tilde{P}| \leq \epsilon\} \vee \nicefrac{1}{\sqrt{\alpha}}\) is a necessary condition to ensure that \(E(T) \geq \nicefrac{1}{\epsilon}\) with confidence level \(1 - \alpha\).
\end{corollary}

Notably, for non-degenerate, i.e., ergodic, fuzzers, the evaluation provides an inverse way to infer \(\delta_R\).
\begin{corollary}[Information about \(\delta_R\)]\label{coro:3}
We obtain that $\delta_R \;\ge\; \nicefrac{T}{4}(|\tilde{P}_R(T)-\tilde{P}|-\sqrt{\nicefrac{\log T}{T}})_{+}^2.$
\end{corollary}

\textbf{Fuzzer design.} As a by-product of the evaluation framework, we take an inverse perspective: using the proxy for \(D_R(T)\), we bound the spectral parameter \(\delta_R\) and reveal its variation across fuzzers. One can then use causal analysis to identify fuzzer components (features) that influence \(\delta_R\), thereby guiding practitioners in designing more effective fuzzers without tracking \(P_R^Z\) or the hidden state \(G_t\). 
%






\section{Numerical Experiments}\label{sec:exp}


In this section, we present a \textbf{comprehensive numerical evaluation of existing fuzzers} that builds on the insights from Section~\ref{4.2.3}, without estimating \(\delta_R\) or assuming that it is non-degenerate. We first assess the soundness of the proposed metric. We then demonstrate practical evaluation using the approximations \(\tilde{P}_R(T)\) and \(\tilde{\pi}(1)\). We further show that the evaluation procedure generalizes across fuzzers and benchmark programs and remains robust to randomness. Finally, we provide practical guidance on selecting the horizon \(T\). In addition, we present a \textbf{numerical study on 4 non-fuzzing domains} in Appendix~\ref{app:non-fuzzing}, where estimating \(\delta_R\) is feasible. We provide these estimates and evaluate accordingly, following both Sections~\ref{4.2.2} and~\ref{4.2.3}; the results align with the established theory. We also discuss the broader impacts in Appendix~\ref{app:impacts}. The code is available at \url{https://anonymous.4open.science/r/Feedback-Driven-Black-Box}.













We evaluate 9 widely used mutation-based fuzzers following~\cite{schloegel2024sok}: AFL~\cite{afl}, AFLFast~\cite{bohme2016coverage}, AFL++~\cite{AFL++}, AFLSmart~\cite{AFLSmart}, Entropic~\cite{Entropic}, FairFuzz~\cite{lemieux2018fairfuzz}, MOpt~\cite{lyu2019mopt}, LibFuzzer~\cite{LibFuzzer}, and Honggfuzz~\cite{honggfuzz}. We run them on 18 well-known program benchmarks from \textbf{FuzzBench}~\cite{fuzzbench} via Algorithm~\ref{alg:mc}: arrow, aspell, ffmpeg, grok, harfbuzz, libhevc, libgit2, libhtp, libxml2, matio, njs, openh264, php, poppler, quickjs, stb, systemd, and wireshark, each for \(T=72\) hours. These fuzzers and benchmarks are widely used in practice and fuzzer evaluation studies~\cite{schloegel2024sok}, representing real-world settings; moreover, some benchmarks include multiple programs, further enhancing experimental coverage. We also report results on \textbf{Magma}~\citep{hazimeh2020magma}, a benchmark suite with ground truth, in Appendix~\ref{app:magma}.

Based on the output, i.e., real fuzzing data, we construct \(\tilde{P}_R(t)\), track its evolution over \(T\), and compute the evaluation metric in Section~\ref{4.2.3}. This approach avoids computing the spectral parameter \(\delta_R\), which is intractable for fuzzers with internal states, and may even inform \(\delta_R\). We further validate our framework against existing metrics, demonstrating its advantages in convergence behavior, fuzzer discrimination, and benchmark-specific assessment of \(T\). For clarity, we present results for a subset of benchmarks and defer the complete results to Appendix~\ref{app:exp}. Experiment details, e.g., bug deduplication, seed, and computing resources, are provided in Appendices~\ref{app:seed_bug} and~\ref{app:exp}.

Through numerical experiments, we study five research questions for fuzzing evaluation: whether the proposed metric (\textbf{RQ1}) \emph{captures bug discovery efficiency as predicted by theory}, (\textbf{RQ2}) \emph{reliably distinguishes between fuzzers}, (\textbf{RQ3}) \emph{generalizes across fuzzers and benchmarks}, (\textbf{RQ4}) \emph{is robust to fuzzing randomness}, and (\textbf{RQ5}) \emph{indicates whether commonly used fuzzing durations are sufficient.}


%


\subsection{RQ1: Representativeness}

We implement the fuzzers on the selected benchmarks and compute the proposed estimator alongside standard metrics, e.g., code coverage and bug counts, at hourly intervals from \(t=6\) to \(t=72\), excluding the initial \(t\in[0,6]\) period due to high fluctuations.

\textbf{Bug detection ratio.} Figures~\ref{exp_fig}(a,b) show \(\tilde{P}_R(t)\) from 6-72 hours, with time \(t\) on the \(x\)-axis and the bug detection ratio \(\tilde{P}_R(t)\) on the \(y\)-axis. The curves converge for all fuzzers, e.g.,  Honggfuzz and AFL++, revealing long-tail behavior and supporting our theory; the dashed line denotes the proxy \(\tilde{P}\). 


\textbf{Bug counts.} Figures~\ref{exp_fig}(c,d) report cumulative bug counts for \textit{libhtp} and \textit{poppler}. While Honggfuzz and AFL++ continue discovering bugs while fluctuating, others plateau earlier, reflecting stagnation. These results show that bug counts alone neither imply convergence nor capture a universal fuzzing objective~\cite{hazimeh2020magma}, and that benchmark characteristics influence discovery dynamics.

\textbf{Code coverage.} Code coverage remains largely flat (Figures~\ref{exp_fig}(e,f)), while bug-related metrics continue to evolve, indicating that coverage-based metrics are less aligned with bug discovery.  

Similar trends appear on other benchmarks (Appendix~\ref{app:exp}), with varying tail behaviors, demonstrating the representativeness of the proposed estimator relative to existing metrics.

\subsection{RQ2: Distinguishment}

In this section, we evaluate fuzzers using multiple metrics, including the proposed \(D_R(T)\), which is based on but distinct from \(\tilde{P}_R(t)\), and existing metrics, e.g., code coverage and bug counts. 
\begin{figure}[htb]
    \centering
\vspace{-11pt}
    \begin{minipage}[b]{0.85\linewidth}
        \centering
        \includegraphics[width=\linewidth]{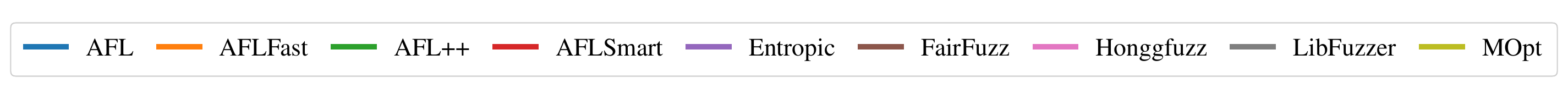}
    \end{minipage}

\vspace{-6pt}

    \figpanel{(a)}{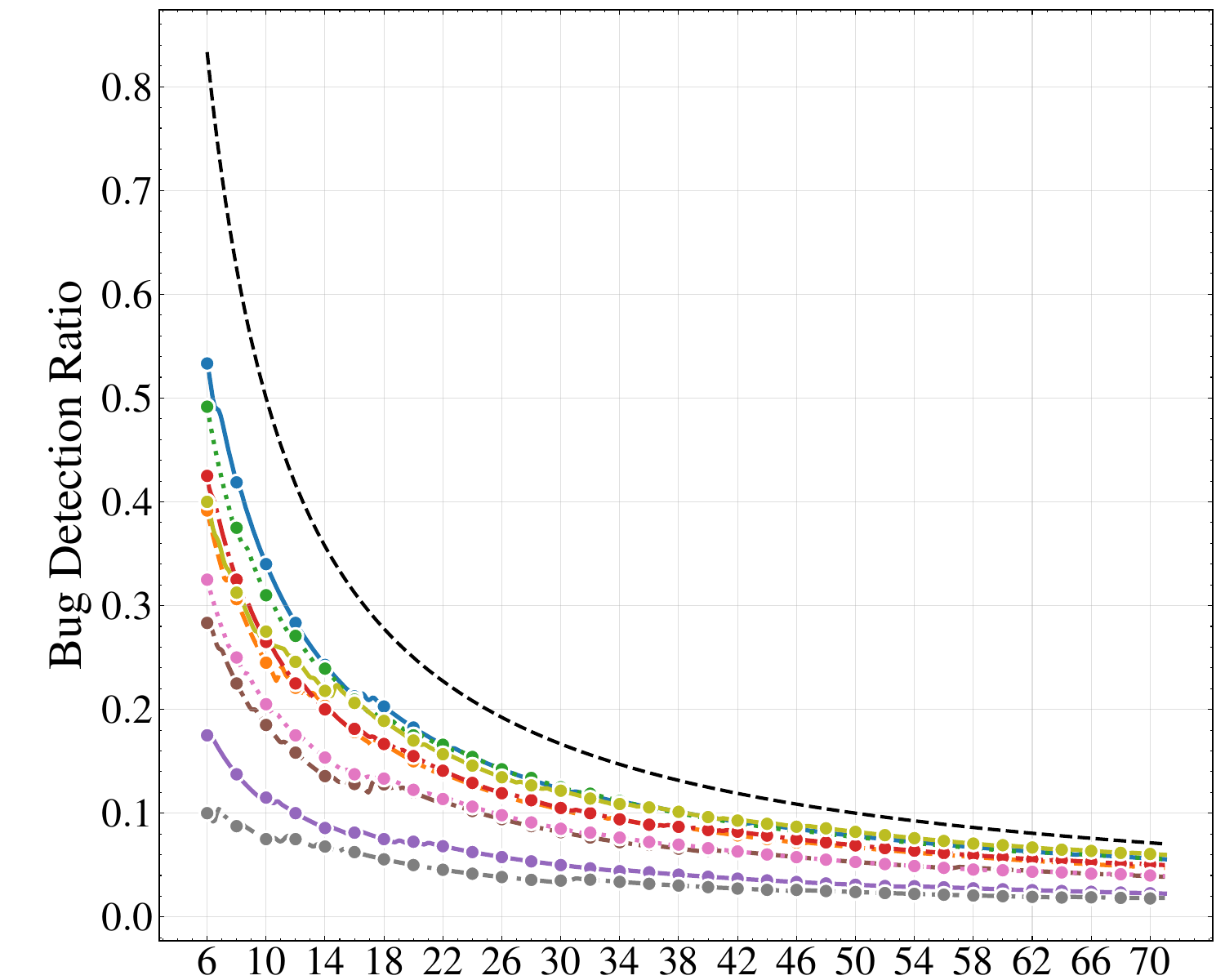}
    \hfill
    \figpanel{(b)}{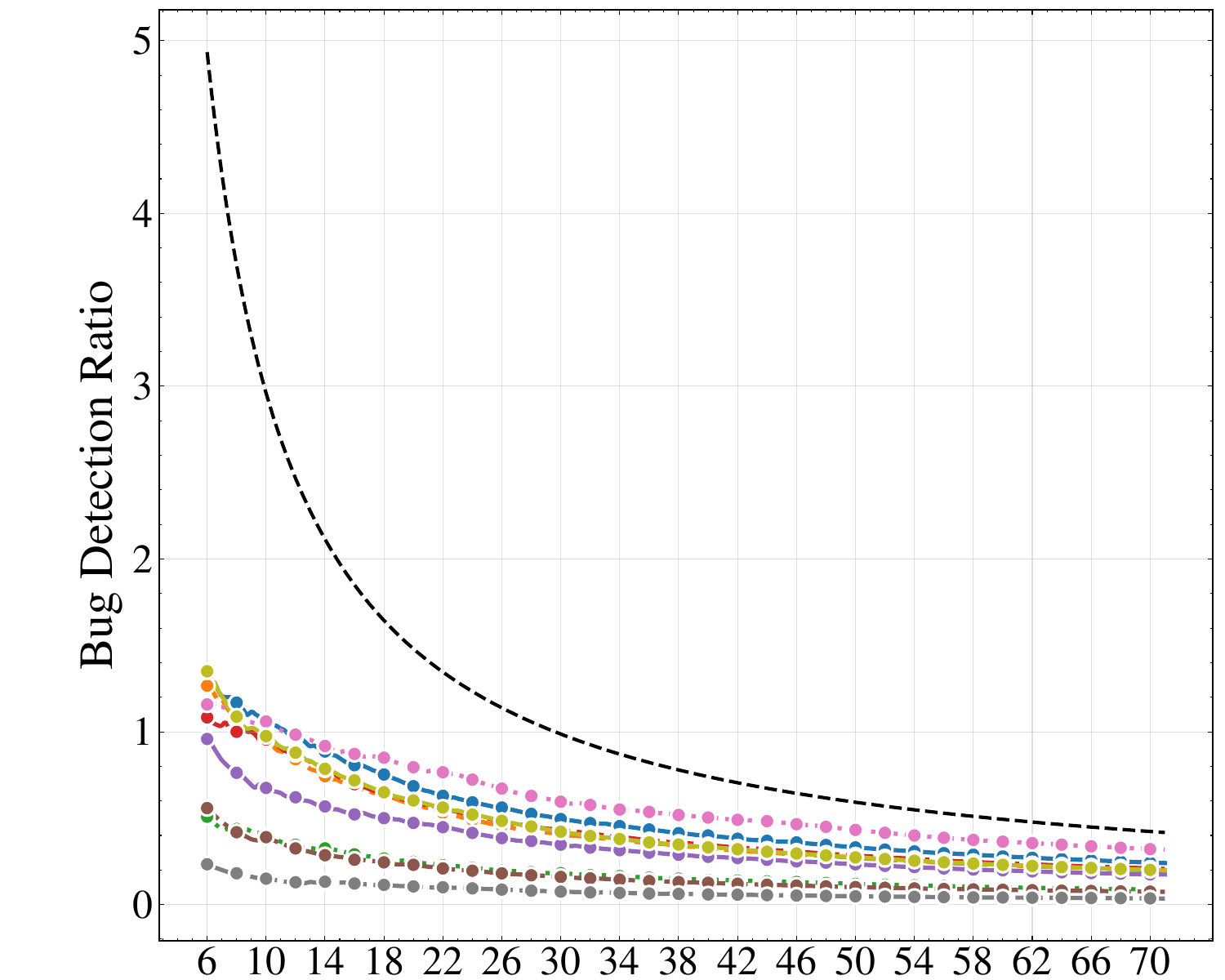}
    \hfill
    \figpanel{(c)}{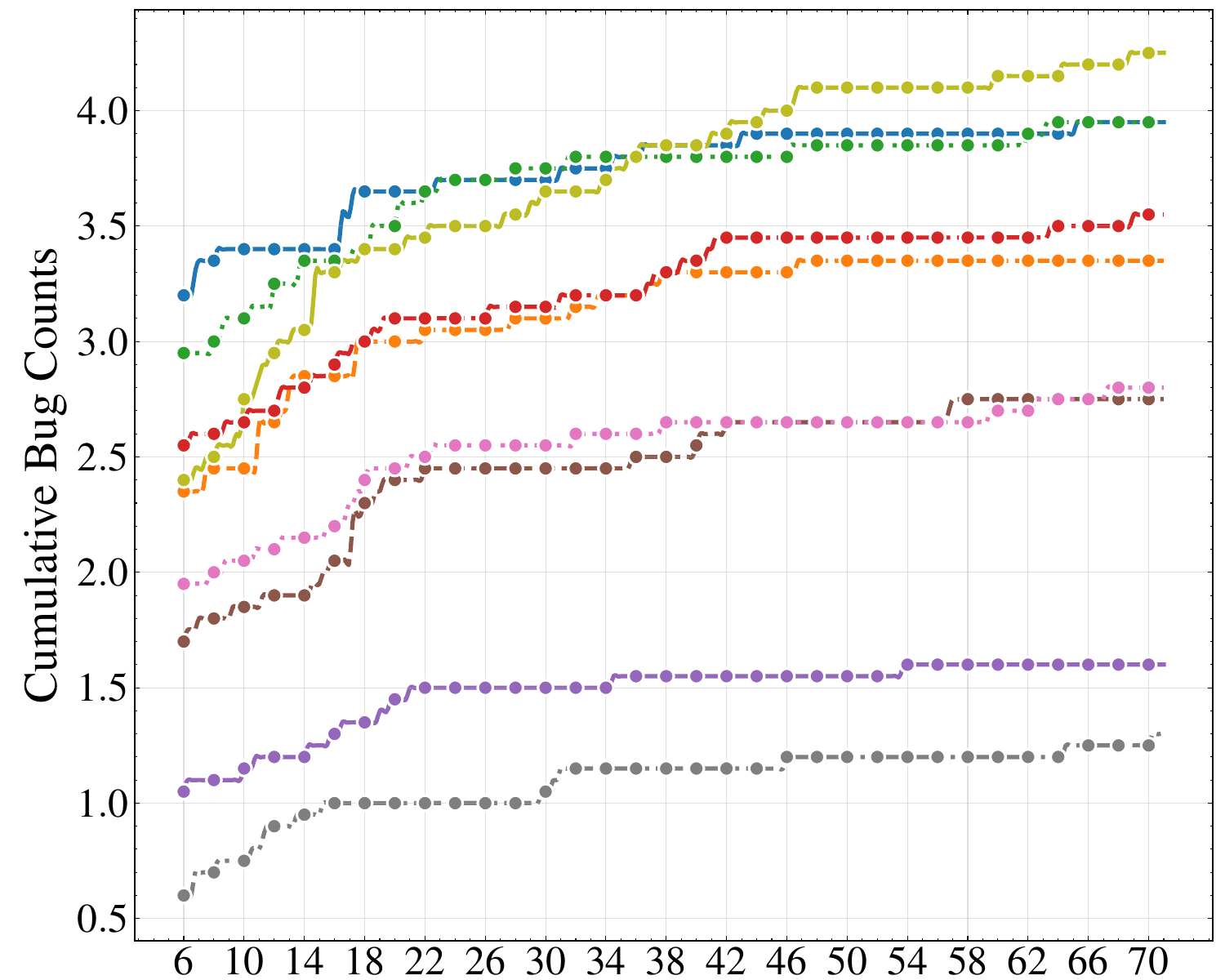}

    \figpanel{(d)}{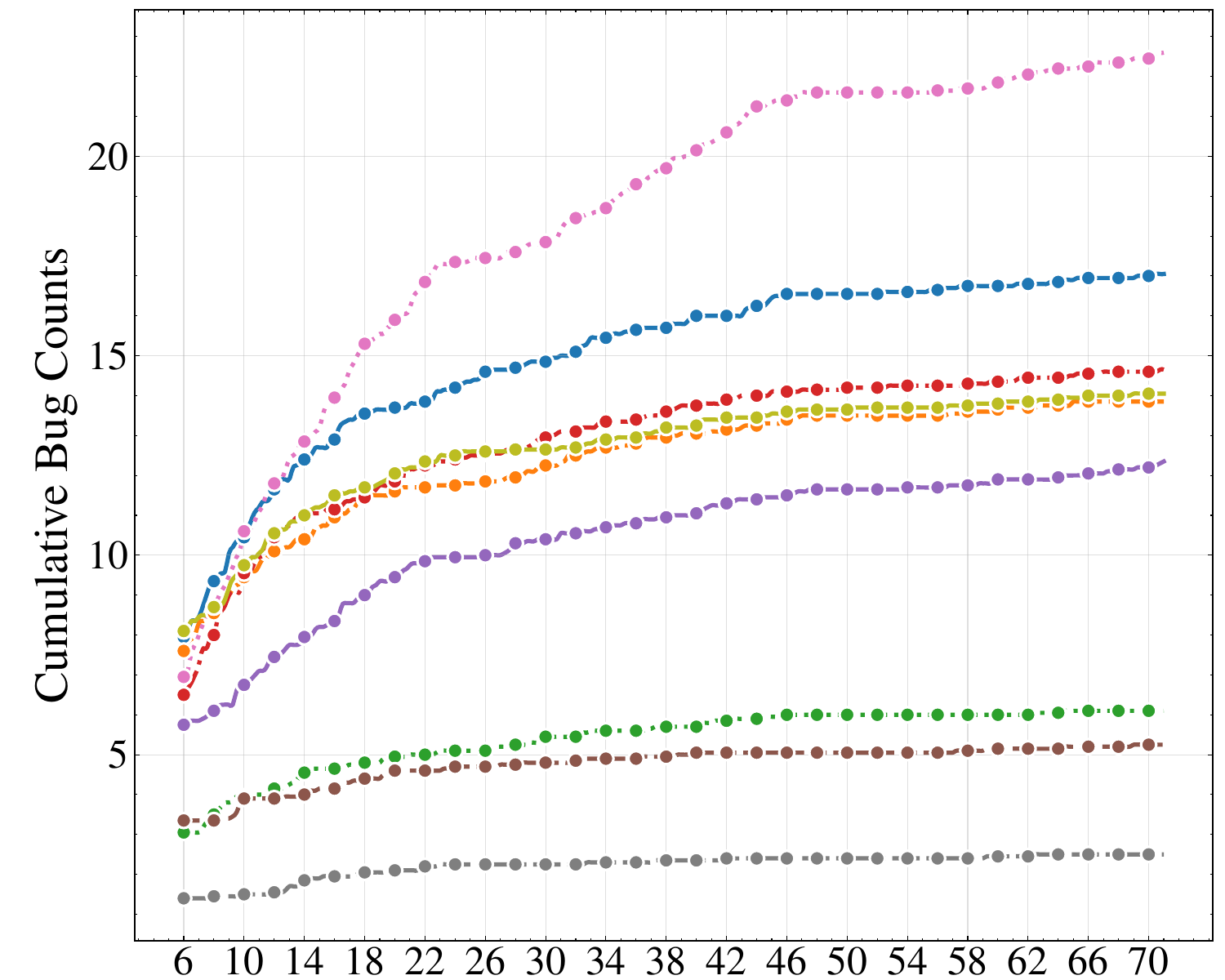}
    \hfill
    \figpanel{(e)}{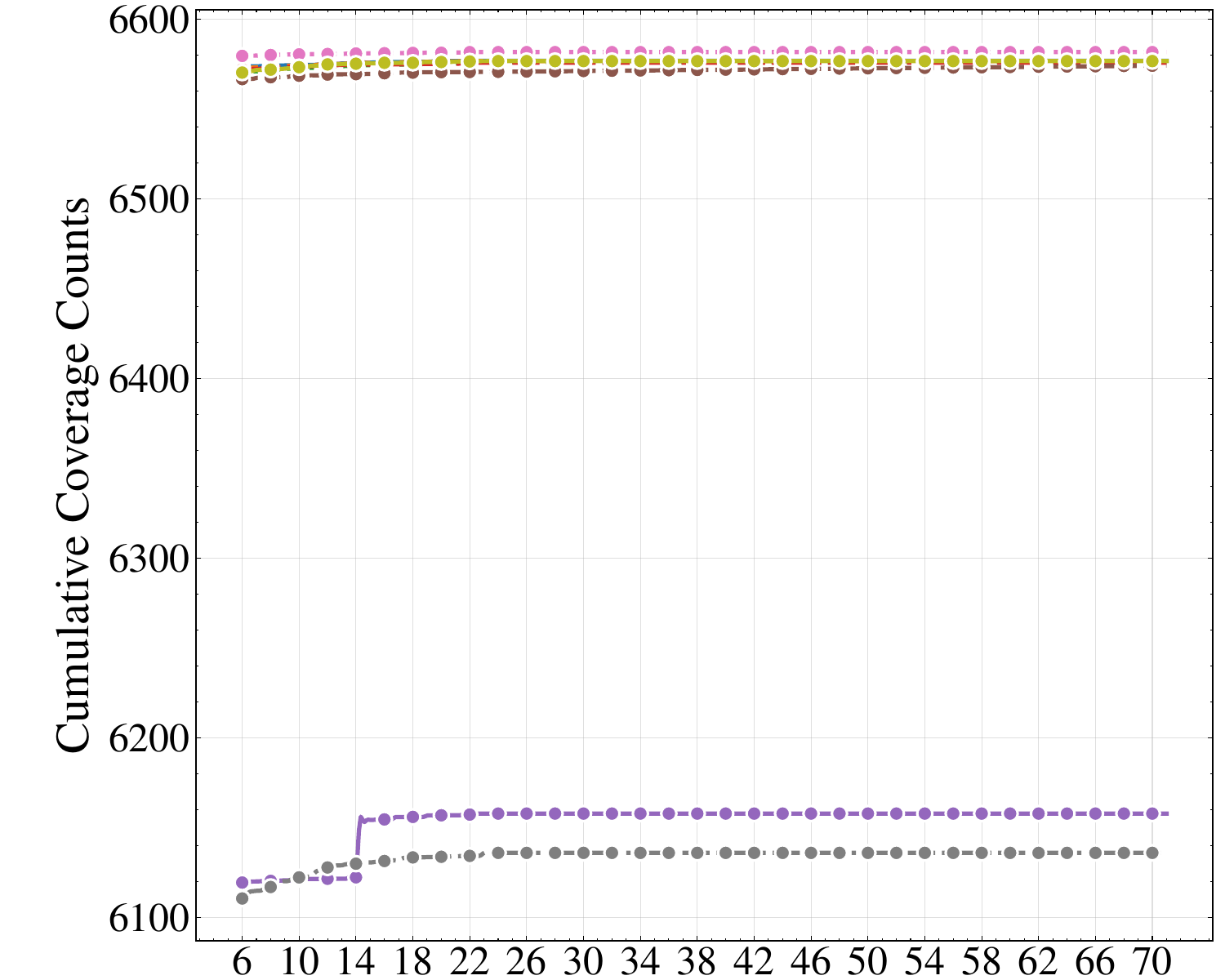}
    \hfill
    \figpanel{(f)}{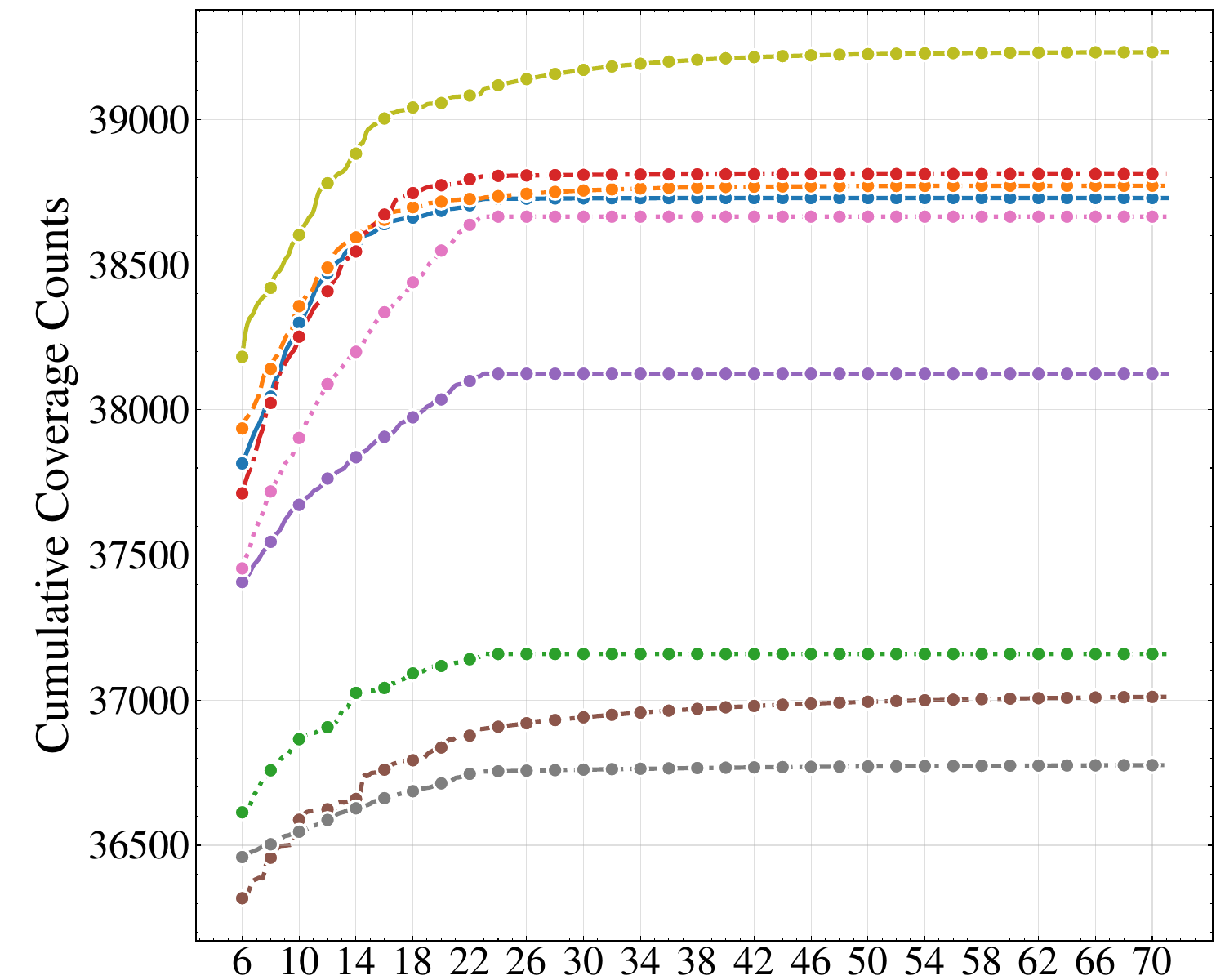}

    \figpanel{(g)}{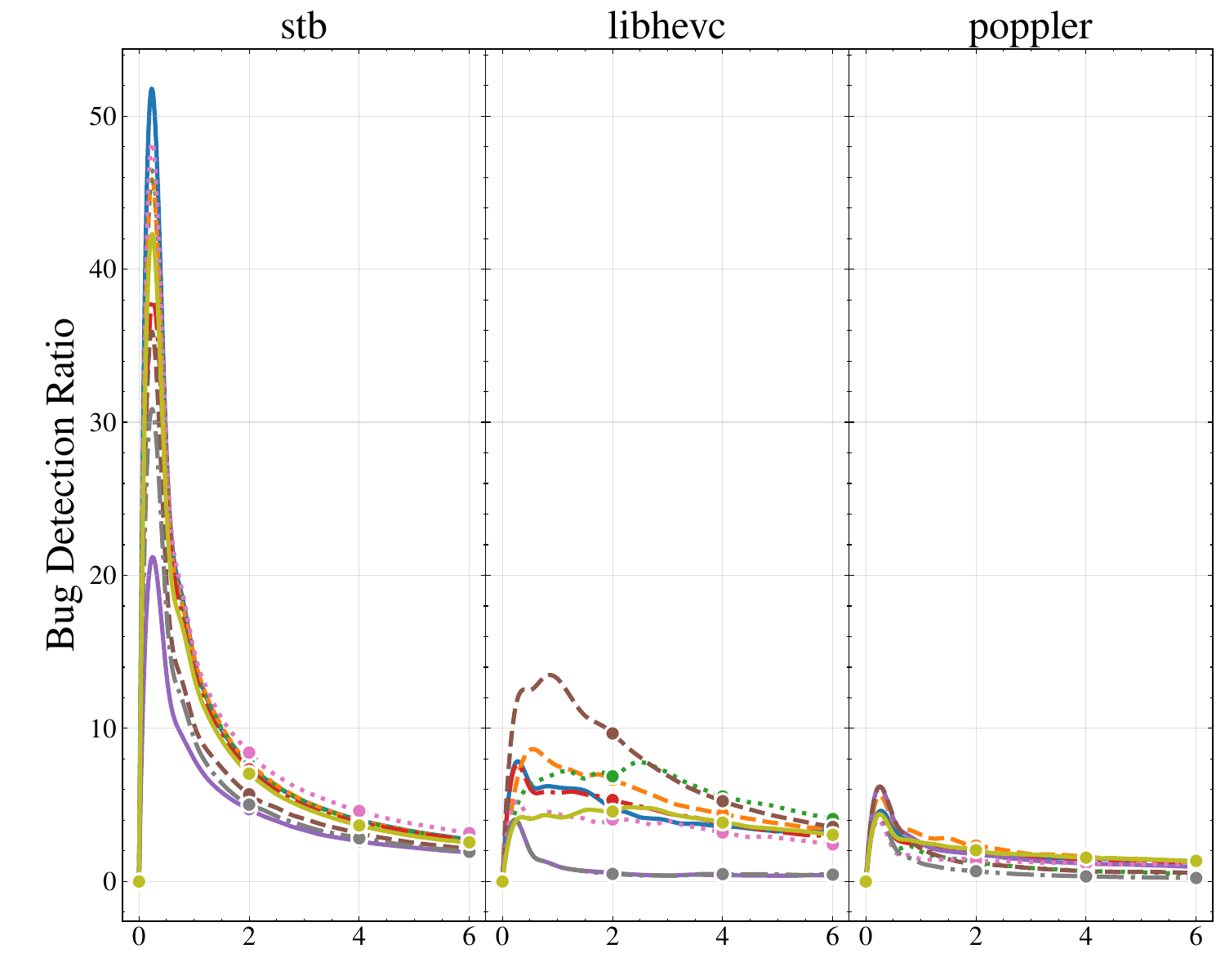}
    \hfill
    \figpanel{(h)}{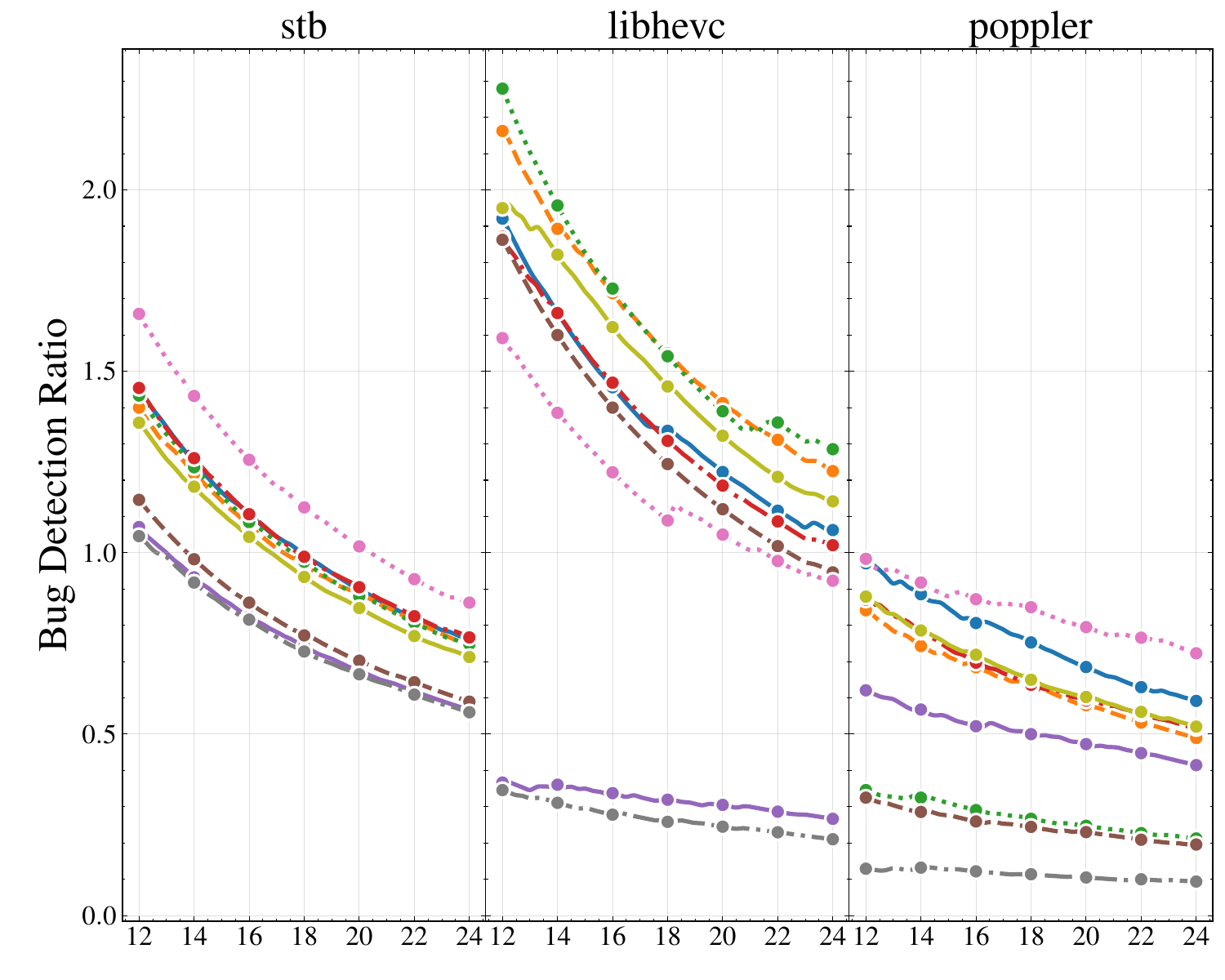}
    \hfill
    \figpanel{(i)}{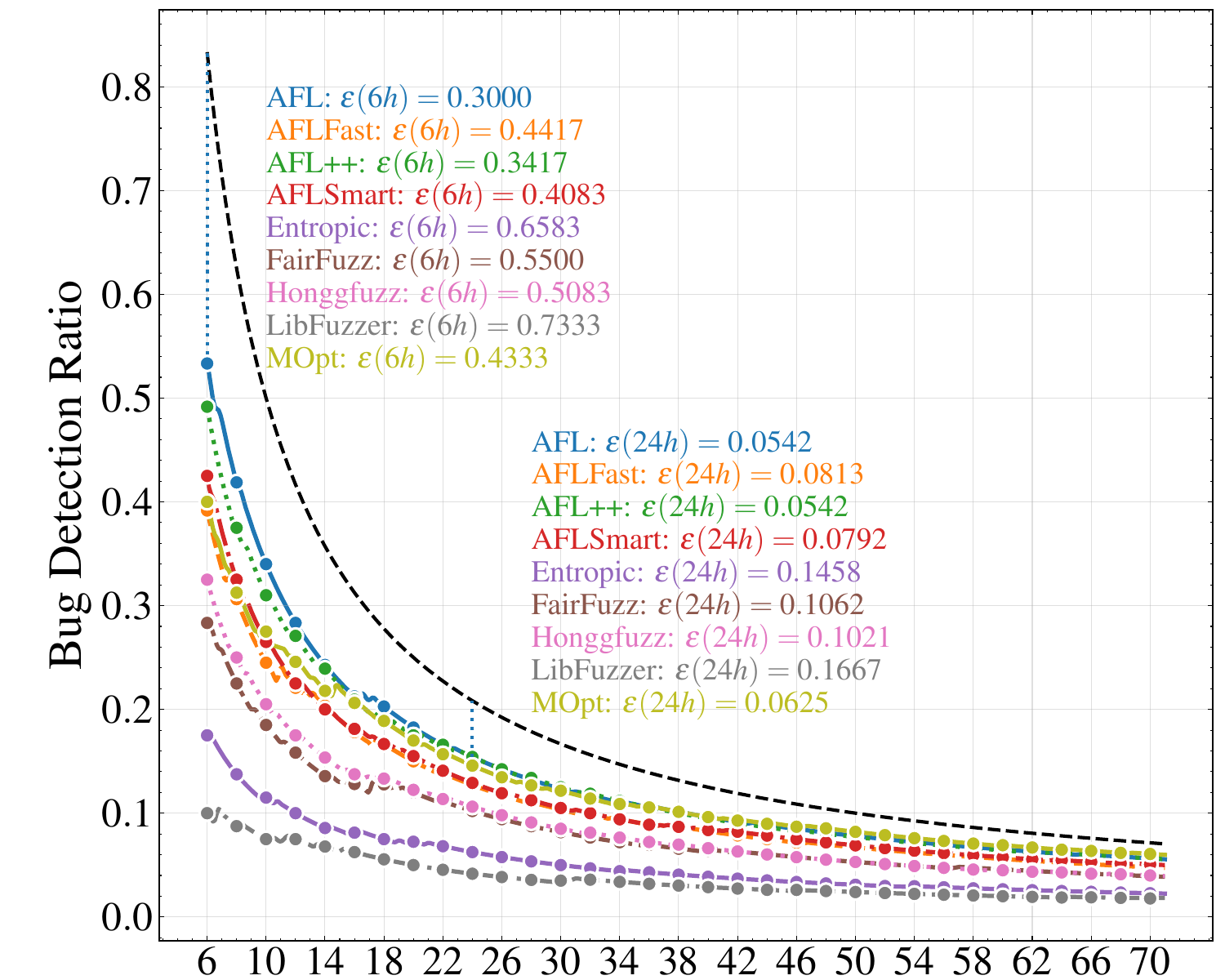}

    \figpanel{(j)}{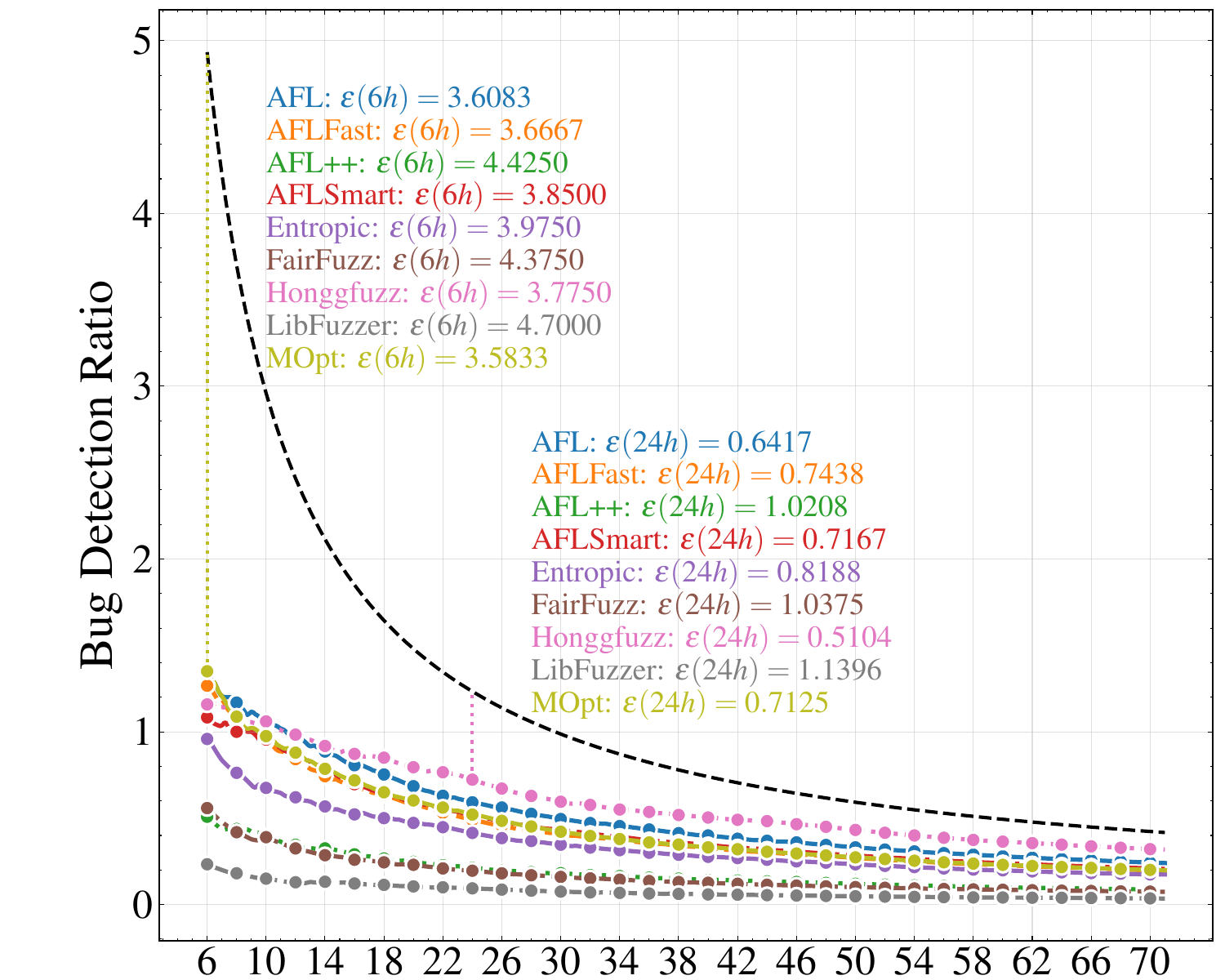}
    \hfill
    \figpanel{(k)}{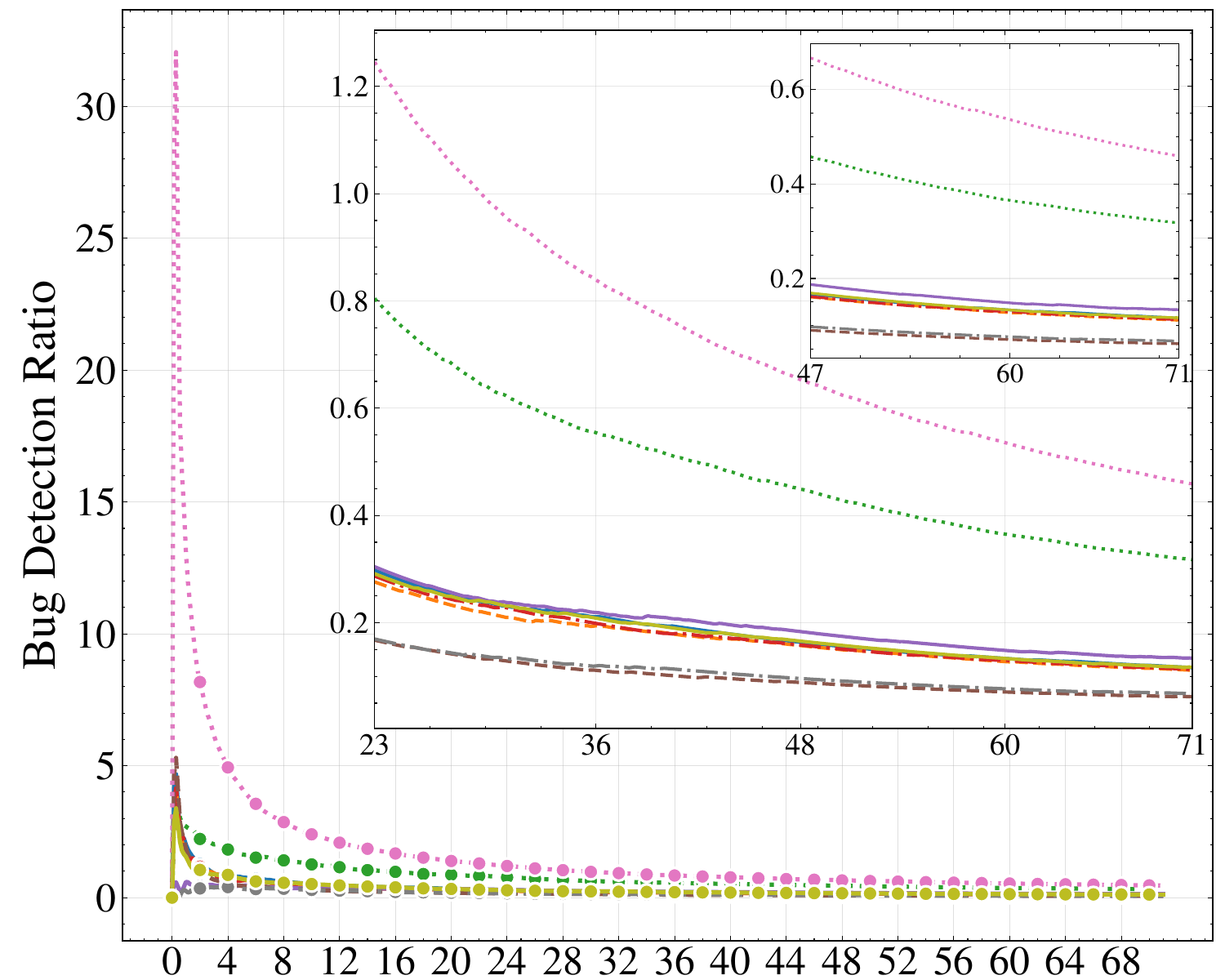}
    \hfill
    \figpanel{(l)}{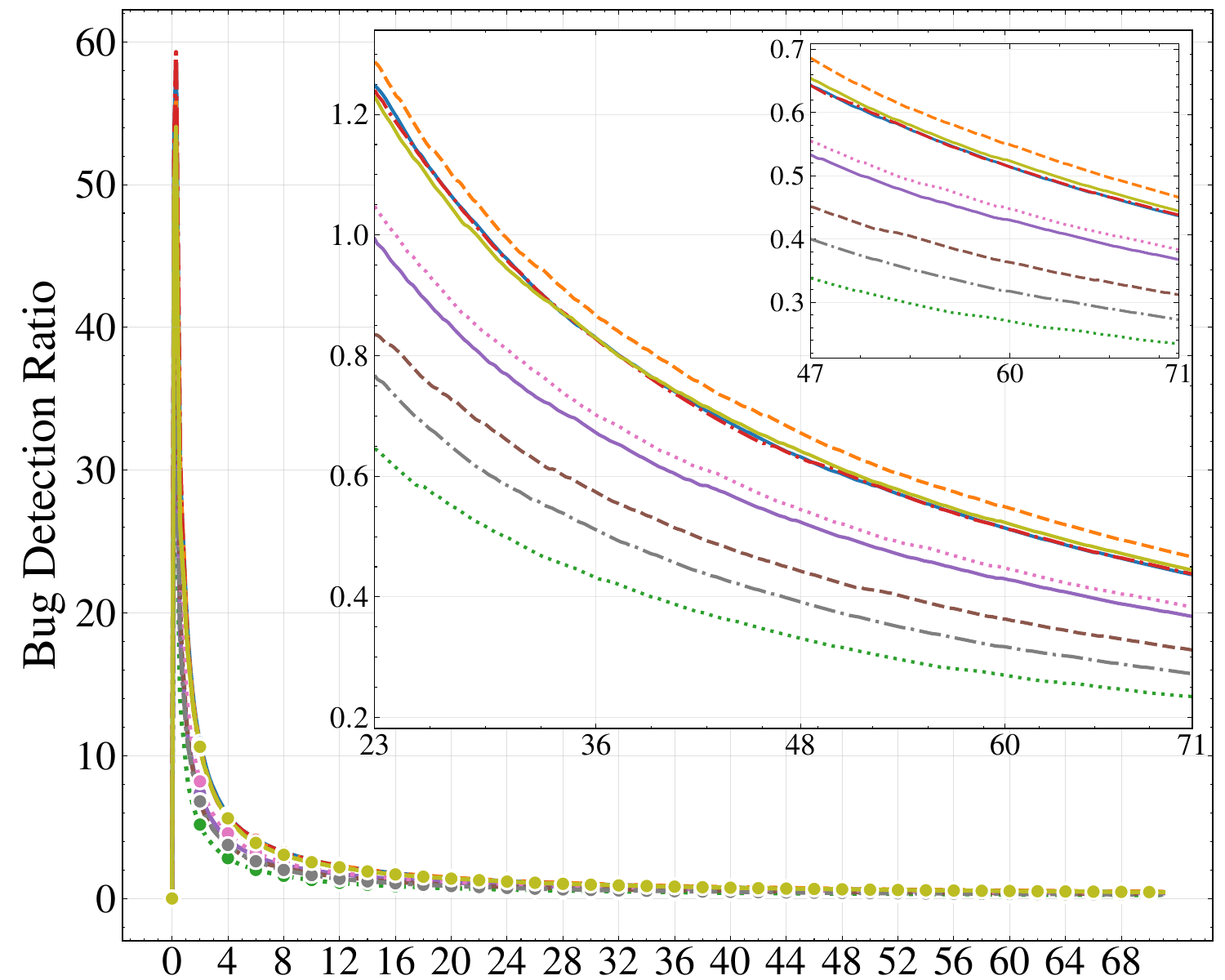}

    \vspace{-6pt}

    \caption{Panels (a),(b) show bug ratio; (c),(d) bug counts; (e),(f) code coverage; (g),(h) convergence across time segments; (i),(j) \(D_R(T)\) and \(\epsilon\); and (k),(l) time sufficiency on \emph{ffmpeg} and \emph{harfbuzz}.}
    \label{exp_fig}
\end{figure}

\vspace{-11pt}

\(\mathbf{D_R(T)}\). We report \(D_R(T)\) at the end of the run (\(T=71\) hours) on two benchmarks, \textit{libhtp} and \textit{poppler}. Here, \(\tilde{P}_R(T)\) is computed as defined earlier, and we also compute the ideal curve \(\tilde{P}\) (dashed line) at 71 hours, following Section~\ref{4.2.3}. Recall that \(\tilde{P}=\nicefrac{\tilde{N}}{T}\), where \(\tilde{N}\) is the union of all bugs detected across fuzzers by time \(T\). As proven in Corollary \ref{coro:1} in Section~\ref{4.2.3}, we use \(|\tilde{P}_R(T)-\tilde{P}|\) as a proxy for \(D_R(T)\). The resulting \(D_R(T)\) values are as follows: on \textit{libhtp}, AFL/AFL++/Honggfuzz/LibFuzzer score \(0.01/0.01/0.03/0.05\); on \textit{poppler}, they score \(0.18/0.33/0.10/0.38\). Full results are in Appendix~\ref{app:exp}. These values clearly distinguish fuzzers by bug detection effectiveness.

\textbf{Spectral parameter.} For non-degenerate fuzzers, i.e., those with a meaningful spectral parameter \(\delta_R<\infty\), we apply Corollary~\ref{coro:3} at \(T=71\) to infer information about \(\delta_R\). The lower bounds on the inverse spectral parameter \(\delta_R\) are \(0\) for AFL/AFL++/Honggfuzz/LibFuzzer on \textit{libhtp}, and \(0,0.128,0,0.323\) for AFL/AFL++/Honggfuzz/LibFuzzer on \textit{poppler}, respectively. These results indicate that, at the current horizon \(T\), the observed \(D_R(T)\) can be insufficient to certify a nontrivial lower bound on \(\delta_R\), especially on \textit{libhtp}. This provides another \textbf{subtle and novel insight from our evaluation}: while \(T=71\) is adequate for evaluating fuzzers, characterizing the inverse spectral parameter \(\delta_R\) may require a larger \(T\) such that \(\sqrt{\nicefrac{\log T}{T}} \leq |\tilde{P}_R(T)-\tilde{P}|\).

\textbf{Bug counts.} Raw bug counts are useful for reporting discoveries but are less reliable for comparing fuzzers. They remain unnormalized (substantially different scales across fuzzers in Figure~\ref{exp_fig}(d)); have larger variance than the bug detection ratio; and depend strongly on benchmark-specific bug density (different patterns in Figures~\ref{exp_fig}(c) and~\ref{exp_fig}(d)). They may plateau (most fuzzers do in Figure~\ref{exp_fig}(c)), or continue increasing (Honggfuzz does in Figure~\ref{exp_fig}(d)), without indicating convergence to a limiting performance measure. In contrast, our bug-ratio metric, shown in Figures~\ref{exp_fig}(a) and~\ref{exp_fig}(b), normalizes discoveries over executions, is more stable, and provides a principled basis for comparing fuzzers and even informing the spectral parameter as discussed above. 

\textbf{Code coverage.} We compare our metric with code coverage~\cite{li2021unifuzz} (Figures~\ref{exp_fig}(e)-(f)). Although the coverage ratio exhibits a longer tail, most fuzzers plateau early and remain close, limiting its ability to distinguish performance. Its scale also changes substantially across benchmarks ((e) and (f)), and higher coverage does not necessarily imply higher bug discovery efficiency (Figures~\ref{exp_fig}(a)-(b)).

\subsection{RQ3: Generalization}
We evaluate each fuzzer across all benchmarks and visualize convergence on representative ones to assess the framework’s generality. For each benchmark, we also derive a difficulty score from the evaluation metrics and compare these scores across benchmarks to validate its discriminative power.

Based on Figures~\ref{exp_fig}(g)-(h), fuzzers exhibit distinct yet stabilizing convergence behaviors across benchmarks, reflecting inherent benchmark-fuzzer differences. These patterns emerge without assuming fuzzers are ergodic on all benchmarks, further supporting the generality of our framework.

We compute \(D_R(T)\) for all fuzzers and benchmarks and define each benchmark’s difficulty as the average \(D_R(T)\) across fuzzers. The illustrative scores for \textit{arrow}, \textit{ffmpeg}, \textit{harfbuzz}, \textit{libhtp}, \textit{poppler}, and \textit{systemd} are \(0.43\), \(0.43\), \(0.27\), \(0.03\), \(0.25\), and \(0.00\), respectively; full results are in Appendix~\ref{app:exp}. These scores vary widely \emph{on a common scale} (unlike code coverage or bug counts), e.g., from \(0\) to \(0.43\), reflecting differences in bug-detection difficulty across benchmarks.



\subsection{RQ4: Robustness to Randomness}

We conduct experiments with 5 and 20 trials, representing small and large trial regimes, to examine the robustness of the metrics. We compute the evaluation metrics and benchmark scores as before.

Tables~\ref{tab:71h_libhtp_poppler_5} and~\ref{tab:avg_71h_benchmarks_5} in Appendix~\ref{app:exp} show that reducing the trial count from 20 to 5 does not significantly affect benchmark scores. Although metric values vary slightly on individual benchmarks, e.g., \textit{libhtp} and \textit{poppler}, the relative differences between fuzzers and conclusions about performance remain consistent. This indicates the evaluation's robustness to the trial count and fuzzing randomness.


\subsection{RQ5: Duration Sufficiency}
We extend the analysis to the choice of \(T\) and assess the adequacy of current practices. Specifically, we use the metrics at \(T=6\) and \(T=24\) as tolerance thresholds \(\epsilon\), set \(\alpha=0.025\), and identify the minimum duration based on Corollary~\ref{coro:2}. 

For \emph{libhtp} and \emph{poppler}, Figure~\ref{exp_fig}(i,j) reports the \(\epsilon\) values for all fuzzers at \(6\) and \(24\) hours. The smallest differences yield \(\epsilon=(0.3,3.6)\) and \((0.05,0.51)\) for \(T=6\) and \(24\) on \emph{libhtp} and \emph{poppler}, respectively, while the largest are \(\epsilon=(0.73,4.70)\) and \((0.17,1.14)\). These correspond to approximate $E(T) = 1/\epsilon$ of \((1.37,0.21)\) and \((5.88,0.88)\). Thus, \(6\) hours is needed when \(\epsilon>0.73\) and \(\epsilon>4.70\) for \emph{libhtp} and \emph{poppler}, respectively, while \(24\) hours is needed when \(\epsilon>0.17\) and \(\epsilon>1.14\). Here, we present representative \(\epsilon\) analyses; the same applies to any value of \(\epsilon\). The necessary duration varies across benchmarks, reflecting differences in bug patterns. In Figures~\ref{exp_fig}(k,l), many fuzzers on \emph{ffmpeg} require $24$ hours, whereas for \emph{harfbuzz}, even \(47\) hours is insufficient.

Notably, existing metrics suggest quite different choices of \(T\): code
coverage suggests a much smaller \(T\) (\(14\) in
Figure~\ref{exp_fig}(e)), whereas bug counts suggest a larger \(T\) (\(71\) in Figure~\ref{exp_fig}(d)). This is risky: an insufficient
\(T\) weakens evaluation claims, while an excessive \(T\) wastes computational
resources. 

%





%
\bibliographystyle{abbrv}
\bibliography{refs}

\begin{thebibliography}{10}

\bibitem{belluzzo_pydtmc}
{PyDTMC}: A library for discrete-time markov chains analysis.
\newblock \url{https://github.com/TommasoBelluzzo/PyDTMC}.

\bibitem{arcuri2010black}
A.~Arcuri, M.~Z. Iqbal, and L.~Briand.
\newblock Black-box system testing of real-time embedded systems using random and search-based testing.
\newblock In {\em IFIP International Conference on Testing Software and Systems}, pages 95--110. Springer, 2010.

\bibitem{auer2002nonstochastic}
P.~Auer, N.~Cesa-Bianchi, Y.~Freund, and R.~E. Schapire.
\newblock The nonstochastic multiarmed bandit problem.
\newblock {\em SIAM Journal on Computing}, 32(1):48--77, 2002.

\bibitem{auger2005restart}
A.~Auger and N.~Hansen.
\newblock A restart cma evolution strategy with increasing population size.
\newblock In {\em 2005 IEEE congress on evolutionary computation}, volume~2, pages 1769--1776. IEEE, 2005.

\bibitem{belete2022grid}
D.~M. Belete and M.~D. Huchaiah.
\newblock Grid search in hyperparameter optimization of machine learning models for prediction of hiv/aids test results.
\newblock {\em International Journal of Computers and Applications}, 44(9):875--886, 2022.

\bibitem{bendkowski2022tuning}
M.~Bendkowski, O.~Bodini, and S.~Dovgal.
\newblock Tuning as convex optimisation: a polynomial tuner for multi-parametric combinatorial samplers.
\newblock {\em Combinatorics, Probability and Computing}, 31(5):765--811, 2022.

\bibitem{Entropic}
M.~B\"{o}hme, V.~J.~M. Man\`{e}s, and S.~K. Cha.
\newblock Boosting fuzzer efficiency: an information theoretic perspective.
\newblock In {\em ESEC/FSE}, 2020.

\bibitem{bohme2017directed}
M.~B{\"o}hme, V.-T. Pham, M.-D. Nguyen, and A.~Roychoudhury.
\newblock Directed greybox fuzzing.
\newblock In {\em Proceedings of the 2017 ACM SIGSAC conference on computer and communications security}, 2017.

\bibitem{bohme2016coverage}
M.~B{\"o}hme, V.-T. Pham, and A.~Roychoudhury.
\newblock Coverage-based greybox fuzzing as markov chain.
\newblock In {\em Proceedings of the 2016 ACM SIGSAC Conference on Computer and Communications Security}, 2016.

\bibitem{braiek2020testing}
H.~B. Braiek and F.~Khomh.
\newblock On testing machine learning programs.
\newblock {\em Journal of Systems and Software}, 164:110542, 2020.

\bibitem{cha2015program}
S.~K. Cha, M.~Woo, and D.~Brumley.
\newblock Program-adaptive mutational fuzzing.
\newblock In {\em IEEE Symposium on Security and Privacy}. IEEE, 2015.

\bibitem{chatterjee2023spectral}
S.~Chatterjee.
\newblock Spectral gap of nonreversible markov chains.
\newblock {\em Annals of Applied Probability}, 2025.

\bibitem{chen2021black}
X.~Chen and E.~Hazan.
\newblock Black-box control for linear dynamical systems.
\newblock In {\em Conference on Learning Theory}, pages 1114--1143. PMLR, 2021.

\bibitem{elder2022really}
S.~Elder, N.~Zahan, R.~Shu, M.~Metro, V.~Kozarev, T.~Menzies, and L.~Williams.
\newblock Do i really need all this work to find vulnerabilities? an empirical case study comparing vulnerability detection techniques on a java application.
\newblock {\em Empirical Software Engineering}, 2022.

\bibitem{esteva2017dermatologist}
A.~Esteva, B.~Kuprel, R.~A. Novoa, J.~Ko, S.~M. Swetter, H.~M. Blau, and S.~Thrun.
\newblock Dermatologist-level classification of skin cancer with deep neural networks.
\newblock {\em nature}, 542(7639):115--118, 2017.

\bibitem{feldman2007budget}
J.~Feldman, S.~Muthukrishnan, M.~Pal, and C.~Stein.
\newblock Budget optimization in search-based advertising auctions.
\newblock In {\em Proceedings of the 8th ACM conference on Electronic commerce}, pages 40--49, 2007.

\bibitem{fill1991eigenvalue}
J.~A. Fill.
\newblock Eigenvalue bounds on convergence to stationarity for nonreversible markov chains, with an application to the exclusion process.
\newblock {\em Annals of Applied Probability}, pages 62--87, 1991.

\bibitem{AFL++}
A.~Fioraldi, D.~C. Maier, H.~Ei{\ss}feldt, and M.~Heuse.
\newblock {AFL++} : Combining incremental steps of fuzzing research.
\newblock In {\em {USENIX} Workshop on Offensive Technologies}, 2020.

\bibitem{freedman2017convergence}
A.~Freedman.
\newblock Convergence theorem for finite markov chains.
\newblock {\em Proc. Reu}, 2017.

\bibitem{godefroid2020fuzzing}
P.~Godefroid.
\newblock Fuzzing: Hack, art, and science.
\newblock {\em Communications of the ACM}, 63(2):70--76, 2020.

\bibitem{godefroid2008grammar}
P.~Godefroid, A.~Kiezun, and M.~Y. Levin.
\newblock Grammar-based whitebox fuzzing.
\newblock In {\em Proceedings of the 29th ACM SIGPLAN conference on programming language design and implementation}, 2008.

\bibitem{guidotti2018survey}
R.~Guidotti, A.~Monreale, S.~Ruggieri, F.~Turini, F.~Giannotti, and D.~Pedreschi.
\newblock A survey of methods for explaining black box models.
\newblock {\em ACM computing surveys (CSUR)}, 51(5):1--42, 2018.

\bibitem{guo2018dlfuzz}
J.~Guo, Y.~Jiang, Y.~Zhao, Q.~Chen, and J.~Sun.
\newblock Dlfuzz: Differential fuzzing testing of deep learning systems.
\newblock In {\em ESEC/FSE}, 2018.

\bibitem{hazimeh2020magma}
A.~Hazimeh, A.~Herrera, and M.~Payer.
\newblock Magma: A ground-truth fuzzing benchmark.
\newblock {\em Proceedings of the ACM on Measurement and Analysis of Computing Systems}, 4(3):1--29, 2020.

\bibitem{he2025llm}
J.~He, C.~Treude, and D.~Lo.
\newblock Llm-based multi-agent systems for software engineering: Literature review, vision, and the road ahead.
\newblock {\em ACM Transactions on Software Engineering and Methodology}, 34(5):1--30, 2025.

\bibitem{hsu2019mixing}
D.~Hsu, A.~Kontorovich, D.~A. Levin, Y.~Peres, C.~Szepesv{\'a}ri, and G.~Wolfer.
\newblock Mixing time estimation in reversible markov chains from a single sample path.
\newblock 2019.

\bibitem{jero2019leveraging}
S.~Jero, M.~L. Pacheco, D.~Goldwasser, and C.~Nita-Rotaru.
\newblock Leveraging textual specifications for grammar-based fuzzing of network protocols.
\newblock In {\em Proceedings of the AAAI}, 2019.

\bibitem{jiang2023evaluating}
L.~Jiang, H.~Yuan, M.~Wu, L.~Zhang, and Y.~Zhang.
\newblock Evaluating and improving hybrid fuzzing.
\newblock In {\em ICSE}. IEEE, 2023.

\bibitem{klees2018evaluating}
G.~Klees, A.~Ruef, B.~Cooper, S.~Wei, and M.~Hicks.
\newblock Evaluating fuzz testing.
\newblock In {\em Proceedings of the 2018 ACM SIGSAC conference on computer and communications security}, pages 2123--2138, 2018.

\bibitem{krishnamoorthy2023diffusion}
S.~Krishnamoorthy, S.~M. Mashkaria, and A.~Grover.
\newblock Diffusion models for black-box optimization.
\newblock In {\em International Conference on Machine Learning}, pages 17842--17857. PMLR, 2023.

\bibitem{leithner2021hydra}
M.~Leithner, B.~Garn, and D.~E. Simos.
\newblock Hydra: Feedback-driven black-box exploitation of injection vulnerabilities.
\newblock {\em Information and Software Technology}, 140:106703, 2021.

\bibitem{lemieux2018fairfuzz}
C.~Lemieux and K.~Sen.
\newblock Fairfuzz: A targeted mutation strategy for increasing greybox fuzz testing coverage.
\newblock In {\em Proceedings of the 33rd ACM/IEEE international conference on automated software engineering}, pages 475--485, 2018.

\bibitem{li2021unifuzz}
Y.~Li, S.~Ji, Y.~Chen, S.~Liang, W.-H. Lee, et~al.
\newblock {UNIFUZZ}: A holistic and pragmatic metrics-driven platform for evaluating fuzzers.
\newblock In {\em USENIX Security}, 2021.

\bibitem{li2021openbox}
Y.~Li, Y.~Shen, W.~Zhang, Y.~Chen, H.~Jiang, M.~Liu, J.~Jiang, J.~Gao, W.~Wu, Z.~Yang, et~al.
\newblock Openbox: A generalized black-box optimization service.
\newblock In {\em Proceedings of the 27th ACM SIGKDD conference on knowledge discovery \& data mining}, pages 3209--3219, 2021.

\bibitem{LibFuzzer}
{LLVM Project}.
\newblock Libfuzzer: A library for coverage-guided fuzz testing.
\newblock \url{https://llvm.org/docs/LibFuzzer.html}.

\bibitem{luo2016forepost}
Q.~Luo, D.~Poshyvanyk, A.~Nair, and M.~Grechanik.
\newblock Forepost: A tool for detecting performance problems with feedback-driven learning software testing.
\newblock In {\em Proceedings of the 38th International Conference on Software Engineering Companion}, pages 593--596, 2016.

\bibitem{luo2024distribution}
Y.~Luo, W.~W. Sun, and Y.~Liu.
\newblock Distribution-free contextual dynamic pricing.
\newblock {\em Mathematics of Operations Research}, 49(1):599--618, 2024.

\bibitem{lyu2019mopt}
C.~Lyu, S.~Ji, C.~Zhang, Y.~Li, et~al.
\newblock {MOPT}: Optimized mutation scheduling for fuzzers.
\newblock In {\em USENIX security}, 2019.

\bibitem{mcmahan2017survey}
H.~B. McMahan.
\newblock A survey of algorithms and analysis for adaptive online learning.
\newblock {\em Journal of Machine Learning Research}, 18(90):1--50, 2017.

\bibitem{meng2024large}
R.~Meng, M.~Mirchev, M.~B{\"o}hme, and A.~Roychoudhury.
\newblock Large language model guided protocol fuzzing.
\newblock In {\em Proceedings of the Annual Network and Distributed System Security Symposium}, 2024.

\bibitem{fuzzbench}
J.~Metzman, L.~Szekeres, L.~Simon, R.~Sprabery, and A.~Arya.
\newblock Fuzzbench: an open fuzzer benchmarking platform and service.
\newblock In {\em ESEC/FSE}, 2021.

\bibitem{moukahal2021vulnerability}
L.~J. Moukahal, M.~Zulkernine, and M.~Soukup.
\newblock Vulnerability-oriented fuzz testing for connected autonomous vehicle systems.
\newblock {\em IEEE transactions on reliability}, 70(4):1422--1437, 2021.

\bibitem{nagy2019full}
S.~Nagy and M.~Hicks.
\newblock Full-speed fuzzing: Reducing fuzzing overhead through coverage-guided tracing.
\newblock In {\em 2019 IEEE Symposium on Security and Privacy (SP)}, pages 787--802. IEEE, 2019.

\bibitem{nelson2010stochastic}
B.~L. Nelson.
\newblock {\em Stochastic modeling: analysis \& simulation}.
\newblock 2010.

\bibitem{nelson2021foundations}
B.~L. Nelson and L.~Pei.
\newblock {\em Foundations and methods of stochastic simulation}.
\newblock Springer, 2021.

\bibitem{openai_clip_2021}
{OpenAI}.
\newblock {CLIP}: Connecting text and images.
\newblock \url{https://openai.com/index/clip/}, Jan. 2021.

\bibitem{petsios2017nezha}
T.~Petsios, A.~Tang, S.~Stolfo, A.~D. Keromytis, and S.~Jana.
\newblock Nezha: Efficient domain-independent differential testing.
\newblock In {\em IEEE Symposium on security and privacy}. IEEE, 2017.

\bibitem{petsios2017slowfuzz}
T.~Petsios, J.~Zhao, A.~D. Keromytis, and S.~Jana.
\newblock Slowfuzz: Automated domain-independent detection of algorithmic complexity vulnerabilities.
\newblock In {\em Proceedings of the 2017 ACM SIGSAC conference on computer and communications security}, pages 2155--2168, 2017.

\bibitem{AFLSmart}
V.~Pham, M.~B{\"{o}}hme, A.~E. Santosa, A.~R. Caciulescu, and A.~Roychoudhury.
\newblock Smart greybox fuzzing.
\newblock {\em CoRR}, 2018.

\bibitem{provodin2024bandits}
D.~Provodin, J.~Joudioux, and E.~Duryev.
\newblock Bandits for sponsored search auctions under unknown valuation model: Case study in e-commerce advertising.
\newblock In {\em Joint European Conference on Machine Learning and Knowledge Discovery in Databases}, pages 263--279. Springer, 2024.

\bibitem{ren2021empirical}
M.~Ren, Z.~Yin, F.~Ma, Z.~Xu, Y.~Jiang, C.~Sun, H.~Li, and Y.~Cai.
\newblock Empirical evaluation of smart contract testing: What is the best choice?
\newblock In {\em Proceedings of the 30th ACM SIGSOFT international symposium on software testing and analysis}, pages 566--579, 2021.

\bibitem{rubino2009rare}
G.~Rubino, B.~Tuffin, et~al.
\newblock {\em Rare event simulation using Monte Carlo methods}, volume~73.
\newblock Wiley Online Library, 2009.

\bibitem{schloegel2024sok}
M.~Schloegel, N.~Bars, N.~Schiller, L.~Bernhard, T.~Scharnowski, A.~Crump, A.~Ale-Ebrahim, N.~Bissantz, M.~Muench, and T.~Holz.
\newblock Sok: Prudent evaluation practices for fuzzing.
\newblock In {\em 2024 IEEE Symposium on Security and Privacy (SP)}, pages 1974--1993. IEEE, 2024.

\bibitem{shalev2025online}
S.~Shalev-Shwartz.
\newblock Online learning and online convex optimization.
\newblock {\em Foundations and Trends{\textregistered} in Machine Learning}, 4(2):107--194, 2025.

\bibitem{she2024fox}
D.~She, A.~Storek, Y.~Xie, S.~Kweon, P.~Srivastava, and S.~Jana.
\newblock Fox: Coverage-guided fuzzing as online stochastic control.
\newblock In {\em Proceedings of the 2024 on ACM SIGSAC Conference on Computer and Communications Security}, pages 765--779, 2024.

\bibitem{sherlaw1995estimating}
C.~Sherlaw-Johnson, S.~Gallivan, and J.~Burridge.
\newblock Estimating a markov transition matrix from observational data.
\newblock {\em JORS}, 1995.

\bibitem{sjoberg1995nonlinear}
J.~Sj{\"o}berg, Q.~Zhang, L.~Ljung, A.~Benveniste, B.~Delyon, P.-Y. Glorennec, H.~Hjalmarsson, and A.~Juditsky.
\newblock Nonlinear black-box modeling in system identification: a unified overview.
\newblock {\em Automatica}, 31(12):1691--1724, 1995.

\bibitem{Sousi2020}
P.~Sousi.
\newblock Mixing times of markov chains, 2020.

\bibitem{honggfuzz}
R.~Swiecki.
\newblock honggfuzz: Security oriented, feedback-driven, evolutionary fuzzer.
\newblock \url{https://github.com/google/honggfuzz}.

\bibitem{vaisnor2024approximate}
C.~Vaisnor.
\newblock Approximate bayesian computation as an informed fuzzing-inference system.
\newblock {\em arXiv preprint arXiv:2404.04303}, 2024.

\bibitem{wang2019superion}
J.~Wang, B.~Chen, L.~Wei, and Y.~Liu.
\newblock Superion: Grammar-aware greybox fuzzing.
\newblock In {\em ICSE}, pages 724--735. IEEE, 2019.

\bibitem{wilson1984variance}
J.~R. Wilson.
\newblock Variance reduction techniques for digital simulation.
\newblock {\em American Journal of Mathematical and Management Sciences}, 1984.

\bibitem{wolfer2020mixing}
G.~Wolfer.
\newblock Mixing time estimation in ergodic markov chains from a single trajectory with contraction methods.
\newblock In {\em Algorithmic Learning Theory}, pages 890--905. PMLR, 2020.

\bibitem{wolfer2024improved}
G.~Wolfer and A.~Kontorovich.
\newblock Improved estimation of relaxation time in nonreversible markov chains.
\newblock {\em Ann. Appl. Probab.}, 2024.

\bibitem{wu2022one}
M.~Wu, L.~Jiang, J.~Xiang, Y.~Huang, H.~Cui, L.~Zhang, and Y.~Zhang.
\newblock One fuzzing strategy to rule them all.
\newblock In {\em ICSE}, 2022.

\bibitem{wu2022evaluating}
M.~Wu, L.~Jiang, J.~Xiang, Y.~Zhang, G.~Yang, H.~Ma, S.~Nie, S.~Wu, H.~Cui, and L.~Zhang.
\newblock Evaluating and improving neural program-smoothing-based fuzzing.
\newblock In {\em ICSE}, pages 847--858, 2022.

\bibitem{xia2024fuzz4all}
C.~S. Xia, M.~Paltenghi, J.~Le~Tian, M.~Pradel, and L.~Zhang.
\newblock Fuzz4all: Universal fuzzing with large language models.
\newblock In {\em ICSE}, 2024.

\bibitem{you2019profuzzer}
W.~You, X.~Wang, S.~Ma, J.~Huang, X.~Zhang, X.~Wang, and B.~Liang.
\newblock Profuzzer: On-the-fly input type probing for better zero-day vulnerability discovery.
\newblock In {\em 2019 IEEE symposium on security and privacy (SP)}, pages 769--786. IEEE, 2019.

\bibitem{yue2020ecofuzz}
T.~Yue, P.~Wang, Y.~Tang, E.~Wang, B.~Yu, K.~Lu, and X.~Zhou.
\newblock $\{$EcoFuzz$\}$: Adaptive $\{$Energy-Saving$\}$ greybox fuzzing as a variant of the adversarial $\{$Multi-Armed$\}$ bandit.
\newblock In {\em USENIX Security}, 2020.

\bibitem{afl}
M.~Zalewski.
\newblock American fuzzy lop.
\newblock \url{http://lcamtuf.coredump.cx/afl/}.

\bibitem{zhang2026instruction}
S.~Zhang, L.~Dong, X.~Li, S.~Zhang, X.~Sun, S.~Wang, J.~Li, R.~Hu, T.~Zhang, G.~Wang, et~al.
\newblock Instruction tuning for large language models: A survey.
\newblock {\em ACM Computing Surveys}, 58(7):1--36, 2026.

\bibitem{ziegler2016synthesis}
M.~M. Ziegler, H.-Y. Liu, G.~Gristede, B.~Owens, R.~Nigaglioni, and L.~P. Carloni.
\newblock A synthesis-parameter tuning system for autonomous design-space exploration.
\newblock In {\em 2016 Design, Automation \& Test in Europe Conference \& Exhibition (DATE)}, pages 1148--1151. IEEE, 2016.

\bibitem{zolfagharian2023search}
A.~Zolfagharian, M.~Abdellatif, L.~C. Briand, M.~Bagherzadeh, et~al.
\newblock A search-based testing approach for deep reinforcement learning agents.
\newblock {\em IEEE Transactions on Software Engineering}, 49(7):3715--3735, 2023.

\end{thebibliography}

\newpage 
\appendix 
\section{Technical appendices and supplementary material}
\subsection{Broader Impacts}\label{app:impacts}

One application of this research is to develop evaluation metrics for mutation-based fuzzing and adversarial attacks from a security perspective. Our empirical study on fuzzing uses established fuzzers and standard public benchmarks, including FuzzBench, and is restricted to stable, previously released benchmark programs. The experiments are conducted only in controlled, offline testing settings. The experiment on adversarial attacks uses an existing public dataset that has been well cited by the community.

The work does not target live systems, disclose new vulnerabilities in deployed software, or introduce foreseeable risks to individuals. As a result, no responsible-disclosure process is triggered. The study is designed to minimize potential harm to users, organizations, and operational infrastructure, and it adheres to responsible research practices throughout.

\subsection{Discussions}\label{app:discussion}
We justify modeling fuzzing from a single initial seed/input string and propose
a new performance measure that, rather than directly adopting existing fuzzing
practices, provides a principled formulation while still connecting to those
practices.

\subsubsection{Modeling}
We start with seed selection. 
\begin{remark}[Seed]
Notably, here we focus on the case of starting from a single input seed; the
setting with multiple seeds can be interpreted as follows. If the seeds are run
simultaneously/independently, this is equivalent to running multiple independent
trials, which is already covered by our modeling. Alternatively, if the seeds
are pre-fixed from a seed corpus and switched over time, this amounts to
modifying the transition matrix by assigning probability one to a transition to
the new seed, after which the Markov chain restarts with the original transition
matrix. 
\begin{center}
\emph{Essentially, changing the seed can be viewed as a mutation by
\(\mathcal{R}\) based on \(S_t\), so it can be incorporated into the
single-seed formulation.} 
\end{center} 
\end{remark}
\vspace{6pt}

\begin{remark}[Bug Deduplication] 
We next justify the proposed performance measure by comparing it with existing
fuzzing metrics and discussing bug deduplication.

\textbf{Why not unique bug rate/counts?} We first explain why we use the bug
detection rate, rather than unique bug rate/counts, as the objective.
Discovering new unique bugs and triggering bug-inducing behaviors are
related but distinct goals. This paper intentionally studies the latter.
Formally, \(Y_t^R\) indicates whether the input executed at time \(t\) triggers
a bug, and
\[
P_R=\lim_{T\to\infty}\frac{1}{T}\sum_{t=1}^T Y_t^R
\]
measures the long-run fraction of bug-triggering inputs among all executed
inputs. We choose this objective because it provides a steady-state asymptotic
target for each fuzzer, supports the finite-time and spectral analysis
developed in this paper, and reflects the fuzzer's bug detection performance
both rigorously and practically.

In contrast, unique-bug discovery is inherently history-dependent: whether an
execution at time \(t\) counts depends on which bugs have already been observed
earlier in the run, i.e., the corresponding unique-bug indicator depends not
only on \(X_t^R\), but also on the entire history. For a finite bug set \(B\), the
asymptotic long-run average rate of new unique-bug discoveries per mutation is
zero.  Therefore it does not play the same role as the steady-state objective
analyzed here. More specifically, convergence of the unique bug rate to zero does not
necessarily imply convergence to the true performance of a fuzzer: the count may
plateau simply because the fuzzer fails to find new bugs and is stuck in certain
regions, rather than because it has exhausted the relevant bug space. 

More precisely, unique bug discovery can still be modeled by augmenting the state with the
execution history. Specifically, if \(G_t\) denotes the set of bugs discovered
before time \(t\), then the unique-bug indicator can be written as
\(Y_t^{\mathrm{uniq}}=\mathbf 1\{g_f(X_t)\in B_f,\; g_f(X_t)\notin G_t\}\),
with \(G_{t+1}=G_t\cup\{g_f(X_t):g_f(X_t)\in B_f\}\). This formulation,
however, makes the history dependence explicit: the objective of unique bug rate  depends on the
augmented state \((X_t,G_t)\), and for a finite bug set \[
\frac{1}{T}\sum_{t=1}^T Y_t^{\mathrm{uniq}}
=
\frac{U_R(T)}{T}
\leq \frac{|B|}{T}
\to 0. 
\] If a fuzzer fails to find new bugs and is stuck in certain regions, $Y_t^{\mathrm{uniq}}$ is $0$, and the target even converges faster to the objective, which is contradictory to practice.  

Using unique bug counts, instead of the unique bug rate, does not solve the
problem, as the objective, \(|B|\), is no longer a long-run asymptotic objective.
Unique bug counts, \(\sum_{t=1}^T Y_t^{\mathrm{uniq}}\), often evolve through
long plateaus followed by occasional jumps, making them less suitable as a
smooth performance measure for convergence and asymptotic behavior as $|\sum_{t=1}^T Y_t^{\mathrm{uniq}} -|B||$ is discrete. This structure does not have an explicit analytical form, such as the cumulative
reward in reinforcement learning, which is typically provable only under strong
structural assumptions on the system; such assumptions, however, do not hold for
fuzzing. \emph{It also does not fit the theme of black-box processes considered
herein, where the target is the rate of certain events.} They also
suffer from higher randomness because the outcomes depend on the entire
history, so differences between two replications can be amplified. Moreover,
they do not indicate how long an evaluation should run to obtain trustworthy
results, nor do they account for the effort \(T\) and thus the efficiency of a
fuzzer. Finally, they are sensitive to initial conditions, if a fuzzer runs with a different seed corpus, then their performance can be quite different, as early outcomes
accumulate over the long run. The true mutation power is thus mixed with the seed selection, thereby less desirable. 

Our goal is therefore not to claim that unique-bug discovery is unimportant,
but to formalize a different quantity, bug-triggering efficiency, that is
well-defined under sequential stochastic dynamics and admits rigorous
convergence and evaluation guarantees that reliably reflects the fuzzer performance.

\textbf{Connection between bug detection rate and unique bug discovery.}
Moreover, in practice, the bug detection rate and unique bug counts are oftentimes 
positively related: a fuzzer with a higher bug detection rate is more likely to
explore bug-sensitive regions and eventually trigger distinct bugs. Thus, the
bug detection rate provides a rigorous analytical objective while still offering
practical insight into unique-bug discovery. 

We acknowledge that, in practice, there are cases where repeatedly triggering the
same bug may make the unique bug rate or bug rate favor certain fuzzers, while
the unique bug count remains unchanged. However, when \(T\) is sufficiently
large, such repeated triggers have limited impact on \(P_R(T)\) because the
denominator is large. Moreover, unique bug counts suffer from several
limitations, as discussed above. These corner cases may not generalize, and a formal discussion of them is an
interesting direction for future exploration. One possible extension is to
account for both the bug detection rate and unique bug counts in a
multi-objective framework; one can always incorporate unique bug counts as a supplementary empirical measure. Since the bug detection rate remains a necessary
component in such a framework, our work provides a foundation for understanding
it rigorously.

\textbf{Why not raw bug counts?} Next, we do not use raw bug counts as the
performance measure. First, raw bug counts inherit a similar limitation from the
perspective of convergence: \(\sum_{t=1}^T Y_t^R\) does not converge to a fixed
long-run objective as \(T\to\infty\). Second, raw counts are not normalized by
the computational effort \(T\), making comparisons across different evaluation
horizons difficult and causing their variance to grow with \(T\). Third, raw
counts do not provide a principled basis for choosing an appropriate evaluation
horizon. These issues motivate our use of the normalized bug detection rate as
the primary performance measure.

\textbf{Connection between bug detection rate and raw bug counts.} By
definition, these two quantities are positively related: a higher bug detection
rate means more raw bug counts, and vice versa. As such, using our new measure
also reflects an objective that practitioners often care about. 
\end{remark}

\subsubsection{Experiments}\label{app:seed_bug}

\begin{remark}[Seed]
\end{remark}
 
All fuzzers are initialized from the same default benchmark seed corpus (the benchmark-provided OSS-Fuzz/FuzzBench seeds), so the comparison is not confounded by unequal initial seeds.

\begin{remark}[Bug Deduplication]
\end{remark}

For bug detection, we use FuzzBench’s default sanitizer-based setup, which is also common practice in the fuzzing domain nowadays. In our experiments, the targets are instrumented with AddressSanitizer (ASan) and UndefinedBehaviorSanitizer (UBSan), and a fatal sanitizer report is treated as evidence of bug-triggering behavior.

To classify repeated sanitizer reports in the experiments, we use the sanitizer-reported error class together with a canonicalized stack-trace signature, so that repeated identical reports within a trial are not redundantly presented as separate failures.






\subsection{Numerical Experiments on Magma}\label{app:magma}

We complement the FuzzBench evaluation with a study on
Magma~\citep{hazimeh2020magma}, the only widely used fuzzing benchmark
that ships verifiable ground-truth bugs. This lets us check the central
claim of Section~\ref{4.2.3} directly: that the proxy
\(\tilde{P}=\nicefrac{\tilde{N}}{T}\) is a faithful drop-in for the
unknown stationary rate \(\pi(1)\) when ranking fuzzers through
\(D_R(T)=|\tilde{P}_R(T)-\tilde{P}|\); it also lets us estimate the
spectral parameter \(\delta_R\) along both paths in
Sections~\ref{4.2.2} and~\ref{4.2.3}.

\paragraph{Setup.} Targets: \texttt{openssl} (asn1parse harness);
trackable Magma bugs (\texttt{SSL001}, \texttt{SSL003}, \texttt{SSL010}, \texttt{SSL016})
and \texttt{poppler} (pdftoppm harness); trackable Magma bugs (\texttt{PDF001},~\texttt{PDF002},~\texttt{PDF004}--\texttt{PDF012},~\texttt{PDF014},~\texttt{PDF016},~\texttt{PDF018},~\texttt{PDF019},~\texttt{PDF021},~\texttt{PDF022}). Fuzzers: \textsc{AFL}, \textsc{AFLFast},
\textsc{Honggfuzz}, \textsc{MOpt} in stand-alone mode. Each pair runs
\(T=12\)\,h with \(M=10\) independent trials seeded from the
benchmark-provided corpora. Magma's reach/trigger oracle is queried at
one-minute resolution. \(\tilde{P}_R(T)\) is the unique-bug rate,
\(\tilde{P}(T)=\nicefrac{\tilde{N}(T)}{T}\) is the union of bugs
\emph{triggered} across all fuzzers/trials, and the ground-truth analog
\(\pi_{\mathrm{GT}}(T)=\nicefrac{\tilde{N}_{\mathrm{GT}}(T)}{T}\) is the
union of bugs \emph{reached}. All numbers are reported as mean
\(\pm\,1\,\sigma\) across the \(M=10\) trials (\emph{error bars}).
We do not report whole-program code coverage because the Magma
reach/trigger oracle used in this dataset only logs per-bug reach and
trigger counts.

\paragraph{Proxy \(\tilde{P}\) validates against ground-truth
\(\pi(1)\).} Recall \(D_R(T)=|\tilde{P}_R(T)-\tilde{P}|\); on Magma we additionally compute its ground-truth analog \(D_R^{\mathrm{GT}}(T)=|\tilde{P}_R(T)-\pi_{\mathrm{GT}}|\). On \texttt{openssl} we have \(\tilde{P}=0.167\) and
\(\pi_{\mathrm{GT}}=0.333\); on \texttt{poppler}, \(\tilde{P}=0.500\)
and \(\pi_{\mathrm{GT}}=1.417\). The shared bias
\(|\tilde{P}-\pi_{\mathrm{GT}}|\) is fuzzer-independent
(\(0.167\) and \(0.917\) respectively), so the rankings induced by
\(D_R(T)\) and the ground-truth analog
\(D_R^{\mathrm{GT}}(T)=|\tilde{P}_R(T)-\pi_{\mathrm{GT}}|\) coincide
exactly: \textsc{Honggfuzz} is best on \texttt{openssl},
\textsc{MOpt} is best on \texttt{poppler} by either metric. The sandwich
\(\tilde{P}_R(T)\le\tilde{P}\le\pi_{\mathrm{GT}}\) of
Section~\ref{4.2.3} holds in every row of Table~\ref{tab:magma-proxy}.

\begin{table}[!htbp]
\centering
\setlength{\tabcolsep}{4pt}
\renewcommand{\arraystretch}{0.92}
\footnotesize
\begin{tabular}{ll cccc}
\toprule
Target & Fuzzer & \(\tilde{P}_R(T)\) & \(D_R(T)\) & \(D_R^{\mathrm{GT}}(T)\) & rank \\
\midrule
\texttt{openssl} & \textsc{AFL}       & \(0.083\pm0.000\) & \(0.083\) & \(0.250\) & 2-4 \\
                 & \textsc{AFLFast}   & \(0.083\pm0.000\) & \(0.083\) & \(0.250\) & 2-4 \\
                 & \textsc{Honggfuzz} & \(0.117\pm0.041\) & \(\mathbf{0.050}\) & \(\mathbf{0.217}\) & \textbf{1} \\
                 & \textsc{MOpt}      & \(0.083\pm0.000\) & \(0.083\) & \(0.250\) & 2-4 \\
\midrule
\texttt{poppler} & \textsc{AFL}       & \(0.125\pm0.042\) & \(0.375\) & \(1.292\) & 4 \\
                 & \textsc{AFLFast}   & \(0.167\pm0.000\) & \(0.333\) & \(1.250\) & 3 \\
                 & \textsc{Honggfuzz} & \(0.183\pm0.033\) & \(0.317\) & \(1.233\) & 2 \\
                 & \textsc{MOpt}      & \(0.292\pm0.067\) & \(\mathbf{0.208}\) & \(\mathbf{1.125}\) & \textbf{1} \\
\bottomrule
\end{tabular}
\caption{Proxy validation on Magma. Smallest \(D_R\) in bold.}
\label{tab:magma-proxy}
\end{table}

\paragraph{Estimating \(\delta_R\).} We estimate \(\delta_R\) along
both paths developed in the paper:
\(\hat\delta_R\) is the data-driven \texttt{pydtmc}~\citep{belluzzo_pydtmc}
estimate from Section~\ref{4.2.2}, applied to the two-state output
chain induced by \(Y_R(\omega,t)\); \(\delta_R^{\mathrm{LB,GT}}\) is
the Corollary~\ref{coro:3} lower bound, instantiated on Magma with
the ground-truth analog \(\pi_{\mathrm{GT}}\) in place of \(\tilde{P}\).
Both paths produce the same ranking as \(D_R(T)\):
\textsc{Honggfuzz} smallest on \texttt{openssl},
\textsc{MOpt} smallest on \texttt{poppler}.

\begin{table}[!htbp]
\centering
\setlength{\tabcolsep}{4pt}
\renewcommand{\arraystretch}{0.92}
\footnotesize
\begin{tabular}{ll cc}
\toprule
Target & Fuzzer & \(\hat\delta_R\) (\texttt{pydtmc}) & \(\delta_R^{\mathrm{LB,GT}}\) \\
\midrule
\texttt{openssl} & \textsc{AFL}       & \(3.30\pm0.27\) & \(4.29\) \\
                 & \textsc{AFLFast}   & \(2.92\pm0.54\) & \(4.29\) \\
                 & \textsc{Honggfuzz} & \(\mathbf{1.03\pm0.05}\) & \(\mathbf{2.64}\) \\
                 & \textsc{MOpt}      & \(1.32\pm0.15\) & \(4.29\) \\
\midrule
\texttt{poppler} & \textsc{AFL}       & \(1.62\pm0.07\) & \(257.5\) \\
                 & \textsc{AFLFast}   & \(1.62\pm0.10\) & \(239.9\) \\
                 & \textsc{Honggfuzz} & \(1.88\pm2.53\) & \(233.0\) \\
                 & \textsc{MOpt}      & \(\mathbf{1.59\pm0.11}\) & \(\mathbf{190.7}\) \\
\bottomrule
\end{tabular}
\caption{Spectral-parameter estimates on Magma. Smallest in bold.
\(\hat\delta_R\) is the data-driven \texttt{pydtmc} estimate of
\(\delta_R\) (Section~\ref{4.2.2}); \(\delta_R^{\mathrm{LB,GT}}\) is
the Corollary~\ref{coro:3} lower bound using the ground-truth analog
\(\pi_{\mathrm{GT}}\).}
\label{tab:magma-delta}
\end{table}

\paragraph{Unique-bug counts and error bars.} 
Table~\ref{tab:magma-detection} reports the two raw detection metrics
the FuzzBench evaluation also considers --- unique bug count and bug
count (\(=\)\,total trigger events) --- with error bars. Trial-to-trial
spread is small for AFL/AFLFast/MOpt across both targets; \textsc{Honggfuzz}
on \texttt{poppler} is the only fuzzer/target with a large bug-count
\(\sigma\) (one trial saw a long bug-free stretch).

\begin{table}[!htbp]
\centering
\setlength{\tabcolsep}{4pt}
\renewcommand{\arraystretch}{0.92}
\footnotesize
\begin{tabular}{ll cc}
\toprule
Target & Fuzzer & Unique bugs & Bug count (trigger events) \\
\midrule
\texttt{openssl} & \textsc{AFL}       & \(1.00\pm0.00\) & \(4683.4\pm254.2\) \\
                 & \textsc{AFLFast}   & \(1.00\pm0.00\) & \(5875.5\pm1408.9\) \\
                 & \textsc{Honggfuzz} & \(\mathbf{1.40\pm0.49}\) & \(1066.9\pm193.9\) \\
                 & \textsc{MOpt}      & \(1.00\pm0.00\) & \(\mathbf{20298.0\pm2591.4}\) \\
\midrule
\texttt{poppler} & \textsc{AFL}       & \(1.50\pm0.50\) & \(130.1\pm6.5\) \\
                 & \textsc{AFLFast}   & \(2.00\pm0.00\) & \(162.1\pm3.2\) \\
                 & \textsc{Honggfuzz} & \(2.20\pm0.40\) & \(1198.8\pm399.2\) \\
                 & \textsc{MOpt}      & \(\mathbf{3.50\pm0.81}\) & \(\mathbf{2482.5\pm175.3}\) \\
\bottomrule
\end{tabular}
\caption{Detection metrics on Magma: unique bug count and bug count
(trigger events), mean \(\pm\,1\,\sigma\) across \(M=10\) trials.
Largest in bold per target.}
\label{tab:magma-detection}
\end{table}

\begin{figure}[!htbp]
\centering
\includegraphics[width=0.99\linewidth]{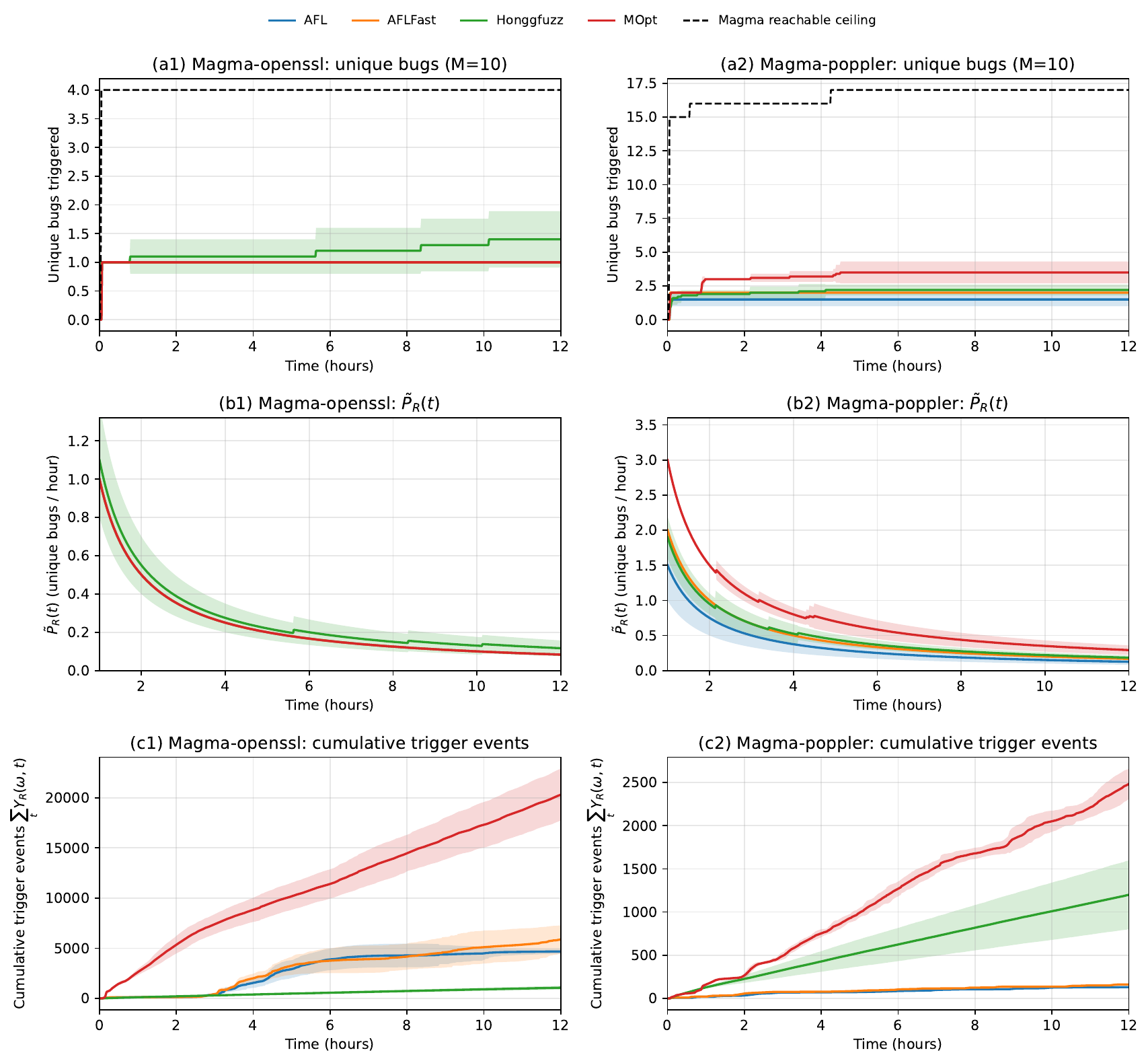}
\caption{Magma diagnostic figure (\(M=10\), \(T=12\)\,h). Top row: unique
bugs triggered, with the \emph{Magma reachable ceiling} (black dashed)
bounding all per-fuzzer trajectories from above. Middle row:
bug-detection rate \(\tilde{P}_R(t)\). Bottom row: cumulative trigger
events \(\sum_t Y_R(\omega,t)\), the raw signal underlying the
\texttt{pydtmc} estimator of \(\delta_R\). Shaded bands are mean
\(\pm\,1\sigma\) across the ten trials.}
\label{fig:magma-combined}
\end{figure}

Figure~\ref{fig:magma-combined} shows, for each fuzzer, the cumulative
unique bugs, the rate \(\tilde{P}_R(t)\), and the cumulative trigger
events, all with shaded \(\pm\,1\sigma\) bands across the \(M=10\)
trials. Trial dispersion is tight throughout, with the lone exception
of \textsc{Honggfuzz} on \texttt{poppler}, where one bug-free trial
inflates the cumulative-event spread.

\paragraph{Takeaway.} On a benchmark with ground truth: (i) the proxy
\(\tilde{P}\) recovers the exact \(\pi(1)\) ranking; (ii) the
\texttt{pydtmc} estimator \(\hat\delta_R\) and the ground-truth-based
Corollary~\ref{coro:3} lower bound \(\delta_R^{\mathrm{LB,GT}}\) agree
with each other and with \(D_R(T)\); (iii) ten-trial replication
produces tight error bars on every reported quantity. This jointly
substantiates Theorem~\ref{thm:unknown_delta_U},
Corollary~\ref{coro:1}, and Corollary~\ref{coro:3} on Magma.

\subsection{Numerical Experiments on Non-fuzzing Domains}\label{app:non-fuzzing}

We instantiate the framework on four feedback-driven black-box processes outside fuzzing, demonstrating that the same Markov-chain analysis, performance estimator $\tilde{P}_R(T)$, and evaluation metric $D_R(T)=|\tilde{P}_R(T)-P_R|$ apply more broadly. The four domains are: \textbf{(i)} random walks on graphs, \textbf{(ii)} image-classifier adversarial input search, \textbf{(iii)} vision--language model adversarial input search, and \textbf{(iv)} mutation-based prompt search on small instruction-tuned language models. In every domain the policy $R$ is a transition kernel over an input state space, the system $g$ produces binary feedback $Y_R(\omega,t)\in\{0,1\}$, and the estimator $\tilde{P}_R(T)=\nicefrac{1}{MT}\sum_{m,t}Y_R^m(t)$ is computed as in Algorithm~\ref{alg:mc}. We report each $D_R(T)$ in the per-domain tables with its $\pm 1$ standard error of the mean across the $M$ replications, $\mathrm{SE}=\sigma_{P_R(T)}/\sqrt{M}$.

\textbf{Time-homogeneity.} Every reported policy is time-homogeneous \emph{by construction}: the mutation rule is a fixed function of the current state with no dependence on the time index $t$ or trajectory history. We additionally provide an empirical confirmation by partitioning each chain trajectory into halves $[1,T/2]$ and $[T/2+1,T]$ and comparing the empirical transition statistics on the discrete signature; in every domain the paired two-sided $\chi^{2}$ test does not reject the null at the $95\%$ level for any policy, supporting the assumption.

\textbf{Spectral parameter estimation.} The relaxation time $\delta_R=1/\gamma_R=1/(1-\lambda_{R,2})$ of each chain is estimated via the \texttt{pydtmc} package~\cite{belluzzo_pydtmc} per Section~\ref{4.2.2}: the package returns the spectral gap $\gamma_R$, and we report $\delta_R$ by inversion. The estimator is applied directly to the discrete chain in domain (i), and to a small discrete signature of the chain in domains (ii)--(iv) (sign-pattern of the perturbation in (ii) and (iii); model output class in (iv)).

\textbf{Performance reference $P_R$.} For ergodic chains $P_R=\pi(1)$. In domain (i), $\pi(1)=|T|/|\mathcal{X}|$ is known by construction. In domains (ii)--(iv), the policies are symmetric reversible random walks with the \emph{uniform} stationary distribution on a bounded state space, so $\pi(1)$ is the probability of positive feedback under uniform sampling and is directly estimated by independent uniform random sampling (no overlap with chain trajectories): we draw $N_\pi$ independent random states, evaluate $g$, and report $\hat{\pi}(1)$ with binomial standard error $\mathrm{SE}(\hat{\pi})=\sqrt{\hat{\pi}(1-\hat{\pi})/N_\pi}$. Across domains (ii)--(iv), $\mathrm{SE}(\hat{\pi})$ is between $3\times 10^{-6}$ and $4\times 10^{-4}$, several orders of magnitude smaller than the inter-policy gaps in $D_R(T)$, so $\hat{\pi}$ is operationally equivalent to a known $P_R$ in the sense of Section~\ref{4.2.1}.

\textbf{What the theorems predict.} With $P_R$ known (analytically in (i), empirically via independent sampling in (ii)--(iv)) and $\delta_R$ estimated via \texttt{pydtmc}, Theorem~\ref{thm7} states that the policy with the smallest $\delta_R$ also achieves the smallest $D_R(T)$ with probability at least $1-\nicefrac{1}{T^2}$. The per-domain tables in Sections~\ref{app:non-fuzzing-mc}--\ref{app:non-fuzzing-llm} give the underlying numbers; Figure~\ref{fig:non-fuzzing-combined} at the end of the section provides a one-shot visual summary across all four domains.

\subsubsection{Random walks on graphs}\label{app:non-fuzzing-mc}
\textbf{Setup.} The state space $\mathcal{X}=\{0,1,\dots,N-1\}$ is the vertex set of a finite graph; the system reports membership in a target subset $T\subseteq\mathcal{X}$, $Y_R(\omega,t)=\mathds{1}\{X_R(\omega,t)\in T\}$, and the stationary mass $\pi(1)=|T|/|\mathcal{X}|$ is known by construction. We consider seven graph structures (chain, cluster, sparse, bimodal, hierarchical, mixed) and evaluate four time-homogeneous policies that share the uniform stationary distribution: \textsc{RandomJump} (uniform restart), \textsc{LocalWalk} (single-step neighbor move), \textsc{StickyWalk} (neighbor move with self-loop probability $0.5$), \textsc{LevyWalk} (heavy-tailed step length).

\textbf{Results.} We run $T=50{,}000$ steps with $M=200$ replications. Table~\ref{tab:non-fuzzing-mc} reports $D_R(T)=|\tilde{P}_R(T)-\pi(1)|$ and $\delta_R$. The smallest $\delta_R$ is achieved by \textsc{RandomJump} on every graph, and the smallest $D_R(T)$ is also achieved by \textsc{RandomJump} on every graph, jointly verifying Theorem~\ref{thm7}.

\begin{table}[H]
\centering
\setlength{\tabcolsep}{2pt}
\footnotesize
\caption{Random walks on graphs. $T=50{,}000$, $M=200$. $\pi(1)$ known by construction. $D_R(T)=|\tilde{P}_R(T)-\pi(1)|$ reported with $\pm 1$ SE of the mean. $\delta_R=1/\gamma_R$ (\texttt{pydtmc}). Smallest entry per row in bold.}
\label{tab:non-fuzzing-mc}
\begin{tabular}{llcccc}
\toprule
Graph ($\pi$) & Metric & \textsc{RandomJump} & \textsc{LocalWalk} & \textsc{StickyWalk} & \textsc{LevyWalk} \\
\midrule
chain\_200 ($0.1$) & $D_R(T)$ & $\mathbf{1.16{\scriptstyle\pm0.09}\!\times\!10^{-4}}$ & $1.60{\scriptstyle\pm2.6}\!\times\!10^{-3}$ & $3.18{\scriptstyle\pm2.5}\!\times\!10^{-3}$ & $1.23{\scriptstyle\pm13.5}\!\times\!10^{-4}$ \\
                   & $\delta_R$ & $\mathbf{1.05}$ & $999.4$ & $1243.9$ & $275.5$ \\
\midrule
chain\_1000 ($0.05$) & $D_R(T)$ & $\mathbf{1.35{\scriptstyle\pm0.07}\!\times\!10^{-4}}$ & $1.86{\scriptstyle\pm1.5}\!\times\!10^{-3}$ & $9.98{\scriptstyle\pm12.7}\!\times\!10^{-4}$ & $1.03{\scriptstyle\pm2.3}\!\times\!10^{-3}$ \\
                     & $\delta_R$ & $\mathbf{1.13}$ & $1255.8$ & $1082.8$ & $2777.8$ \\
\midrule
cluster\_500 ($0.2$) & $D_R(T)$ & $\mathbf{1.34{\scriptstyle\pm1.3}\!\times\!10^{-4}}$ & $3.88{\scriptstyle\pm4.5}\!\times\!10^{-3}$ & $3.54{\scriptstyle\pm4.1}\!\times\!10^{-3}$ & $6.78{\scriptstyle\pm4.0}\!\times\!10^{-3}$ \\
                     & $\delta_R$ & $\mathbf{1.09}$ & $1338.3$ & $1264.2$ & $977.0$ \\
\midrule
sparse\_500 ($0.05$) & $D_R(T)$ & $\mathbf{4.66{\scriptstyle\pm0.65}\!\times\!10^{-5}}$ & $2.46{\scriptstyle\pm1.4}\!\times\!10^{-3}$ & $8.04{\scriptstyle\pm13.0}\!\times\!10^{-4}$ & $8.47{\scriptstyle\pm14.0}\!\times\!10^{-4}$ \\
                     & $\delta_R$ & $\mathbf{1.09}$ & $1289.5$ & $1195.3$ & $976.6$ \\
\midrule
bimodal\_400 ($0.1$) & $D_R(T)$ & $\mathbf{3.91{\scriptstyle\pm9.1}\!\times\!10^{-5}}$ & $8.29{\scriptstyle\pm14.7}\!\times\!10^{-4}$ & $4.88{\scriptstyle\pm12.3}\!\times\!10^{-4}$ & $1.66{\scriptstyle\pm1.4}\!\times\!10^{-3}$ \\
                     & $\delta_R$ & $\mathbf{1.08}$ & $1473.3$ & $1209.6$ & $721.0$ \\
\midrule
hierarchical ($0.3333$) & $D_R(T)$ & $\mathbf{1.21{\scriptstyle\pm1.5}\!\times\!10^{-4}}$ & $5.95{\scriptstyle\pm6.9}\!\times\!10^{-3}$ & $4.22{\scriptstyle\pm5.8}\!\times\!10^{-3}$ & $4.92{\scriptstyle\pm6.1}\!\times\!10^{-3}$ \\
                        & $\delta_R$ & $\mathbf{1.10}$ & $1471.5$ & $1181.3$ & $1059.8$ \\
\midrule
mixed ($0.1$) & $D_R(T)$ & $\mathbf{2.51{\scriptstyle\pm0.97}\!\times\!10^{-4}}$ & $1.35{\scriptstyle\pm1.5}\!\times\!10^{-3}$ & $2.24{\scriptstyle\pm1.3}\!\times\!10^{-3}$ & $8.41{\scriptstyle\pm15.4}\!\times\!10^{-4}$ \\
              & $\delta_R$ & $\mathbf{1.09}$ & $1451.2$ & $1072.6$ & $879.2$ \\
\bottomrule
\end{tabular}
\end{table}

\subsubsection{Image-classifier adversarial input search}\label{app:non-fuzzing-adv}
\textbf{Setup.} For a pretrained CIFAR-10 classifier $g$ and a correctly-classified image $\mathbf{x}$, the perturbation $\boldsymbol{\delta}_t$ in the $L_\infty$ ball $B_\infty(\epsilon)$ with $\epsilon=16/255$ is the chain state, and $Y_R(\omega,t)=\mathds{1}\{g(\mathbf{x}+\boldsymbol{\delta}_t)\neq y\}$. Four time-homogeneous symmetric reversible policies share the uniform stationary on $B_\infty(\epsilon)$: \textsc{SmallMut} ($\sigma=0.02\epsilon$), \textsc{MedMut} ($\sigma=0.1\epsilon$), \textsc{LargeMut} ($\sigma=0.4\epsilon$), and \textsc{Random} (uniform sample on $B_\infty(\epsilon)$). $\delta_R$ is estimated on a discrete sign-pattern signature of the perturbation. $\hat{\pi}(1)$ is estimated by drawing $N_\pi=2{,}000$ independent uniform perturbations per image and averaging $\mathds{1}\{g(\mathbf{x}+\boldsymbol{\delta})\neq y\}$ across images.

\textbf{Results.} We evaluate \textsc{wrn28\_10}, \textsc{resnet50}, \textsc{densenet169}, \textsc{vgg16bn} at $T=5000$, $M=10$, $n=600$ images per model. Table~\ref{tab:non-fuzzing-adv} reports $D_R(T)=|\tilde{P}_R(T)-\hat{\pi}(1)|$ and $\delta_R$. The smallest $\delta_R$ and the smallest $D_R(T)$ are both achieved by \textsc{Random} on every model, verifying Theorem~\ref{thm7}.

\begin{table}[H]
\centering
\setlength{\tabcolsep}{3pt}
\footnotesize
\caption{Image-classifier adversarial input search. $T=5000$, $M=10$, $n=600$ images per model, $\epsilon=16/255$. $\hat{\pi}(1)$ from $N_\pi=2{,}000$ uniform perturbations per image. $D_R(T)$ reported with $\pm 1$ SE. $\delta_R$ via \texttt{pydtmc}. Smallest entry per row in bold.}
\label{tab:non-fuzzing-adv}
\begin{tabular}{llcccc}
\toprule
Model ($\pi$) & Metric & \textsc{SmallMut} & \textsc{MedMut} & \textsc{LargeMut} & \textsc{Random} \\
\midrule
wrn28\_10 ($0.159$) & $D_R(T)$ & $0.142{\scriptstyle\pm0.108}$ & $0.102{\scriptstyle\pm0.104}$ & $0.178{\scriptstyle\pm0.109}$ & $\mathbf{0.070{\scriptstyle\pm0.101}}$ \\
                     & $\delta_R$ & $45.45$ & $10.04$ & $3.14$ & $\mathbf{1.01}$ \\
\midrule
resnet50 ($0.176$) & $D_R(T)$ & $0.175{\scriptstyle\pm0.114}$ & $0.140{\scriptstyle\pm0.111}$ & $0.232{\scriptstyle\pm0.116}$ & $\mathbf{0.100{\scriptstyle\pm0.108}}$ \\
                   & $\delta_R$ & $56.29$ & $10.27$ & $3.13$ & $\mathbf{1.00}$ \\
\midrule
densenet169 ($0.174$) & $D_R(T)$ & $0.186{\scriptstyle\pm0.117}$ & $0.156{\scriptstyle\pm0.116}$ & $0.238{\scriptstyle\pm0.120}$ & $\mathbf{0.122{\scriptstyle\pm0.113}}$ \\
                       & $\delta_R$ & $51.29$ & $10.14$ & $3.15$ & $\mathbf{1.00}$ \\
\midrule
vgg16bn ($0.144$) & $D_R(T)$ & $0.141{\scriptstyle\pm0.101}$ & $0.100{\scriptstyle\pm0.097}$ & $0.191{\scriptstyle\pm0.105}$ & $\mathbf{0.064{\scriptstyle\pm0.093}}$ \\
                   & $\delta_R$ & $52.99$ & $10.14$ & $3.15$ & $\mathbf{1.01}$ \\
\bottomrule
\end{tabular}
\end{table}

\subsubsection{Vision--language model adversarial input search}\label{app:non-fuzzing-vlm}
\textbf{Setup.} The state, the policies (\textsc{SmallMut}, \textsc{MedMut}, \textsc{Random}), the discrete signature, and the stationary distribution are as in Section~\ref{app:non-fuzzing-adv}; only the system $g$ changes from a CIFAR-10 classifier to a CLIP zero-shot classifier scoring image--text similarity against ten templated prompts of the form ``a photo of a \{class\}''. $\hat{\pi}(1)$ is estimated as in (ii) with $N_\pi=2{,}000$ independent uniform perturbations per image.

\textbf{Results.} We evaluate \textsc{CLIP-ViT-B32} at $T=3000$, $M=5$, $n=300$ images, at two perturbation budgets $\epsilon\in\{8/255,\,16/255\}$. Table~\ref{tab:non-fuzzing-vlm} reports $D_R(T)$ and $\delta_R$. On every cell the smallest $\delta_R$ and the smallest $D_R(T)$ are both achieved by \textsc{Random}, verifying Theorem~\ref{thm7}.

\begin{table}[H]
\centering
\setlength{\tabcolsep}{3pt}
\footnotesize
\caption{Vision--language model (CLIP) adversarial input search. $T=3000$, $M=5$, $n=300$ images. $\hat{\pi}(1)$ from $N_\pi=2{,}000$ uniform perturbations per image. $D_R(T)$ reported with $\pm 1$ SE. $\delta_R$ via \texttt{pydtmc}. Smallest entry per row in bold.}
\label{tab:non-fuzzing-vlm}
\begin{tabular}{llccc}
\toprule
Cell ($\pi$) & Metric & \textsc{SmallMut} & \textsc{MedMut} & \textsc{Random} \\
\midrule
CLIP-ViT-B32, $\epsilon=8/255$ ($0.067$) & $D_R(T)$ & $4.45{\scriptstyle\pm9.5}\!\times\!10^{-3}$ & $1.20{\scriptstyle\pm9.6}\!\times\!10^{-3}$ & $\mathbf{1.08{\scriptstyle\pm9.6}\!\times\!10^{-4}}$ \\
                                         & $\delta_R$ & $39.08$ & $19.71$ & $\mathbf{1.01}$ \\
\midrule
CLIP-ViT-B32, $\epsilon=16/255$ ($0.094$) & $D_R(T)$ & $5.04{\scriptstyle\pm10.7}\!\times\!10^{-3}$ & $1.29{\scriptstyle\pm10.6}\!\times\!10^{-3}$ & $\mathbf{1.24{\scriptstyle\pm10.6}\!\times\!10^{-5}}$ \\
                                          & $\delta_R$ & $39.08$ & $19.71$ & $\mathbf{1.01}$ \\
\bottomrule
\end{tabular}
\end{table}

\subsubsection{Mutation-based prompt search on small instruction-tuned language models}\label{app:non-fuzzing-llm}
\textbf{Setup.} A mutable suffix of $L=16$ tokens is appended to a fixed instruction prefix and evolves as a Markov chain on $\mathcal{V}^{L}$, where $\mathcal{V}$ is the model vocabulary. The system $g$ is \textsc{SmolLM-360M-Instruct}; the binary feedback is $Y_R(\omega,t)=\mathds{1}\{\text{first response token to (prefix$+$suffix$_t$)} = \text{ref token}\}$, where the ref token is the model's first response to a fixed initial suffix. Four time-homogeneous symmetric token-level mutation policies are evaluated: \textsc{RandomFull} (uniform redraw of all $L$ positions), \textsc{SwapOne}, \textsc{SwapTwo}, \textsc{SmallMix} ($70\%$ \textsc{SwapOne}, $30\%$ \textsc{SwapTwo}). All four share the uniform stationary on $\mathcal{V}^{L}$, so $\pi(1)$ is the probability of first-token match under uniform suffix sampling. $\hat{\pi}(1)$ is estimated by drawing $N_\pi=10^{7}$ independent uniform suffixes per prompt (no chain overlap) and averaging the binary indicator. $\delta_R=1/\gamma_R$ is estimated via \texttt{pydtmc} on the projected output-class trajectory.

\textbf{Results.} We run $T=4000$, $M=10$, on four instruction prompts. Table~\ref{tab:non-fuzzing-llm} reports $D_R(T)=|\tilde{P}_R(T)-\hat{\pi}(1)|$ and $\delta_R$. The smallest $\delta_R$ and the smallest $D_R(T)$ are both achieved by \textsc{RandomFull} on every prompt, verifying Theorem~\ref{thm7}.

\begin{table}[H]
\centering
\setlength{\tabcolsep}{3pt}
\footnotesize
\caption{Mutation-based prompt search on \textsc{SmolLM-360M-Instruct}. $T=4000$, $M=10$, $L=16$. $\hat{\pi}(1)$ from $N_\pi=10^{7}$ uniform suffixes per prompt. $D_R(T)$ reported with $\pm 1$ SE. $\delta_R$ via \texttt{pydtmc}. Smallest entry per row in bold.}
\label{tab:non-fuzzing-llm}
\begin{tabular}{llcccc}
\toprule
Prompt ($\pi$) & Metric & \textsc{RandomFull} & \textsc{SwapOne} & \textsc{SwapTwo} & \textsc{SmallMix} \\
\midrule
P1 ($8.89\!\times\!10^{-5}$) & $D_R(T)$ & $\mathbf{2.36{\scriptstyle\pm0.46}\!\times\!10^{-4}}$ & $9.36{\scriptstyle\pm3.7}\!\times\!10^{-4}$ & $3.86{\scriptstyle\pm0.77}\!\times\!10^{-4}$ & $6.19{\scriptstyle\pm2.2}\!\times\!10^{-4}$ \\
                             & $\delta_R$ & $\mathbf{1.01}$ & $2.28$ & $2.00$ & $1.71$ \\
\midrule
P2 ($3.40\!\times\!10^{-4}$) & $D_R(T)$ & $\mathbf{1.52{\scriptstyle\pm0.83}\!\times\!10^{-4}}$ & $9.18{\scriptstyle\pm2.8}\!\times\!10^{-4}$ & $6.60{\scriptstyle\pm3.2}\!\times\!10^{-4}$ & $1.21{\scriptstyle\pm1.2}\!\times\!10^{-3}$ \\
                             & $\delta_R$ & $\mathbf{1.01}$ & $2.83$ & $2.36$ & $2.53$ \\
\midrule
P3 ($3.26\!\times\!10^{-4}$) & $D_R(T)$ & $\mathbf{2.24{\scriptstyle\pm0.92}\!\times\!10^{-4}}$ & $8.24{\scriptstyle\pm2.6}\!\times\!10^{-4}$ & $4.49{\scriptstyle\pm2.0}\!\times\!10^{-4}$ & $4.24{\scriptstyle\pm1.7}\!\times\!10^{-4}$ \\
                             & $\delta_R$ & $\mathbf{1.02}$ & $2.24$ & $2.06$ & $2.01$ \\
\midrule
P4 ($3.32\!\times\!10^{-4}$) & $D_R(T)$ & $\mathbf{1.93{\scriptstyle\pm0.75}\!\times\!10^{-4}}$ & $4.93{\scriptstyle\pm1.8}\!\times\!10^{-4}$ & $5.68{\scriptstyle\pm3.5}\!\times\!10^{-4}$ & $2.43{\scriptstyle\pm1.3}\!\times\!10^{-4}$ \\
                             & $\delta_R$ & $\mathbf{1.02}$ & $1.95$ & $2.52$ & $2.00$ \\
\bottomrule
\end{tabular}
\end{table}

\begin{figure}[H]
\centering
\includegraphics[width=0.95\linewidth]{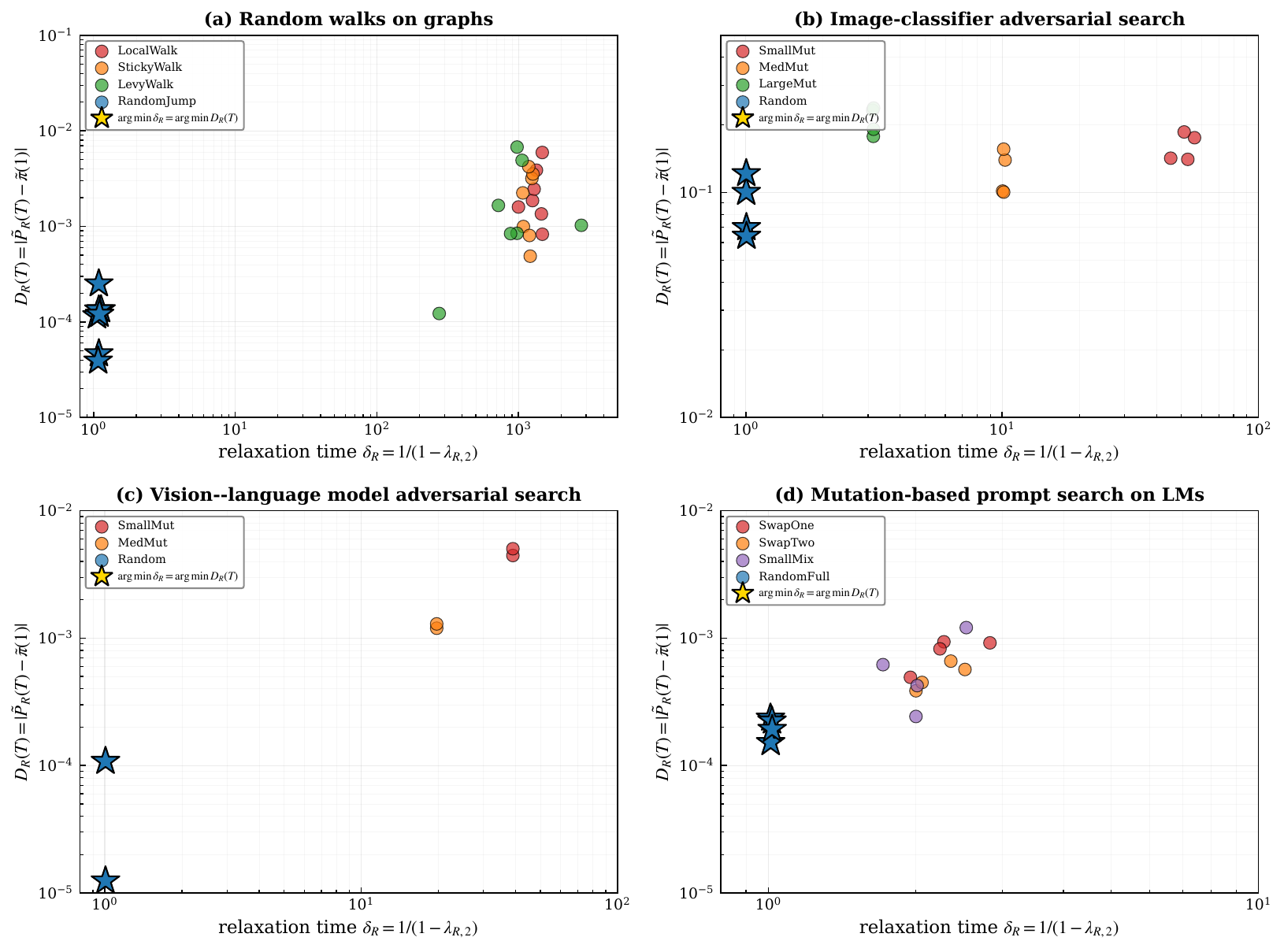}
\caption{Joint visual demonstration of Theorem~\ref{thm7} across all four non-fuzzing domains. Each panel plots the estimated relaxation time $\delta_R=1/(1-\lambda_{R,2})$ on the horizontal axis (log) against $D_R(T)$ on the vertical axis (log) for every (policy, benchmark) pair. Each star marks the within-cell policy that simultaneously achieves the smallest $\delta_R$ and the smallest $D_R(T)$. On every cell the one-step-mixing policy (\textsc{RandomJump}, \textsc{Random}, \textsc{RandomFull}) is the argmin in both quantities. Numerical $\pm 1$ standard errors of the mean are reported in the per-domain tables (Sections~\ref{app:non-fuzzing-mc}--\ref{app:non-fuzzing-llm}).}
\label{fig:non-fuzzing-combined}
\end{figure}

\paragraph{Summary.}
Across all four non-fuzzing domains and every benchmark, the policy minimizing $\delta_R$ coincides with the policy minimizing $D_R(T)$, verifying Theorem~\ref{thm7}. In every domain the one-step-mixing policy ($\delta_R\!\approx\!1$) jointly achieves the smallest $\delta_R$ and the smallest $D_R(T)$, while local-mutation policies have $\delta_R$ orders of magnitude larger and correspondingly larger $D_R(T)$. Time-homogeneity is satisfied by construction (mutation rules are $t$-independent) and confirmed by the half-trajectory $\chi^{2}$ test described above. Including a one-step uniform sampler in each policy set is a deliberate calibration choice. Since this sampler has relaxation time $\delta_R \approx 1$, it provides a fastest-mixing spectral anchor within each domain. The non-fuzzing experiments therefore test the implication of Theorem~\ref{thm7} that, under the same target event and horizon, local-mutation policies with much larger $\delta_R$ should not achieve smaller finite-time deviation $D_R(T)$ through convergence alone. This comparison isolates the spectral-convergence effect; it does not claim that uniform sampling is generally the best optimization strategy. The joint $\delta_R$-vs-$D_R(T)$ pattern across all four domains is summarized in Figure~\ref{fig:non-fuzzing-combined}.

\subsection{Related Work}\label{app:related_work}

{\bf Black-box processes.} Black-box processes are systems whose internal dynamics are inaccessible, so analysis and decision-making rely solely on observed input–output behavior \citep{guidotti2018survey, sjoberg1995nonlinear}. They arise in settings such as black-box optimization \citep{krishnamoorthy2023diffusion, li2021openbox} and  program testing, and are often studied in static or non-adaptive regimes where inputs are fixed in advance. A more structured subclass comprises \emph{feedback-driven black-box processes} \citep{leithner2021hydra, luo2016forepost}, in which inputs are selected sequentially based on past observations. This feedback-driven structure underlies evolutionary and genetic algorithms \citep{auger2005restart}, parameter tuning systems  in machine learning \citep{ziegler2016synthesis} and mutation-based fuzzing \citep{xia2024fuzz4all, you2019profuzzer, klees2018evaluating}, where policies interact with an unknown system over time. Prior work has largely emphasized algorithm design and empirical performance, with evaluation relying on problem-specific or surrogate metrics. In contrast, principled performance objectives and evaluation methodologies for feedback-driven black-box processes remain relatively underexplored. Our work addresses this gap by developing a unified stochastic modeling and evaluation framework, with theoretical analysis and empirical validation carried out in the context of mutation-based fuzzing. 


{\bf Fuzzing.} The key component in fuzzing is test case (input) generation, as the inputs are fed into a program, which directly determines whether they trigger bugs or vulnerabilities. Historically, there have been two main approaches to input generation: grammar-based methods and mutation-based methods combined with initial input selection. Grammar-based fuzzing~\cite{godefroid2008grammar, wang2019superion, jero2019leveraging} generates inputs based on grammar constraints unique to the target program. However, it becomes ineffective when the program's grammar is not well understood upfront, thus limiting its practical application. In contrast, mutation-based fuzzing mutates existing inputs to generate new ones, which is guided by the program's feedback, such as bug signals~\cite{moukahal2021vulnerability}, total execution time~\cite{petsios2017slowfuzz}, differential signals~\cite{petsios2017nezha, guo2018dlfuzz}, or code coverage~\cite{nagy2019full, lemieux2018fairfuzz,she2024fox}. Notably, the emergence of Large Language Models (LLMs) has inspired recent work on LLM-guided mutation, which designs prompts for an LLM and uses its responses as mutation rules~\cite{meng2024large, xia2024fuzz4all}. Mutation-based fuzzing naturally remains the most widely used technique, and thus, we focus on mutation-based fuzzing throughout, highlighting the broad applicability of our framework. While existing work proposes empirical mutation rules aimed at practical success, these approaches may not rigorously capture the underlying dynamics of the fuzzing process. In contrast, we focus on reliably modeling fuzzing and evaluating existing fuzzers, both of which are crucial for improving their effectiveness and guiding scientific design.

{\bf Fuzzing evaluation.} Recently, fuzzing evaluation has been drawing increasing attention from the community as more fuzzers are developed. Without a proper evaluation framework, it becomes infeasible to assess different fuzzers for deployment or to inform the development of genuinely improved ones. In~\cite{klees2018evaluating}, the authors implement a wide range of existing fuzzers and evaluate their performance using heuristic metrics (e.g.\ code coverage), showing that these metrics may not always accurately reflect the fuzzers’ true capabilities and are sensitive to the inherent randomness of fuzzing. For example, high code coverage does not imply that more bugs are detected. Consequently, they emphasize using statistical testing through multiple trials, which, however, are only empirical guidelines. Our evaluations, grounded in both theory and practice, validate some of these guidelines while also offering new insights for practitioners. For example, increasing the number of mutation rounds can substantially reduce fluctuations caused by randomness, although conducting multiple trials remains a viable option. Alternatively,~\cite{li2021unifuzz} propose multiple empirical metrics, ranging from the number of unique bugs to the quality of bugs. Notably, counting total bugs, rather than just unique ones, is more informative, as distinct inputs triggering the same bug may inform distinct bug patterns. Yet,~\cite{hazimeh2020magma} highlights the empirical limitations of counting total bugs, such as high stochasticity and fluctuations, which prevent them from being a reliable evaluation metric. Additionally, these works treat performance metrics, such as bug counts and code coverage, as equivalent to evaluation metrics used to compare fuzzers, which can be problematic. 
In contrast, we propose both performance and evaluation metrics that are directly tied to total bug detection, account for stochasticity, and are grounded in theory, enabling reliable fuzzer evaluation.  

Moreover, it remains unclear which metrics practitioners should prioritize; we note a growing body of work on empirical evaluation and refer readers to~\cite{ren2021empirical, jiang2023evaluating, wu2022evaluating, elder2022really}. Most recently,~\cite{schloegel2024sok} adhere to the guidelines proposed by~\cite{klees2018evaluating} and conduct an empirical examination of numerous fuzzers. They also report that fuzzing evaluation still lacks rigor, an open problem they highlight, yet propose new empirical guidelines. While existing work empirically observes that more mutation rounds improve performance, running the fuzzing process for 24 hours as a standard evaluation remains a heuristic approach and lacks rigor. 
Our work addresses these open problems by introducing a rigorous evaluation framework that reliably guides the choice of metrics and fuzzing duration, while also informing practical evaluation and fuzzer design.

{\bf Stochastic modeling and simulation. }
Given the power of stochastic modeling, several studies have explored its use in developing new fuzzers. For example,~\cite{bohme2016coverage} model the fuzzing process as a Markov chain over all possible code coverages (i.e., indirect program outputs), later extending this to address power scheduling~\cite{bohme2017directed}. However, in general fuzzing processes, these output transitions are not always Markovian. Moreover, modeling indirect outputs alone may not capture the fundamental dynamics of how inputs transition from one to another in the input space which interact with fuzzers, nor does it account for dynamics in direct outputs such as bugs. It may also have limited ability to distinguish fuzzers whose outputs appear similar despite differences in their input mutations. In contrast, we model the input and bug-related output spaces directly as state spaces, with mutations as transitions. Additionally, existing work has not yet explored a theoretical understanding of general fuzzing processes and evaluation. Our work addresses this by integrating and analyzing the new modeling into a theoretical evaluation framework. More recently,~\cite{yue2020ecofuzz, wu2022one} combined Markov chains with multi-armed bandits (MAB)~\cite{auer2002nonstochastic}, but their approach inherits the same limitations from~\cite{bohme2016coverage}. 
In contrast, we begin by constructing a rigorous and reliable stochastic model. Simulation based on stochastic modeling has also been explored for input generation through techniques such as Sequential Monte Carlo~\cite{vaisnor2024approximate} and Markov Chain Monte Carlo~\cite{bohme2017directed}. However, these efforts focus on designing and empirically evaluating new fuzzers. In contrast, our work proposes a simulation framework for modeling fuzzing processes and theoretically evaluating any mutation-based fuzzer.


\subsection{Algorithms}\label{app:alg}

\begin{algorithm}[h]
\caption{Monte Carlo Simulation Algorithm}\label{alg:mc}
\begin{algorithmic}[1]
\REQUIRE The number of trials \(M\); the length of a trial \(T\); the input seed string \(X_0^R\); the fuzzer \(R\); the number of bugs \(N_T\); the sets of executed and buggy inputs, \(I_R^T\) and \(H_R^T\), respectively.
\FOR{$m = 1$ to $M$}
\STATE \(N_T(m) = 0\), \(H_R^{T}(m), I_R^{T}(m)= \emptyset\)
\FOR{$t = 1$ to $T$}
\STATE Mutate \(s_t = X_t^{R}\) with fuzzer \(R\) and generate \(X_{t+1}^{R} = \mathcal{R}(X_t^{R})\) or $\mathcal{R}(X_t^{R}, G_t)$
\STATE Add \(X_{t+1}^{R}\) to \(I_R^T(m)\)
\STATE The program executes \(X_{t+1}^{R}\) and generates the program behavior $g_f(X_{t+1}^{R})$
\STATE Observe \(Y_t^{R}(m) = 1_{g_f(X_{t+1}^{R}) \in B_f}\)
\IF{$Y_t^{R}(m) = 1$}
\STATE Add \(X_{t+1}^{R}\) to \(H_R^T(m)\)
\STATE Update \(N_T(m) = N_T(m)+1\)
\ENDIF
\STATE \(s_t = X_{t+1}^{R}\)
\ENDFOR
\ENDFOR
\ENSURE \({\bigl(H_R^{T}(m)\bigr)}_{m=1}^{M}\), \({\bigl(I_R^{T}(m)\bigr)}_{m=1}^{M}\), \({\bigl(N_T(m)\bigr)}_{m=1}^{M}\), \({\bigl(Y_{t}^R(m)\bigr)}_{m,t}\)
\end{algorithmic}
\end{algorithm}

\begin{algorithm}[h]
\caption{Data-driven estimation \& evaluation}\label{alg:data-driven}
\begin{algorithmic}[1]
\REQUIRE The number of trials \(M\); the length of a trial \(t_1\); the input seed string \(X_0^R\); the fuzzer \(R\)
\FOR{$m = 1$ to $M$}
\STATE \(N_{t_1}(m) = 0\), \(H_R^{t_1}(m), I_R^{t_1}(m)= \emptyset\)
\FOR{$t = 1$ to $t_1$}
\STATE Run Algorithm 1 in \cite{wolfer2024improved}
\ENDFOR
\STATE Collect the history \({\bigl(X_t^{R}\bigr)}_{t=1}^{t_1}\), \({\bigl(Y_t^{R}\bigr)}_{t=1}^{t_1}\)
\STATE Estimator construction of $\delta_{dps}(R)$ via the point estimator in Section 3.2 \cite{hsu2019mixing} or the estimator in Section 3.2 \cite{wolfer2024improved}
\STATE Fuzzer evaluation by \(\hat{U}_R =  \sqrt{\nicefrac{\log{T}}{T}} + \sqrt{\nicefrac{4\delta_{dps}(R)}{T}}\)
\ENDFOR
\ENSURE \(\hat{U}_R\) and \(\delta_{dps}(R)\)
\end{algorithmic}
\end{algorithm}

\subsection{Additional Theoretical Results}\label{app:theory}

\subsubsection{Asymptotic Property}
We begin with the asymptotic case where \(T \to \infty\).
\begin{theorem}\label{thm: Asymptotic Results}
Assume that transition matrix \(P_R^Z\) is ergodic. We have \(\lim_{T \to \infty}\tilde{P}_{R}(T)  = P_R = \pi^{R}(1)\). 

\end{theorem}

\begin{lemma}[Asymptotic bias under rare-event scaling]\label{lemma:0}
Let \(D\) be a finite input space with
\(|D|=N^{\ell_{\max}}\), and let \(B\subseteq D\)
denote the set of bug-triggering inputs. Suppose the software-level reference
distribution \(\pi\) is uniform over \(D\), so that
\(\pi(1)=\pi(B)=|B|/N^{\ell_{\max}}\) and $\pi(x) > 0$ for all $x \in D$. For a fuzzer \(R\),
let \(\pi^R\) denote the stationary distribution of the fuzzer-induced Markov
chain and define \(P_R=\pi^R(B)\). Assume that the evaluation horizon
\(T\) satisfies the rare-event scaling
\(|B|T=o(N^{\ell_{\max}})\). Further assume that the fuzzer-induced
stationary distribution has a bounded density ratio with respect to \(\pi\):
there exists a constant \(\kappa_R<\infty\), independent of \(N^{\ell_{\max}}\),
such that \(\pi^R(x)/\pi(x)\leq \kappa_R\) for all \(x\in B\); this holds for most fuzzers. Then
\(|P_R-\pi(1)|=o(1/T)\).
\end{lemma}

The above asymptotic results show that our constructed estimators converge to \(P_R\) and to $\pi^{R}(1)$ and thus to the true bug detection rate of the program \(\pi(1)\) due to the scale of the bug rate, independent of the fuzzer. This implies that \(\lim_{T \to \infty} D_R(T) \approx 0\), validating the effectiveness of ergodic fuzzers as the sample path length \(T\) approaches infinity. Notably, this finding also aligns with the heuristic proposed in~\cite{klees2018evaluating}: \emph{all fuzzers eventually detect the same bugs and thus achieve equivalent performance when allowed to run indefinitely}. 

\begin{tcolorbox}[colback=custompurple!10, colframe=custompurple, title=Asymptotic Property]
\emph{As \(T\) goes to infinity, the bug detection ratio converges to the underlying true bug detection rate. }
\end{tcolorbox}


\begin{remark}[Discussion on Ergodicity] We assume that the fuzzers are ergodic, a reasonable condition for most well-designed fuzzers. Ergodicity implies a positive probability of transitioning between any two states, ensuring the chain does not get stuck or exhibit periodic revisits. Good fuzzers avoid such cycles by exploring the entire input space. If a fuzzer violates this assumption, it may become stuck and is generally undesirable in practice. Hence, assuming ergodicity naturally filters out less preferable fuzzers. This assumption is also common in Markov chain analyses, and is required only for theoretical analysis rather than the formulation.

\end{remark}

While the asymptotic results are informative in characterizing the representativeness of the performance and evaluation metrics, we further consider two key questions: (1) how to evaluate fuzzers for a finite \(T\), and (2) what minimal \(T\) is required to ensure a trustworthy evaluation (i.e., one that closely approximates the ground truth). The first question informs evaluation under a finite number of mutation rounds, while the second guides simulation design, both addressed via finite-time analysis.

\subsubsection{Finite-time Analysis}
In this section, we examine the statistical properties of the proposed estimators \(\tilde{P}_R(T)\) and \(D_R(T)\) for finite \(T\). Specifically, we establish the convergence rate of the Markov chain \(P_R^Z(x, \cdot)\) to its stationary distribution \(\pi^X\), and, consequently, the convergence behavior of the performance estimator \(\tilde{P}_R(T)\) and the evaluation metric \(D_R(T) = |\tilde{P}_R(T) - \pi(1)| \approx |\tilde{P}_R(T) - P_R| = |\tilde{P}_R(T) - \pi^R(1)|\) due to $|\pi(1) - \pi^R(1)| =  o(1) \approx 0$ for bug discovery. %

By characterizing the total variation distance between \(P_R^Z(x, \cdot)\) and \(\pi^Z\), a standard measure for comparing two distributions, we obtain the following result. 

\begin{theorem}\label{finite-time:matrix}
Let us assume that the Markov transition matrix is ergodic.  
Then we have that \(\max_{x}\lVert (P_{R}^{Z})^{T}(x,\cdot) - \pi^X \rVert_{TV} \leq C_R\alpha_R^{T}\) where \(C_R = \alpha^{-r}, \alpha_R = \theta^{\frac{1}{r}}\) are constants associated with the transition matrix \(P_{R}^Z\), with \(\theta = 1 - \min_{x_1, x_2}(\nicefrac{P_{R}^{Z})^{r}(x_1,x_2)}{\pi(x_2)}, r = \min_{t}(P_{R}^{Z})^{t}(x_1, x_2) > 0\) for any \(x_1, x_2\). Here, \((P_{R}^{Z})^{t}\) denotes \(P_{R}^Z\) to the power of \(t\).
\end{theorem}

\begin{tcolorbox}[colback=custompurple!10, colframe=custompurple, title=Input Distribution Convergence]
\emph{As \(T\) increases, the distribution of inputs generated by a fuzzer approaches the true/stationary input distribution in the input space. }
\end{tcolorbox}

Building on the finite-time convergence of \((P_{R}^{Z})^{T}\), we next establish the finite-time convergence of the estimator \(\tilde{P}_R(T)\).  

\begin{theorem}[Finite-time Convergence]\label{finite-time:estimator}
Let us assume that the Markov transition matrix is ergodic. 
Then we have that with probability at least \(1 - \nicefrac{1}{T^2}\), \(|\tilde{P}_R(T) - \pi(1)|  \leq \sqrt{\nicefrac{\log{T}}{T}} + \nicefrac{1}{T}\nicefrac{C_R}{(1-\alpha_R)}\).\end{theorem}
\begin{remark}[Convergence Rate]
Notably, the convergence rate \(\max\{\sqrt{\nicefrac{\log T}{T}}, \nicefrac{1}{T}\nicefrac{C_R}{(1 - \alpha_R})\}\) lies between sublinear and linear convergence. Different from the asymptotic case, this convergence rate is fuzzer-dependent, as reflected in the parameters \(C_R\) and \(\alpha_R\), whose key components, \(r\) and \(\theta\), depend on the transition matrix \(P_R^{Z}(\cdot, \cdot)\).
\end{remark}

\begin{tcolorbox}[colback=custompurple!10, colframe=custompurple, title=Estimator Convergence]
\emph{The estimator, the bug detection ratio, converges toward the objective, the bug detection rate, with the distance depending on \(T\) and the fuzzer-related parameters \(C_R\) and \(\alpha_R\).}
\end{tcolorbox}

The finite-time convergence depends on \(C_R\) and \(\alpha_R\) of \(P_R^Z(\cdot, \cdot)\) that can be used to evaluate different fuzzers but lack interpretability.  In other words, it is not straightforward to intuitively relate changes in a fuzzer’s transition matrix to its convergence rate and overall performance, let alone to guide fuzzer design. To this end, we next leverage the spectral properties of the transition matrix to establish the convergence rate, as these properties can be explicitly characterized and intuitively interpreted.

Spectral theory has been an active area of research, particularly for reversible Markov chains. A Markov chain is said to be reversible if for any \(i, j \in S\), we have \(\pi(i)P_R^Z(i,j) = \pi(j)P_R^Z(j,i)\). We first focus on the convergence of \(\tilde{P}_R(T)\) when the transition matrix is reversible.

We consider 
that \(\lambda_{R,j}\) is the \(j\)-th eigenvalue of \(P_R^Z(\cdot, \cdot)\), and \(E\) is the rank of \(P_R^Z(\cdot, \cdot)\). We define the spectral gap of \(P_R^Z\) as \(\gamma_R = 1 - \lambda_{R,2}\) and the corresponding relaxation time as \(\delta_R = 1 / \gamma_R\), both spectral properties of \(P_R^Z\). Based on \(\delta_R\), we establish the following finite-time convergence result for the transition matrix \((P_R^Z)^T\) and thus for the estimator \(\tilde{P}_R(T)\).


\begin{theorem}[Reversible MC (Poincare inequality)]\label{finite-time:matrix-spectral-rev}
Let us assume that the Markov transition matrix is reversible. Then we have that \(\lVert \pi_0^X (P_{R}^{Z})^T - \pi^X \rVert_{L_{2}} \leq {\left(1-\gamma_R\right)}^{T}\lVert \pi_0^X - \pi^X \rVert_{L_2}  \leq e^{-\frac{T}{\delta_R}}\lVert \pi_0^X - \pi^X \rVert_{L_2}\)
where \(\pi_0^X\) is the initial distribution of the input Markov chain.
\end{theorem}


\begin{theorem}[Reversible MC (Convergence)]\label{finite-time:estimator-spectral-reversible}
Let us assume that the Markov transition matrix is reversible and ergodic. Then we have that with probability at least \(1 - \nicefrac{1}{T^2}\), \(|\tilde{P}_R(T) - \pi(1)|  \leq \sqrt{\nicefrac{\log{T}}{T}} + {(1-\pi(1))} \nicefrac{1}{T}\nicefrac{1}{\gamma_R} = \sqrt{\nicefrac{\log{T}}{T}} + {(1-\pi(1))} \nicefrac{\delta_R}{T}\).
\end{theorem}


While the order of the convergence rate in \(T\) remains the same, these new results capture differences between fuzzers through \(\delta_R\), facilitating a better understanding of fuzzer performance. However, they apply only to reversible Markov chains.  
Consequently, we extend our analysis to the non-reversible setting and examine the convergence of the estimator. Recently, a generalized definition of the spectral gap for such Markov chains has been developed~\cite{chatterjee2023spectral}, which we adapt to establish the following convergence result. 

\begin{theorem}[Nonreversible MC (Convergence)]\label{finite-time:matrix-spectral-nonrev}
Let us define \(\delta_R = \nicefrac{1}{\lambda_{R}^{\prime}}\) where \(\lambda_{R}^{\prime}\) represents the second-largest eigenvalue of matrix \(I - P_R\) where \(I\) is the identical matrix with all elements in the diagonal being 1 and otherwise 0. Let us assume that \(P_R^Z\) is ergodic. Then we have  \(\left|\sum_{t=1}^{T} P_{R}^t(y,1) - \pi(1)\right| \leq \sqrt{\nicefrac{4\delta_{R}}{T}}\)
and
\(\left|\sum_{t=1}^{T} P_{R}^t(y,0) - \pi(0)\right| \leq \sqrt{\nicefrac{4\delta_{R}}{T}}\)
by specifying \(g(X_t) = 1_{X_t \in B}\) and \(g(X_t) = 1_{X_t \not \in B}\) in Theorem 1.2 in~\cite{chatterjee2023spectral}, respectively. Subsequently, we derive that w.p. \(1 - \nicefrac{1}{T^2}\), \(|\tilde{P}_R(T) - \pi(1)|  \leq \sqrt{\nicefrac{\log{T}}{T}} + \sqrt{\nicefrac{4\delta_{R}}{T}}\)
which again depends on the spectral parameter \(\delta_R\) of fuzzers.

\end{theorem}

\begin{tcolorbox}[colback=custompurple!10, colframe=custompurple, title= Estimator Convergence (Intepretable via Spectrum)]
\emph{The bug detection ratio converges toward the bug detection rate, with the deviation depending on \(T\) and the spectral parameter \(\delta_R\), which measures how slowly the mutations reach stationarity. }
\end{tcolorbox}

\begin{remark}[Comparison]
We note that the convergence rate in \(T\) here is of order \(\sqrt{\nicefrac{1}{T}}\), rather than \(\max\{\sqrt{\nicefrac{1}{T}}, \nicefrac{1}{T}\}\) as in both Theorem~\ref{finite-time:estimator} and Theorem~\ref{finite-time:estimator-spectral-reversible}. But it features an interpretable dependency on the spectral gap and does not require the reversibility assumption for the Markov chain.
\end{remark}

For both reversible and non-reversible Markov chains, the result depends directly on the transition matrix \(P_R^Z(\cdot, \cdot)\) through \(\delta_R\) and \(T\), making it both tractable and interpretable. When the fuzzing path length \(T\) satisfies \(T \leq e^{4\delta_R}\), the convergence rate is dominated by \(\delta_R\); hence, a smaller \(\delta_R\) (i.e., shorter relaxation time) implies faster convergence and, consequently, better fuzzers. However, when \(T > e^{4\delta_R}\), the convergence behavior across fuzzers becomes similar, as the dominant term \(\log T\) is independent of the fuzzer. This again aligns with intuition and~\cite{schloegel2024sok}: given sufficient time, all fuzzers should eventually detect all bugs.

Also, alternatively, we derive computable upper bounds on \(\delta_R\) and \(D_R(T)\) and use them to evaluate fuzzers.

\begin{theorem}[Evaluation of \(D_R(T)\)]\label{thm:unknown_delta} Assume \(T > O(\hat{\tau}^2(\epsilon)\ln(\nicefrac{\hat{\tau}(\epsilon)}{\beta})\ln{\ln{\nicefrac{\hat{\tau}(\epsilon)}{\beta}}})\) for \(0 < \beta < 1 \). The fuzzer with the smallest \(\hat{k}_{R, gen}\) is optimal w.p. \(1-\alpha-\beta\), where \(\hat{\tau}(\epsilon) = \nicefrac{1 + \ln{\frac{1}{\epsilon}}}{1 - \hat{k}_{R, gen}}\) and \(\hat{k}_{R, gen}\) is in~\cite{wolfer2020mixing}.\end{theorem}

We evaluate the effectiveness of \(T\) with \(\hat{\tau}(\epsilon)\) and \(\hat{k}_{R, \text{gen}}\).

\begin{theorem}[Evaluation of \(E(T)\)]\label{thm:unknown_delta_T}
    For any \(T \geq argmax_{R \in \mathcal{R}}argmin_{T \in \mathbb{N}}\{\max\{\sqrt{\nicefrac{\log{T}}{T}} + \sqrt{\nicefrac{4\hat{\delta}_{R}}{T}} \leq \epsilon, 1- \frac{1}{T^2} \geq 1- \alpha, O(\hat{\tau}^2(\epsilon)\ln(\nicefrac{\hat{\tau}(\epsilon)}{\beta})\ln{\ln{\nicefrac{\hat{\tau}(\epsilon)}{\beta}}})\}\}\), \(E(T) \geq \nicefrac{1}{\epsilon}\) w.p. \(1 - \alpha -\beta\) where \(\hat{\delta}_{R} = (\nicefrac{4}{\log{(\nicefrac{2}{1 + 4\epsilon + 2\epsilon^2})}}) \hat{\tau}(\epsilon)\).
\end{theorem}


\subsection{Proof of Results}\label{app:theorems}

This section provides the complete proofs of the theoretical results in Section 2 and 5, whose proofs were not included there.

Here, \(N_T^m\) represents the terminal value of the fuzzing path in the \(m\)-th replication, i.e., \(N_T^m=\sum_{t=1}^T Y_R^m(\omega,t)\), while \(N_T\) represents this value along a single trajectory.

\subsubsection{Lemma \ref{lemma:0}}

\begin{proof}[Proof]
Since \(\pi\) is uniform over \( D\), we have
\(\pi(x)=1/N^{\ell_{\max}}\) for all \(x\in D\). The bounded density
ratio assumption gives
\(\pi^R(x)\leq \kappa_R\pi(x)=\kappa_R/N^{\ell_{\max}}\). Therefore,
\(\pi^R(B)=\sum_{x\in B}\pi^R(x)
\leq \kappa_R|B|/N^{\ell_{\max}}\). Hence,
\(|P_R-\pi(1)|
=|\pi^R(B)-\pi(B)|
\leq \pi^R(B)+\pi(B)
\leq (\kappa_R+1)|B|/N^{\ell_{\max}}\).
Since \(|B|T=o(N^{\ell_{\max}})\), we have
\(|B|/N^{\ell_{\max}}=o(1/T)\). Thus,
\(|P_R-\pi(1)|=o(1/T)\).

\end{proof}

\subsubsection{Proposition \ref{prop:S_t markovian}}

\begin{proof}[Proof]

Let $(S_X, \Sigma_X)$ and $(S_G, \Sigma_G)$ be measurable spaces, and let
\[
(\mathsf S,\mathcal S):=(S_X\times S_G,\Sigma_X\otimes \Sigma_G).
\]

Define the measurable map $F:\mathsf S\to \mathsf S$ by
\[
F(x,g):=\bigl(\mathcal{R}(x,g),\,\Phi(g,G(\mathcal{R}(x,g)))\bigr).
\]
Then, by the update rules,
\[
S_{t+1}^R=F(S_t^R).
\]
Hence for every $A\in\mathcal S$,
\[
\mathbb P(S_{t+1}^R\in A\mid S_0,\dots,S_t^R)
=
\mathbf 1_{\{F(S_t^R)\in A\}}.
\]
Since the right-hand side depends only on $S_t^R$, we obtain
\[
\mathbb P(S_{t+1}^R\in A\mid S_0^R,\dots,S_t^R)
=
\mathbb P(S_{t+1}^R\in A\mid S_t^R),
\qquad \forall A\in\mathcal S.
\]
Therefore, $\{S_t^R\}_{t\ge 0}$ is a Markov chain on $(\mathsf S,\mathcal S)$.

\end{proof}

\subsection{Corollary \ref{coro:1}}
\begin{proof}
Since \(\tilde{P}_R(T)\leq \tilde{P}\leq \pi(1)\), we have \(D_R(T)=|\tilde{P}_R(T)-\pi(1)|=\pi(1)-\tilde{P}_R(T)\) and \(Q_R(T)=|\tilde{P}_R(T)-\tilde{P}|=\tilde{P}-\tilde{P}_R(T)\). Therefore, \(D_R(T)=Q_R(T)+\pi(1)-\tilde{P}\). The term \(\pi(1)-\tilde{P}\) is common to all fuzzers, so it does not affect the ranking. Hence minimizing \(Q_R(T)\) is equivalent to minimizing \(D_R(T)\).

\end{proof}

\subsection{Corollary \ref{coro:2}}

\begin{proof}
By definition, \(E(T)=\nicefrac{1}{\max_{R\in\mathcal R}D_R(T)}\). Thus, \(E(T)\geq \nicefrac{1}{\epsilon}\) implies \(\max_{R\in\mathcal R}D_R(T)\leq \epsilon\), and hence \(D_R(T)\leq \epsilon\) for every \(R\). Since \(\tilde{P}_R(T)\leq \tilde{P}\leq \pi(1)\), we have \(Q_R(T)=\tilde{P}-\tilde{P}_R(T)\leq \pi(1)-\tilde{P}_R(T)=D_R(T)\). Therefore, \(Q_R(T)\leq \epsilon\) for every \(R\), which yields the stated necessary condition on \(T\). Finally, \(1-\nicefrac{1}{T^2}\geq 1-\alpha\) is equivalent to \(T\geq \nicefrac{1}{\sqrt{\alpha}}\).

\end{proof}

\subsection{Corollary \ref{coro:3}}

\begin{proof}
Since \(\tilde{P}_R(T)\leq \tilde{P}\leq \pi(1)\), we have \(Q_R(T)\leq D_R(T)\). Combining this with \(D_R(T)\leq b_T+\sqrt{\nicefrac{4\delta_R}{T}}\) where $b_T =\sqrt{\nicefrac{\log{T}}{T}}$ gives \(Q_R(T)\leq b_T+\sqrt{\nicefrac{4\delta_R}{T}}\). Hence \(\sqrt{\nicefrac{4\delta_R}{T}}\geq Q_R(T)-b_T\). If \(Q_R(T)\leq b_T\), this only gives the trivial bound \(\delta_R\geq 0\). Otherwise, squaring both sides gives \(\delta_R\geq \frac{T}{4}(Q_R(T)-b_T)^2\). Combining the two cases yields \(\delta_R\geq \frac{T}{4}(Q_R(T)-b_T)_+^2\).

\end{proof}

\subsubsection{Theorem~\ref{thm: Asymptotic Results}}

\begin{proof}[Proof]


We define \(\tau^{+}(B,m)\) as the time elapsed until the first bug is detected in the $m^{th}$ trial. It is worth noting that \(\tau^{+}(B,1), \tau^{+}(B,2), \ldots, \tau^{+}(B,M)\) are independently and identically distributed based on the definition, and thus the law of large numbers gives us
\begin{align*}
  \lim_{M \to \infty }\frac{\sum_{m=1}^{M}\tau^{+}(B,m)}{M} = E[\tau^{+}(B,1)] = E[\tau^{+}(B)]
\end{align*}
and subsequently we have
\begin{align*}
     & \lim_{M \to \infty }\frac{\sum_{m=1}^{M}\tau^{+}(B,m)}{M} \\
     & = \frac{1}{P(X\in D: X\in B)}\\
     & = \frac{1}{\pi^R(1)} \approx\frac{1}{\pi(1)}.
\end{align*}


In the meantime, we have that
\begin{align*}
   & E[\tau^{+}(B)] \\
   & = E[\min{\{t \geq 0: X(t) \in B\}}] \\
   & = E[\min{\{t \geq 0: N_{t} = 1\}}].
\end{align*}

We further define
\begin{align*}
    \tau^{+,1}(B) = \min{\{t \geq \tau^{+}(B): N_{t} = 2\}}
\end{align*}
and generally for any \(n \geq 1\)
\begin{align*}
    \tau^{+,n}(B) = \min{\{t \geq \tau^{+,n-1}(B): N_{t} = n\}}.
\end{align*}

Based on the Markov property and the fact that the transition matrix is homogeneous in time, we derive that \(\tau^{+}(B), \tau^{+,1}(B), \ldots, \tau^{+,n}(B)\) are independently and identically distributed.

It is worth noting that \(N_T\) meets that
\begin{align*}
    \sum_{n=1}^{N_T}\tau^{+,n}(B) < T \leq \sum_{n=1}^{N_T+1}\tau^{+,n}(B)
\end{align*}
which implies that
    \begin{align*}
    \frac{\sum_{n=1}^{N_T}\tau^{+,n}(B)}{N_{T}} < \frac{T}{N_T} \leq \frac{\sum_{n=1}^{N_T+1}\tau^{+,n}(B)}{N_T}.
    \end{align*}

By leveraging the law of large number, we obtain that
\begin{align*}
    & \lim_{T \to \infty}\frac{\sum_{n=1}^{N_T}\tau^{+,n}(B)}{N_{T}} = E[\tau^{+}(B)] \\
    & \lim_{T \to \infty}\frac{\sum_{n=1}^{N_T+1}\tau^{+,n}(B)}{N_T} = E[\tau^{+}(B)].
\end{align*}

Consequently, we have
\begin{align*}
    \lim_{T \to \infty}\frac{T}{N_T} = E[\tau^{+}(B)],
\end{align*}
or equivalently,
\begin{align*}
    \lim_{T \to \infty}\frac{N_T}{T} = \frac{1}{E[\tau^{+}(B)]}  = \Pr(X\in B|X\in D)
\end{align*}
which completes last part of the statement, and thus concludes the proof.

\end{proof}

\subsubsection{Theorem~\ref{finite-time:matrix}}

\begin{proof}
    This follows from~\cite{freedman2017convergence}.

\end{proof}

\subsubsection{Theorem~\ref{finite-time:estimator}}

\begin{proof}
We note that
\begin{align*}
   & \tilde{P}_R(T)  \\
   & = \frac{1}{M}\sum_{m=1}^{M}\frac{1}{T}\sum_{t=1}^T1_{X_t^m \in B}
\end{align*}
and if we take the expectation, we derive that
\begin{align*}
   & E[\tilde{P}_R(T)]  = \frac{1}{T} \sum_{t=1}^{T} P_R^t(1).
\end{align*}

Subsequently, we derive that
\begin{align*}
    & \bigl|E[\tilde{P}_R(T)] - \pi(1)\bigr| \\
    & = \left|\frac{1}{T} \sum_{t=1}^{T} P_R^t(1) - \pi(1)\right| \\
    & \leq \frac{1}{T} \sum_{t=1}^{T} C_R \alpha_R^t \\
    & \leq C_R\frac{1}{T}\frac{1}{1-\alpha_R}. 
\end{align*}

Subsequently, we characterize the difference between \(E[\tilde{P}_R(T)\) and \(\tilde{P}_R(T)\). It is worth noting that \(\frac{1}{T}\sum_{t=1}^T1_{X_t^m \in B} \leq 1\) and we repeat the experiment for \(M\) times. Using Hoeffding's inequality, we immediately obtain that
\begin{align*}
    P(|\tilde{P}_R(T) - E[\tilde{P}_R(T)]| \geq \sqrt{\frac{\log{T}}{T}}) \leq \frac{1}{T^2}. 
\end{align*}

This implies that with probability at least \(1 - \frac{1}{T^2}\)
\begin{align*}
   & |\tilde{P}_R(T) - \pi(1)| \\
   & \leq |\tilde{P}_R(T) - E[\tilde{P}_R(T)] + E[\tilde{P}_R(T)] - \pi(1)| \\
   & \leq |\tilde{P}_R(T) - E[\tilde{P}_R(T)]| + |E[\tilde{P}_R(T)] - \pi(1)| \\
   & \leq \sqrt{\frac{\log{T}}{T}} + \frac{1}{T}\frac{C_R}{1-\alpha_R}. 
\end{align*}

\end{proof}

\subsubsection{Theorem~\ref{finite-time:matrix-spectral-rev}}

\begin{proof}
    This follows from~\cite{Sousi2020}.

\end{proof}


We also another alternative of analyzing nonreversible Markov chains. More specifically, we define \(MP_R^Z = \tilde{P}_R P_R^Z\), where
\(\tilde{P}_R = \frac{P_R^Z(x, y) \pi(x)}{\pi(y)}\) and let \(\lambda_{M,R,2}\) and \(\gamma_{M,R}\) denote the second-largest eigenvalue and the spectral gap of \(MP_R^Z\), respectively.

\begin{theorem}[Spectral property for coupling MC (Poincare inequality)]\label{finite-time:matrix-spectral-nonrev-coupling}
The result reads as based on~\cite{fill1991eigenvalue}
 \(\lVert (P_R^Z)^T(x,\cdot) - \pi \rVert_{TV} \leq \frac{\lambda_{M,R, 2}^T}{4\pi(x)} = \frac{{\left(1 - \gamma_{M,R}
\right)}^{T}}{4\pi(x)}\).

In other words, we recover the result in Theorem~\ref{finite-time:matrix} with \(C_R = \max_{\pi_0}\lVert \pi_0 - \pi \rVert_{L_2}\) and \(\alpha_R = 1-\gamma_R\).
\end{theorem}

\subsubsection{Theorem~\ref{finite-time:matrix-spectral-nonrev}}

\begin{proof}
   The proof partly follows from that of Theorem~\ref{finite-time:estimator}. Specifically, we again have the following decomposition.

Denote $P_{R}(x \in I_R^T: x \in H_R^T) = \tilde{P}_R(T)$. This implies that with probability at least \(1 - \frac{1}{T^2}\)
\begin{align*}
   & |P_{R}(x \in I_R^T: x \in H_R^T) - \pi(1)| \\
   & \leq |P_{R}(x \in I_R^T: x \in H_R^T) - E[P_{R}(x \in I_R^T: x \in H_R^T)] \\ &\qquad + E[P_{R}(x \in I_R^T: x \in H_R^T)] - \pi(1)| \\
   & \leq |P_{R}(x \in I_R^T: x \in H_R^T) - E[P_{R}(x \in I_R^T: x \in H_R^T)]| \\ &\qquad + |E[P_{R}(x \in I_R^T: x \in H_R^T)] - \pi(1)|.
\end{align*}

For the first quantity, we again have the Hoeffding's inequality hold, which reads as
\begin{align}\label{eq:6.1}
    & P(|P_{R}(x \in I_R^T: x \in H_R^T) - E[P_{R}(x \in I_R^T: x \in H_R^T)]| \notag \\
    & \geq \sqrt{\frac{\log{T}}{T}}) \leq \frac{1}{T^2  }. 
\end{align}

For the second quantity, we use the result from Theorem 1.2 in~\cite{chatterjee2023spectral} by specifying the target function \(g(X_t)\) as \(g(X_t) = 1_{X_t \in B}\). This immediately gives us that
\begin{align}\label{eq:6.2}
    E[P_{R}(x \in I_R^T: x \in H_R^T)] - \pi(1)| \leq \sqrt{\frac{4\delta_{R}}{T}}.
\end{align}

Combining (\ref{eq:6.1}) and (\ref{eq:6.2}) immediately leads to that  with probability \(1 - \frac{1}{T^2}\), 
\begin{align*}
   & |P_{R}(x \in I_R^T: x \in H_R^T) - \pi(1)| \\
   & \leq |P_{R}(x \in I_R^T: x \in H_R^T) - E[P_{R}(x \in I_R^T: x \in H_R^T)] \\ &\qquad + E[P_{R}(x \in I_R^T: x \in H_R^T)] - \pi(1)| \\
   & \leq \sqrt{\frac{\log{T}}{T}} + \sqrt{\frac{4\delta_{R}}{T}}, 
\end{align*}
which concludes the proof.

\end{proof}

\subsubsection{Theorem~\ref{thm7}}

\begin{proof}
    The proof follows from that in~\cite{fill1991eigenvalue}.

\end{proof}

\subsubsection{Theorem~\ref{thm: E_T_known}}

\begin{proof}
The proof follows directly from the finite-time convergence result of \(\tilde{P}_R(T)\). The first term ensures that \(|\tilde{P}_R(T) - \pi(1)| \leq \epsilon\) with probability at least \(1 - \frac{1}{T^2}\), and the second term guarantees that this probability is at least \(1 - \alpha\). Consequently, we derive both terms and determine the minimal \(T_R\) that satisfies the condition for a specific fuzzer \(R\), then select the maximum \(T_R\) across all fuzzers.

\end{proof}


\subsubsection{Theorem~ \ref{thm:unknown_delta} and  \ref{thm:unknown_delta_T}}

\begin{proof}
The proof follows from Theorem~\ref{thm: E_T_known}, with \(D_R(T)\) replaced by \(U_R(T)\).

\end{proof}

\subsubsection{Theorem~\ref{thm:unknown_delta_U}}

\begin{proof}

The proof first leverages the result in \cite{hsu2019mixing} or \cite{wolfer2024improved}. By the theoretical guarantee established in these work, we have that with probability at least \(1-\beta\), 
\begin{align*}
|\gamma_{dps}(R) - \gamma_R| \leq O(\frac{1}{\sqrt{T}}). 
\end{align*}
Subsequently, we derive that 
\begin{align*}
   &  |\hat{U}_R - U_R| \\
   & \leq |\sqrt{\frac{\log{T}}{T}} + \sqrt{\frac{4}{T\gamma_{dps}(R)}} - \sqrt{\frac{\log{T}}{T}} + \sqrt{\frac{4}{T\gamma_R}}| \\
   & \leq |\sqrt{\frac{4}{T}(\frac{\gamma_{dps}(R) - \gamma_R}{\gamma_{dps}(R)\cdot \gamma_R})}|\\
   & \leq \sqrt{\frac{4}{T}(\frac{O(\frac{1}{\sqrt{T}})}{\gamma_{dps}(R)\cdot \gamma_R})} \\
   & \leq O(\frac{1}{T^{\frac{3}{4}}}) \sqrt{\frac{1}{\gamma_{dps}(R) \cdot \gamma_R}}.
\end{align*}

In other words, with probability \(1-\beta\),
\begin{align*}
    & U_R \leq \hat{U}_R + O(\frac{1}{T^{\frac{3}{4}}}) \sqrt{\frac{1}{\gamma_{dps}(R) \cdot \gamma_R}}  \\
    & \leq \sqrt{\frac{\log{T}}{T}} + \sqrt{\frac{4}{T\gamma_{dps}(R)}} + O(\frac{1}{T^{\frac{3}{4}}}) \sqrt{\frac{1}{\gamma_{dps}(R) \cdot \gamma_R}} \\
    & \leq \sqrt{\frac{\log{T}}{T}} + \sqrt{\frac{4}{T\gamma_{dps}(R)}} + O(\frac{1}{T^{\frac{3}{4}}}) \sqrt{\frac{1}{\gamma_{dps}(R) \cdot \gamma_R}}.  
\end{align*}

Notably, the term \(\sqrt{\frac{1}{T}}\) dominates \(\frac{1}{T^{\frac{3}{4}}}\), allowing us to compare \(\sqrt{\frac{4}{T\gamma_{dps}(R)}}\) across fuzzers for evaluation. This reduces the problem to comparing \(\gamma_{dps}\) or equivalently \(\hat{U}_R\), thereby completing the proof.

\end{proof}

\subsubsection{Theorem~\ref{thm:unknown_delta_U_T}}

\begin{proof}

The proof again follows from Theorem~\ref{thm: E_T_known}, again with \(D_R(T)\) replaced by \(U_R(T)\).

\end{proof}

\subsubsection{Proof of Theorem \ref{finite-time:estimator-spectral-reversible}}
\begin{proof}[Proof of Theorem \ref{finite-time:estimator-spectral-reversible}]
    Notably, \(\lVert \pi_0 (P_R^Z)^T - \pi^Z \rVert_{L_{2}}
       = \sqrt{E\left[|\pi_0^{X}(P_R^Z)^T - \pi^{X}(x)|^2\right]} 
       \leq C_1 \cdot {\left(1-\gamma_R\right)}^{T}\)
    does not imply \(\left|E[P_{R}(x \in I_R^T : x \in H_R^T)] - \pi(1)\right|
       \leq C_1 \cdot {\left(1-\gamma_R\right)}^{T}. \)

    However, we have that
    \begin{align*}
       & |(P_R^Z)^T(x_0,) - \pi(1)|  \leq \frac{\sqrt{E\left[|\pi_0^{X}(P_R^Z)^T - \pi^{X}(x)|^2\right]}}{|\pi(1)|} \\
       & \leq \frac{\lVert \pi_0^Z - \pi^Z \rVert_{L_2}}{\pi(1)} \cdot {\left(1-\gamma_R\right)}^{T} \\
       & \leq  \frac{\left(1-\pi(1)\right)\pi(1)}{\pi(1)} \cdot {\left(1-\gamma_R\right)}^{T}  \leq \left(1-\pi(1)\right) \cdot {\left(1-\gamma_R\right)}^{T}. 
    \end{align*}

   Subsequently, we derive that
\begin{align*}
    & \left|E[P_{R}(x \in I_R^T : x \in H_R^T)] - \pi(1)\right| \\
    & = \left|\frac{1}{T} \sum_{t=1}^{T} P_R^t(1) - \pi(1)\right| \\
    & \leq \frac{1}{T}\sum_{t=1}^T \left(1-\pi(1)\right) \cdot {\left(1-\gamma_R\right)}^{T}\leq \left(1-\pi(1)\right) \frac{1}{T}\frac{1}{\gamma_R}.
\end{align*}

Again, by using the Hoeffding's inequality, we derive that \(P(|P_{R}(x \in I_R^T: x \in H_R^T) - E[P_{R}(x \in I_R^T: x \in H_R^T)]| \geq \sqrt{\frac{\log{T}}{T}}) \leq \frac{1}{T^2  }.\)

Likewise, we obtain that with probability at least \(1 - \frac{1}{T^2}\)
\begin{align*}
   & |P_{R}(x \in I_R^T: x \in H_R^T) - \pi(1)| \\
   & \leq |P_{R}(x \in I_R^T: x \in H_R^T) - E[P_{R}(x \in I_R^T: x \in H_R^T)] \\ &\qquad + E[P_{R}(x \in I_R^T: x \in H_R^T)] - \pi(1)| \\
   & \leq |P_{R}(x \in I_R^T: x \in H_R^T) - E[P_{R}(x \in I_R^T: x \in H_R^T)]| \\ &\qquad+ \left|E[P_{R}(x \in I_R^T : x \in H_R^T)] - \pi(1)\right| \\
   & \leq \sqrt{\frac{\log{T}}{T}} + {1-\pi(1)} \frac{1}{T}\frac{1}{\gamma_R} = \sqrt{\frac{\log{T}}{T}} + {1-\pi(1)} \frac{\delta_R}{T}. 
\end{align*}
\end{proof}

\subsubsection{Proof of Theorem \ref{thm:unknown_delta}}
\begin{proof}[Proof of Theorem \ref{thm:unknown_delta}]

More specifically, we define a new evaluation metric as \(U_R(T) = \sqrt{\frac{\log T}{T}} + \sqrt{\frac{4\delta_R}{T}}\). Based on the result in~\cite{chatterjee2023spectral}, we have for any \(\epsilon\), \(\delta_R \leq (\frac{4}{\log{(\frac{2}{1 + 4\epsilon + 2\epsilon^2})}}) \tau(\epsilon)\) 
where \(\tau(\epsilon)\) is the mixing time defined as \(\min_{t}\lVert P_{R}^t(x,\cdot) - \pi \rVert_{TV} \leq \epsilon\).

Meanwhile, for the mixing time of an ergodic Markov chain, we have \(\tau(\epsilon) \leq \frac{1 + \ln{\frac{1}{\epsilon}}}{1 - k_{R, gen}}.\)

By the estimation procedure for \(k_{R, gen}\) as presented in~\cite{wolfer2020mixing}, we have that \(T > O(\hat{\tau}^2(\epsilon)\ln(\frac{\hat{\tau}(\epsilon)}{\delta})\ln{\ln{\frac{\hat{\tau}(\epsilon)}{\delta}}})\), with probability \(1 - \delta\), \(\tau(\epsilon)  \leq 3\hat{\tau}(\epsilon) = 3\frac{1 + \ln{\frac{1}{\epsilon}}}{1 - \hat{k}_{R, gen}}\), 
where \(\hat{k}_{R, gen}\) is constructed in~\cite{wolfer2020mixing}.

Subsequently, we derive that with probability \(1 -\delta\), \(U_R(T) = \)\begin{align*}
   &\sqrt{\frac{\log T}{T}} + \sqrt{\frac{4\delta_R}{T}}  = \sqrt{\frac{\log T}{T}} + \sqrt{\frac{4(\frac{4}{\log{(\frac{2}{1 + 4\epsilon + 2\epsilon^2})}}) \tau(\epsilon)}{T}} \\
   & \leq \sqrt{\frac{\log T}{T}} + \sqrt{\frac{4(\frac{4}{\log{(\frac{2}{1 + 4\epsilon + 2\epsilon^2})}}) 3\frac{1 + \ln{\frac{1}{\epsilon}}}{1 - \hat{k}_{R, gen}}}{T}}.
\end{align*}\end{proof}

\subsection{Additional Experiments}\label{app:exp}

\paragraph{\textbf{Setup}}
All experiments are  conducted on a server running Ubuntu 24.04.2 LTS (Noble) with 2.0 TB of RAM and Intel Xeon Platinum 8480+ processors (56 cores / 112 threads per socket; 2.00 GHz base frequency). 
Each fuzzer–benchmark pair is evaluated using 20 independent trials, with each trial lasting 72 hours, if not specified otherwise. The total running time is over 233,000 CPU hours. The code is available in the supplementary material and the accompanying code url.

\subsubsection{Tables}
\begin{table}[htbp]
\centering
\setlength{\tabcolsep}{3pt}
\caption{Results on \textit{libhtp} (1st row) and \textit{poppler} (2nd row).}
\label{tab:71h_results}
\begin{tabular}{lcccc}
\toprule
AFL & AFLFast & AFL++ & AFLSmart & Entropic \\
\midrule
0.01 & 0.02 & 0.01 & 0.02 & 0.05 \\
0.18 & 0.22 & 0.33 & 0.21 & 0.24 \\
\midrule
FairFuzz & Honggfuzz & LibFuzzer & MOpt &  \\
\midrule
0.03 & 0.03 & 0.05 & 0.01 &  \\
0.34 & 0.10 & 0.38 & 0.22 &  \\
\bottomrule
\end{tabular}
\end{table}

\begin{table}[htbp]
\centering
\setlength{\tabcolsep}{3pt}
\caption{Benchmark scores across all benchmarks.}
\label{tab:avg_all_71h}
\begin{tabular}{lcccc}
\toprule
arrow & aspell & ffmpeg & grok & harfbuzz \\
\midrule
0.43 & 0.06 & 0.43 & 0.01 & 0.27 \\
\midrule
libhevc & libgit2 & libhtp & libxml2 &  \\
\midrule
0.27 & 0.03 & 0.03 & 0.04 &  \\
\midrule
matio & njs & openh264 & php & poppler \\
\midrule
0.12 & 0.02 & 0.05 & 0.06 & 0.25 \\
\midrule
quickjs & stb & systemd & wireshark &  \\
\midrule
0.02 & 0.19 & 0.00 & 0.03 &  \\
\bottomrule
\end{tabular}
\end{table}

\begin{table}[H]
\centering
\setlength{\tabcolsep}{3pt}
\caption{Results on \textit{libhtp} (1st row) and \textit{poppler} (2nd row) (5 trials).}
\label{tab:71h_libhtp_poppler_5}
\begin{tabular}{lcccc}
\toprule
AFL & AFLFast & AFL++ & FairFuzz & MOpt \\
\midrule
0.03 & 0.06 & 0.05 & 0.04 & 0.05 \\
0.34 & 0.33 & 0.38 & 0.30 & 0.32 \\
\midrule
Honggfuzz & Entropic & LibFuzzer & Eclipser &  \\
\midrule
0.07 & 0.03 & 0.04 & 0.02 &  \\
0.40 & 0.35 & 0.28 & 0.31 &  \\
\bottomrule
\end{tabular}
\end{table}

\begin{table}[H]
\centering
\setlength{\tabcolsep}{3pt}
\caption{Benchmark scores across all benchmarks (5 trials).}
\label{tab:avg_71h_benchmarks_5}
\begin{tabular}{lcccc}
\toprule
arrow & aspell & ffmpeg & grok & harfbuzz \\
\midrule
0.60 & 0.07 & 0.47 & 0.01 & 0.42 \\
\midrule
libhevc & libgit2 & libhtp & libxml2 &  \\
\midrule
0.43 & 0.03 & 0.04 & 0.03 &  \\
\midrule
matio & njs & openh264 & php & poppler \\
\midrule
0.14 & 0.02 & 0.07 & 0.06 & 0.32 \\
\midrule
quickjs & stb & systemd & wireshark &  \\
\midrule
0.02 & 0.23 & 0.00 & 0.03 &  \\
\bottomrule
\end{tabular}
\end{table}

\subsubsection{Figures}

\begin{figure}[H]
\centering\footnotesize
\setlength{\tabcolsep}{1pt}
\begin{tabular}{cc}
\includegraphics[width=0.3\textwidth]{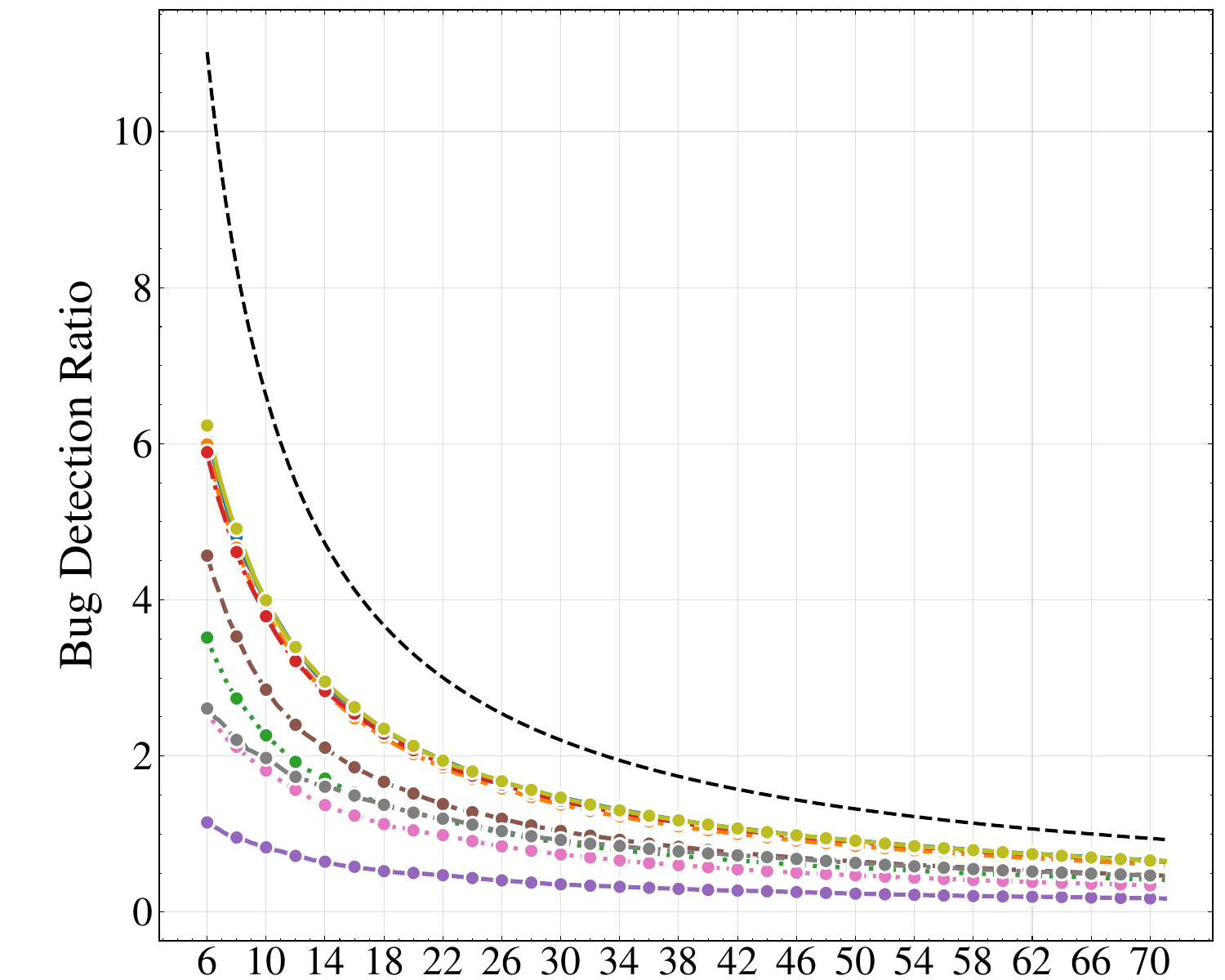} &
\includegraphics[width=0.3\textwidth]{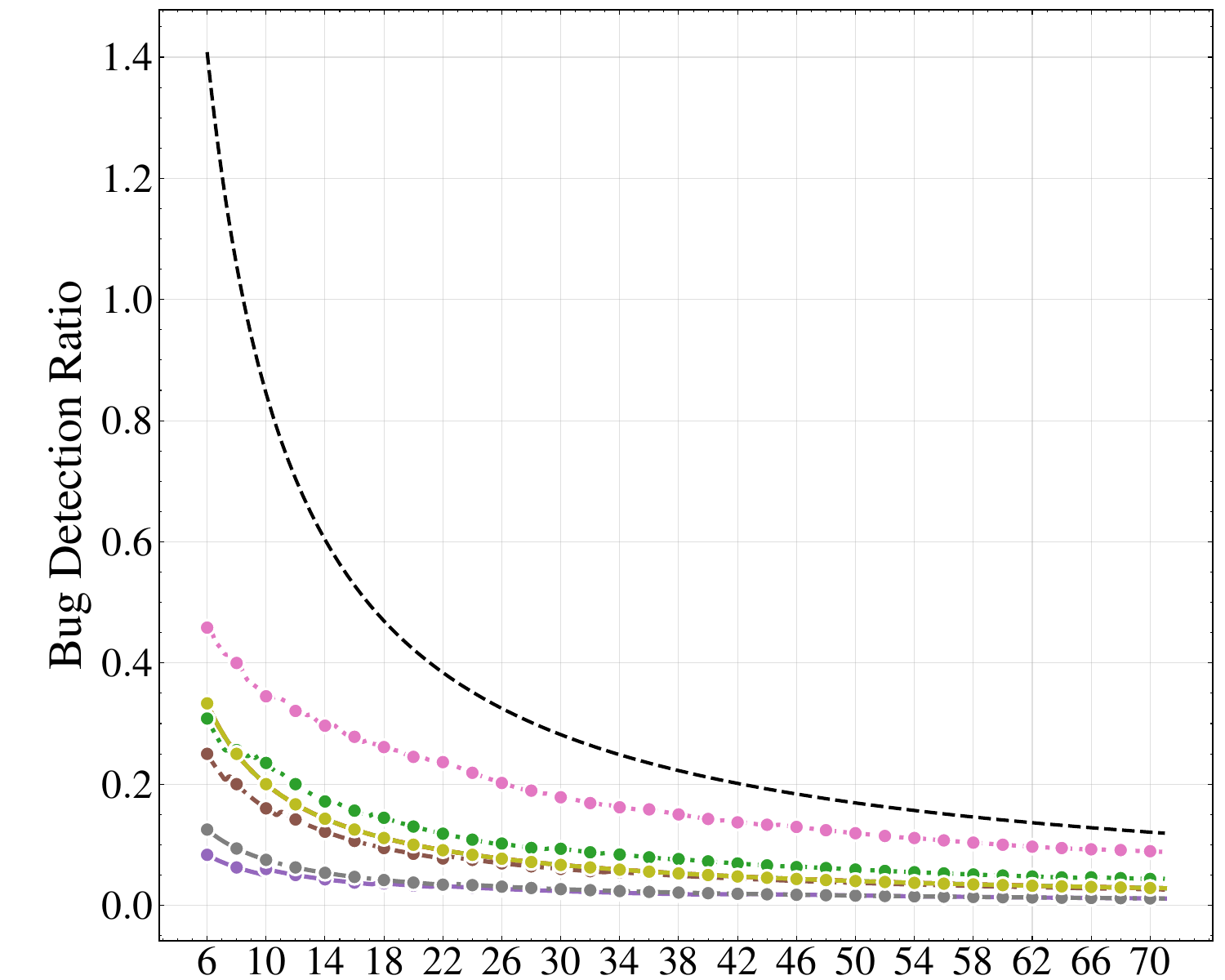} \\
(a) arrow & (b) aspell \\

\includegraphics[width=0.3\textwidth]{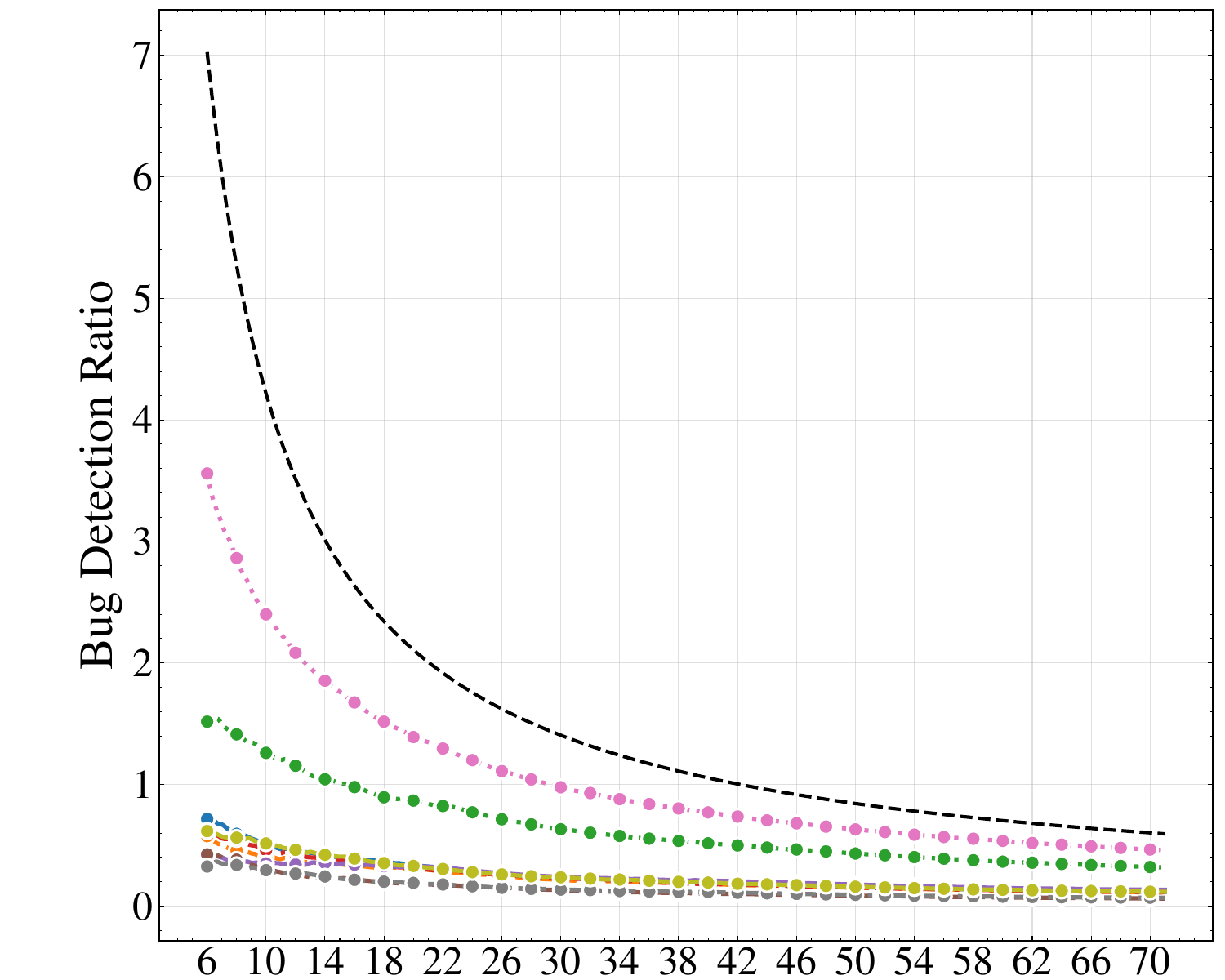} &
\includegraphics[width=0.3\textwidth]{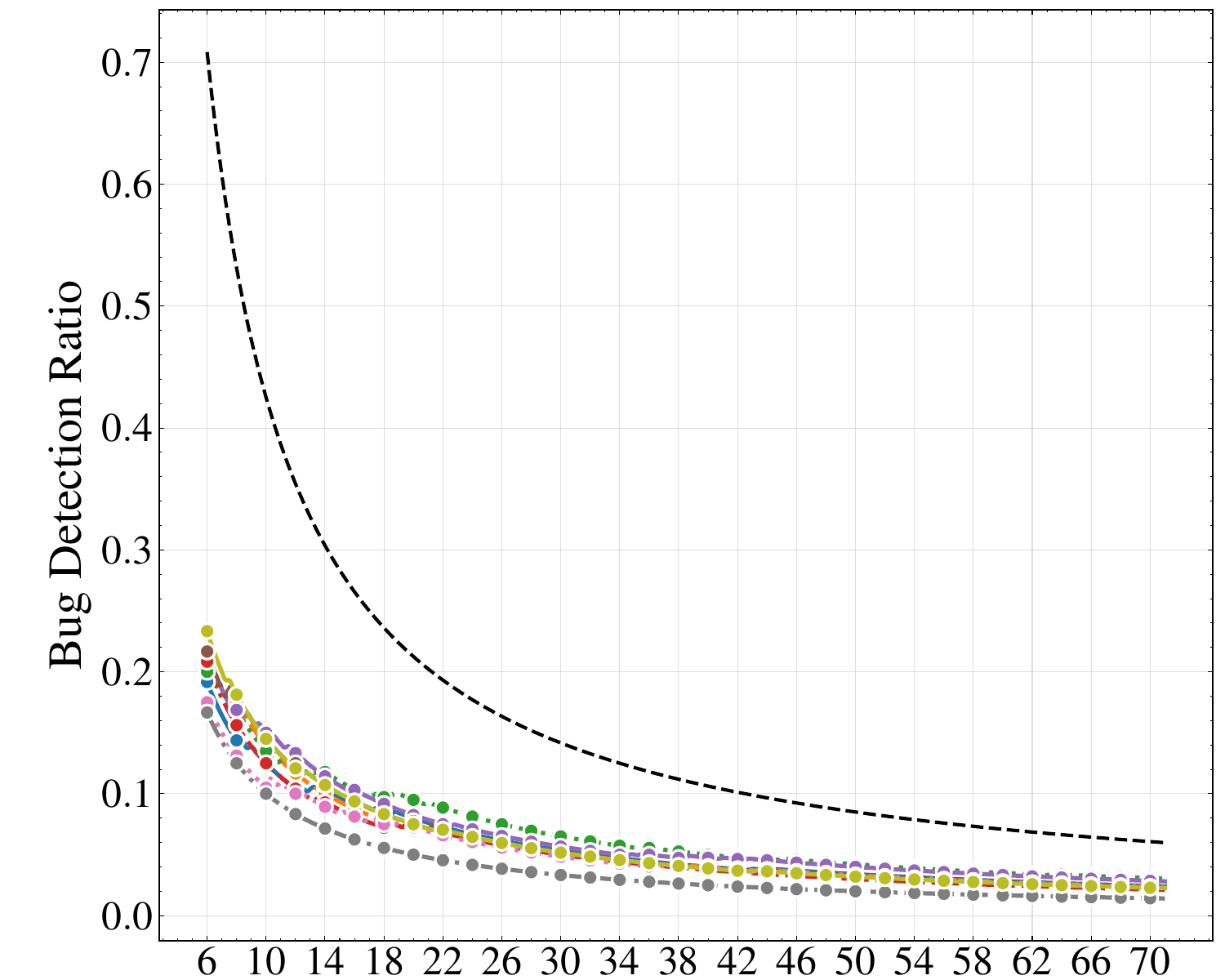} \\
(c) ffmpeg & (d) grok \\

\includegraphics[width=0.3\textwidth]{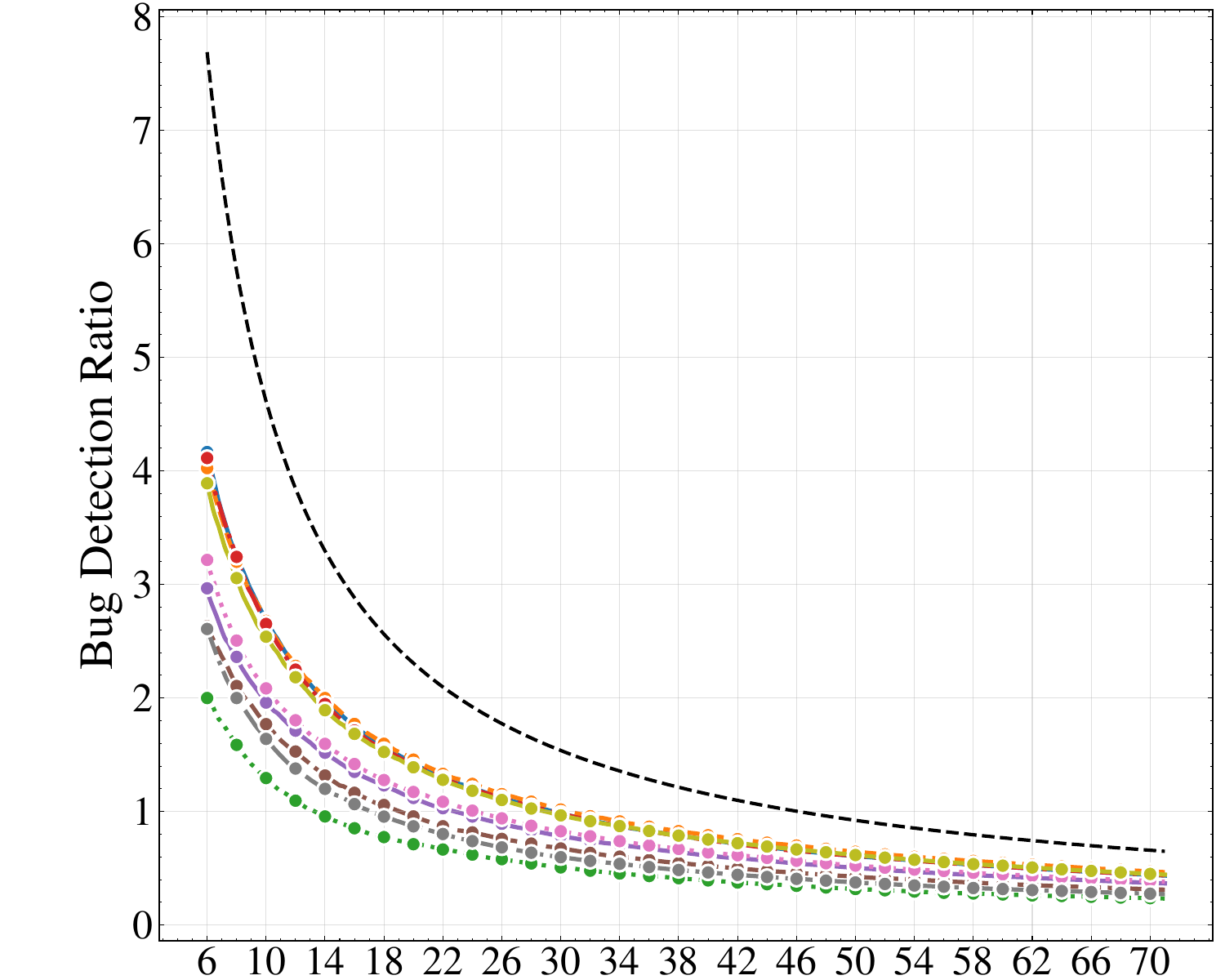} &
\includegraphics[width=0.3\textwidth]{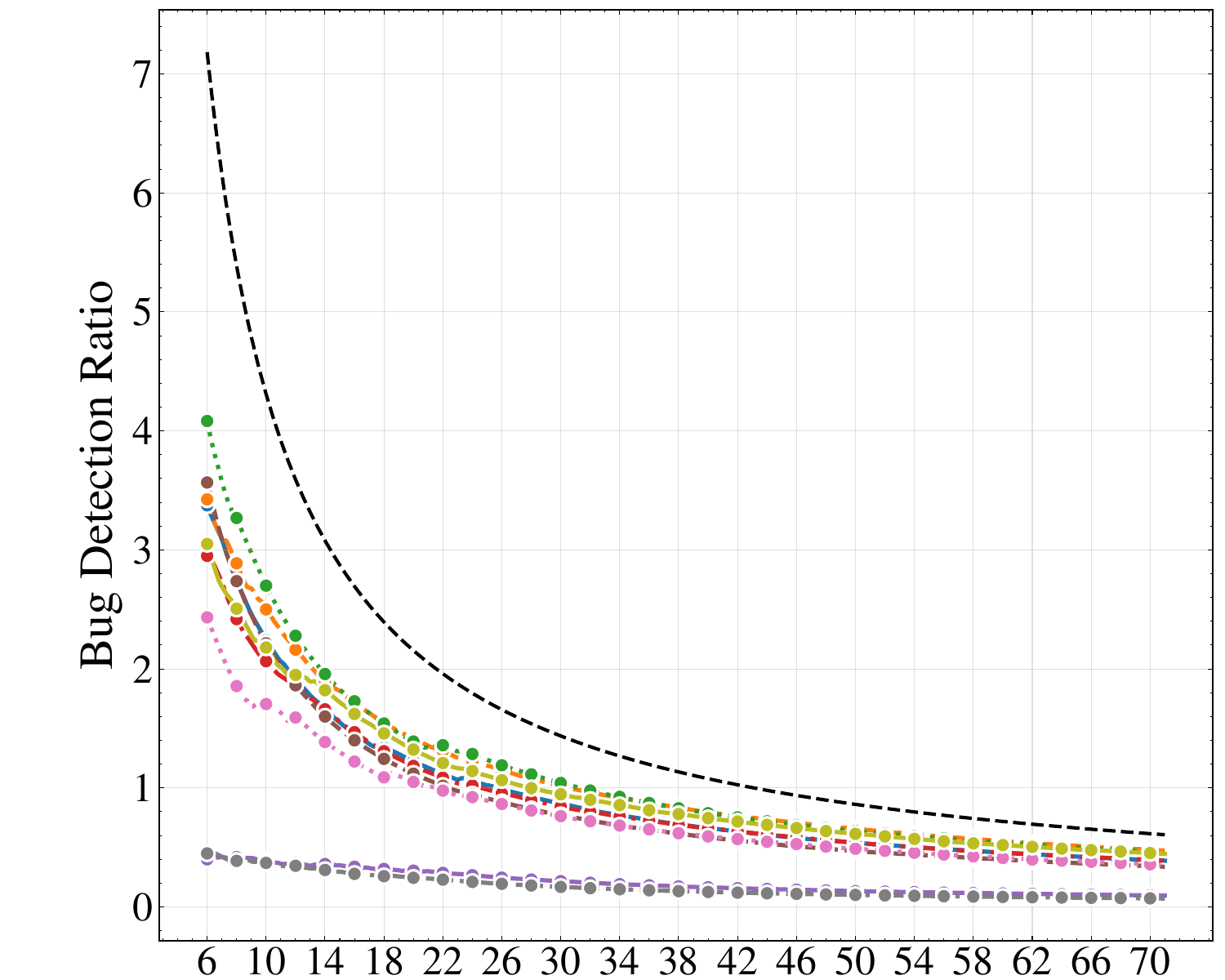} \\
(e) harfbuzz & (f) libhevc \\
\end{tabular}
\caption{Bug detection ratio (3×2 layout).}
\label{fig:rates_3x2}
\end{figure}

\begin{figure}[H]
\centering\footnotesize
\setlength{\tabcolsep}{1pt}
\begin{tabular}{cc}
\includegraphics[width=0.3\textwidth]{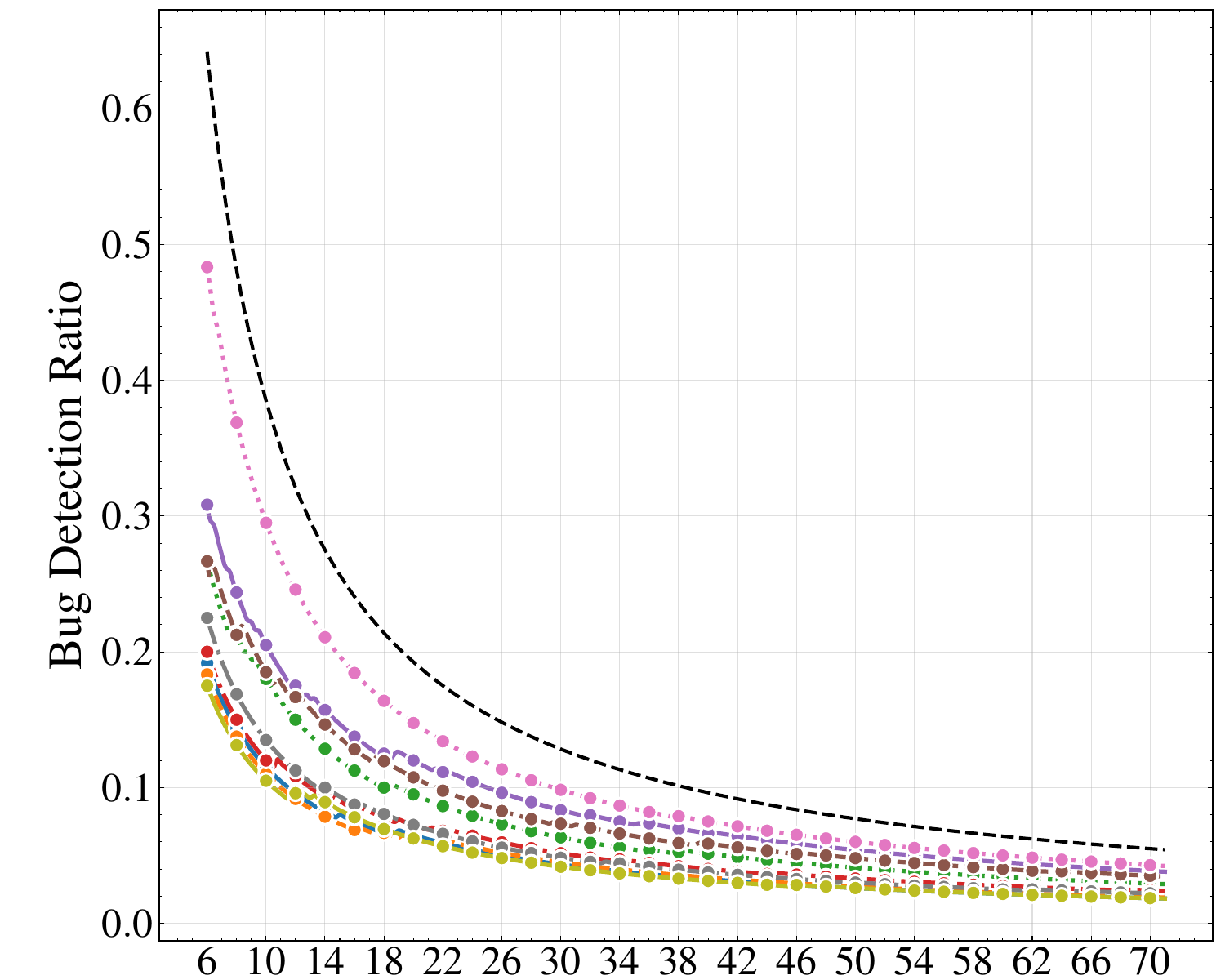} &
\includegraphics[width=0.3\textwidth]{figs/data/combined_rates_libhtp_fuzz_htp.pdf} \\
(a) libgit2 & (b) libhtp \\

\includegraphics[width=0.3\textwidth]{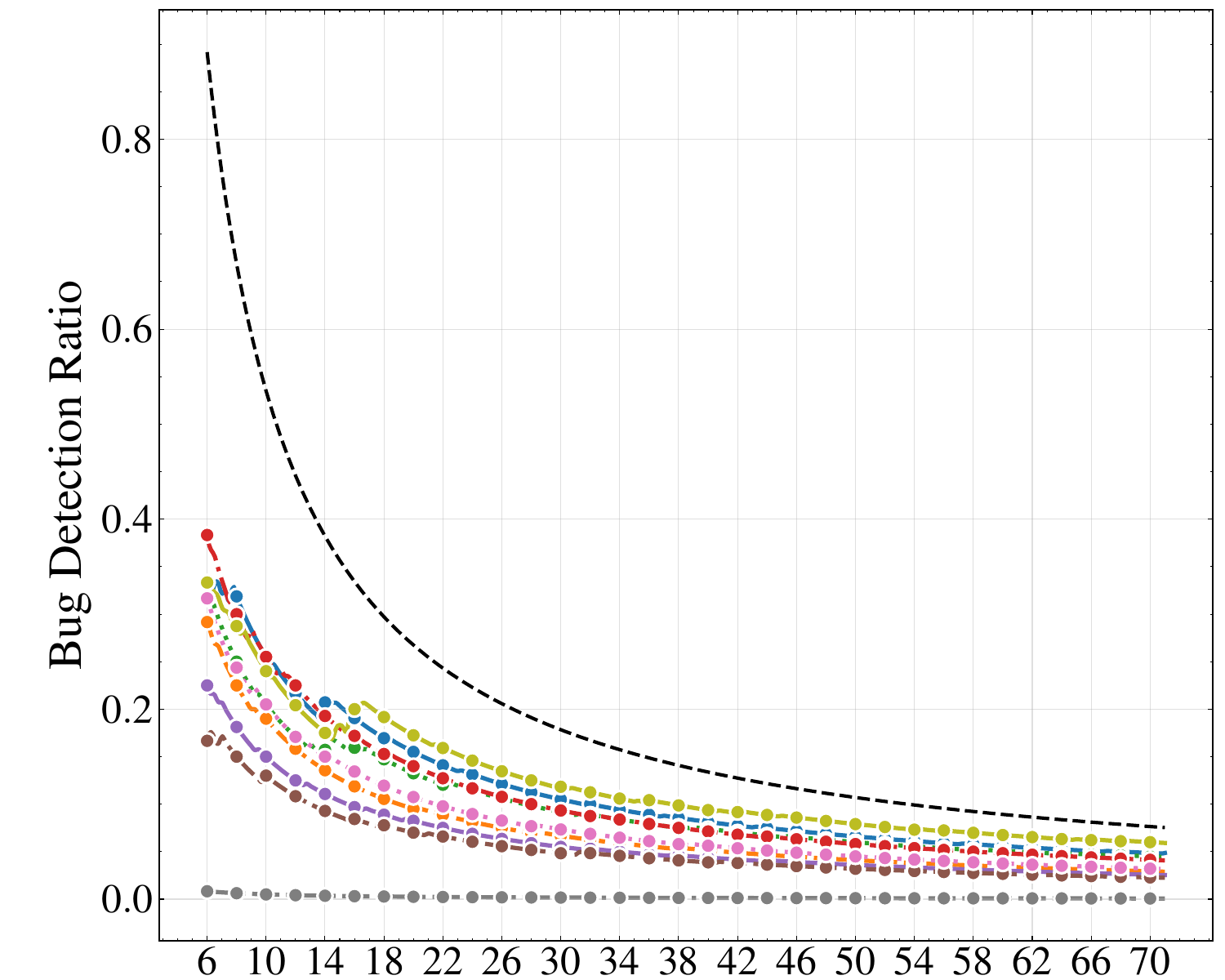} &
\includegraphics[width=0.3\textwidth]{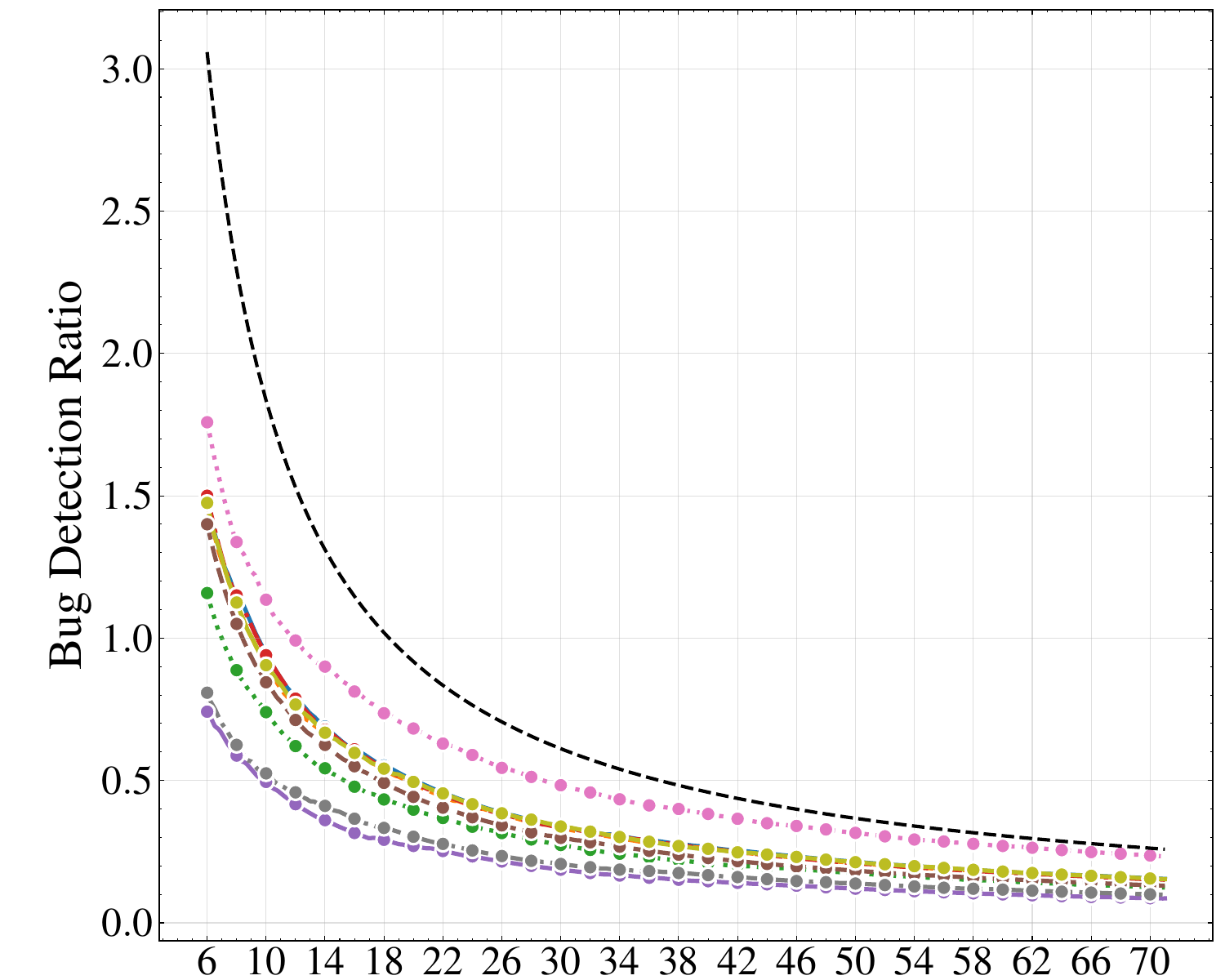} \\
(c) libxml2 & (d) matio \\

\includegraphics[width=0.3\textwidth]{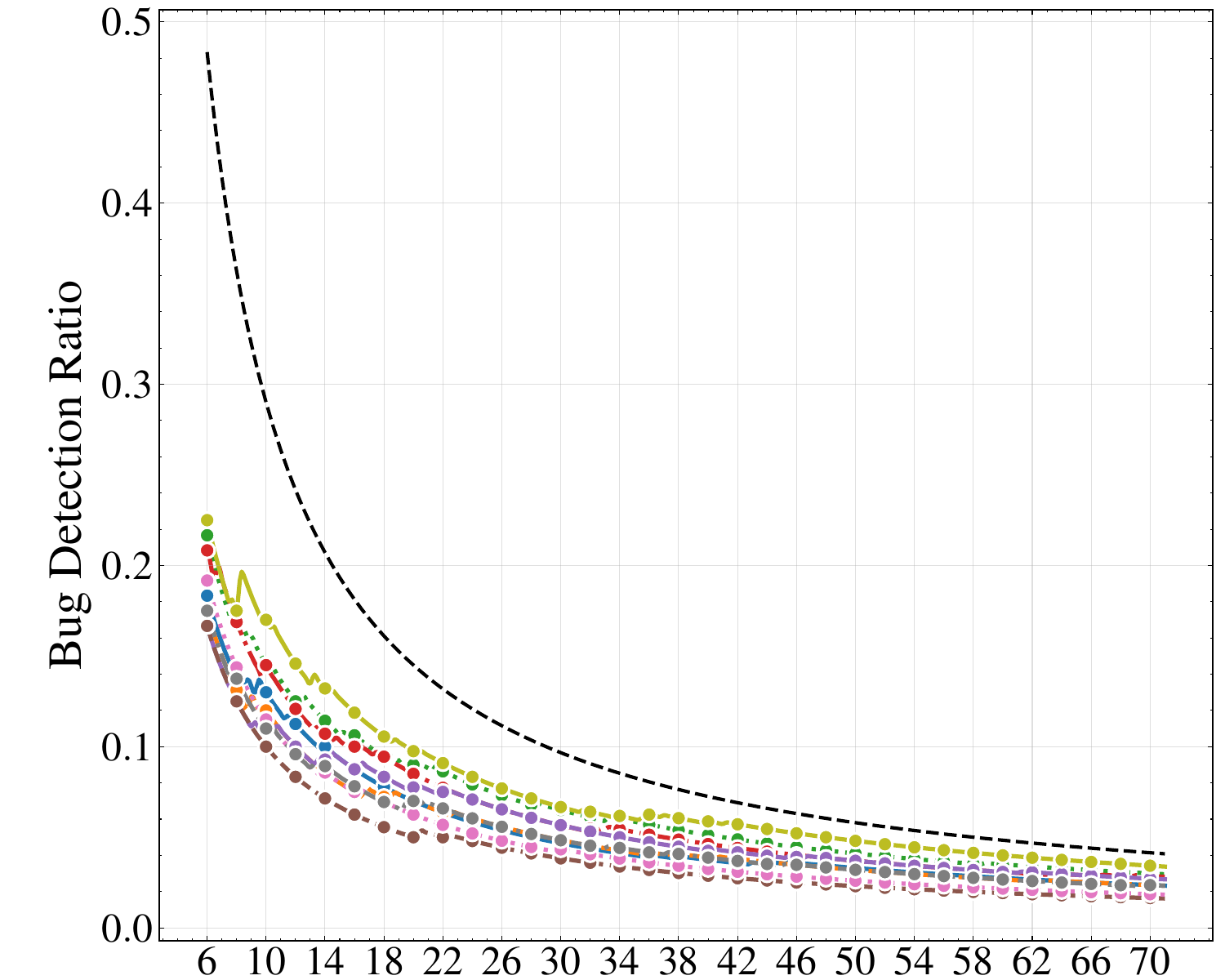} &
\includegraphics[width=0.3\textwidth]{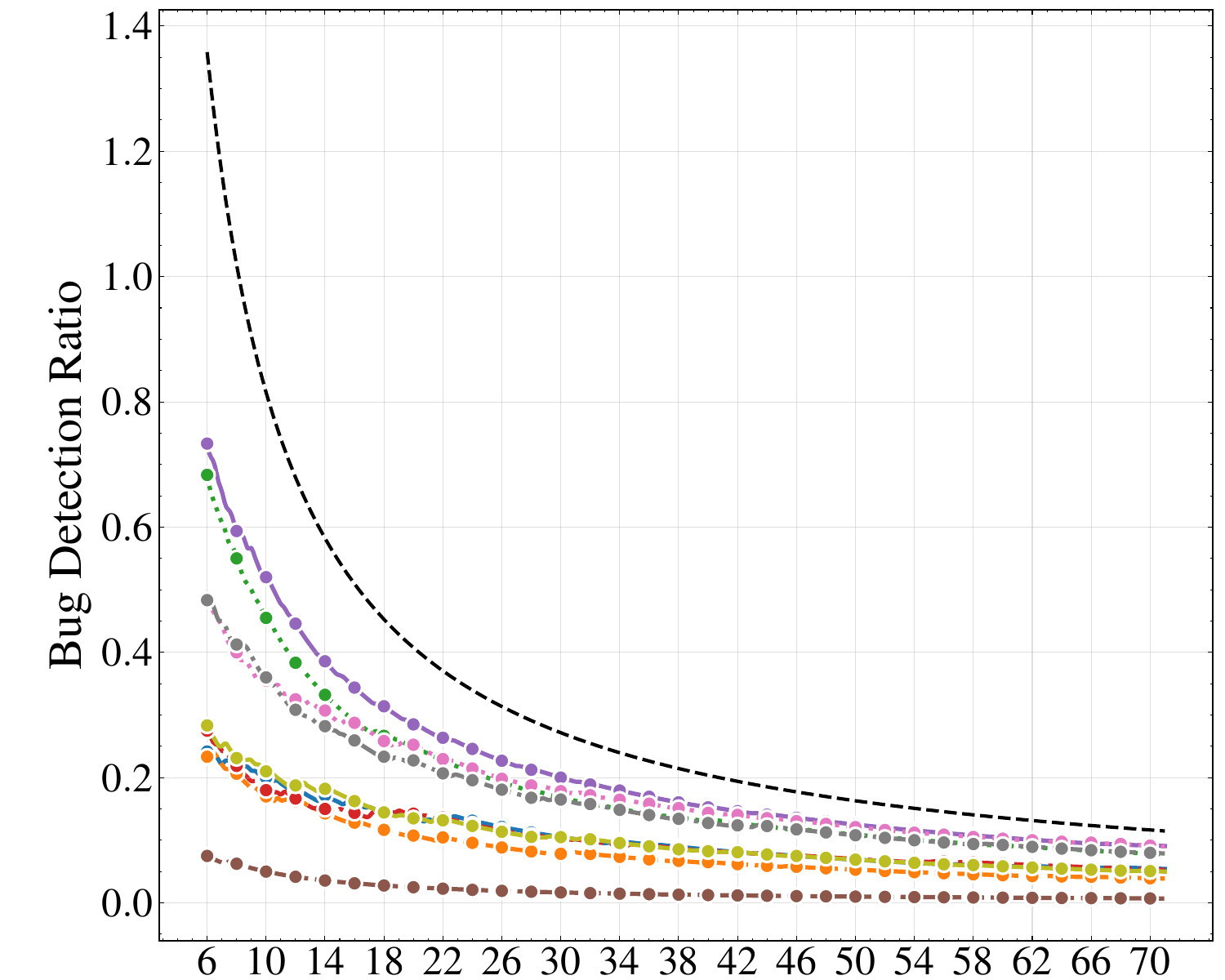} \\
(e) njs & (f) openh264 \\

\includegraphics[width=0.3\textwidth]{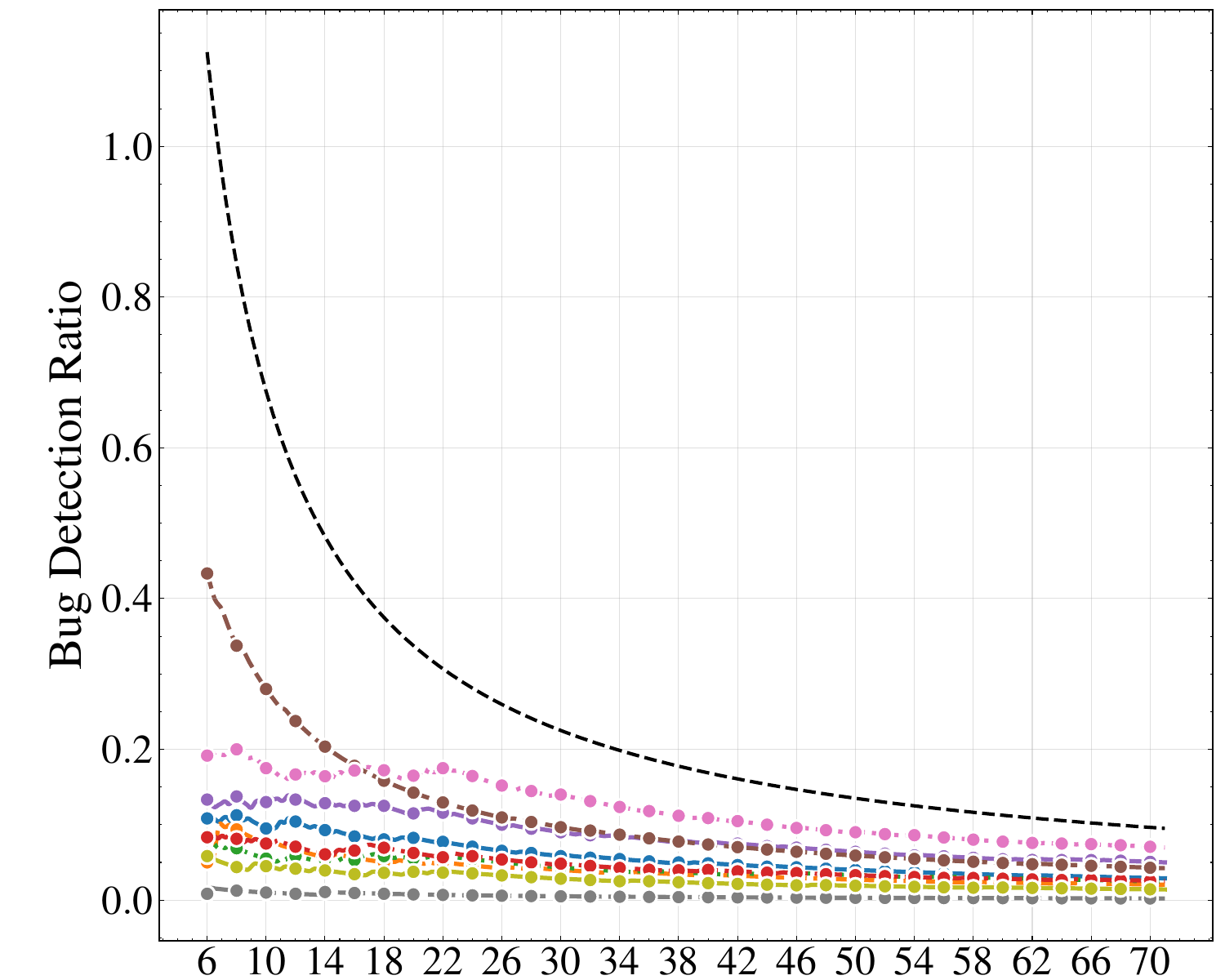} &
\includegraphics[width=0.3\textwidth]{figs/data/combined_rates_poppler_pdf_fuzzer.pdf} \\
(a) php & (b) poppler \\

\includegraphics[width=0.3\textwidth]{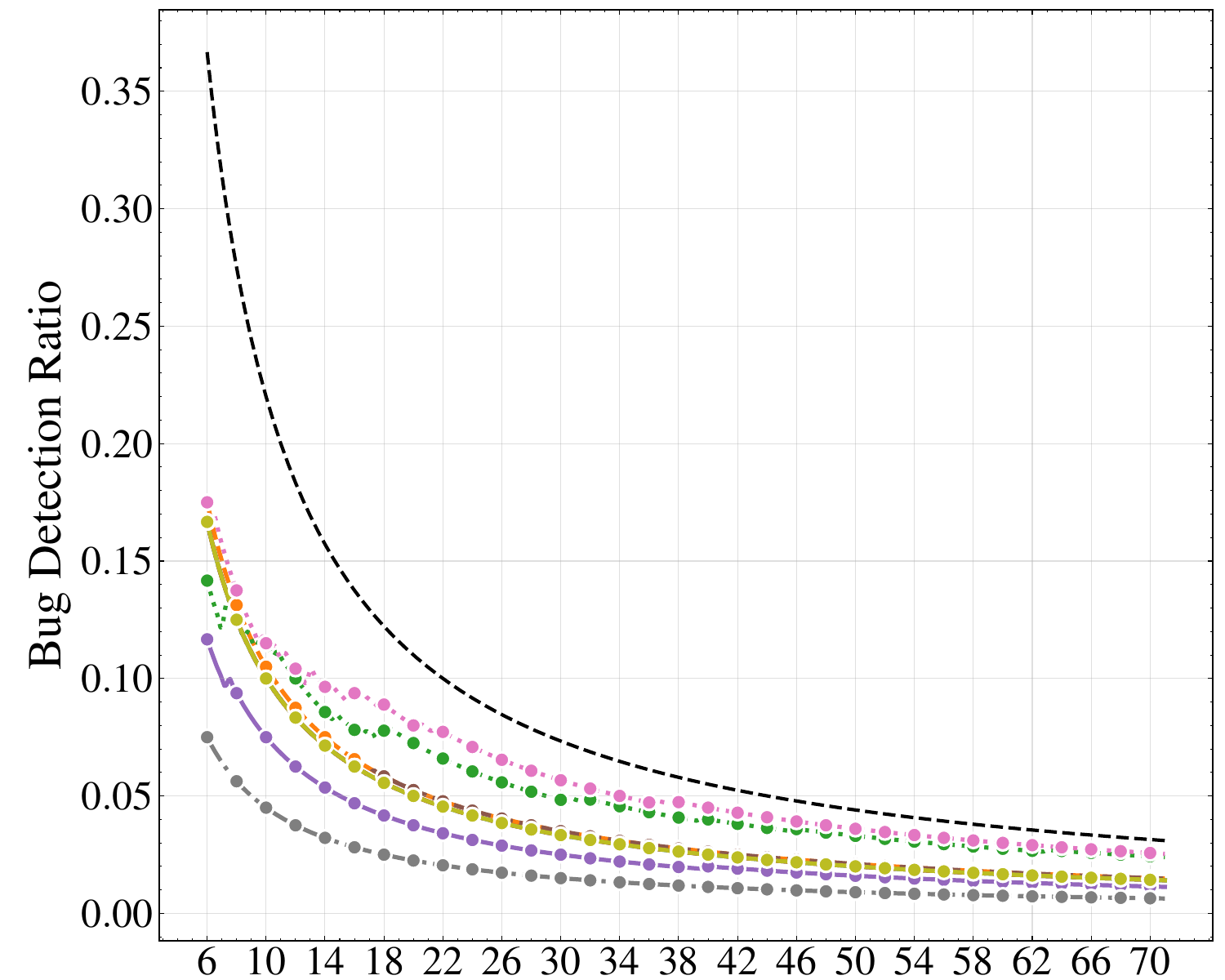} &
\includegraphics[width=0.3\textwidth]{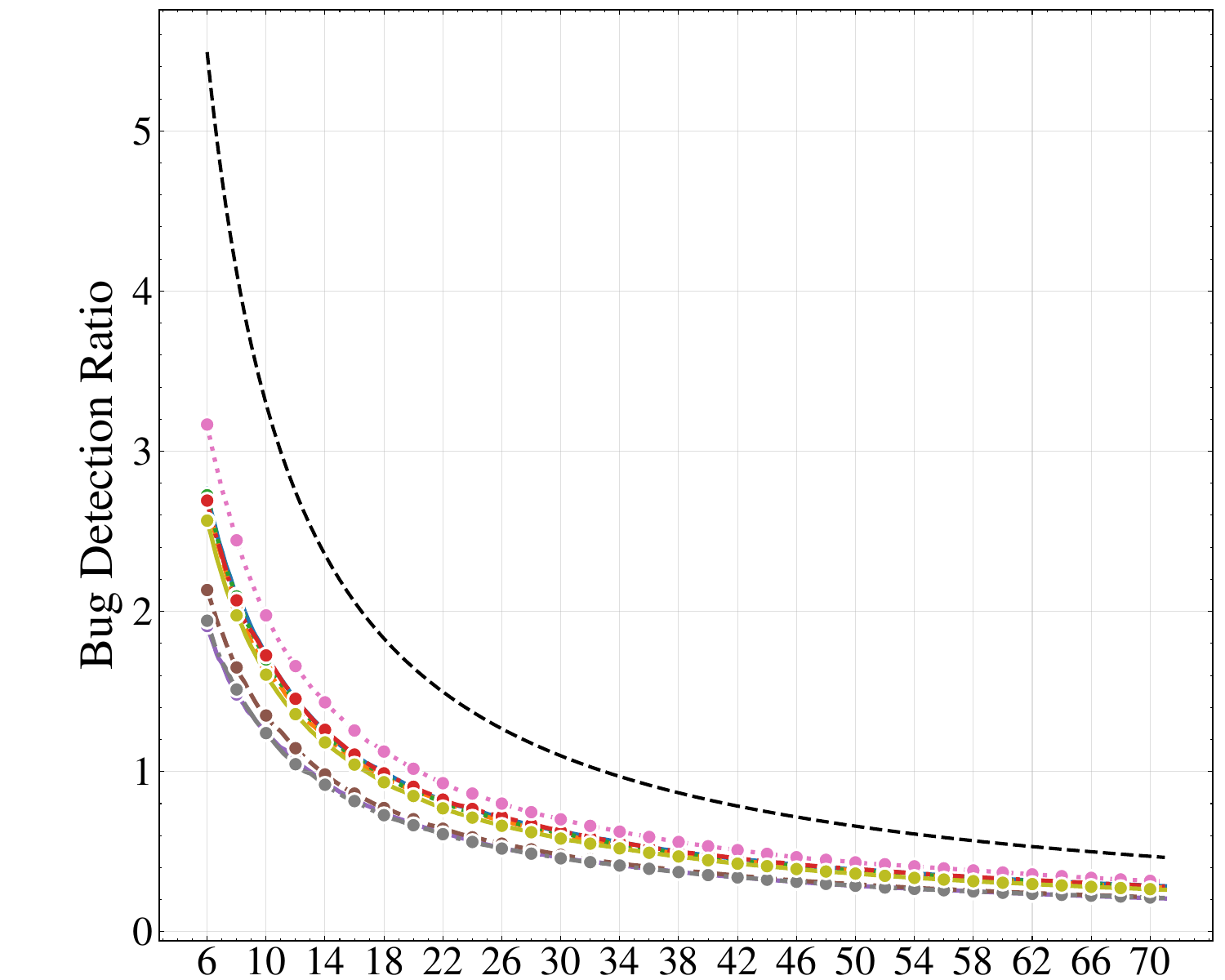} \\
(c) quickjs & (d) stb \\

\includegraphics[width=0.3\textwidth]{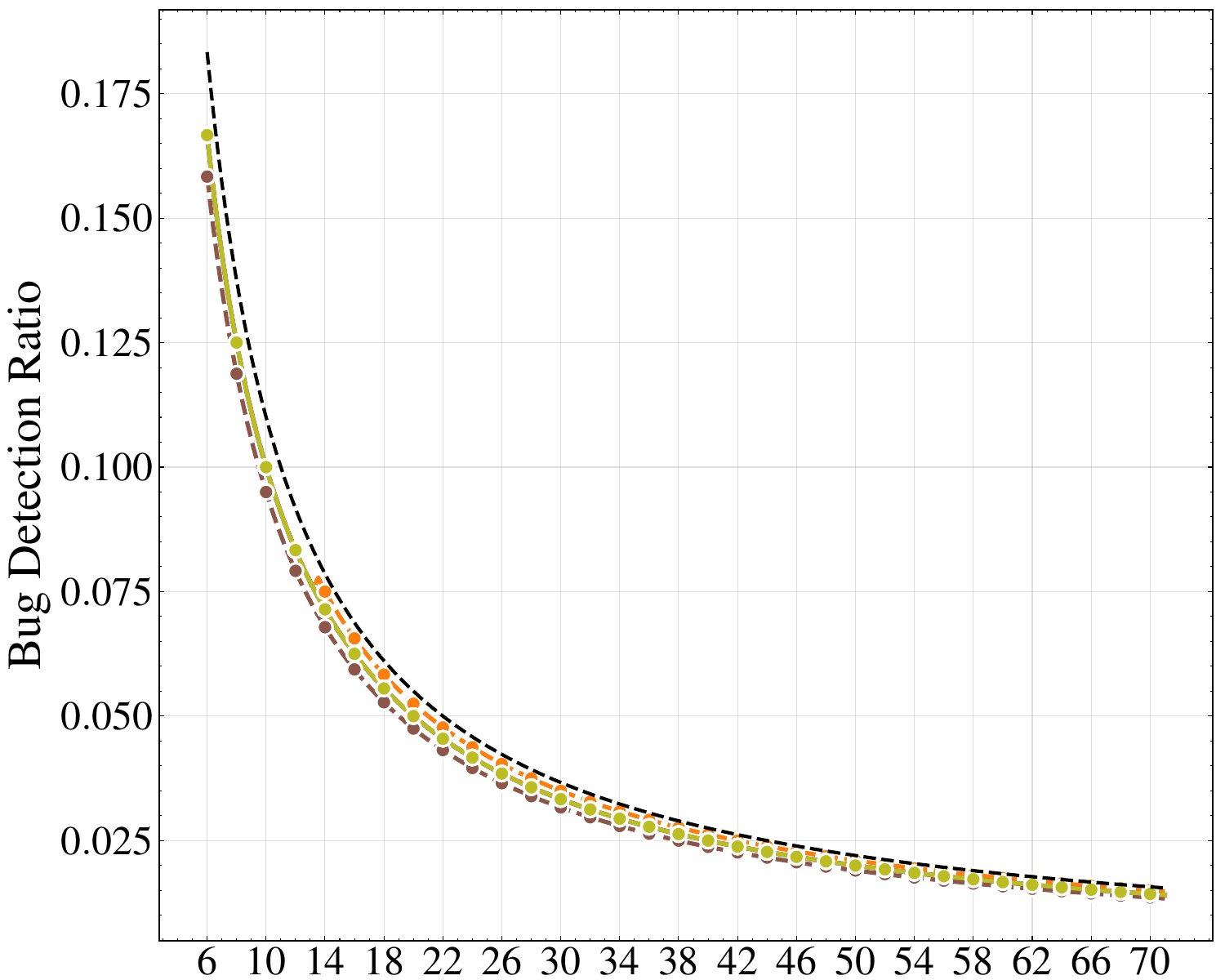} &
\includegraphics[width=0.3\textwidth]{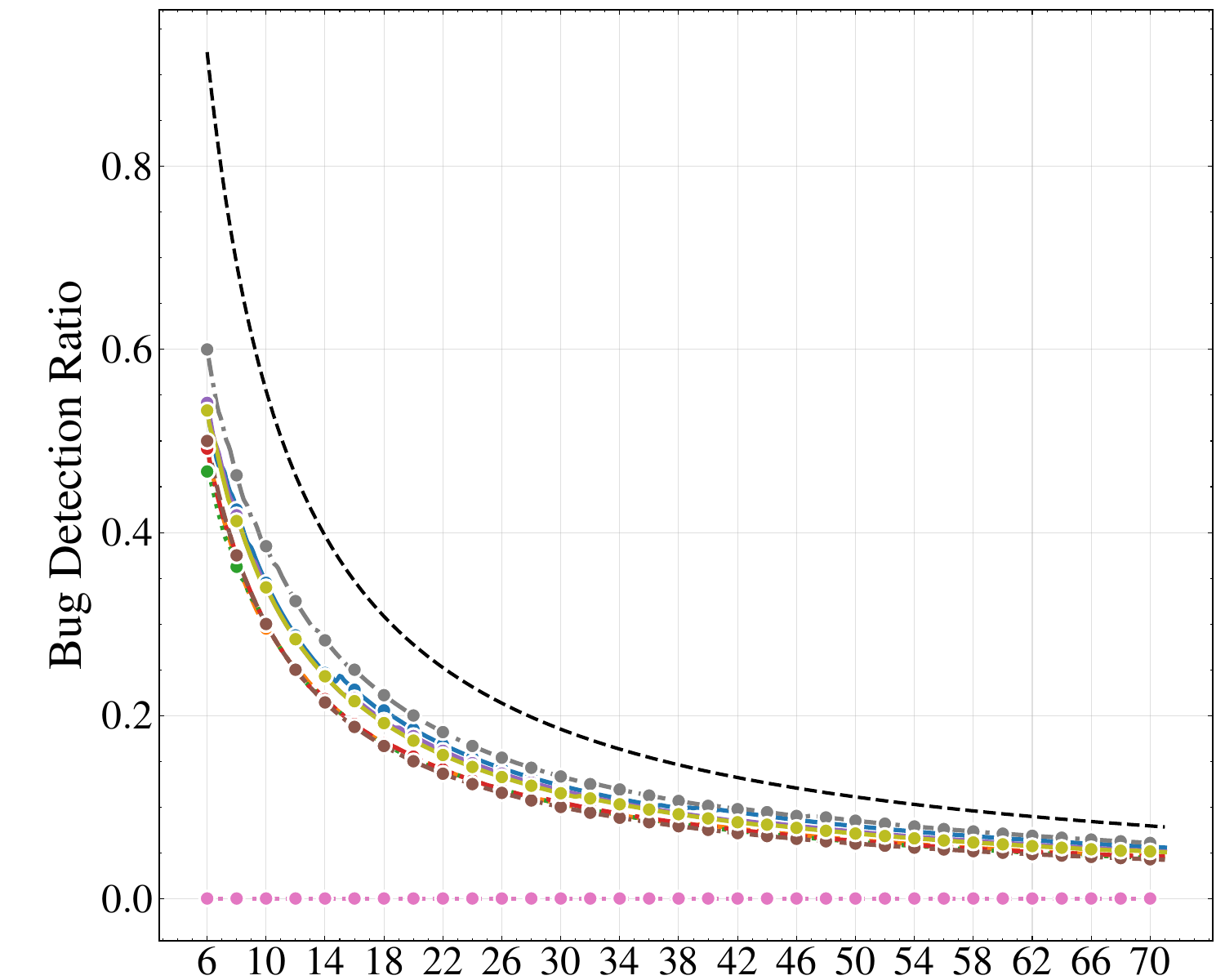} \\
(e) systemd & (f) wireshark \\
\end{tabular}
\caption{Bug detection ratio (6×2 layout).}
\label{fig:rates_6x2}
\end{figure}

\begin{figure}[H]
\centering\footnotesize
\setlength{\tabcolsep}{1pt}

\begin{tabular}{cc}
\includegraphics[width=0.3\textwidth]{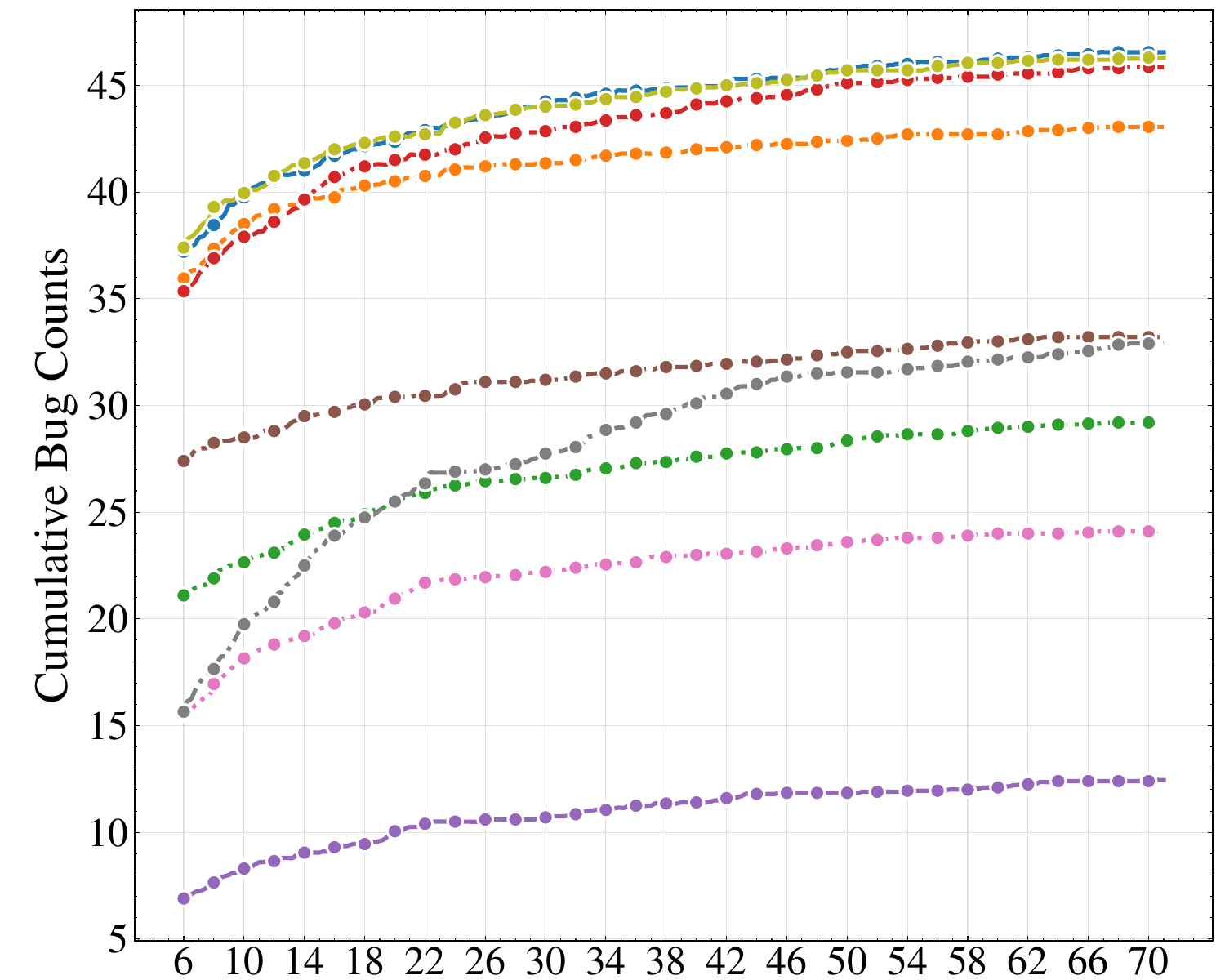} &
\includegraphics[width=0.3\textwidth]{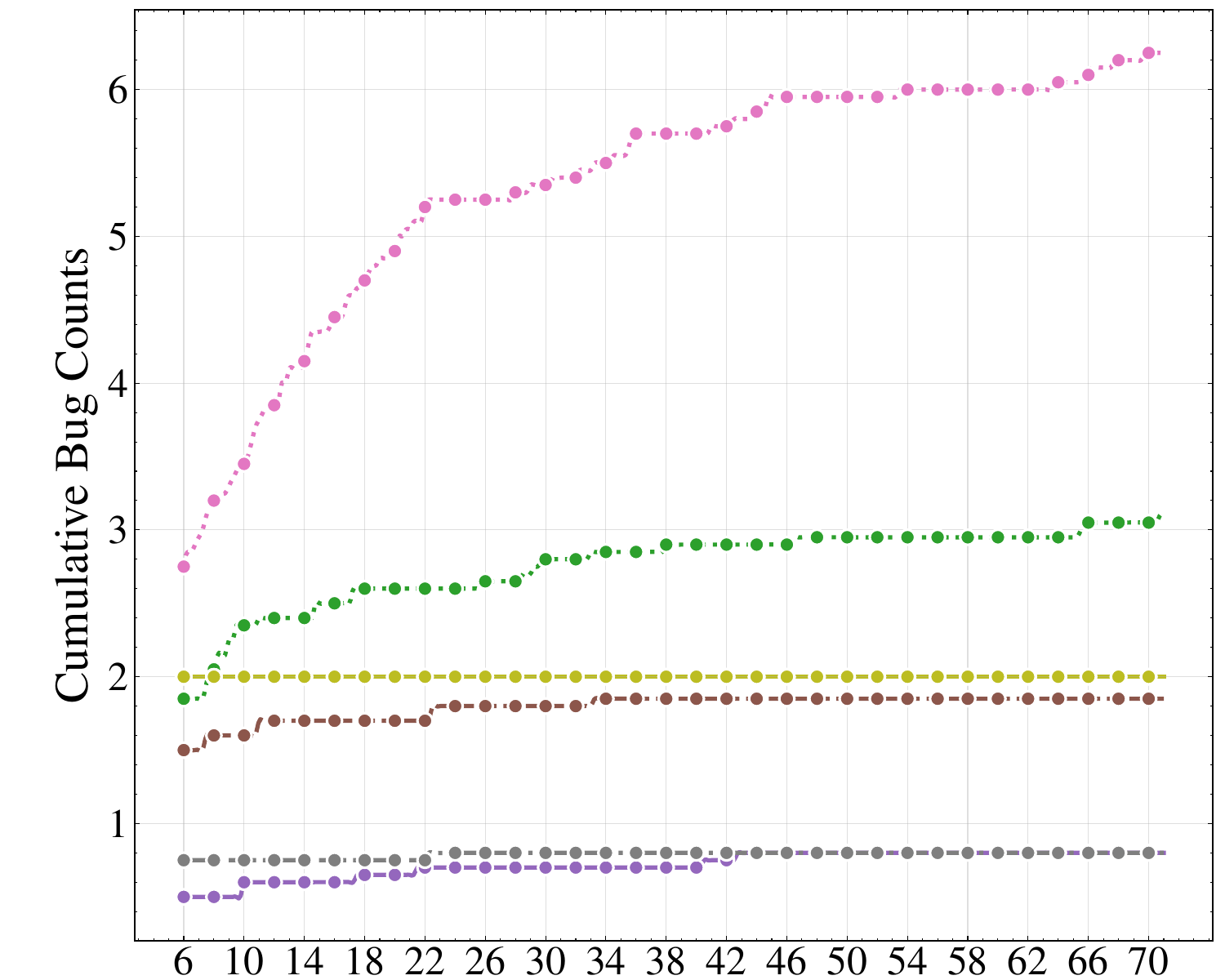} \\
(a) arrow & (b) aspell \\

\includegraphics[width=0.3\textwidth]{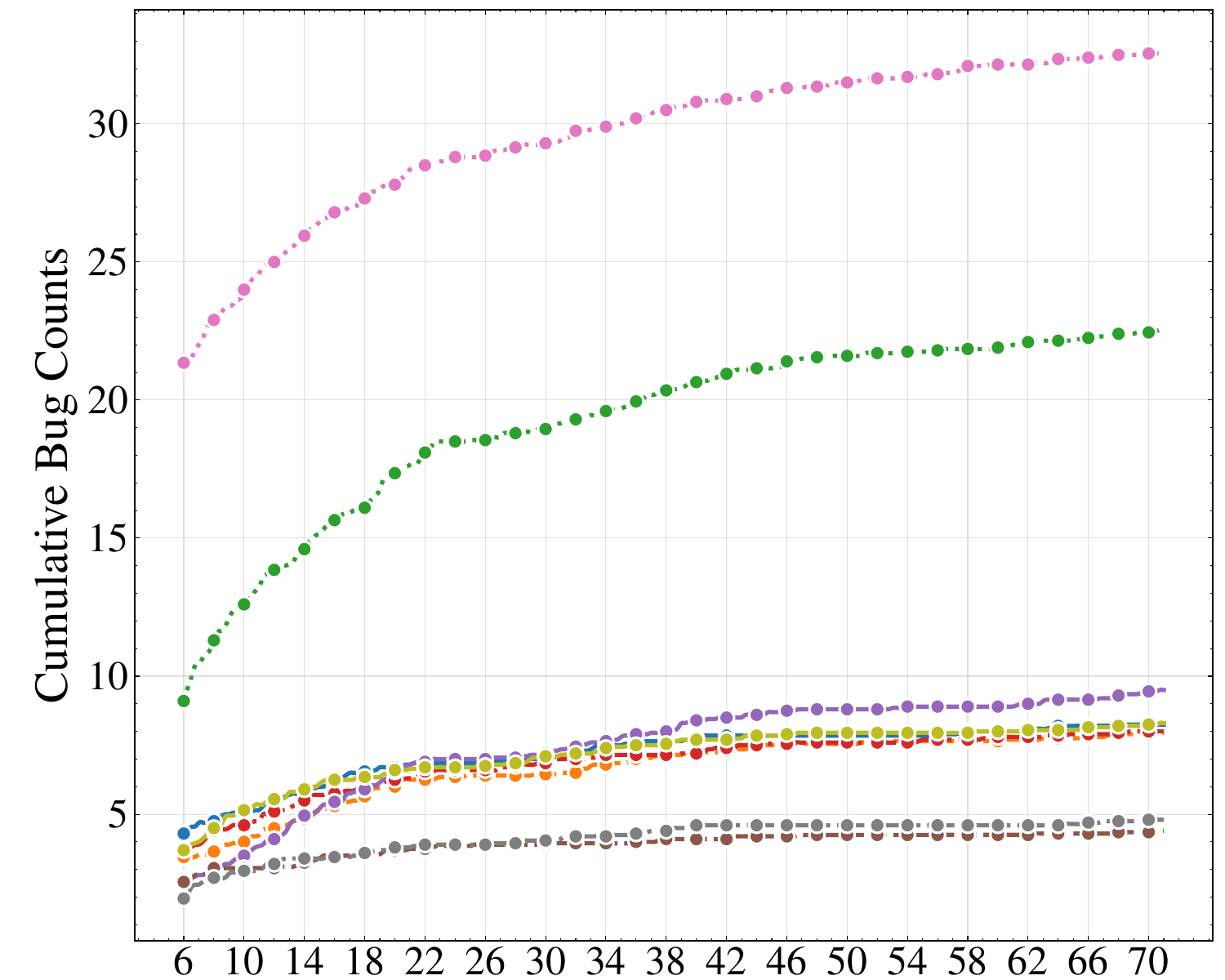} &
\includegraphics[width=0.3\textwidth]{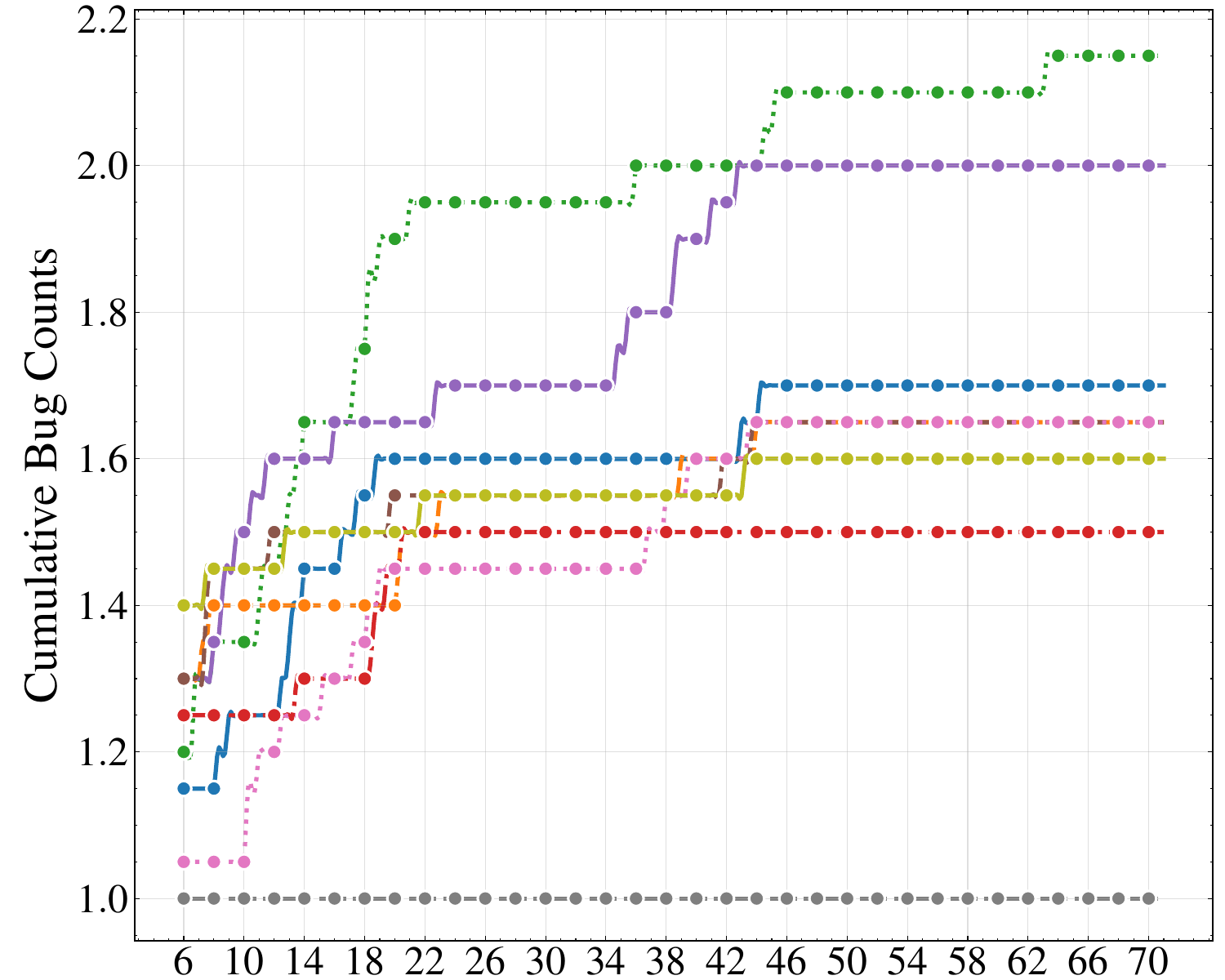} \\
(c) ffmpeg & (d) grok \\

\includegraphics[width=0.3\textwidth]{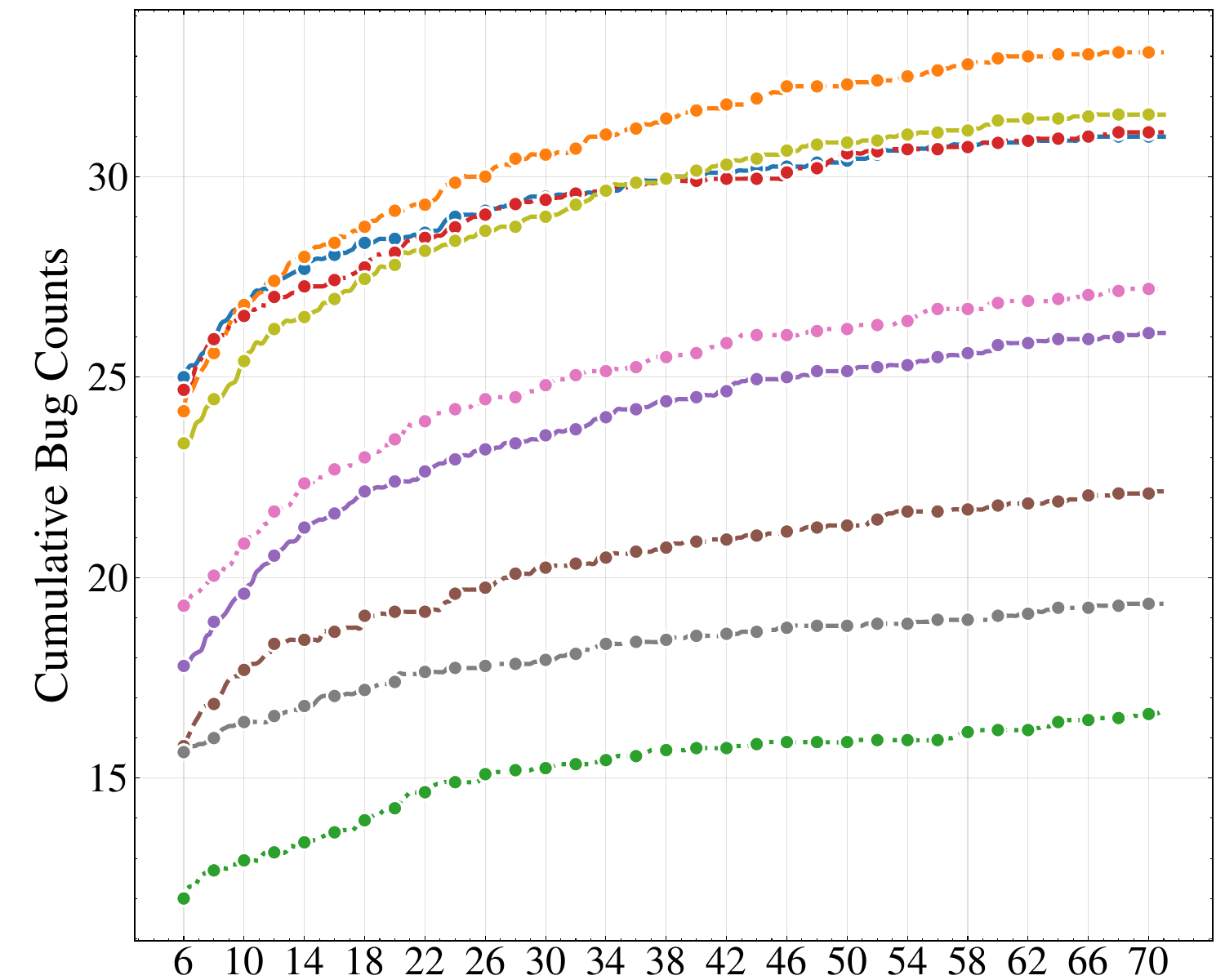} &
\includegraphics[width=0.3\textwidth]{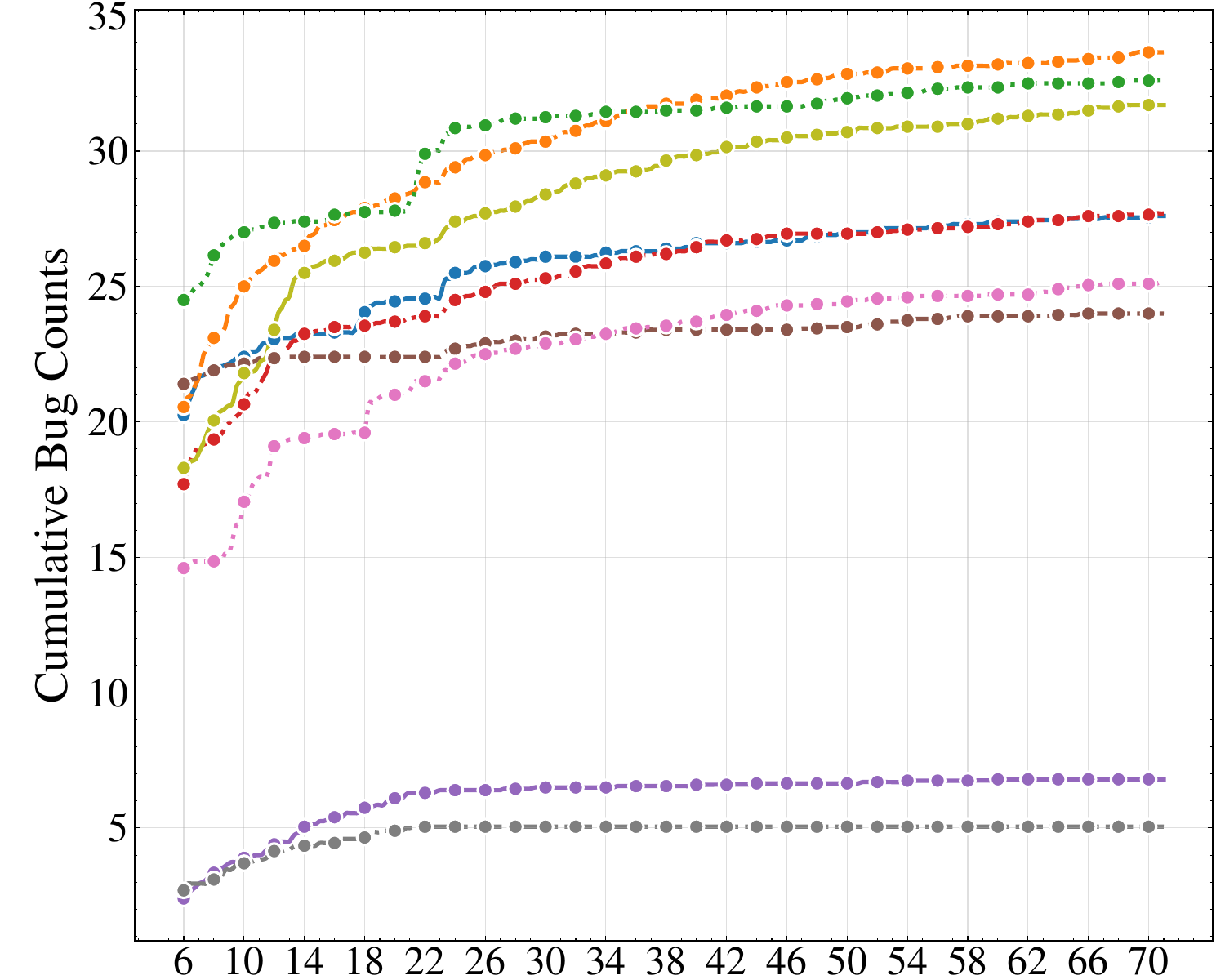} \\
(e) harfbuzz & (f) libhevc \\

\includegraphics[width=0.3\textwidth]{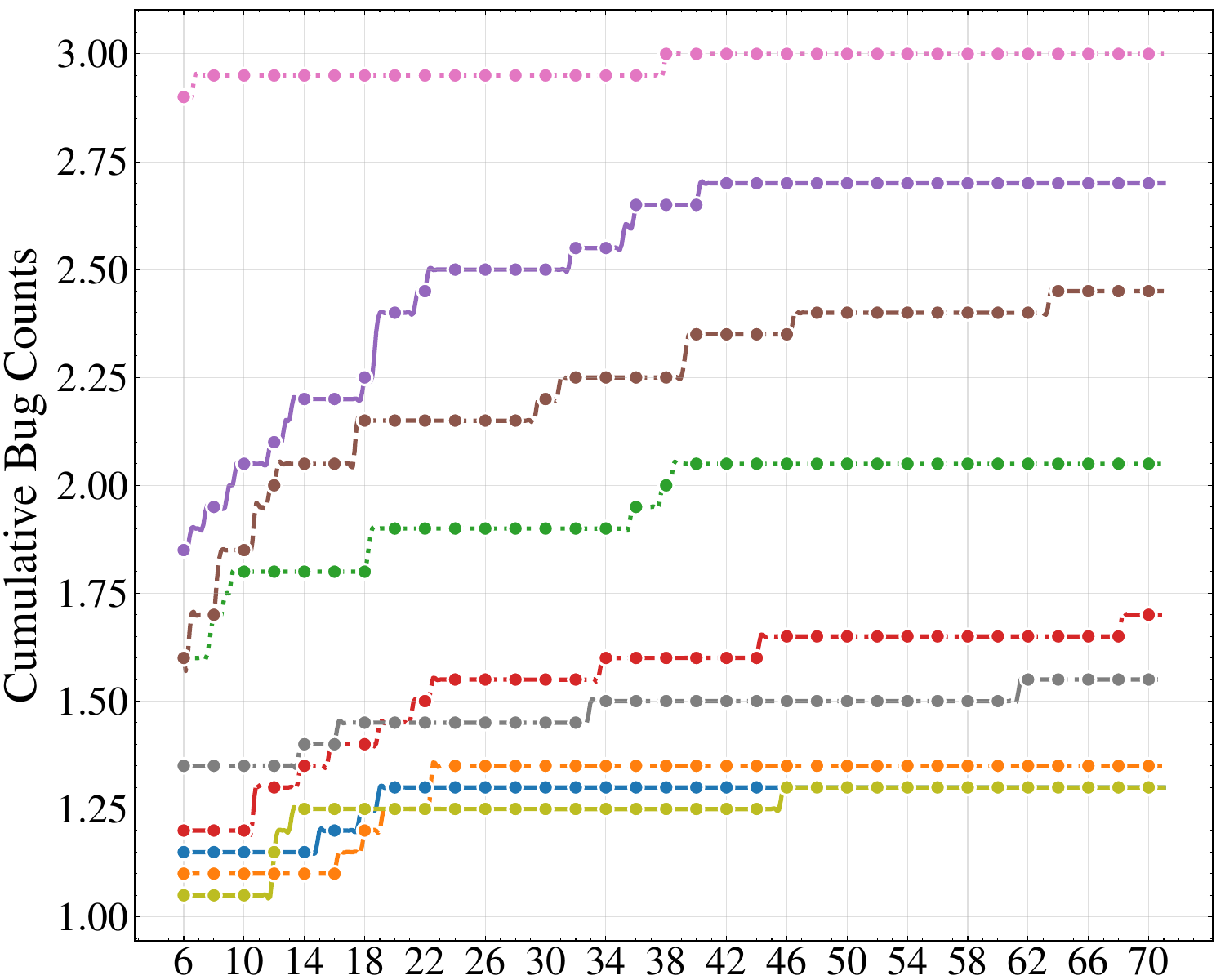} &
\includegraphics[width=0.3\textwidth]{figs/data/combined_cumulative_libhtp_fuzz_htp_06.pdf} \\
(g) libgit2 & (h) libhtp \\

\includegraphics[width=0.3\textwidth]{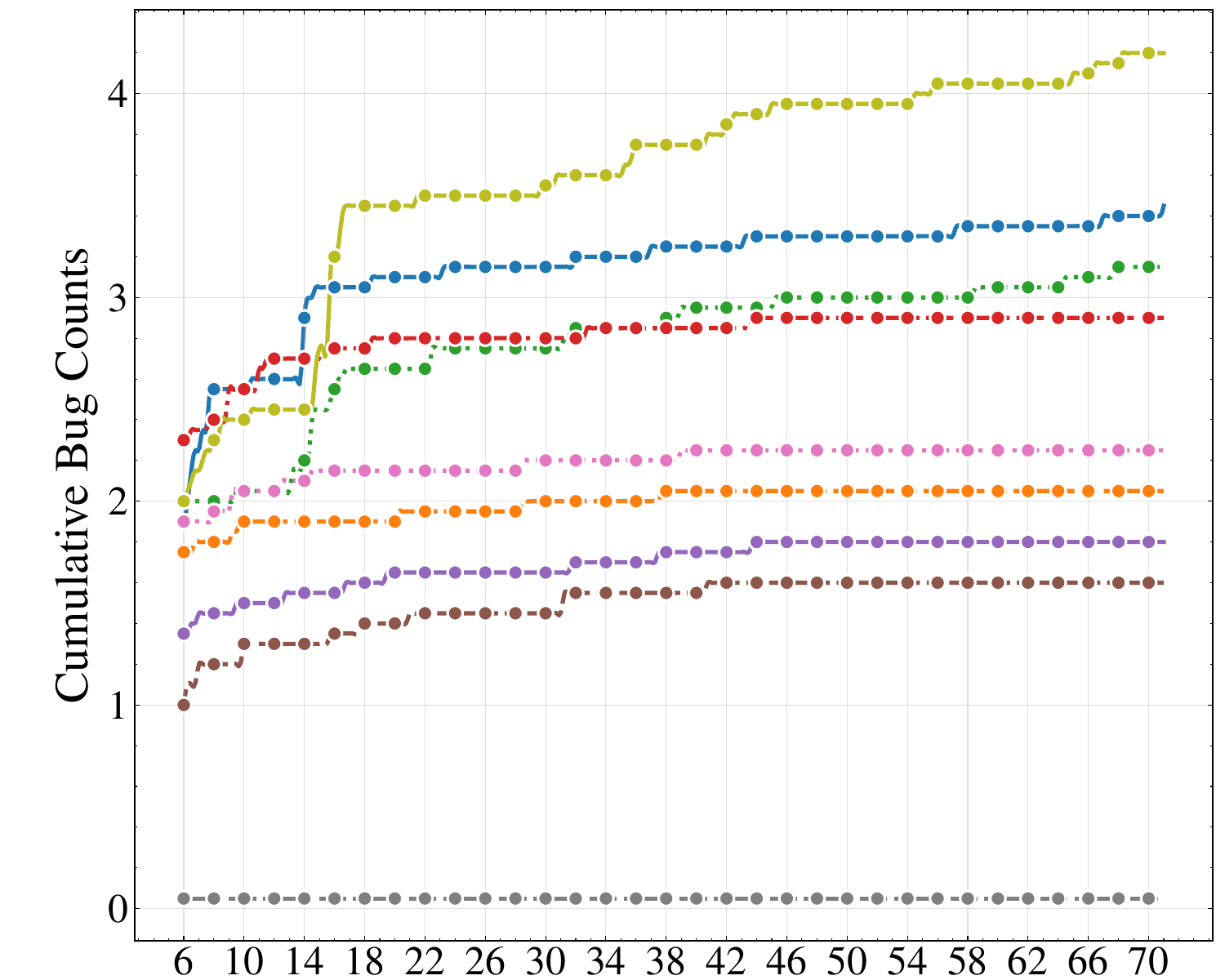} &
\includegraphics[width=0.3\textwidth]{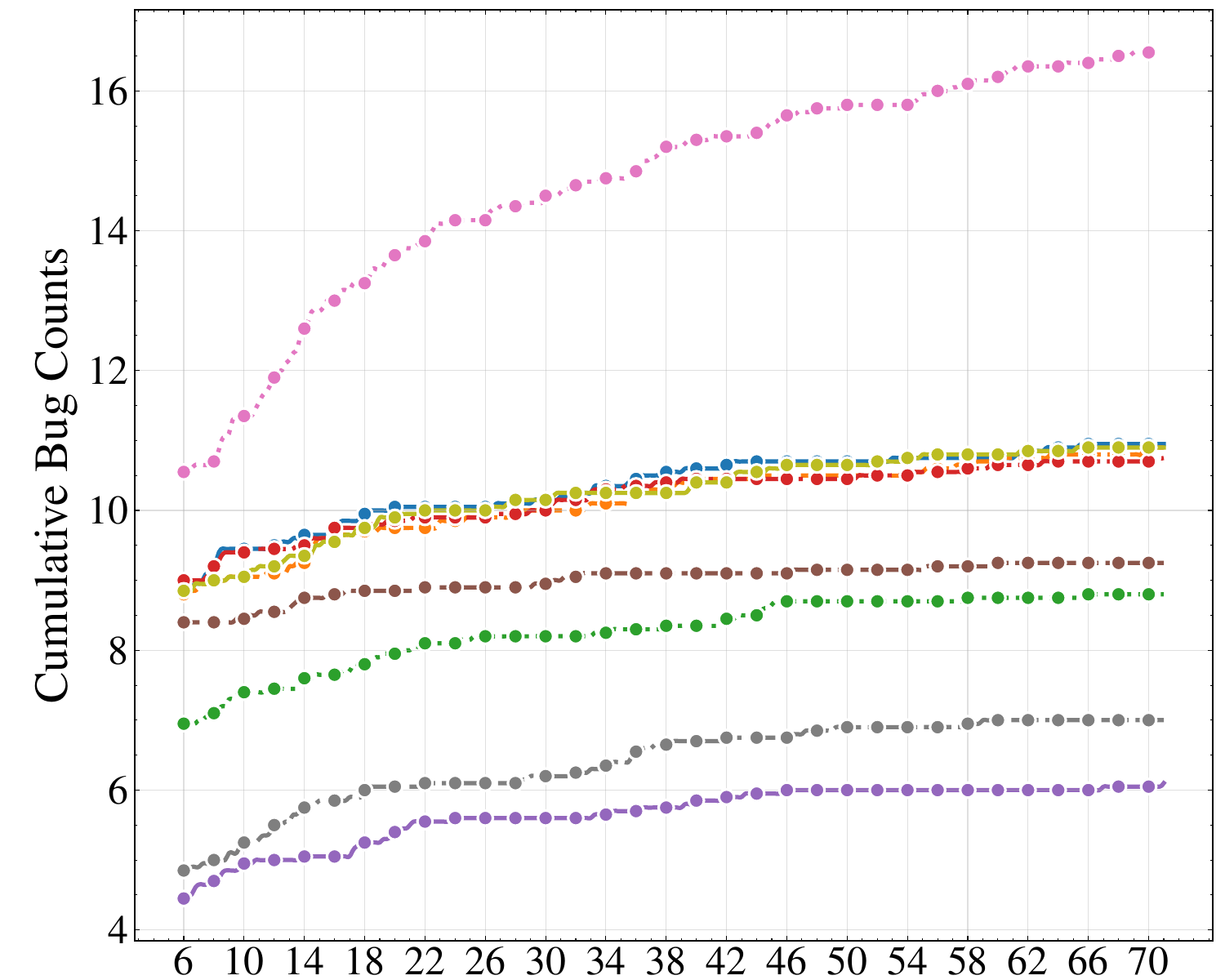} \\
(i) libxml2 & (j) matio \\

\includegraphics[width=0.3\textwidth]{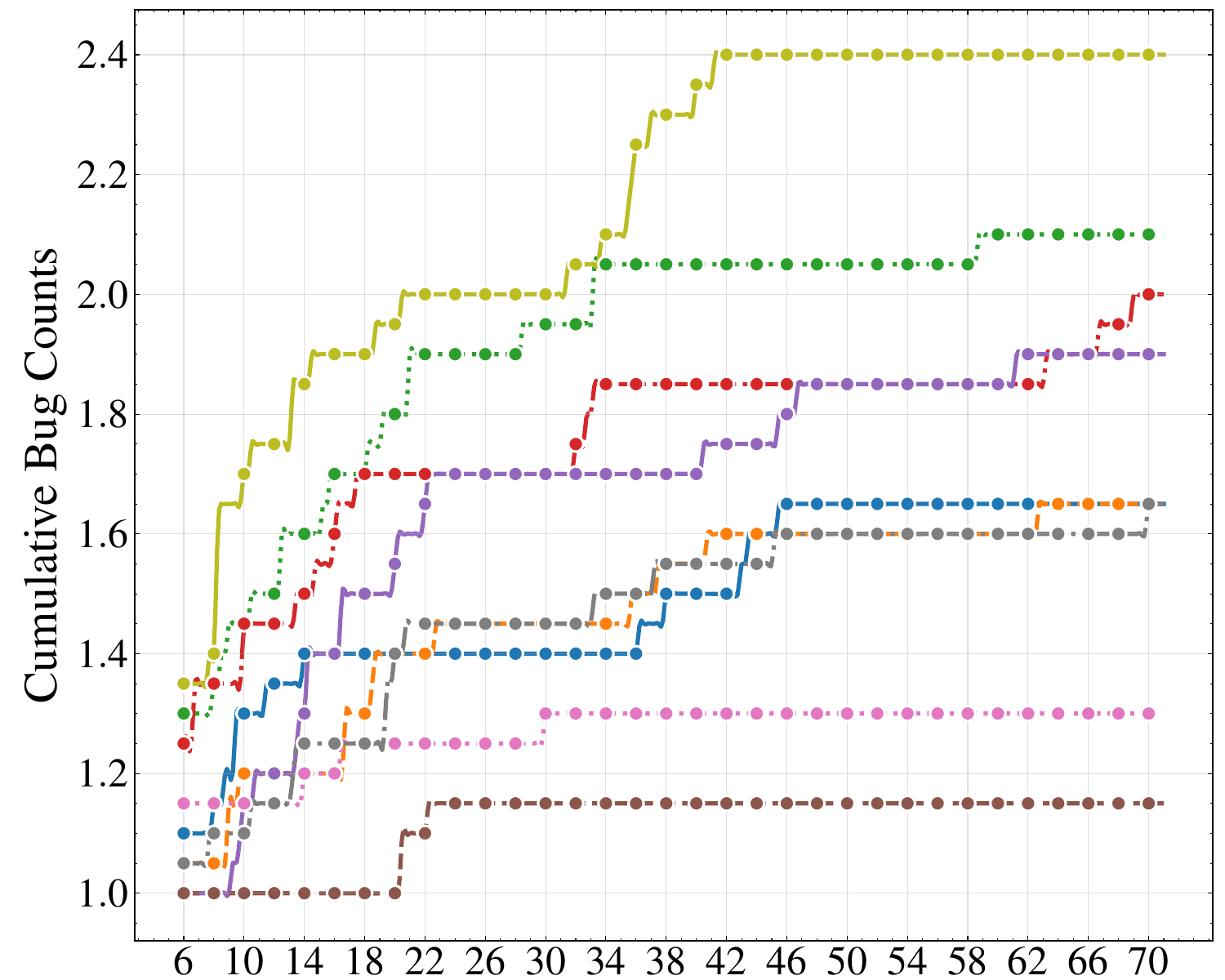} &
\includegraphics[width=0.3\textwidth]{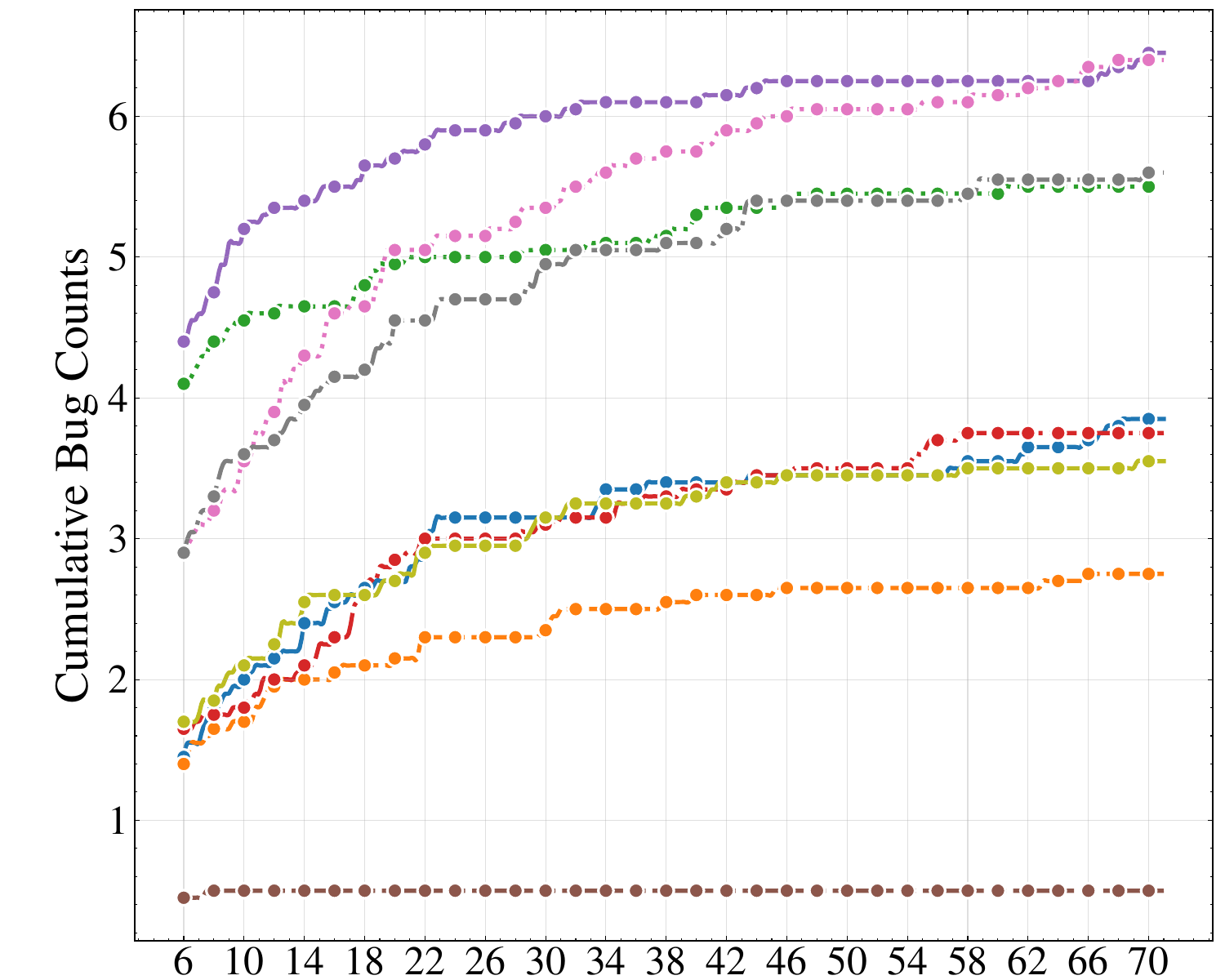} \\
(k) njs & (l) openh264 \\
\end{tabular}

\caption{Bug counts  (6×2 layout). }
\label{fig:cumulative_6x2_final}
\end{figure}

\begin{figure}[H]
\centering\footnotesize
\setlength{\tabcolsep}{1pt}

\begin{tabular}{cc}

\includegraphics[width=0.3\textwidth]{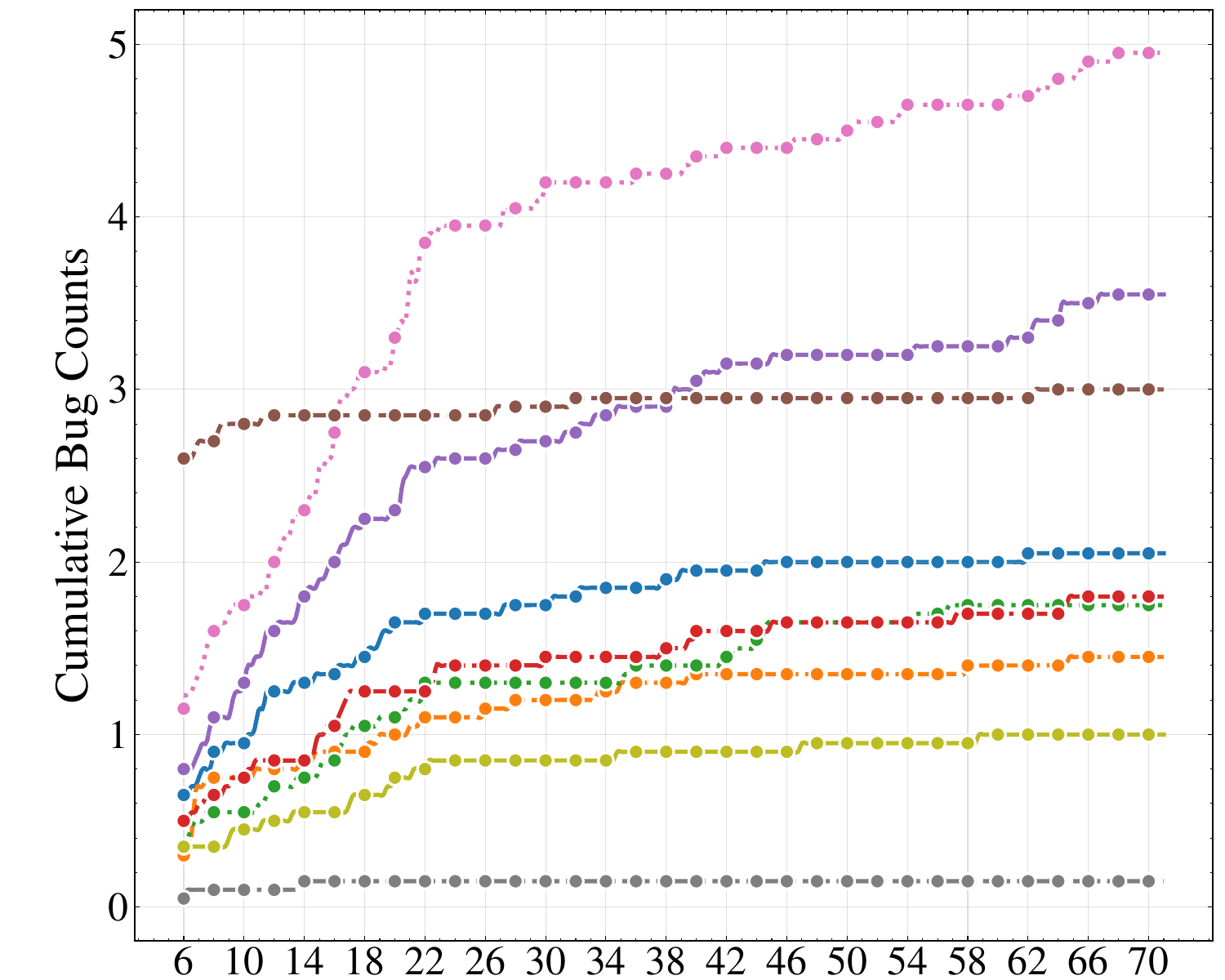} &
\includegraphics[width=0.3\textwidth]{figs/data/combined_cumulative_poppler_pdf_fuzzer_06.pdf} \\
(a) php & (b) poppler \\

\includegraphics[width=0.3\textwidth]{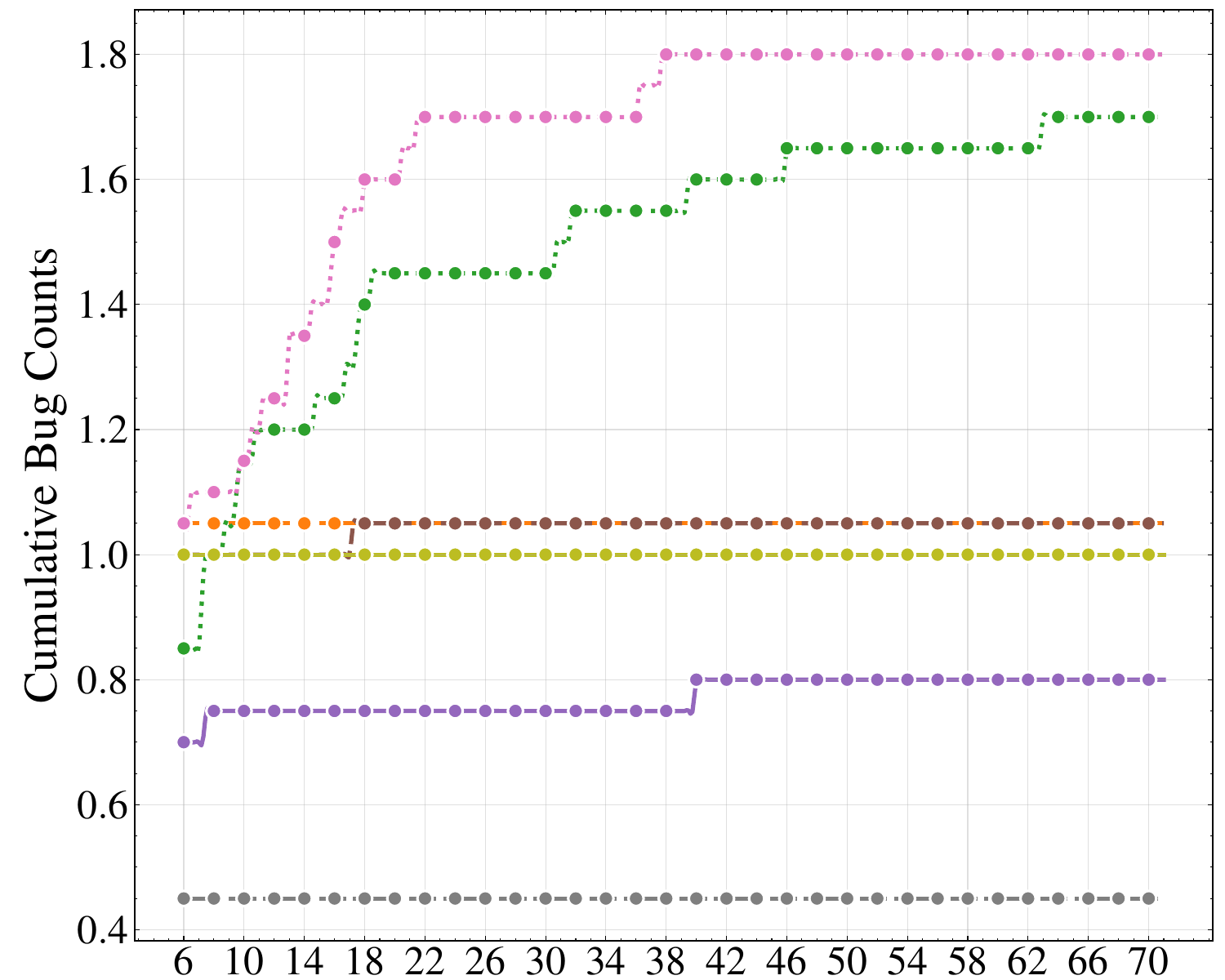} &
\includegraphics[width=0.3\textwidth]{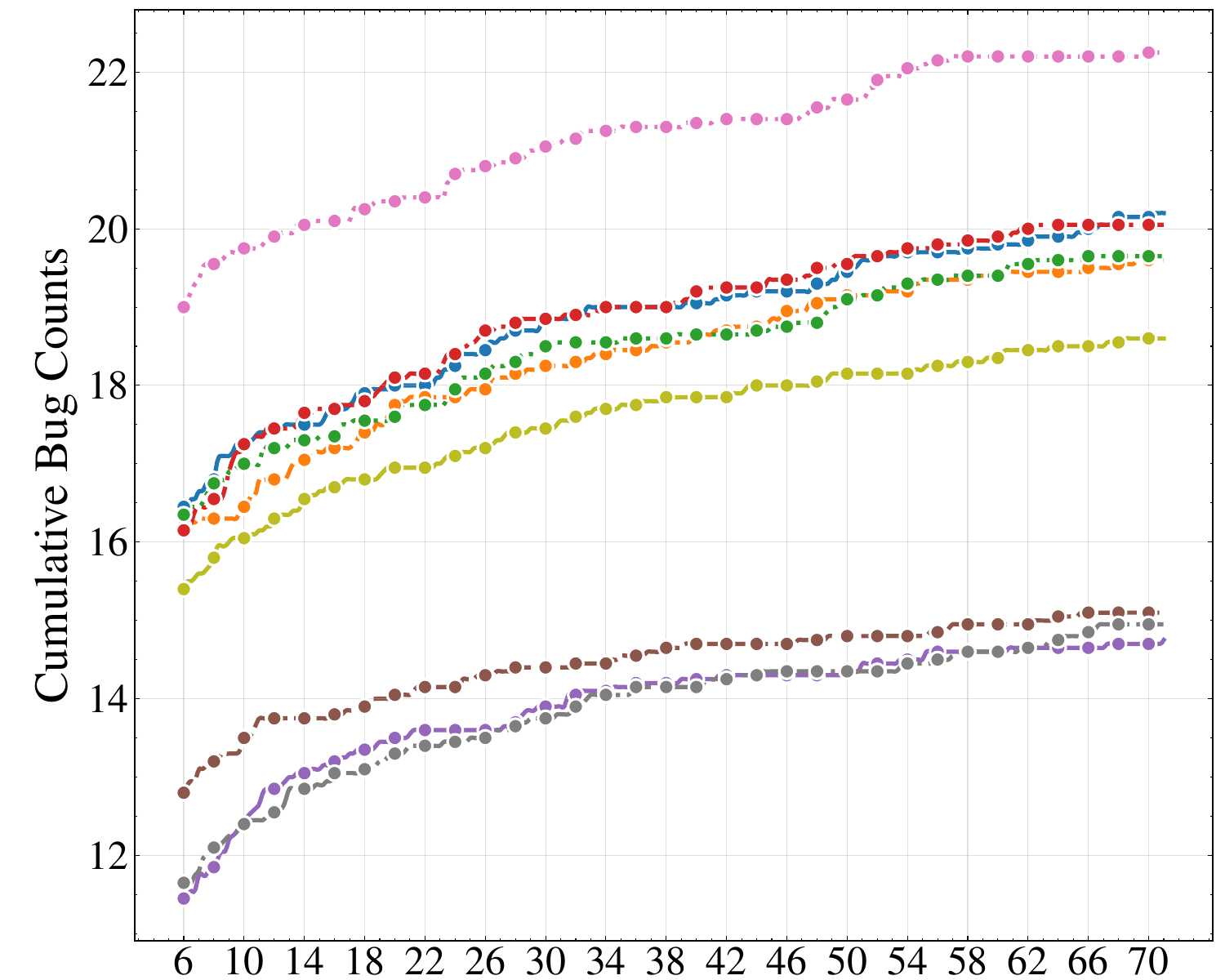} \\
(c) quickjs & (d) stb \\

\includegraphics[width=0.3\textwidth]{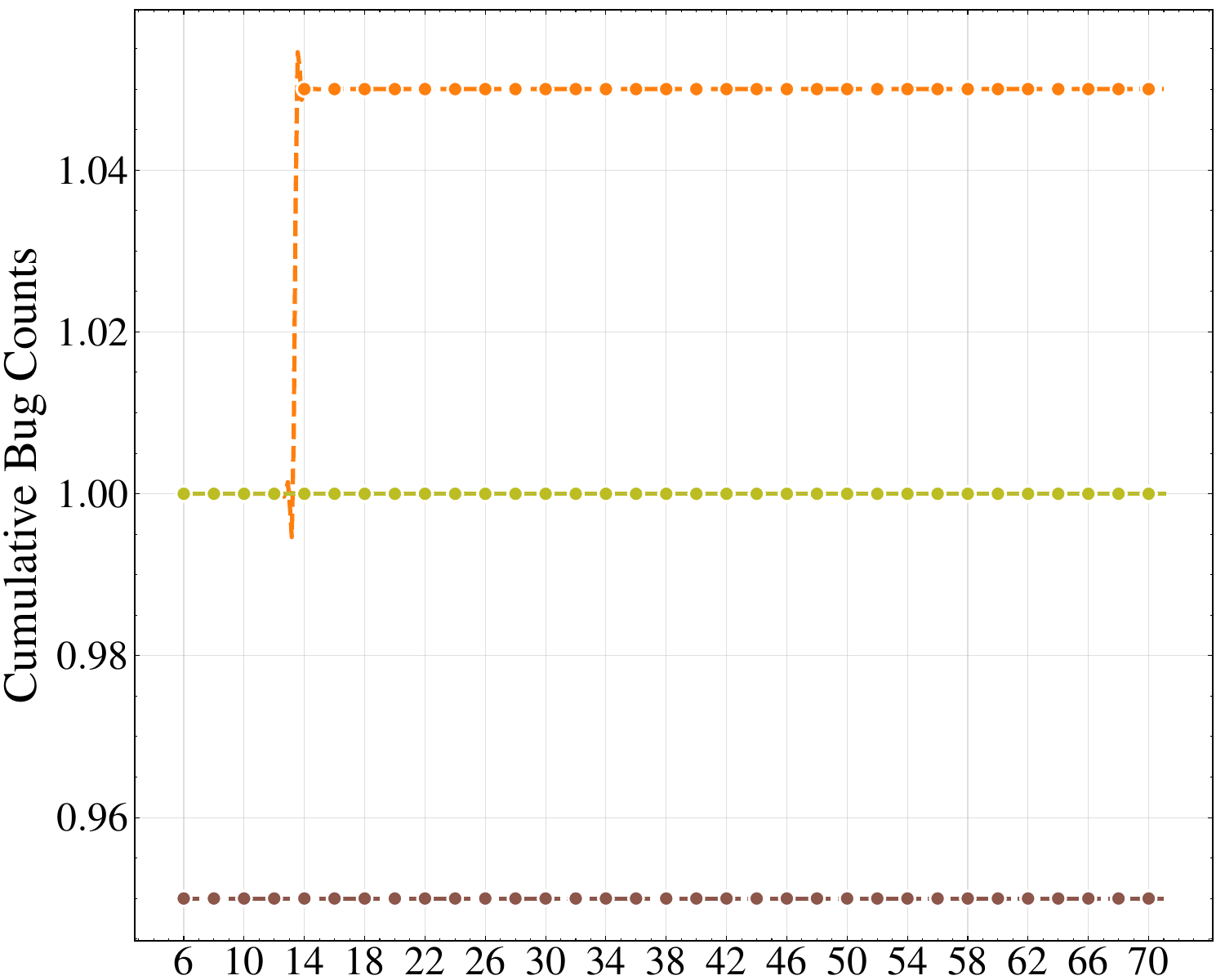} &
\includegraphics[width=0.3\textwidth]{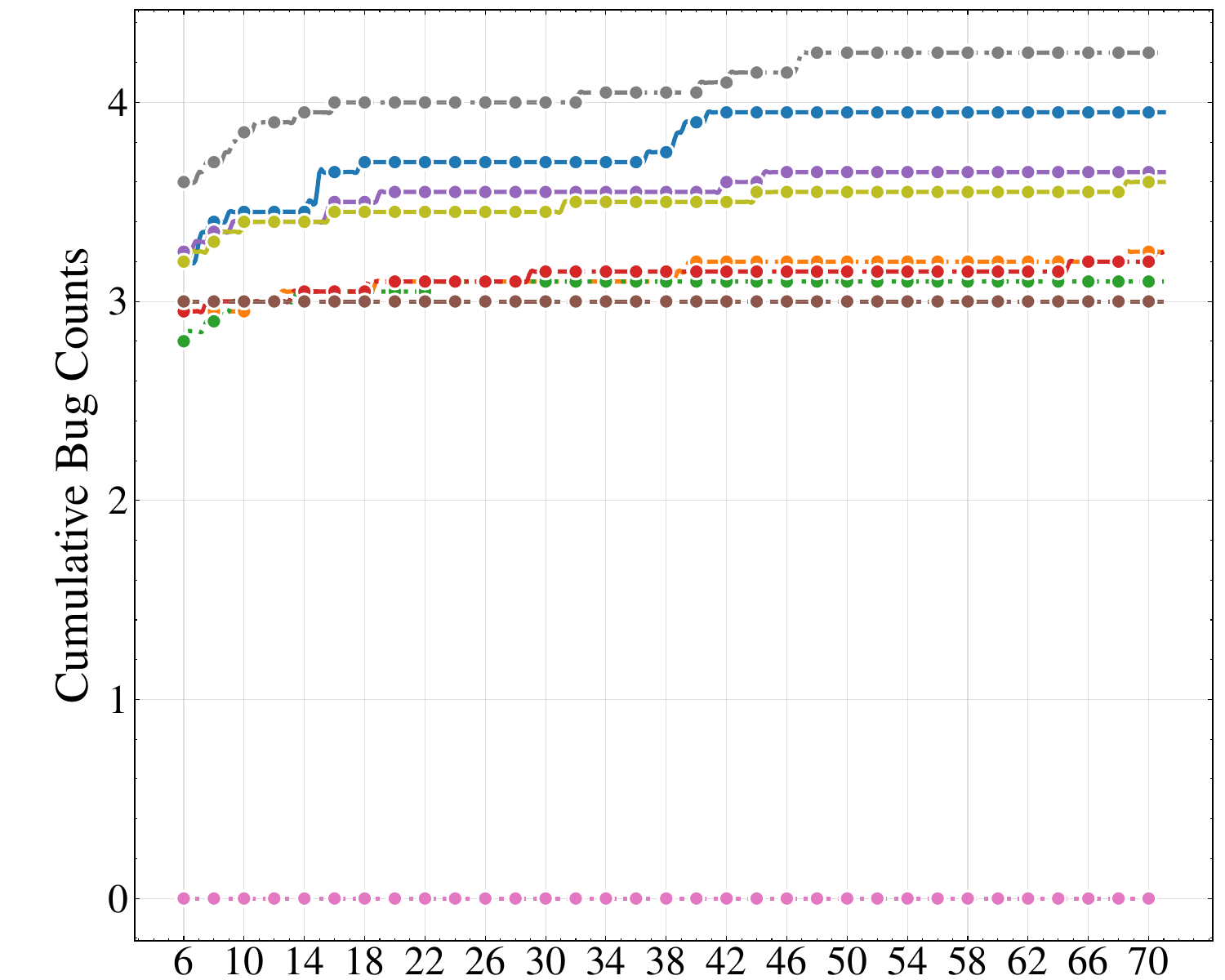} \\
(e) systemd & (f) wireshark \\

\includegraphics[width=0.3\textwidth]{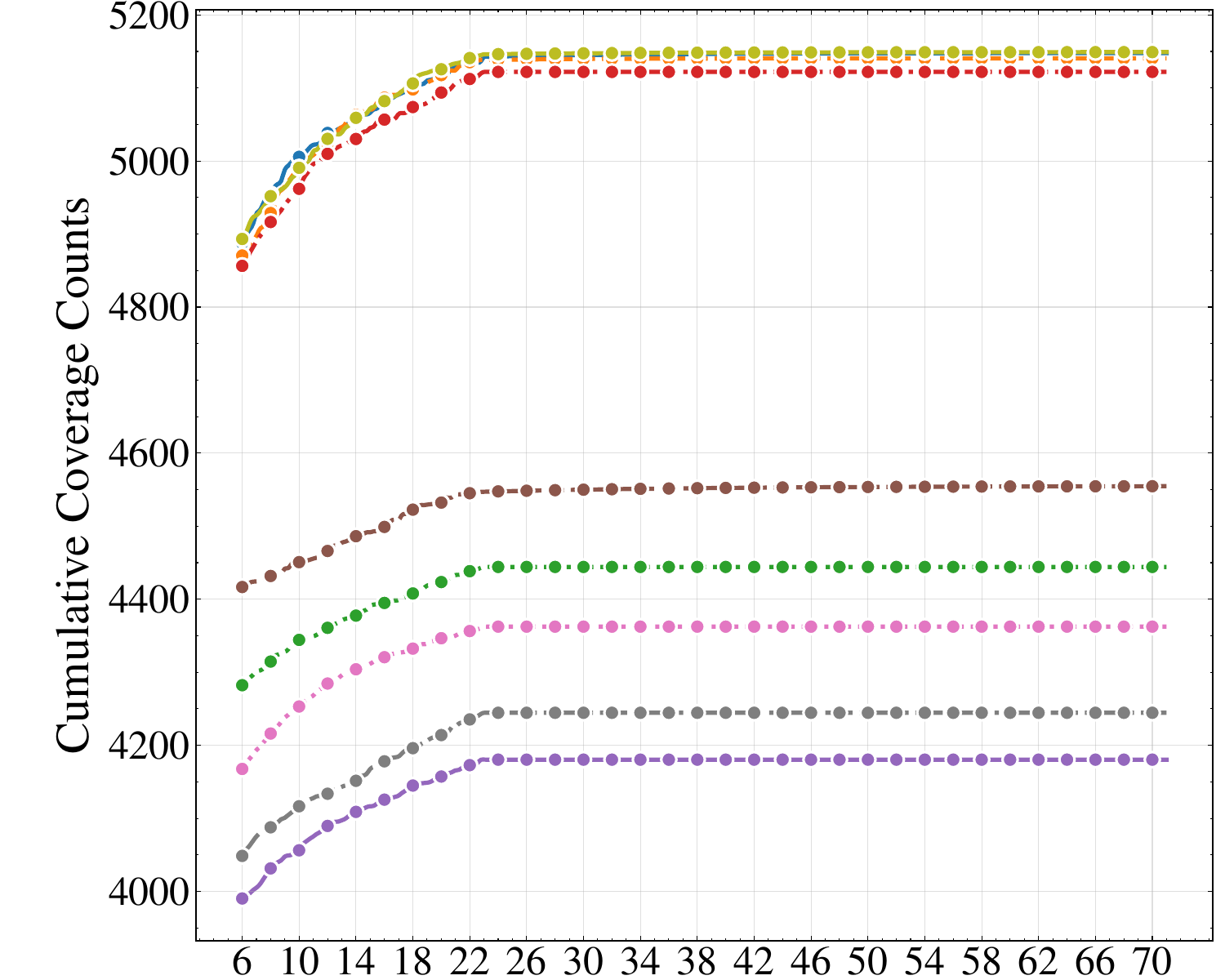} &
\includegraphics[width=0.3\textwidth]{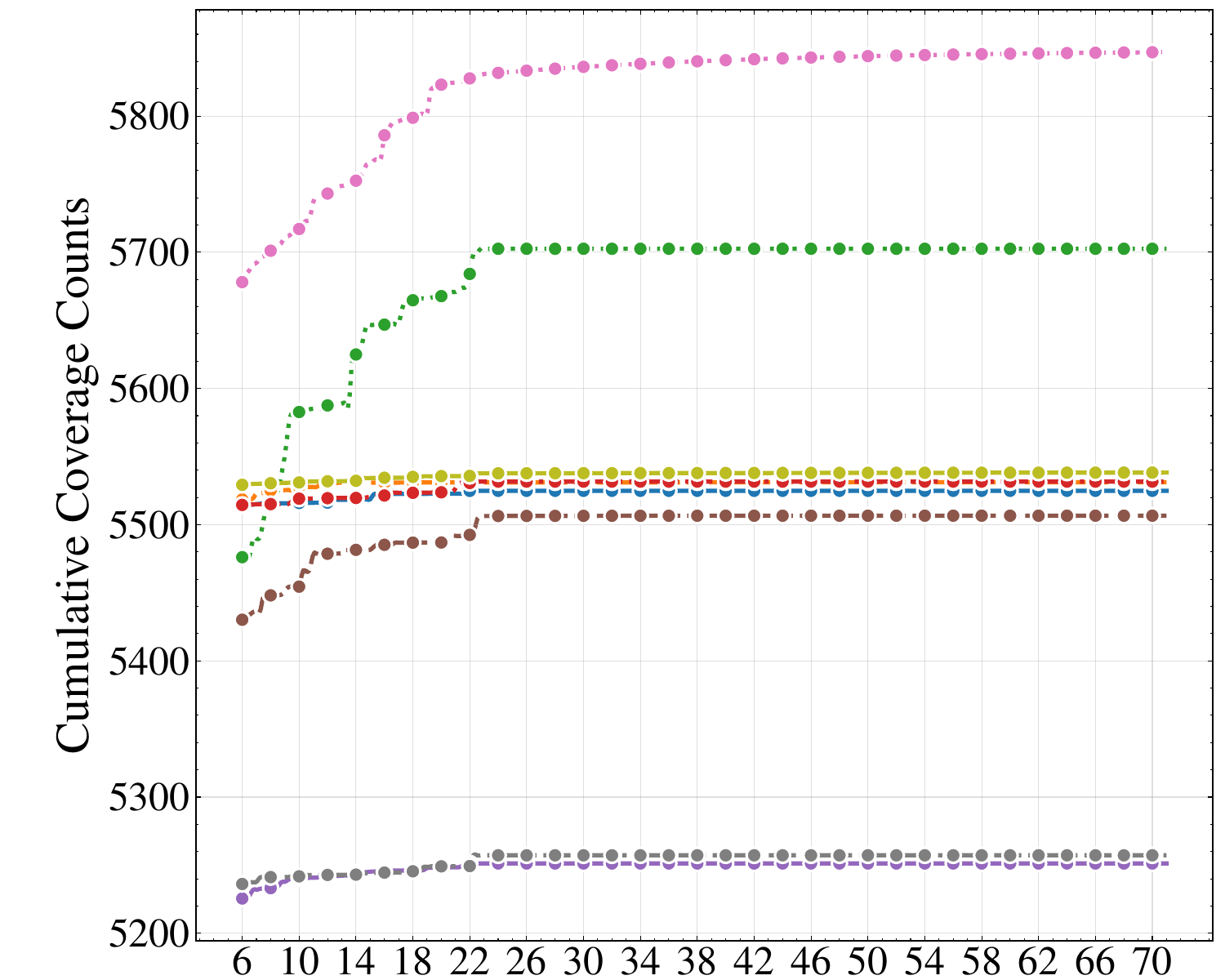} \\
(g) arrow & (h) aspell \\

\includegraphics[width=0.3\textwidth]{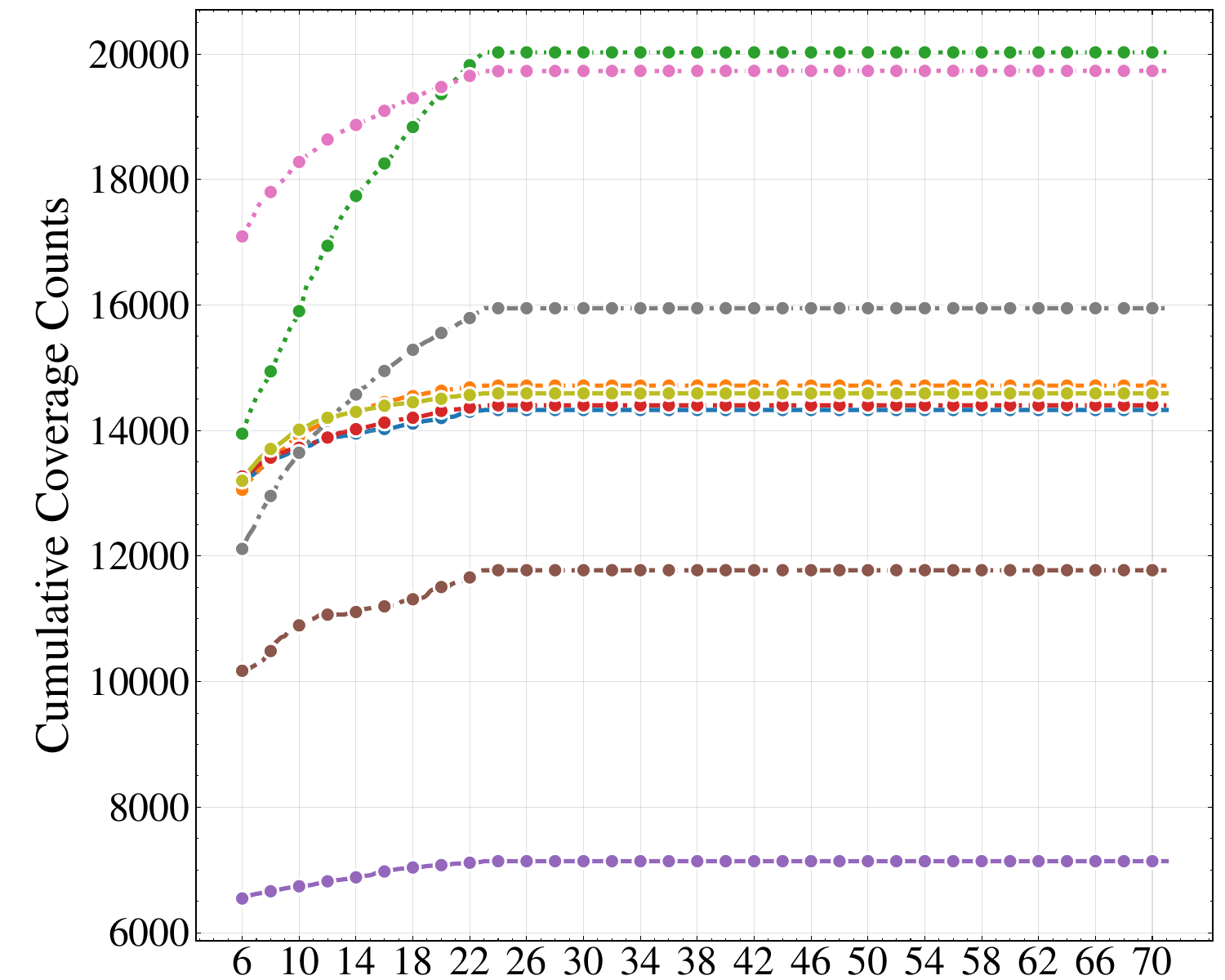} &
\includegraphics[width=0.3\textwidth]{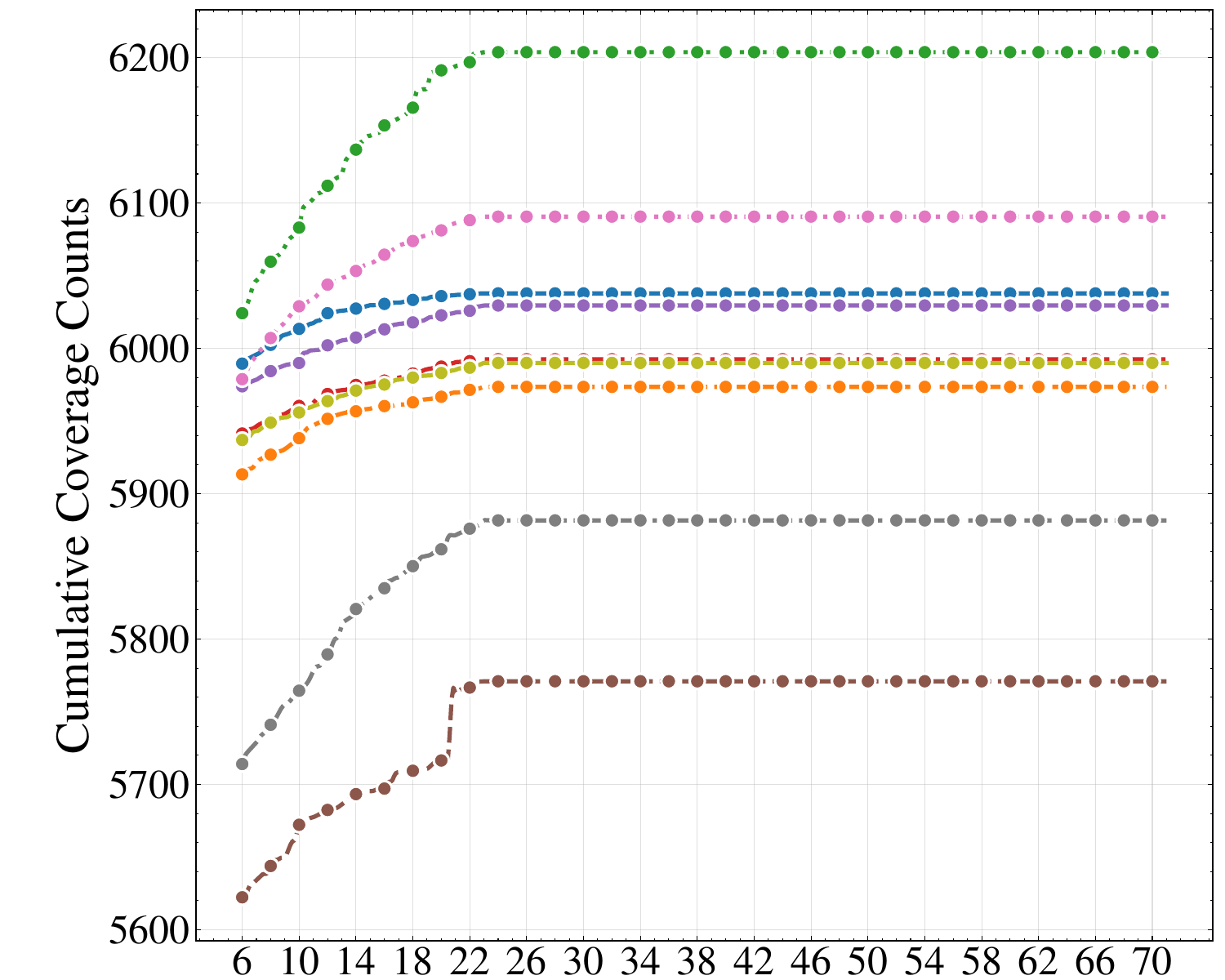} \\
(i) ffmpeg & (j) grok \\

\includegraphics[width=0.3\textwidth]{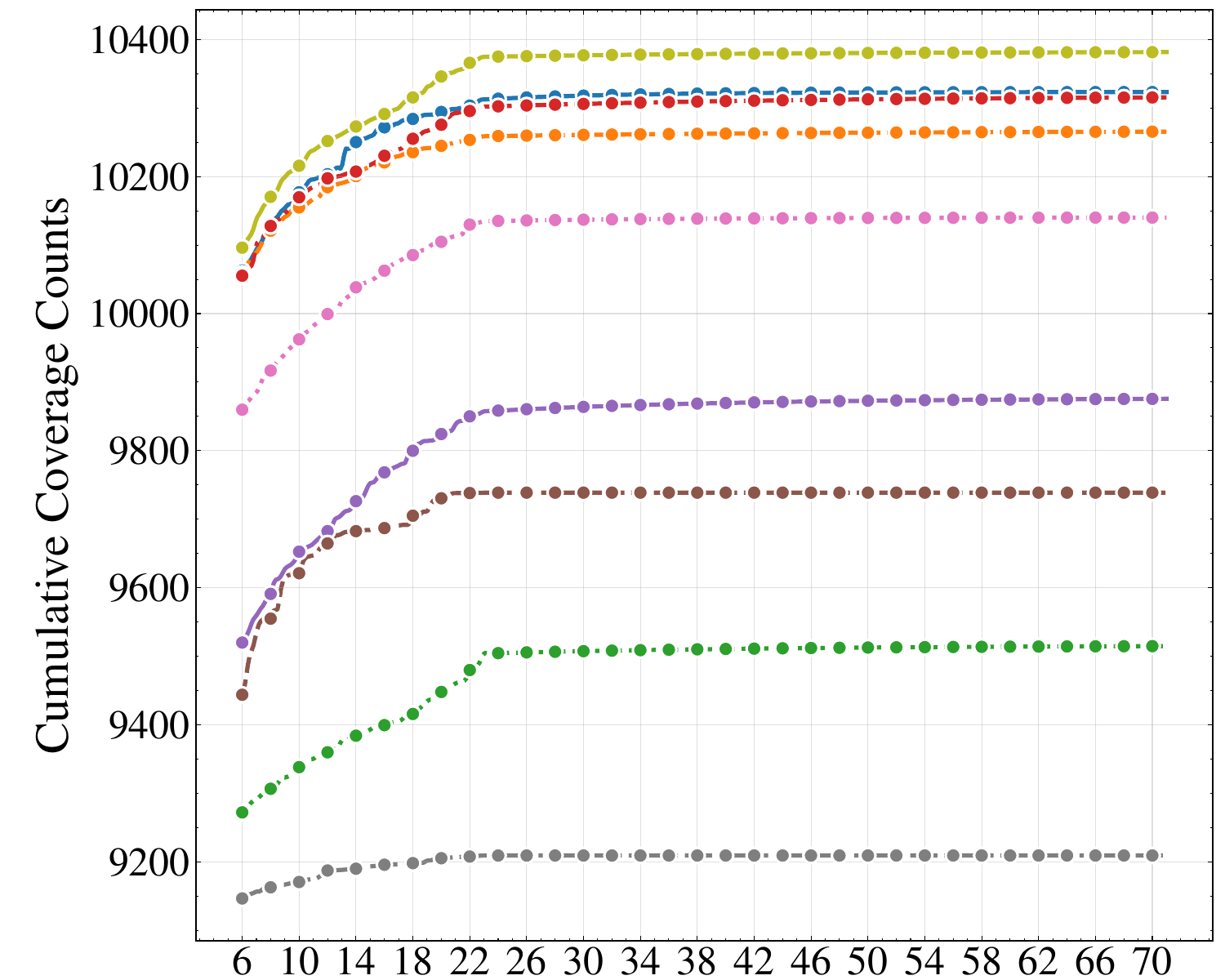} &
\includegraphics[width=0.3\textwidth]{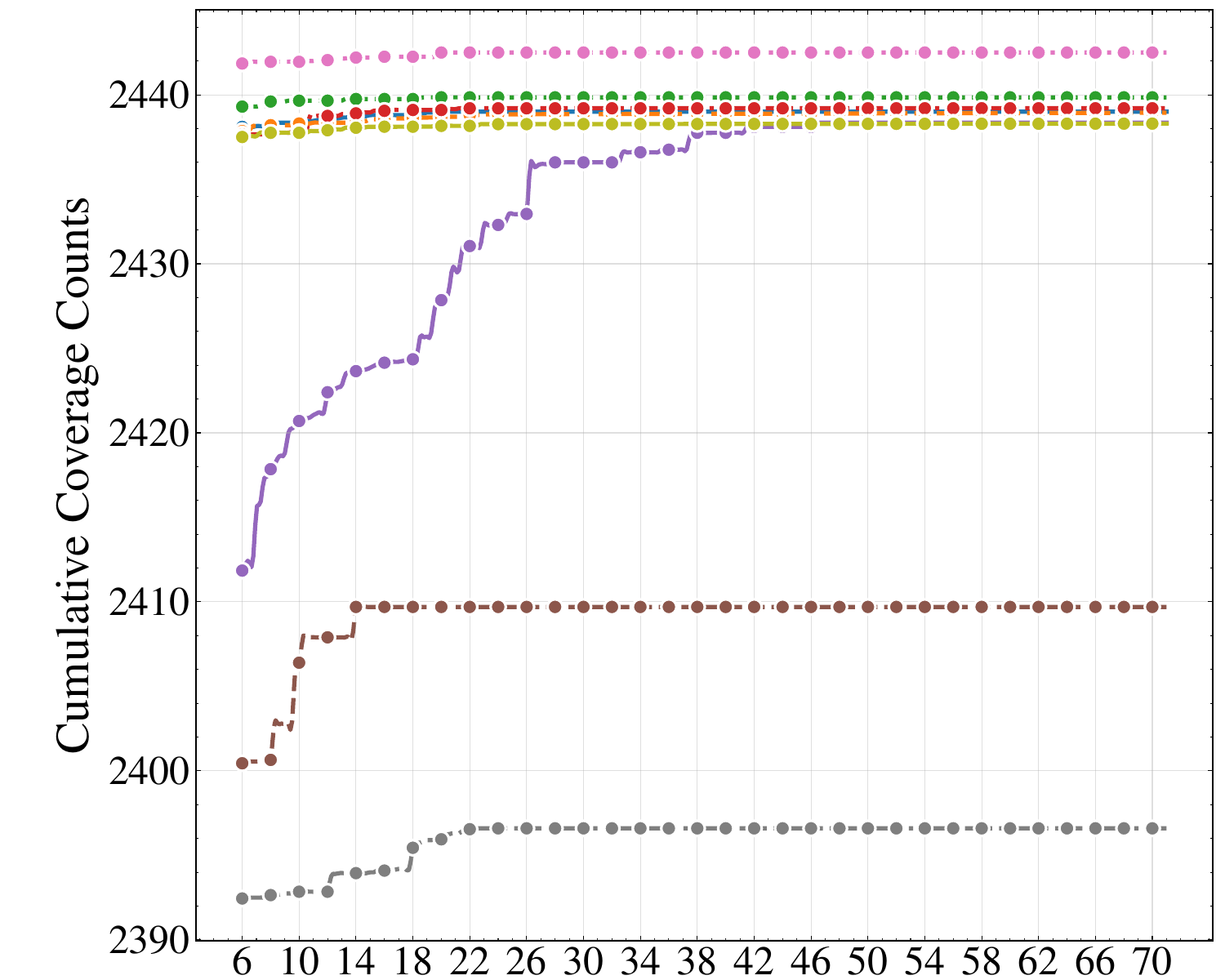} \\
(k) harfbuzz & (l) libgit2 \\
\end{tabular}

\caption{Bug counts (3×2 layout) and code coverage (3×2 layout).}
\label{fig:cumulative_6x2}
\end{figure}

\begin{figure}[H]
\centering\footnotesize
\setlength{\tabcolsep}{1pt}

\begin{tabular}{cc}

\includegraphics[width=0.3\textwidth]{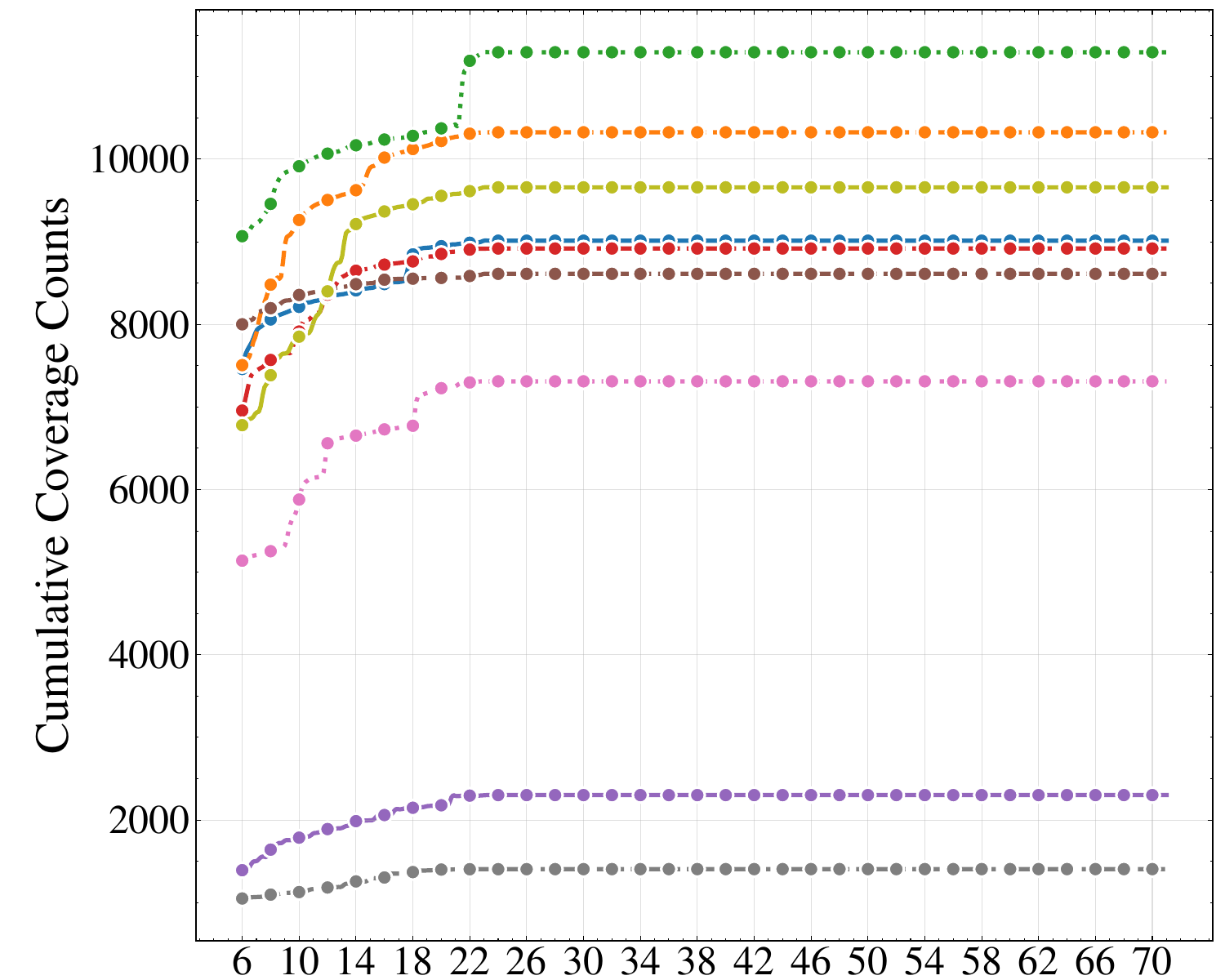} &
\includegraphics[width=0.3\textwidth]{figs/data/combined_cumulative_libhtp_fuzz_htp.pdf} \\
(a) libhevc & (b) libhtp \\

\includegraphics[width=0.3\textwidth]{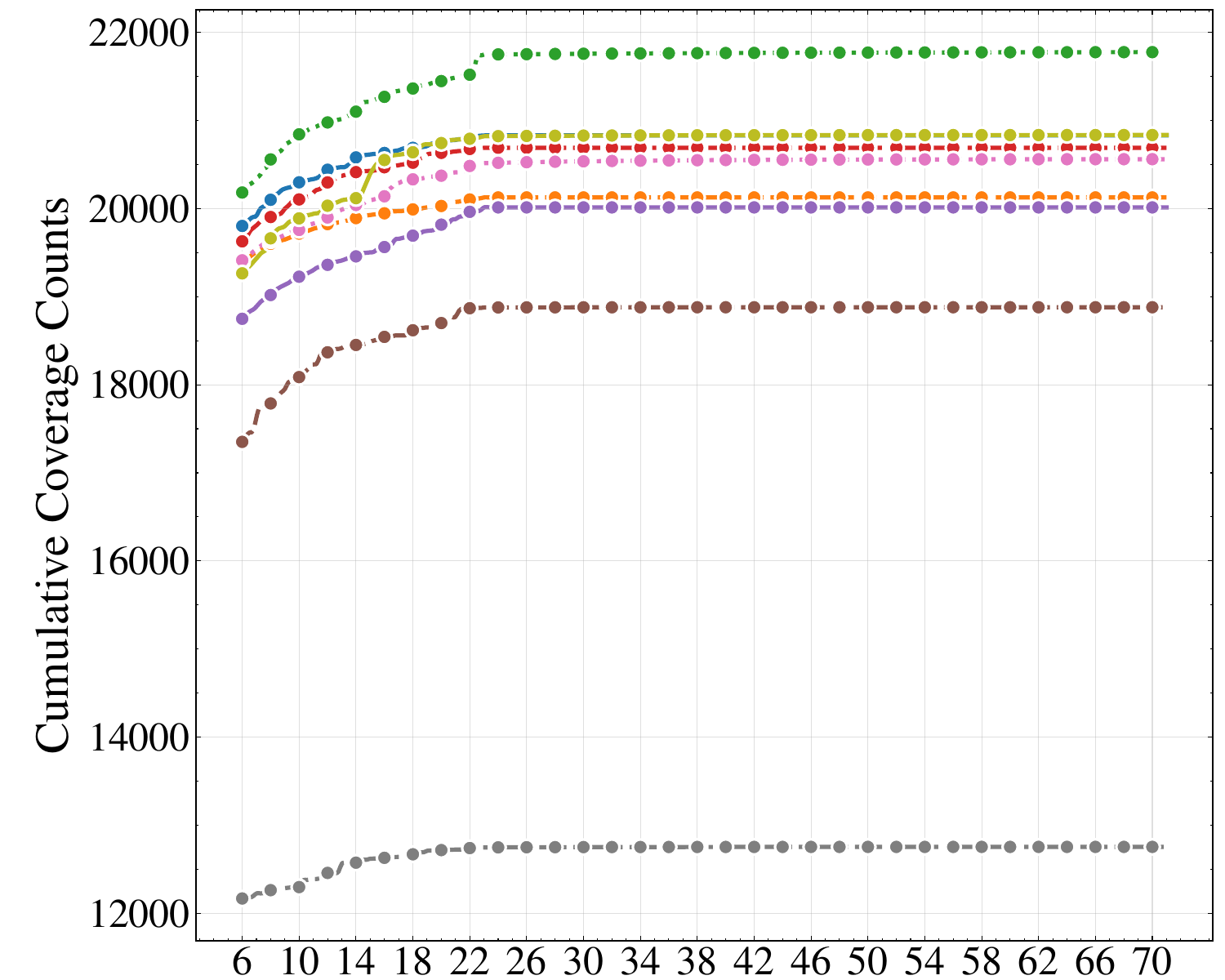} &
\includegraphics[width=0.3\textwidth]{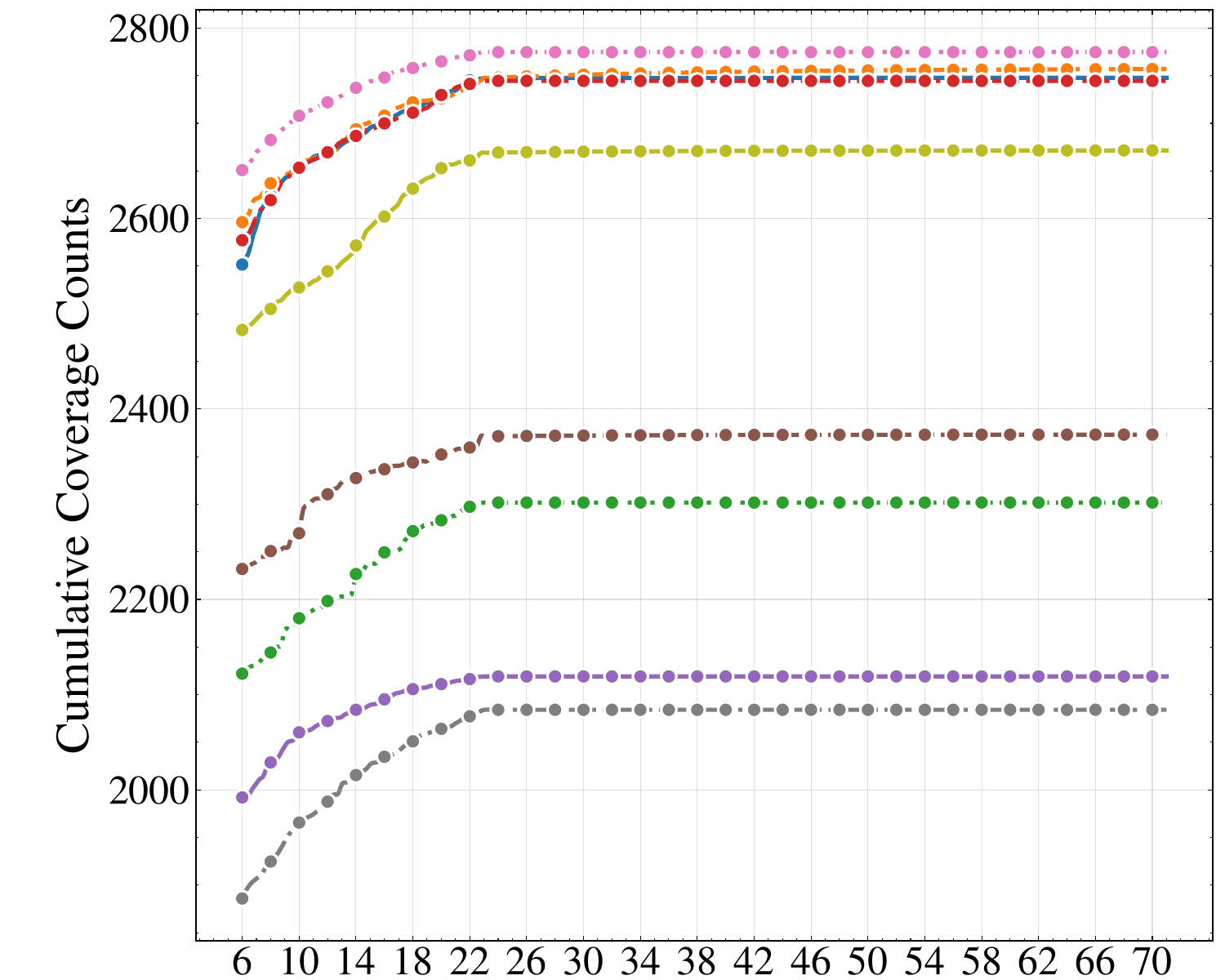} \\
(c) libxml2 & (d) matio \\

\includegraphics[width=0.3\textwidth]{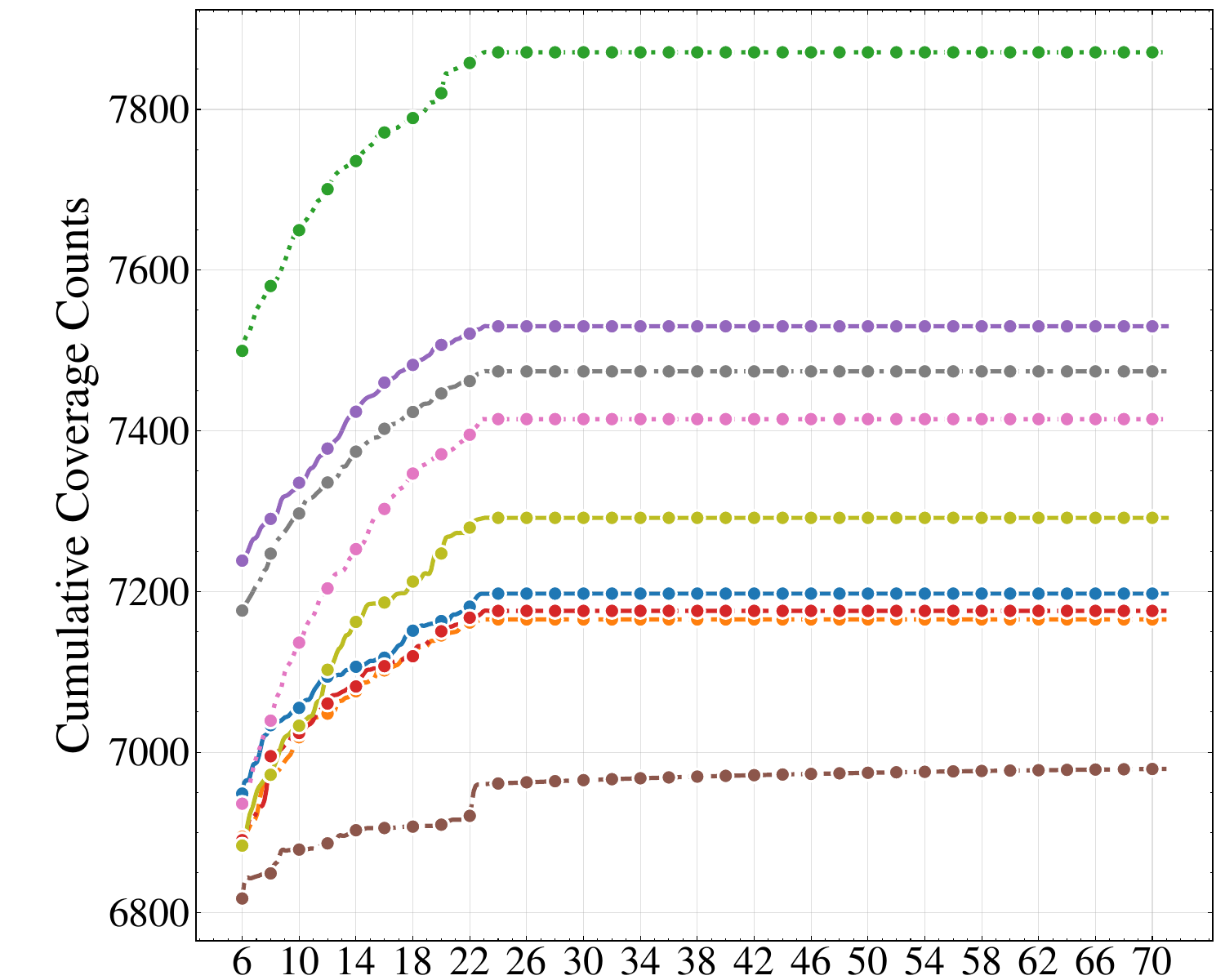} &
\includegraphics[width=0.3\textwidth]{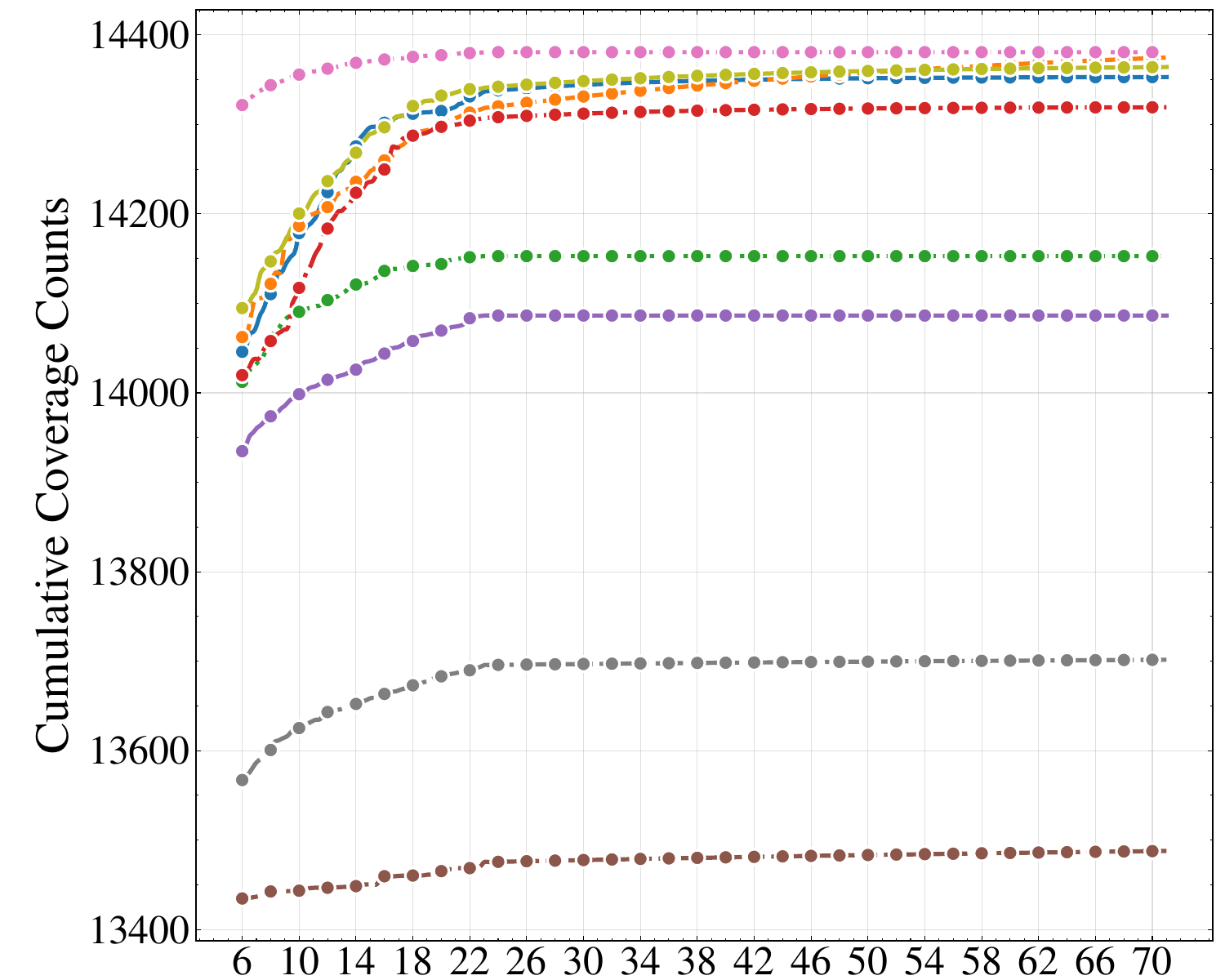} \\
(e) njs & (f) openh264 \\

\includegraphics[width=0.3\textwidth]{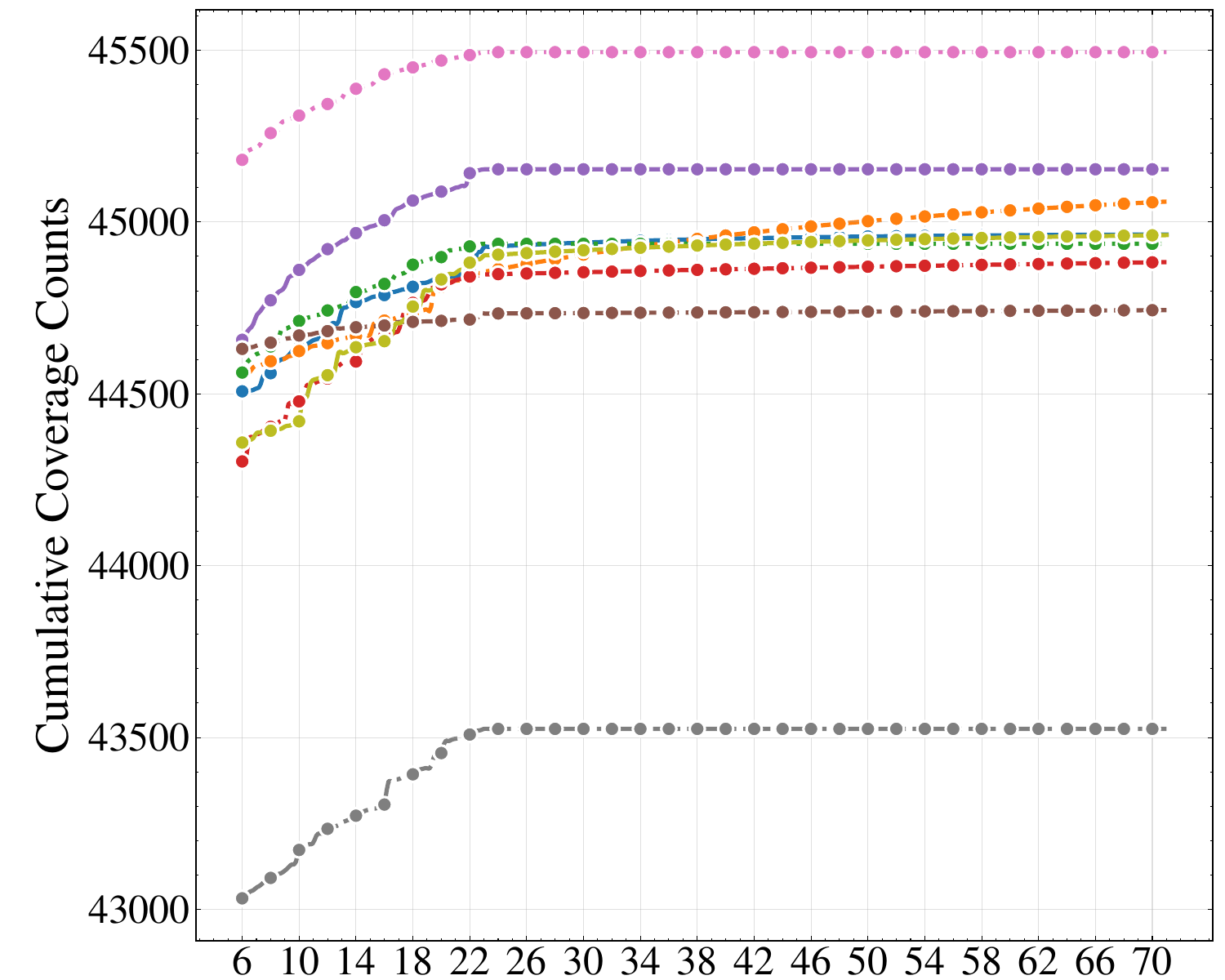} &
\includegraphics[width=0.3\textwidth]{figs/data/combined_cumulative_poppler_pdf_fuzzer.pdf} \\
(g) php & (h) poppler \\

\includegraphics[width=0.3\textwidth]{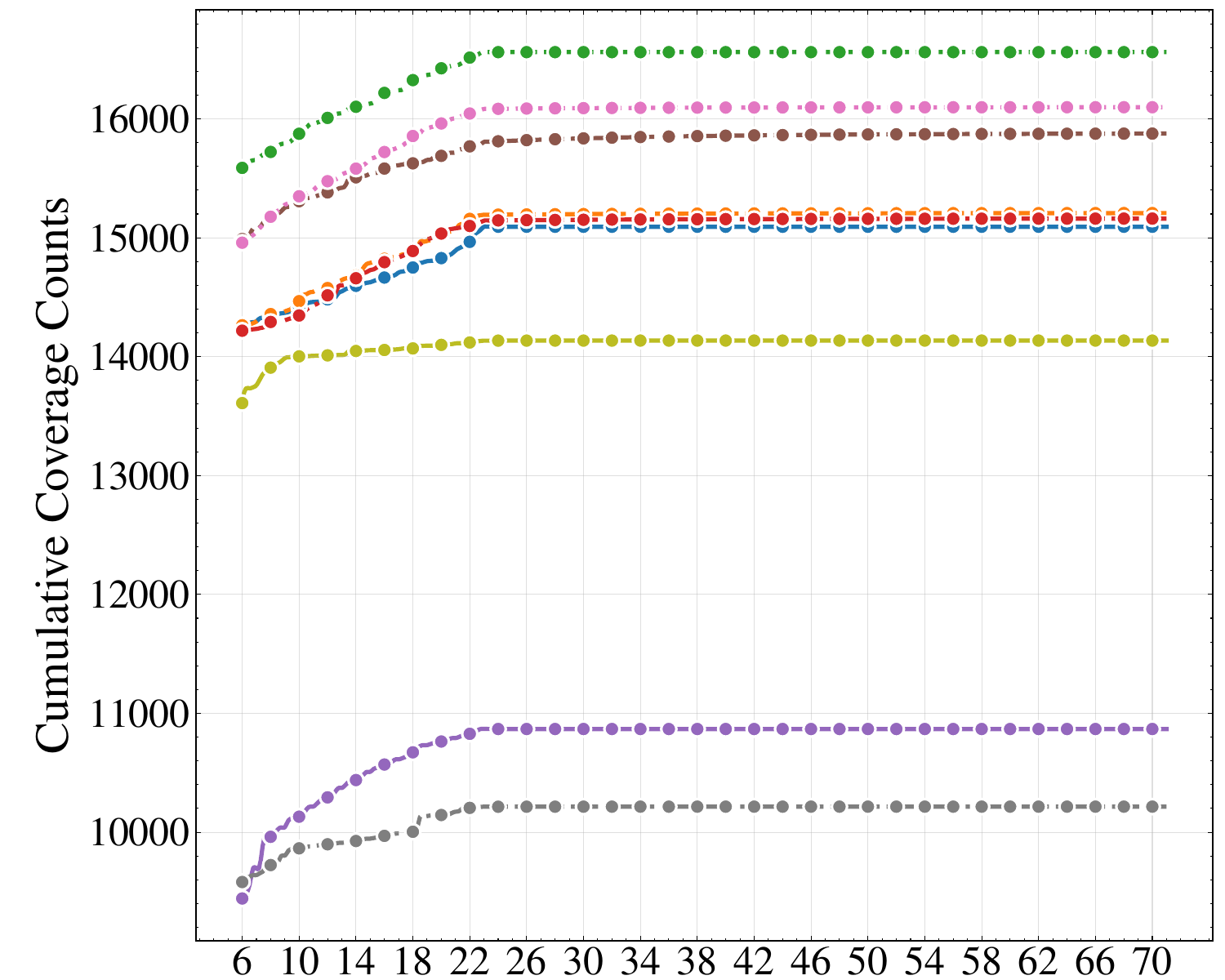} &
\includegraphics[width=0.3\textwidth]{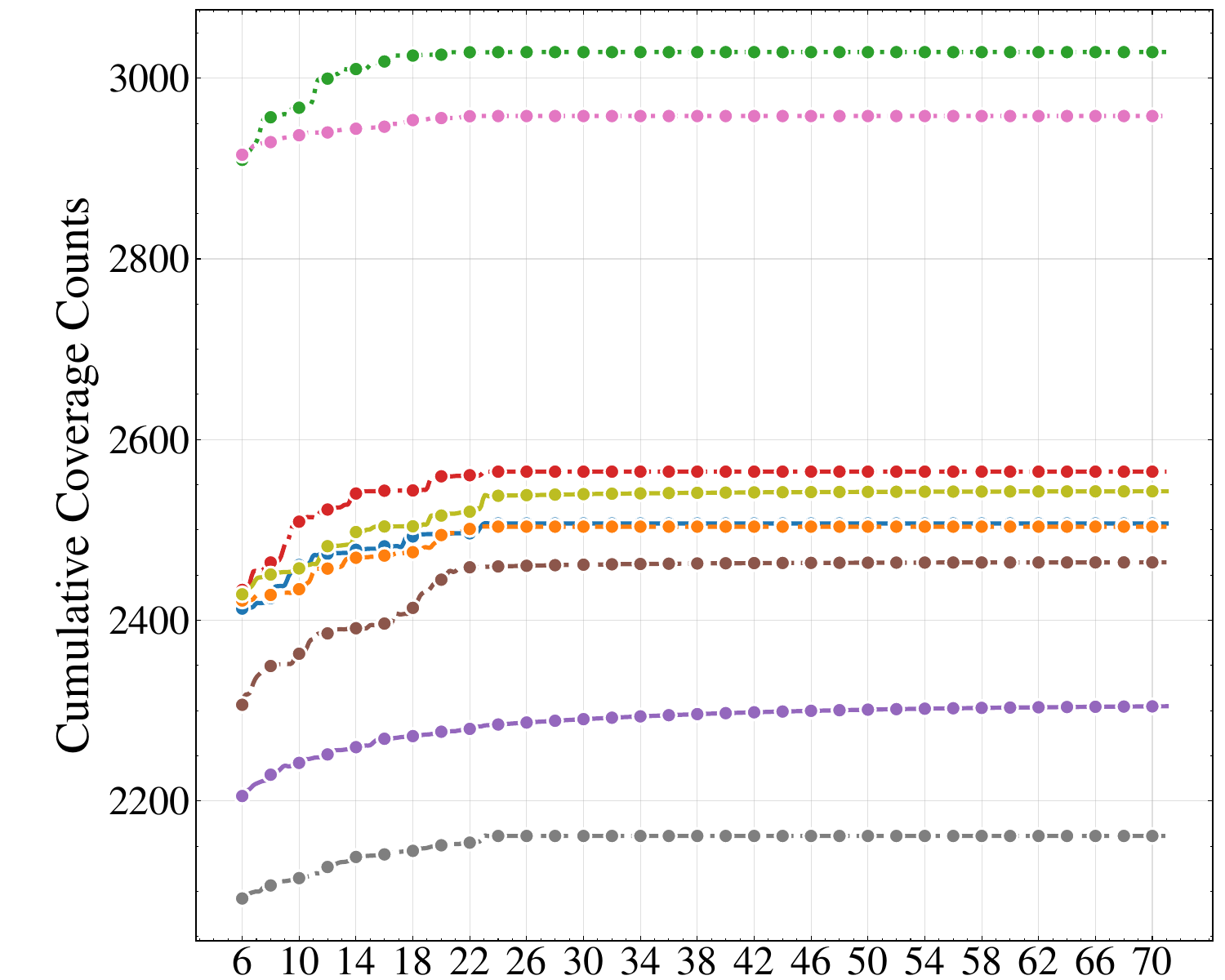} \\
(i) quickjs & (j) stb \\

\includegraphics[width=0.3\textwidth]{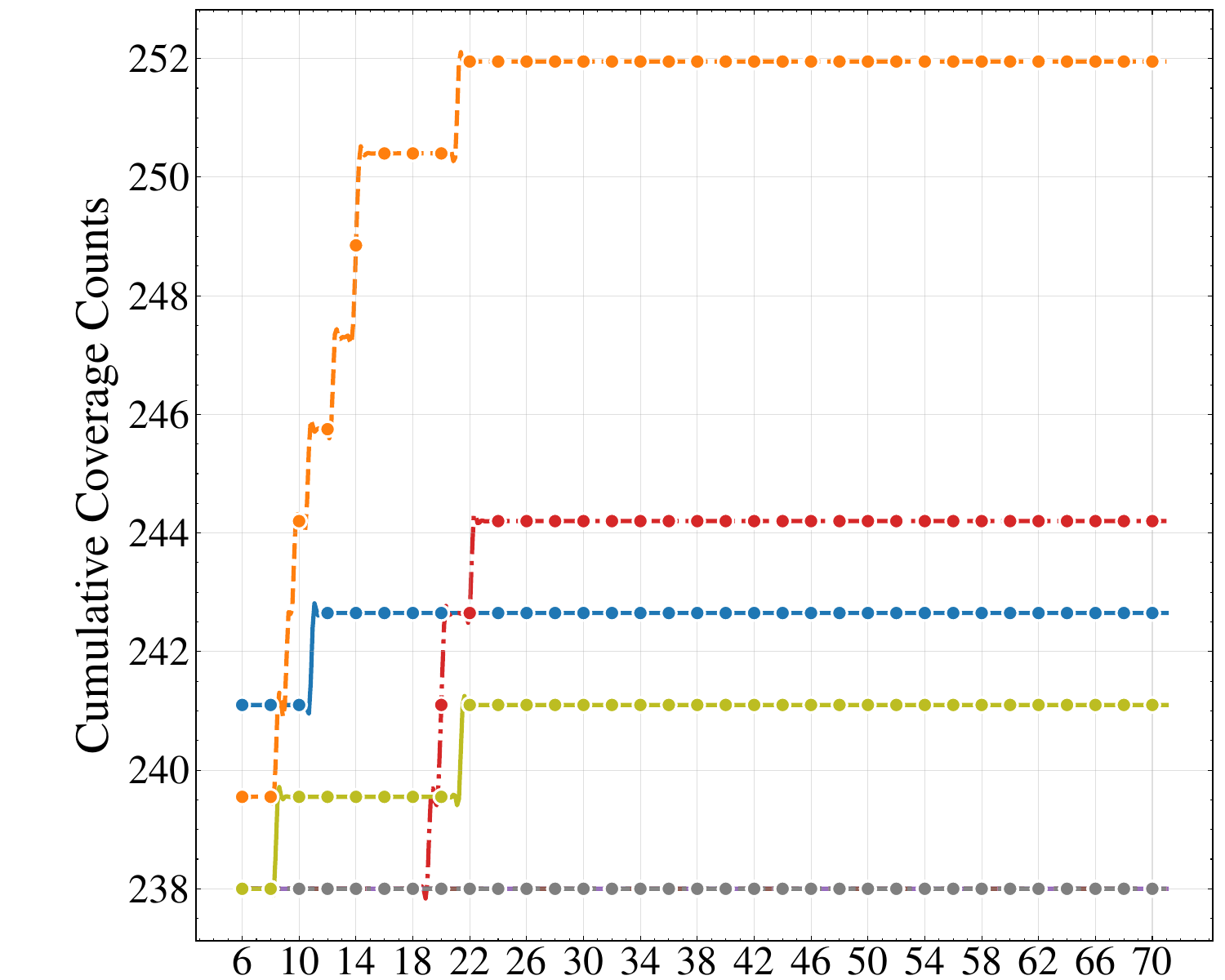} &
\includegraphics[width=0.3\textwidth]{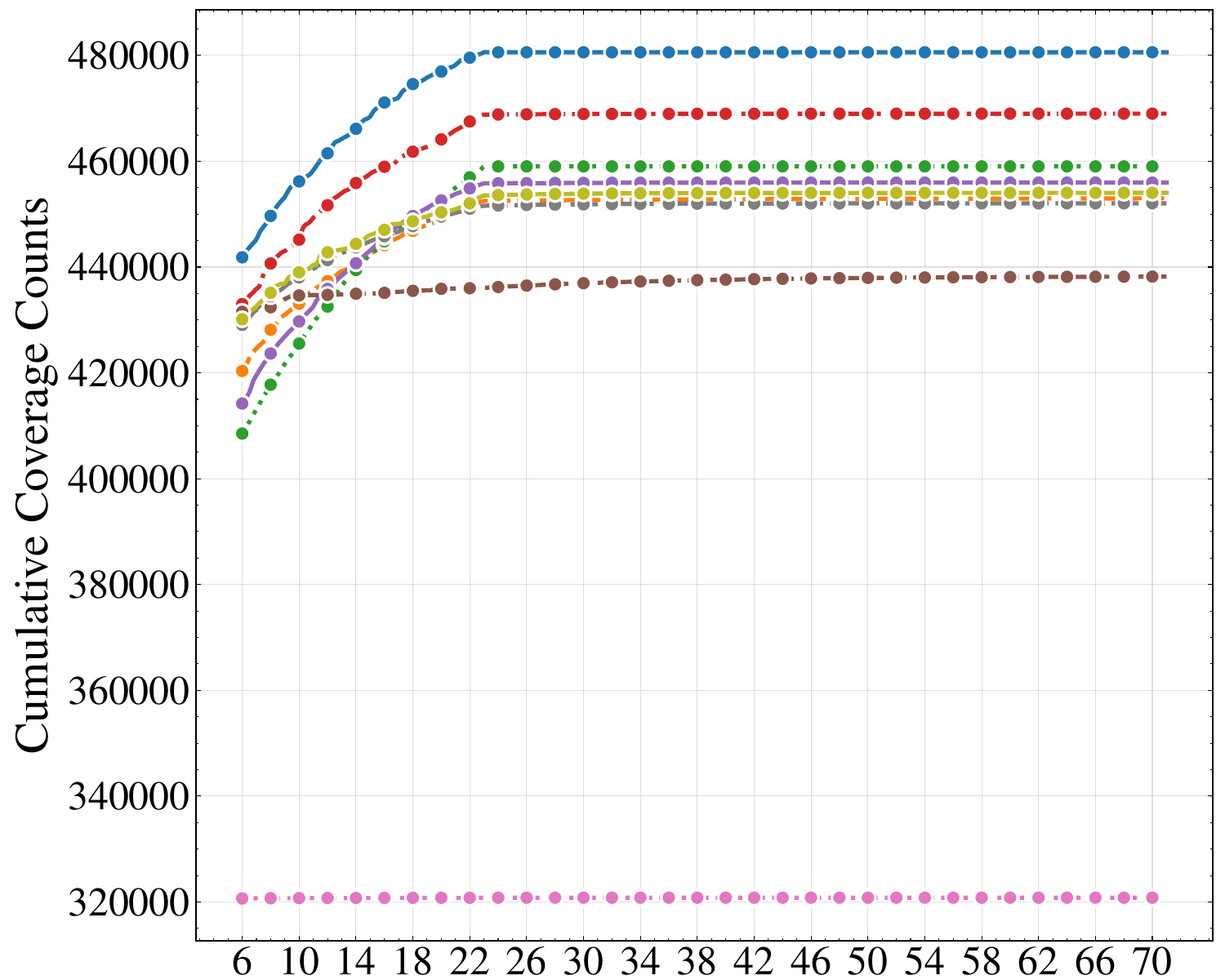} \\
(k) systemd & (l) wireshark \\

\end{tabular}

\caption{Code coverage (6×2 layout). Top: coverage ratio on the first set; Bottom: coverage on the second set.}
\label{fig:combined_cov}
\end{figure}

\begin{figure}[H]
\centering\footnotesize
\setlength{\tabcolsep}{1pt}

\begin{tabular}{cc}

\includegraphics[width=0.3\textwidth]{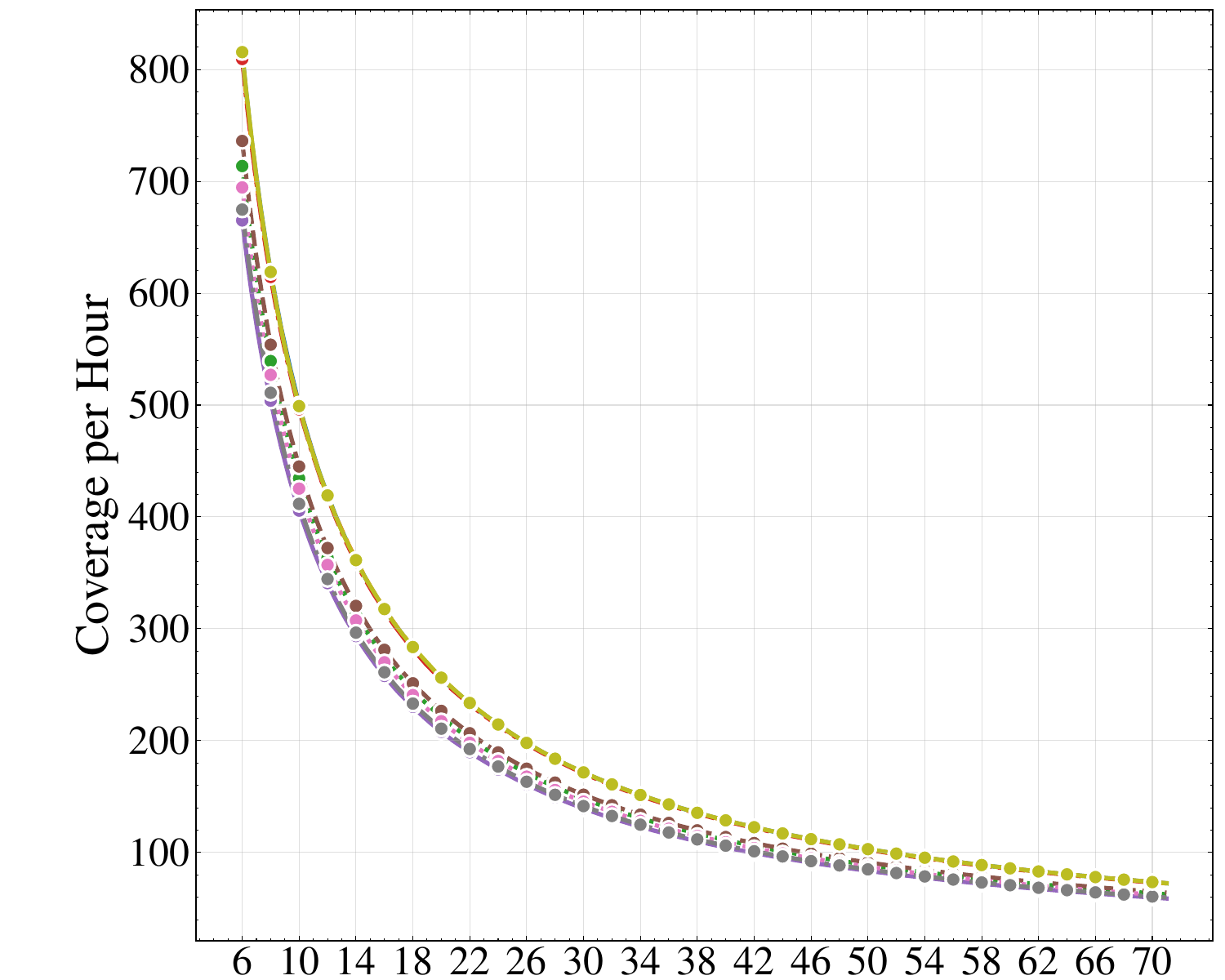} &
\includegraphics[width=0.3\textwidth]{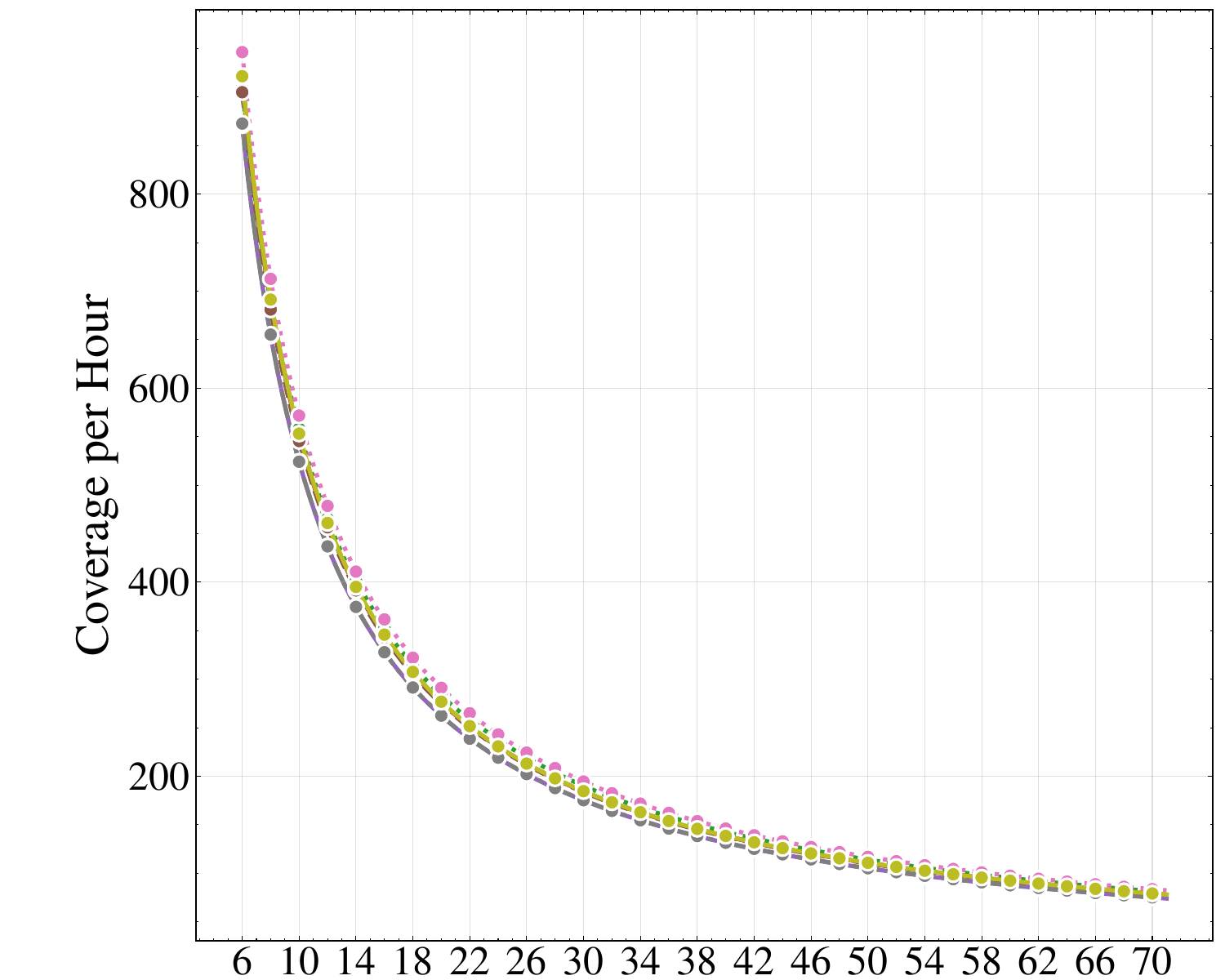} \\
(a) arrow & (b) aspell \\

\includegraphics[width=0.3\textwidth]{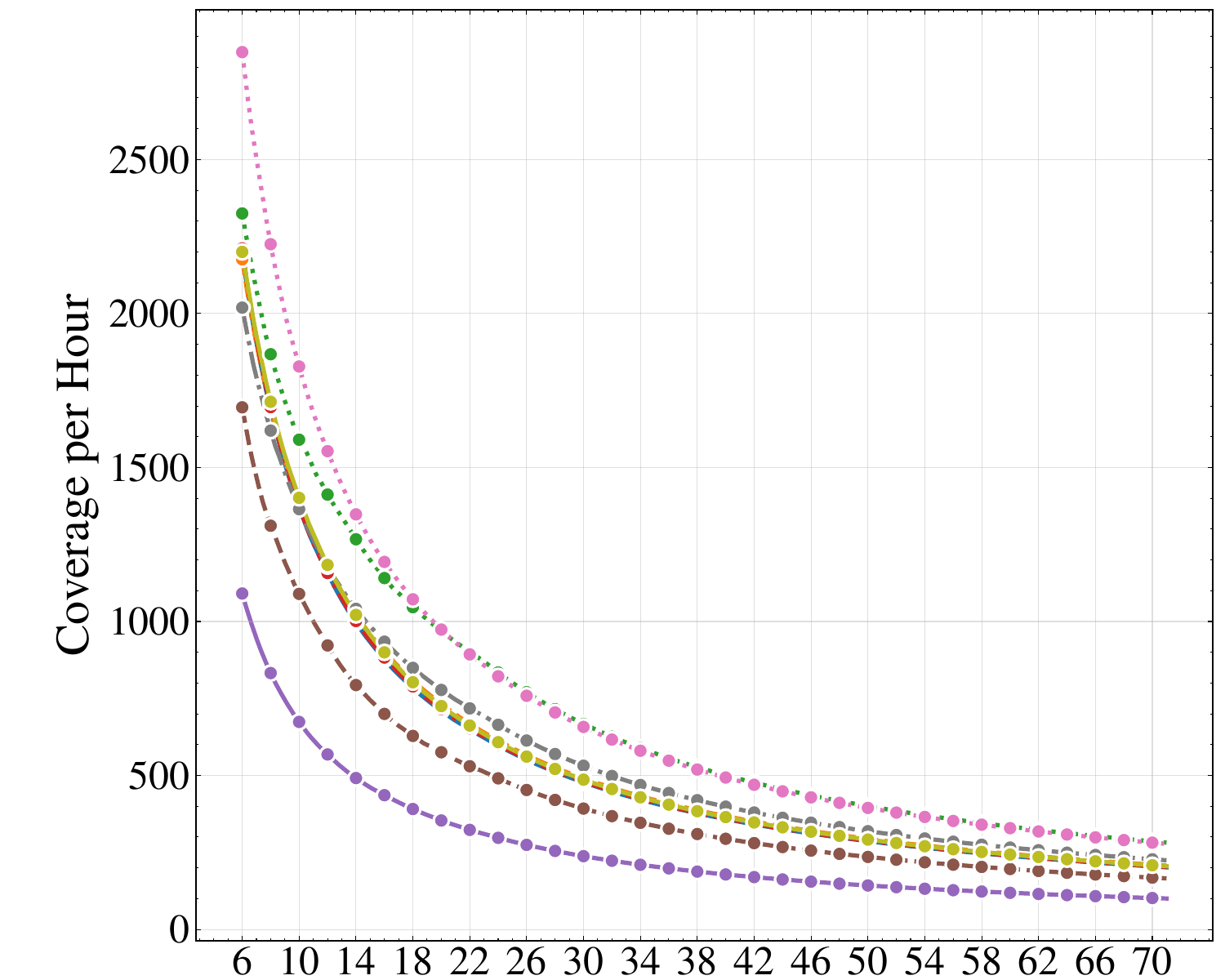} &
\includegraphics[width=0.3\textwidth]{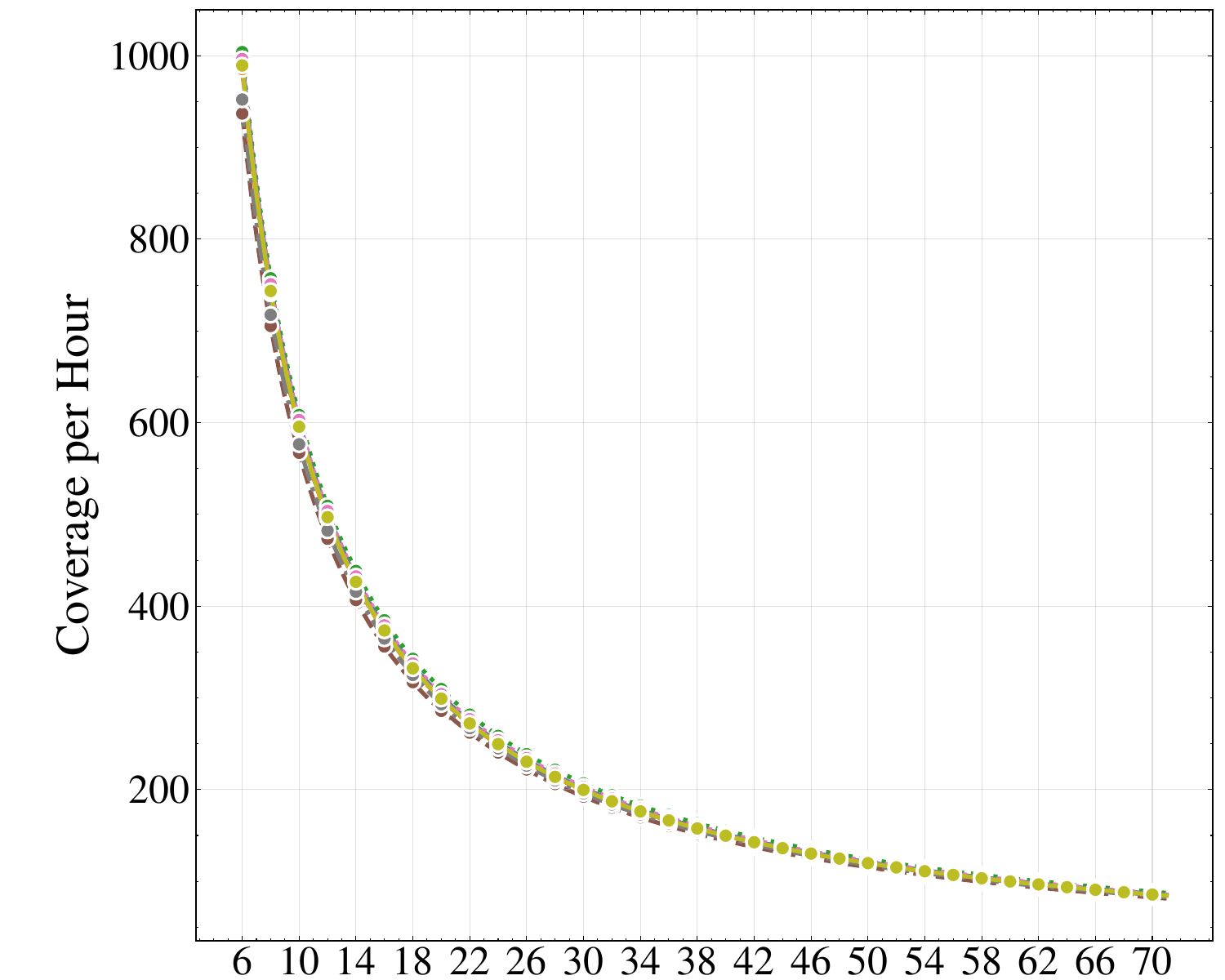} \\
(c) ffmpeg & (d) grok \\

\includegraphics[width=0.3\textwidth]{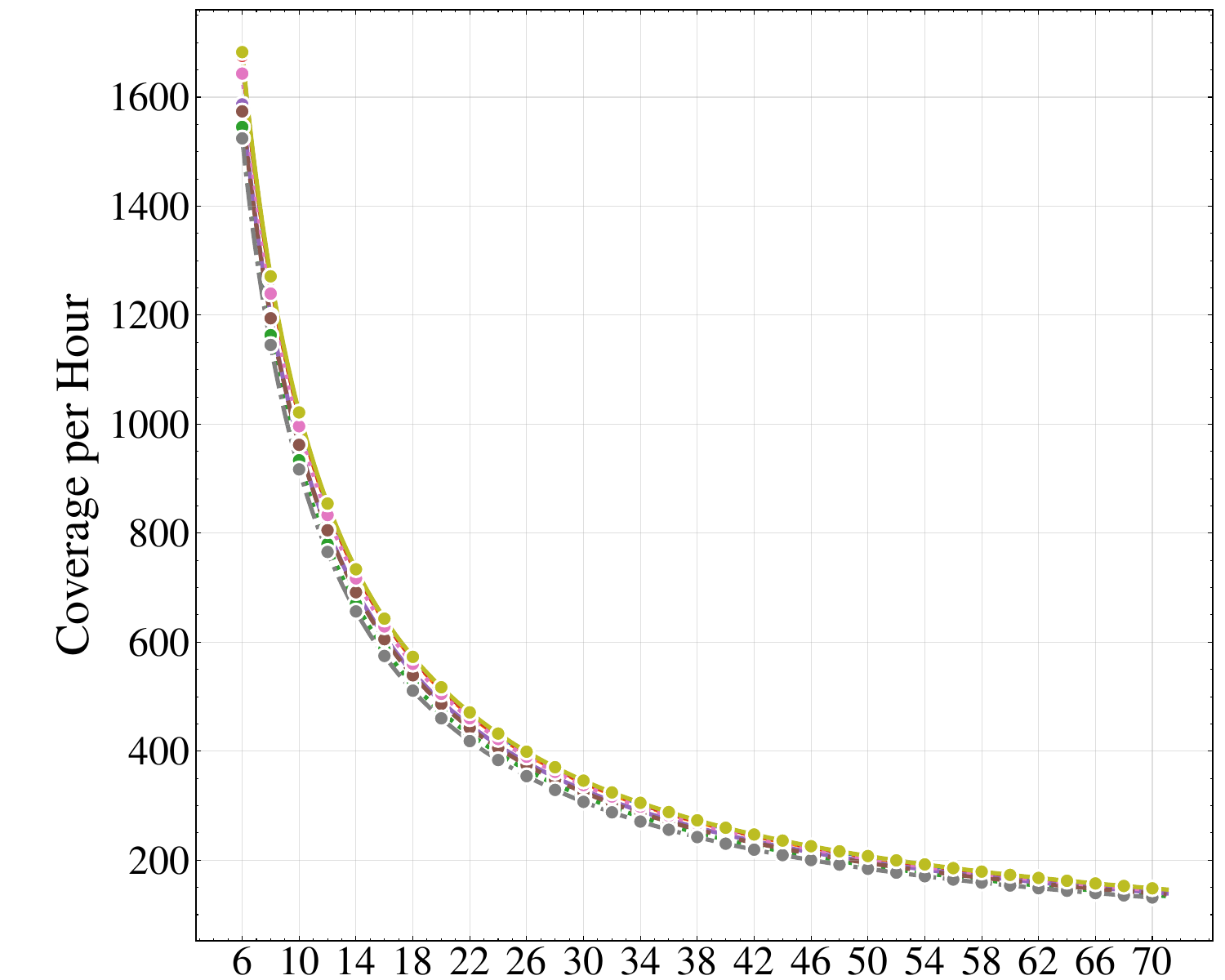} &
\includegraphics[width=0.3\textwidth]{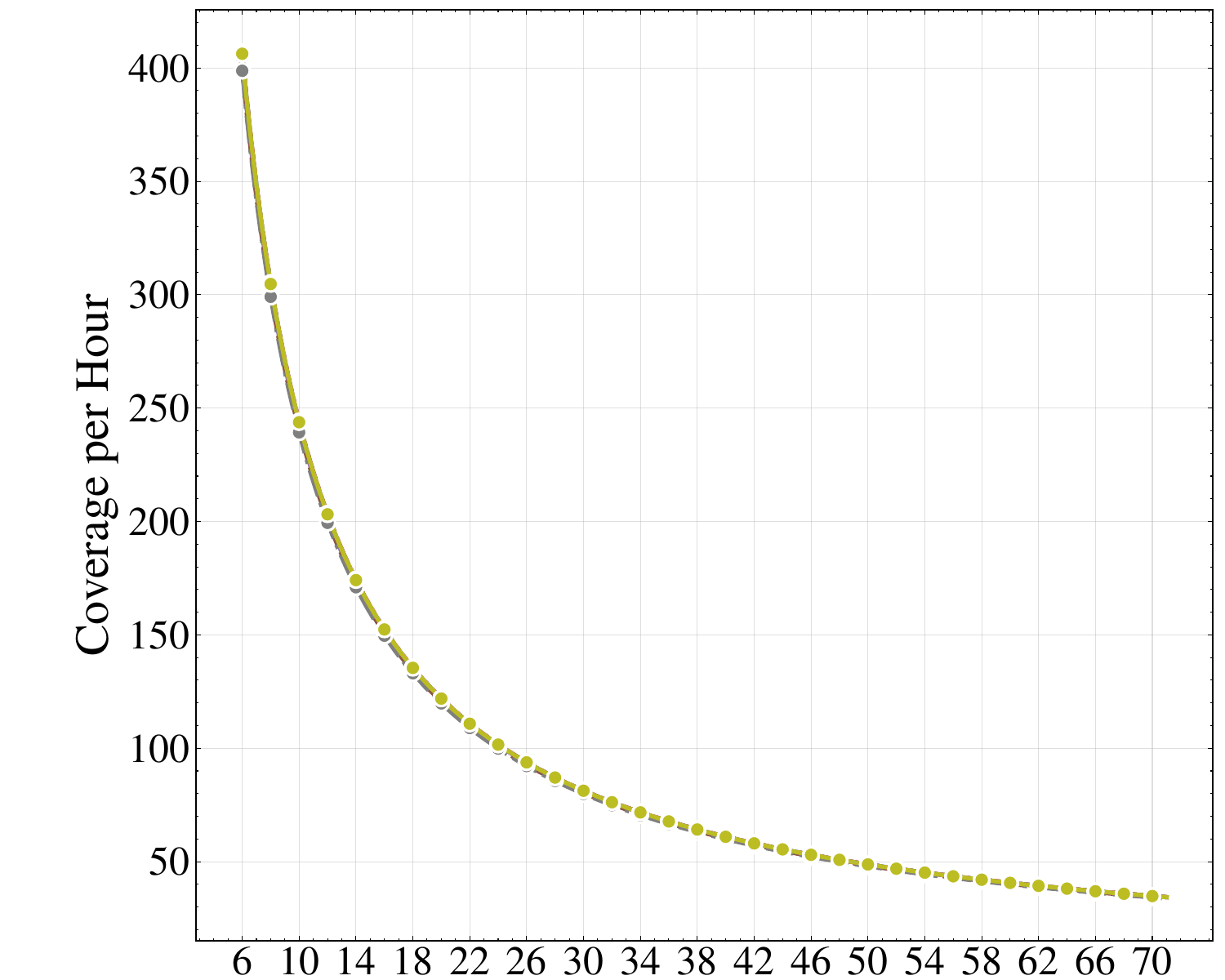} \\
(e) harfbuzz & (f) libgit2 \\

\includegraphics[width=0.3\textwidth]{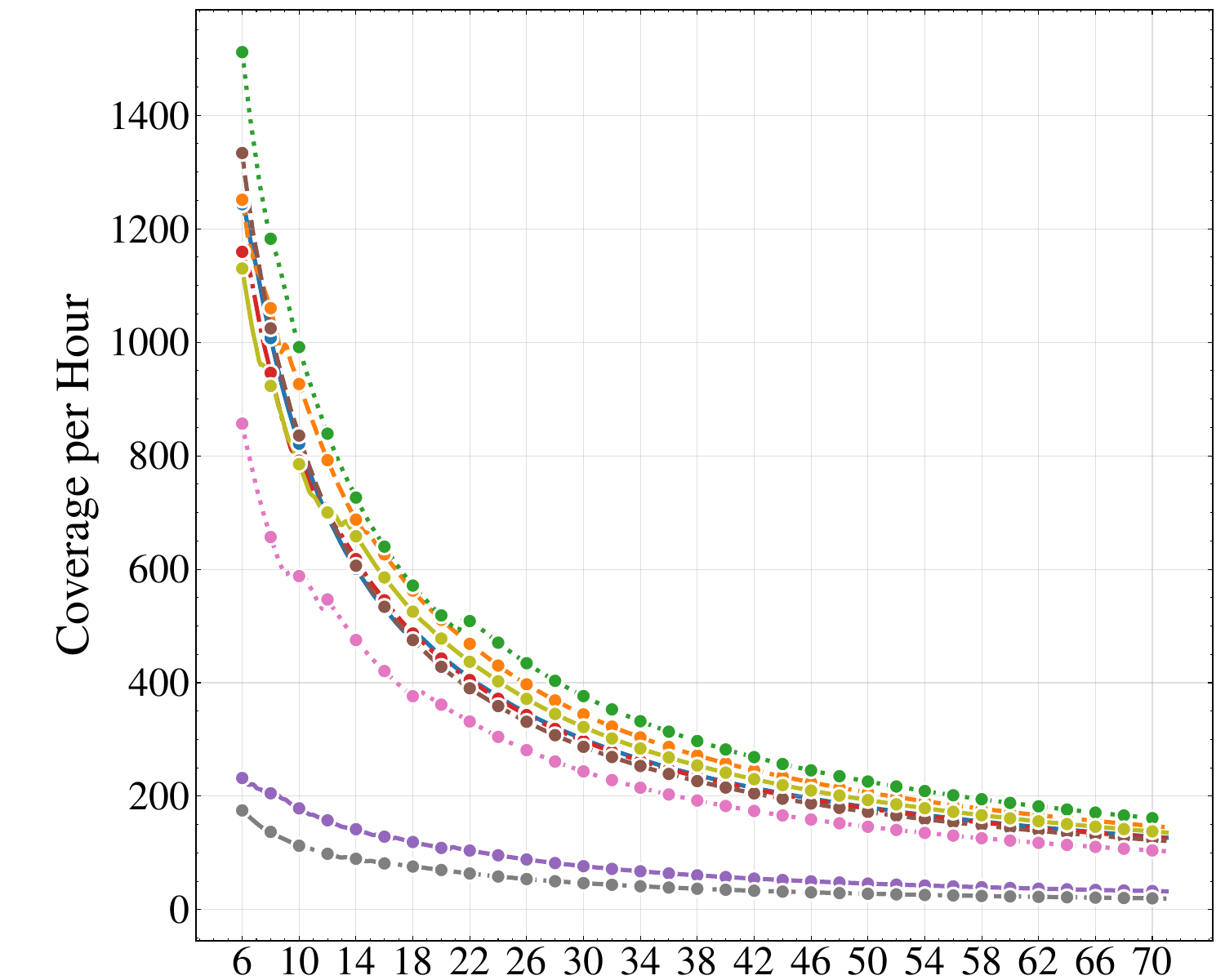} &
\includegraphics[width=0.3\textwidth]{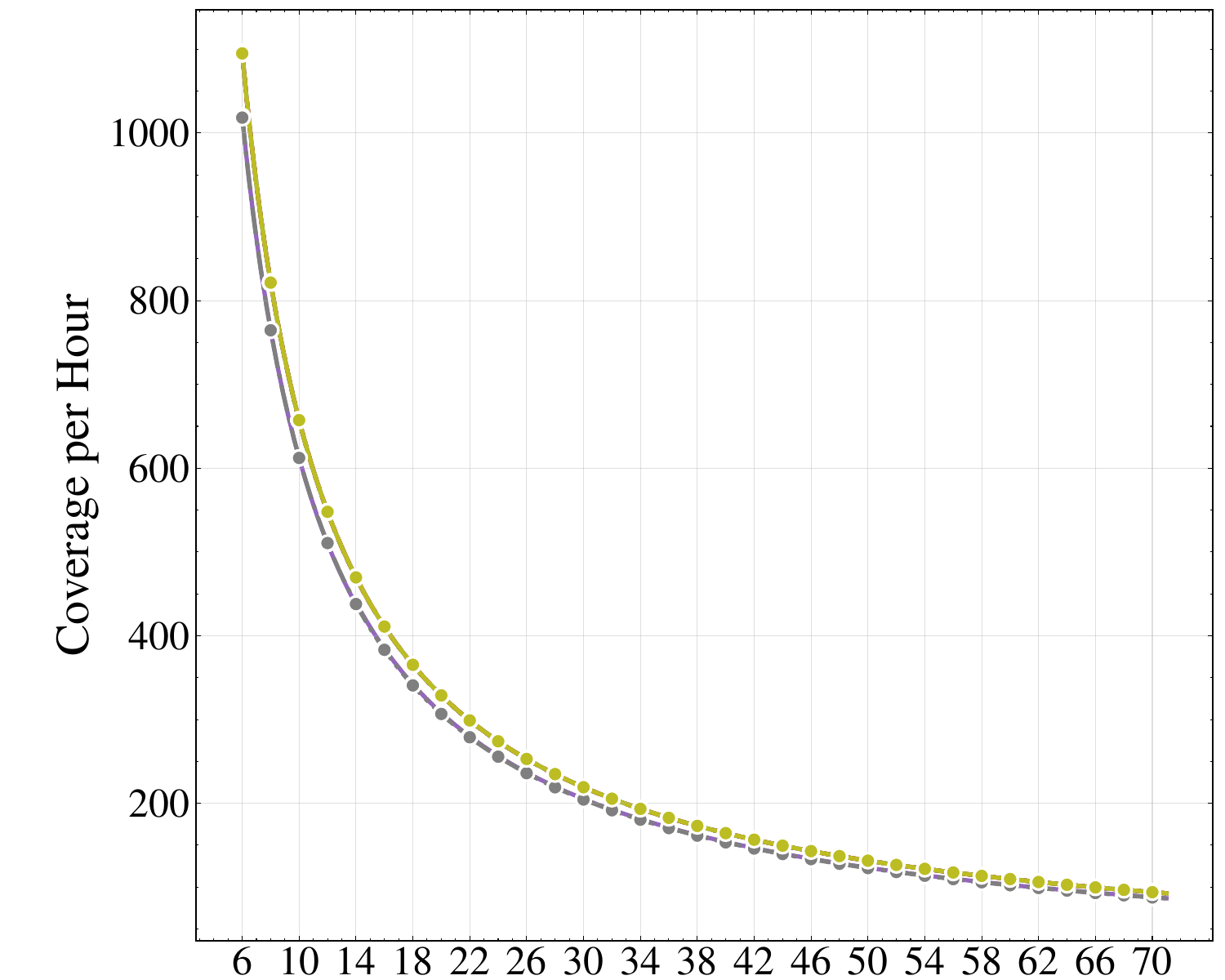} \\
(g) libhevc & (h) libhtp \\

\includegraphics[width=0.3\textwidth]{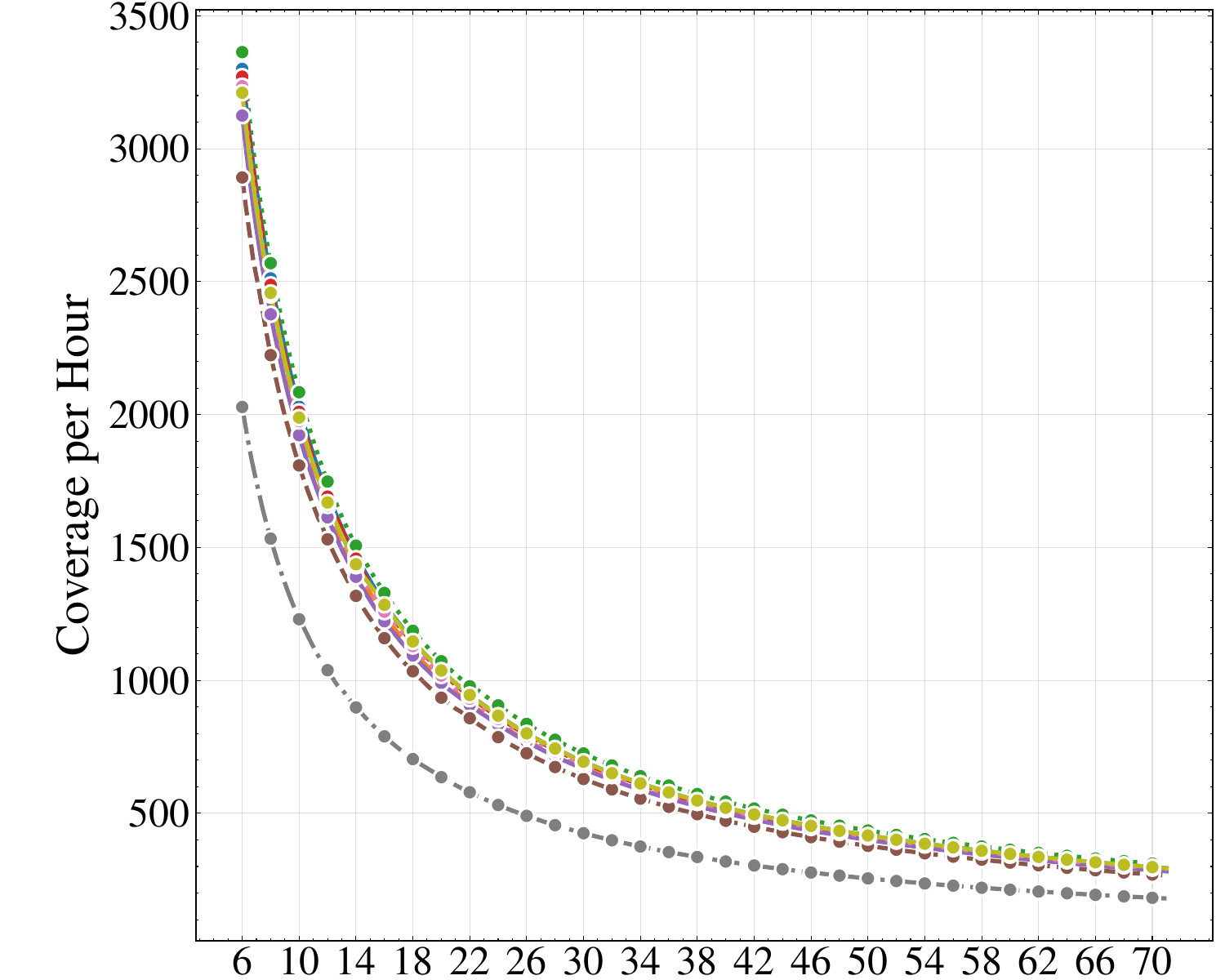} &
\includegraphics[width=0.3\textwidth]{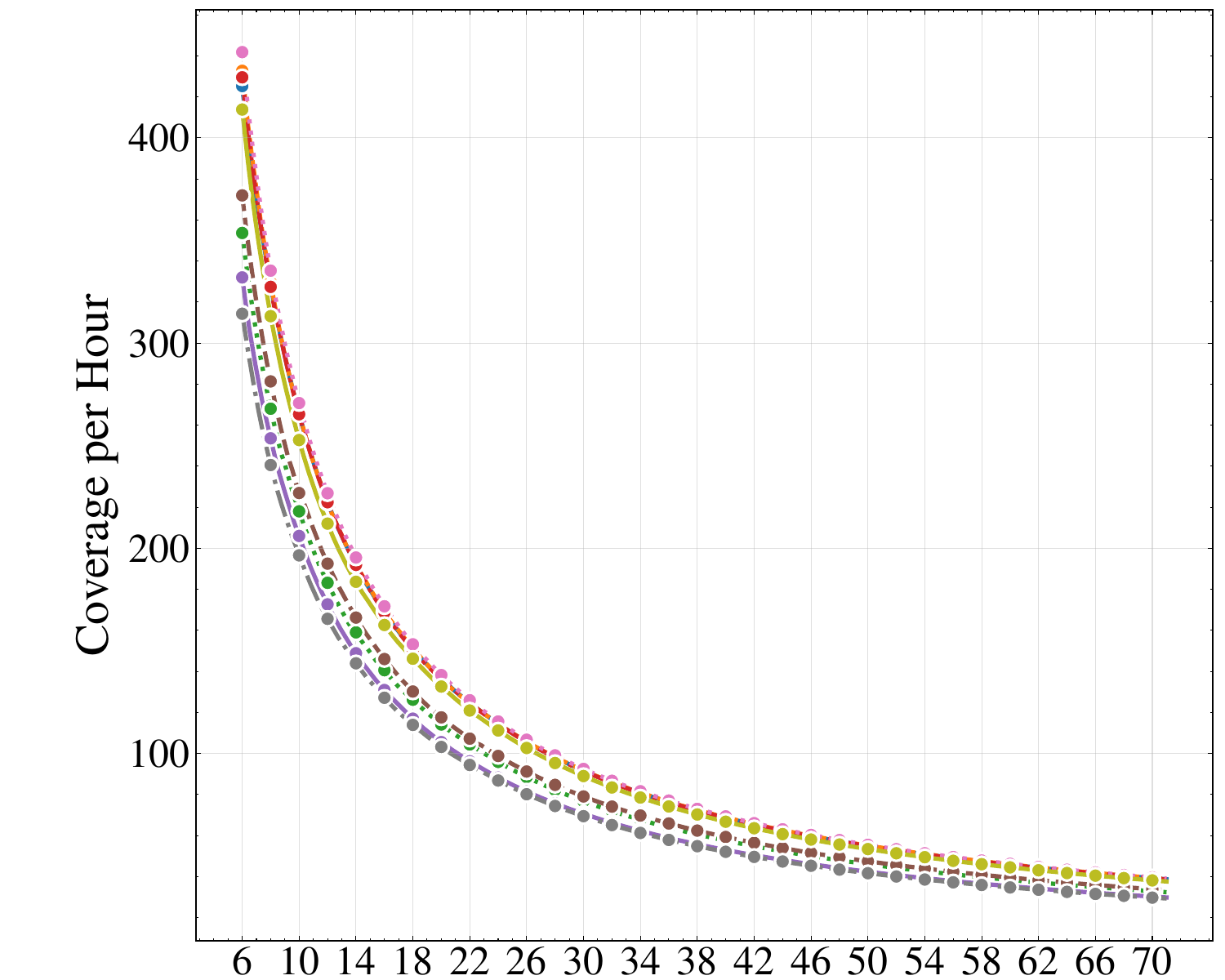} \\
(i) libxml2 & (j) matio \\

\includegraphics[width=0.3\textwidth]{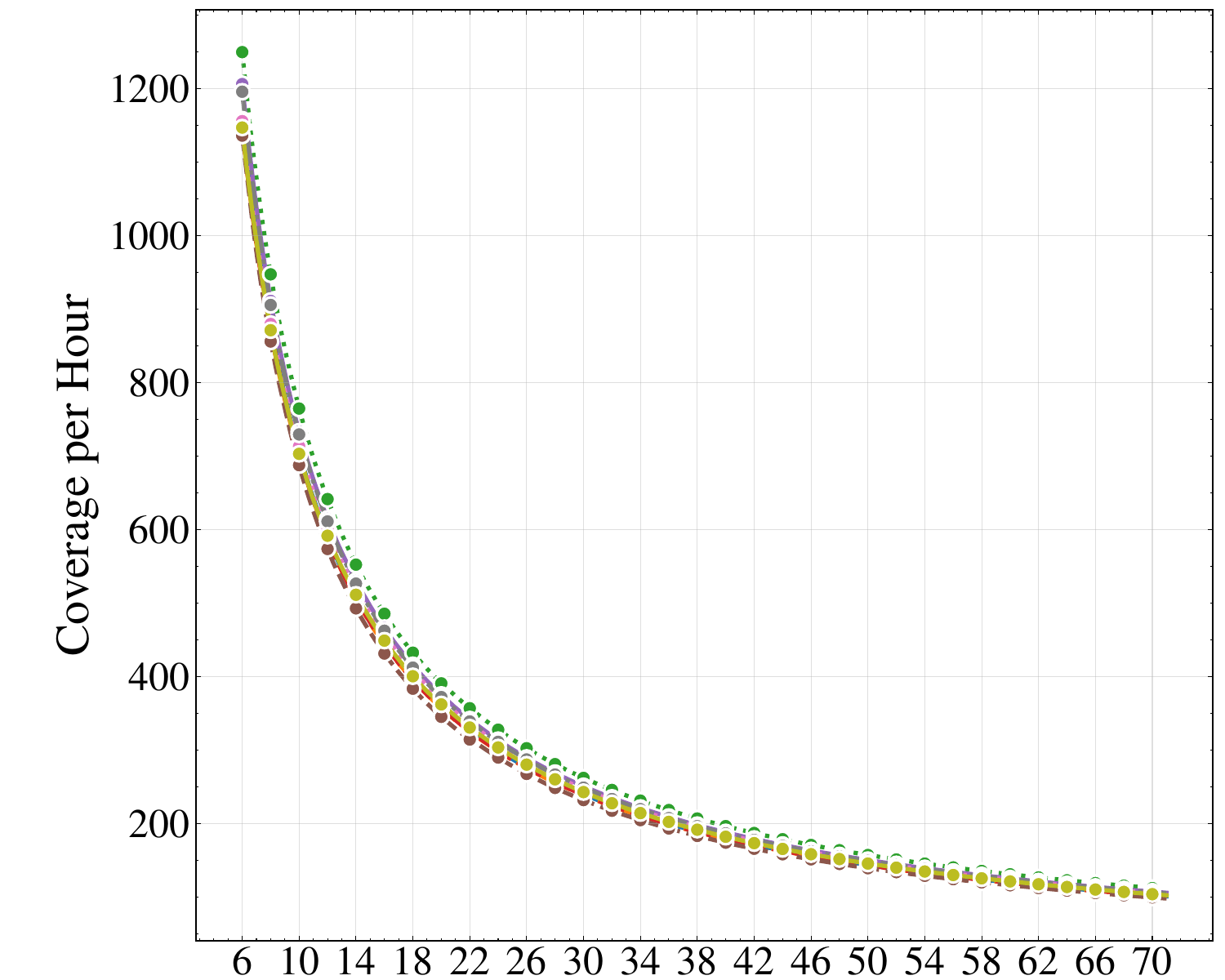} &
\includegraphics[width=0.3\textwidth]{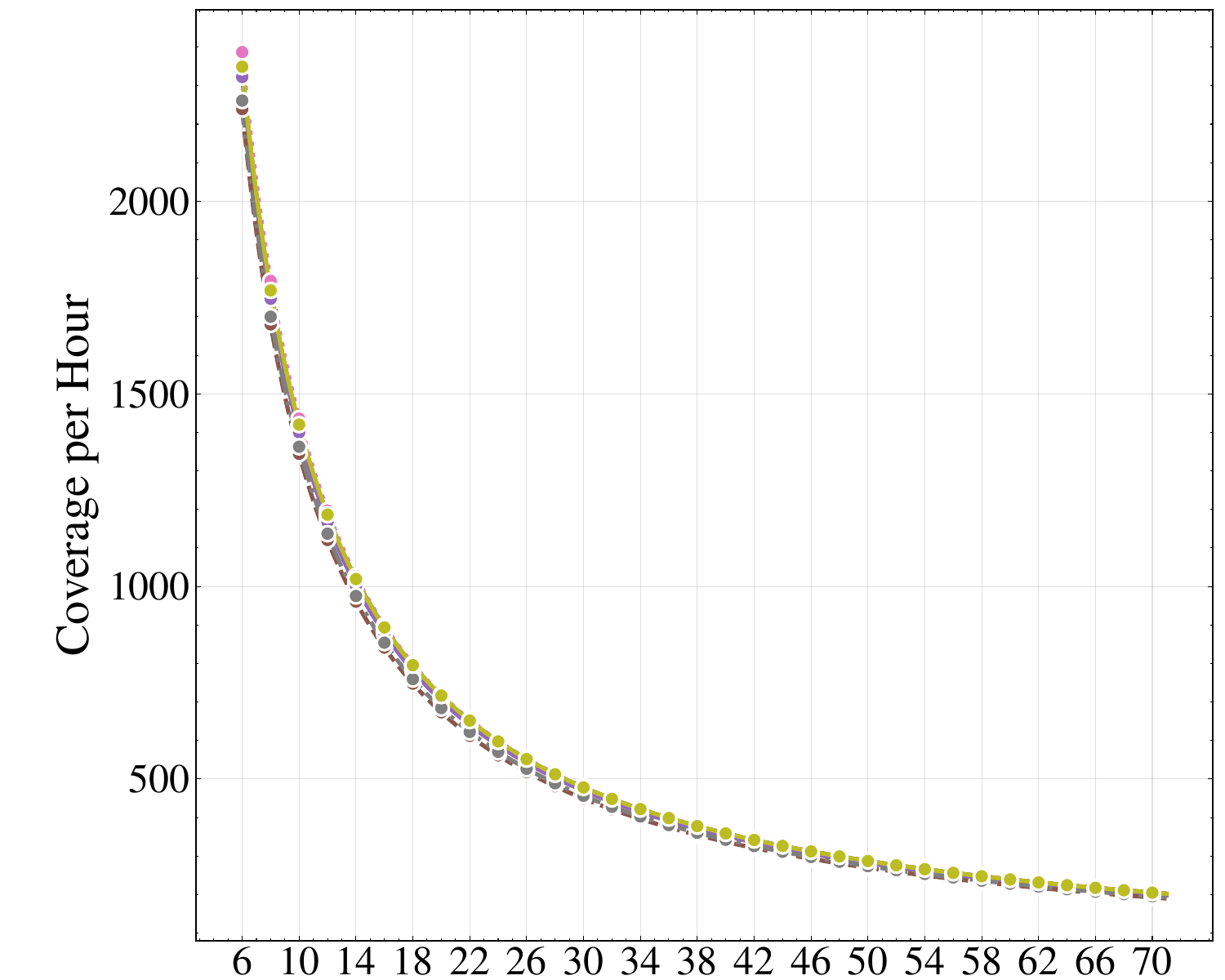} \\
(k) njs & (l) openh264 \\

\end{tabular}

\caption{Code coverage ratio (6×2 layout).}
\label{fig:cov_6x2_combined}
\end{figure}

\begin{figure}[H]
\centering\footnotesize
\setlength{\tabcolsep}{1pt}

\begin{tabular}{cc}

\includegraphics[width=0.3\textwidth]{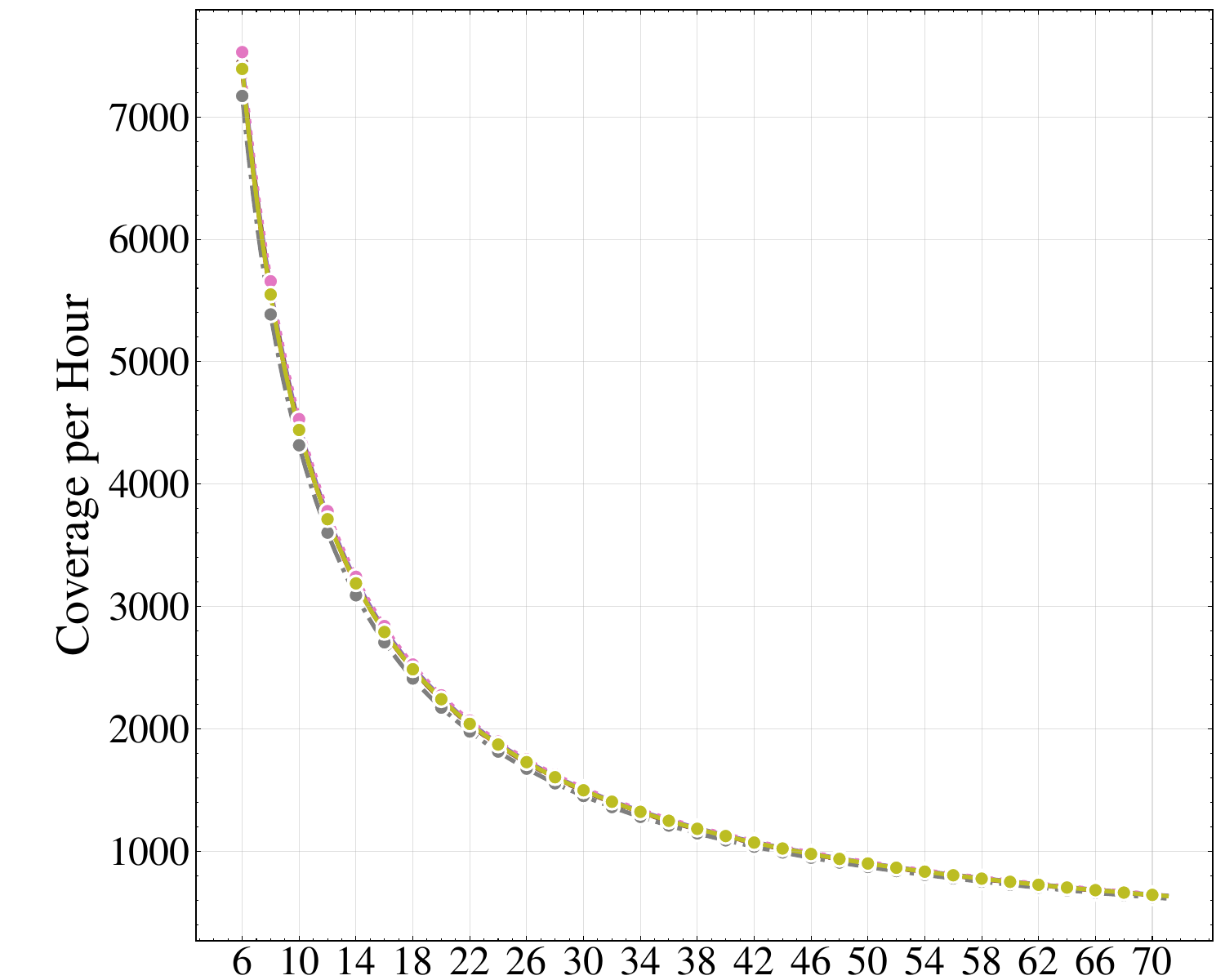} &
\includegraphics[width=0.3\textwidth]{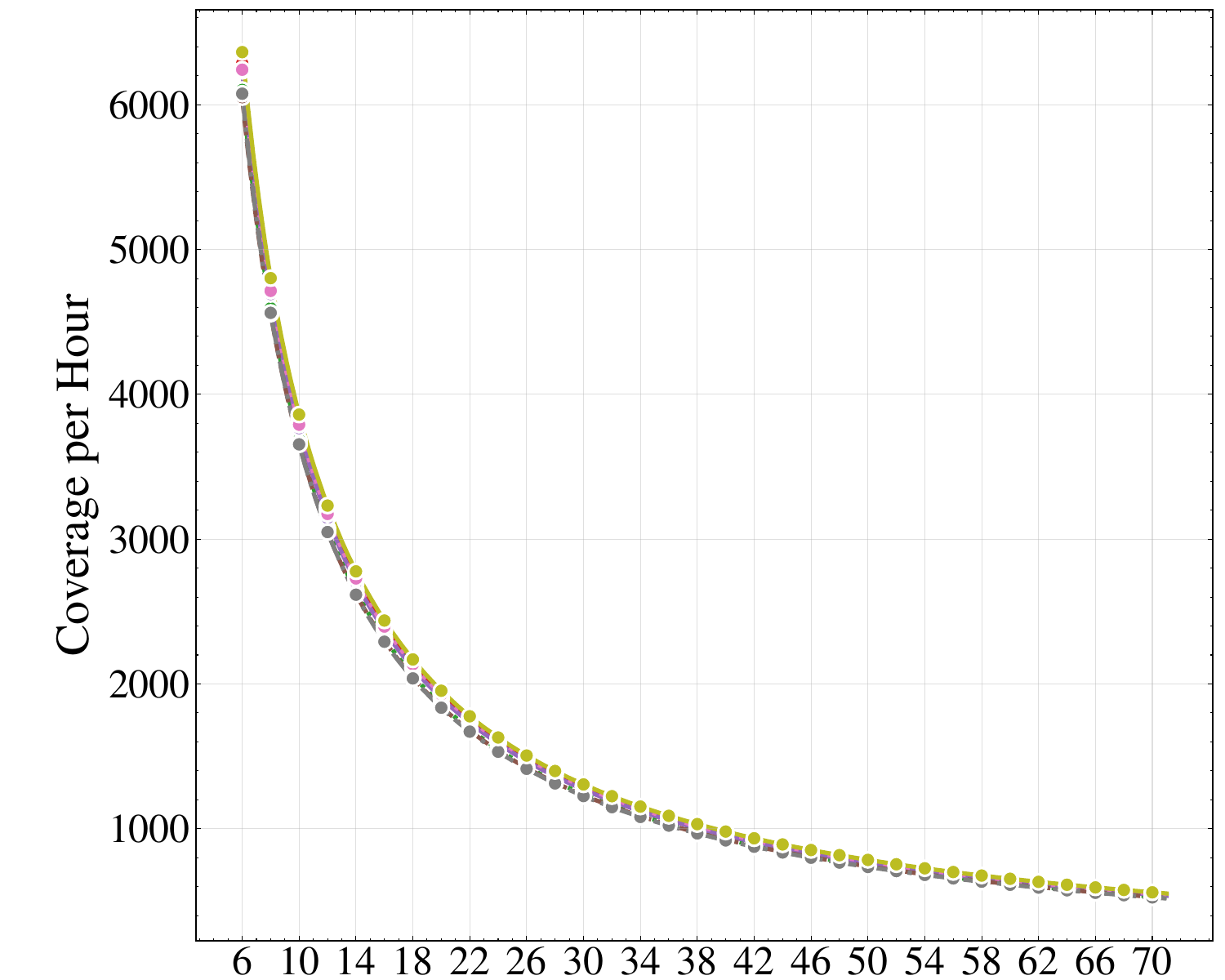} \\
(a) php & (b) poppler \\

\includegraphics[width=0.3\textwidth]{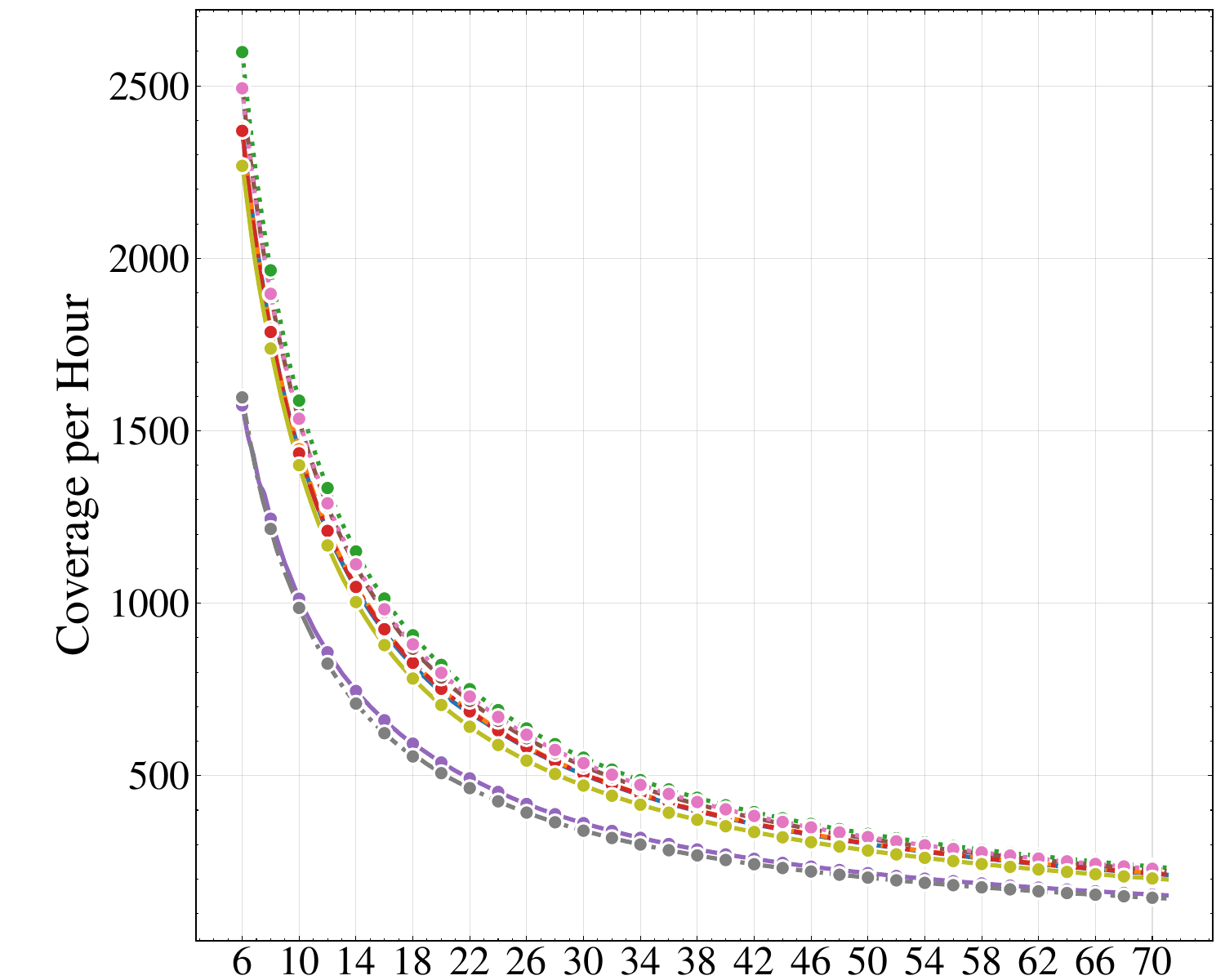} &
\includegraphics[width=0.3\textwidth]{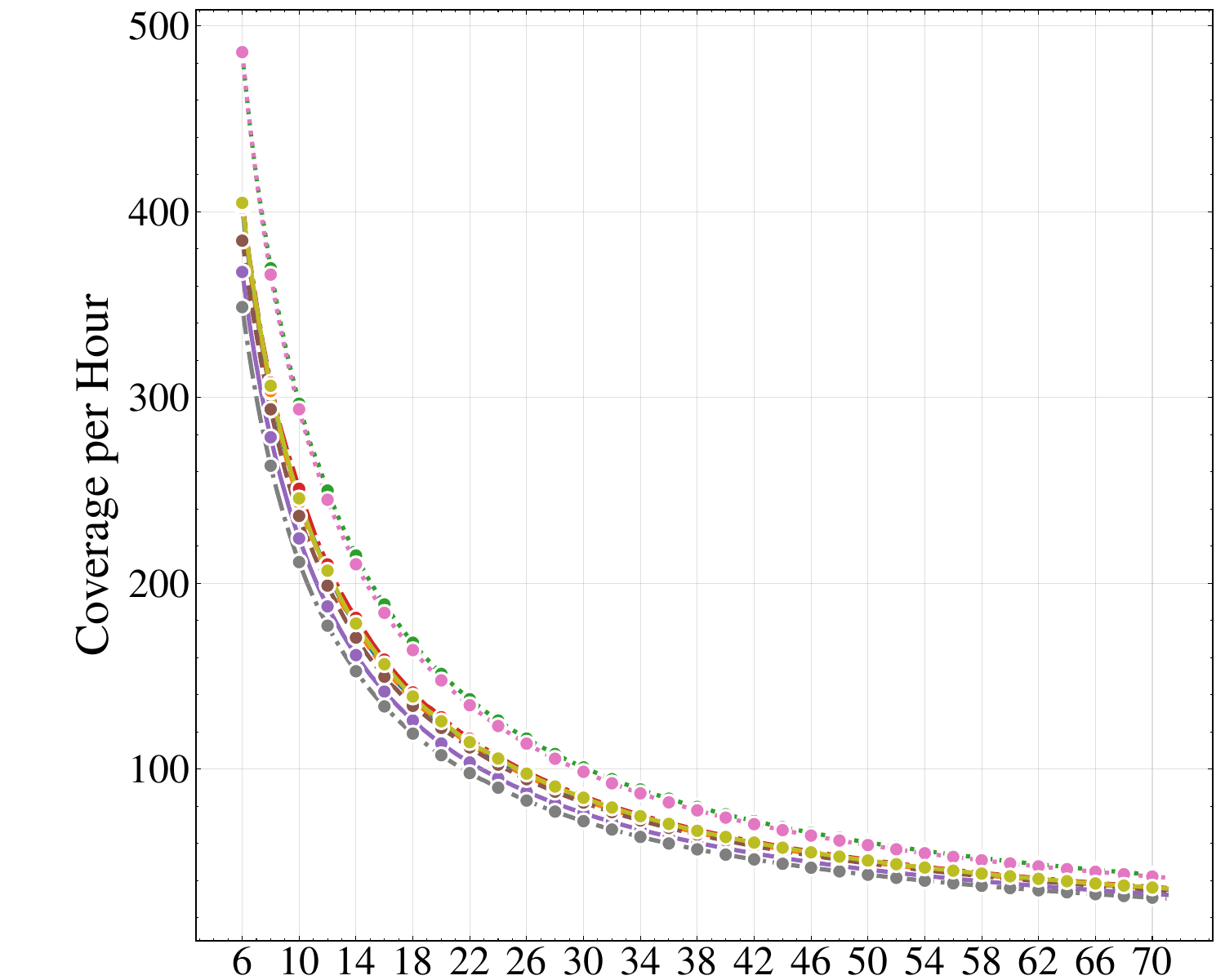} \\
(c) quickjs & (d) stb \\

\includegraphics[width=0.3\textwidth]{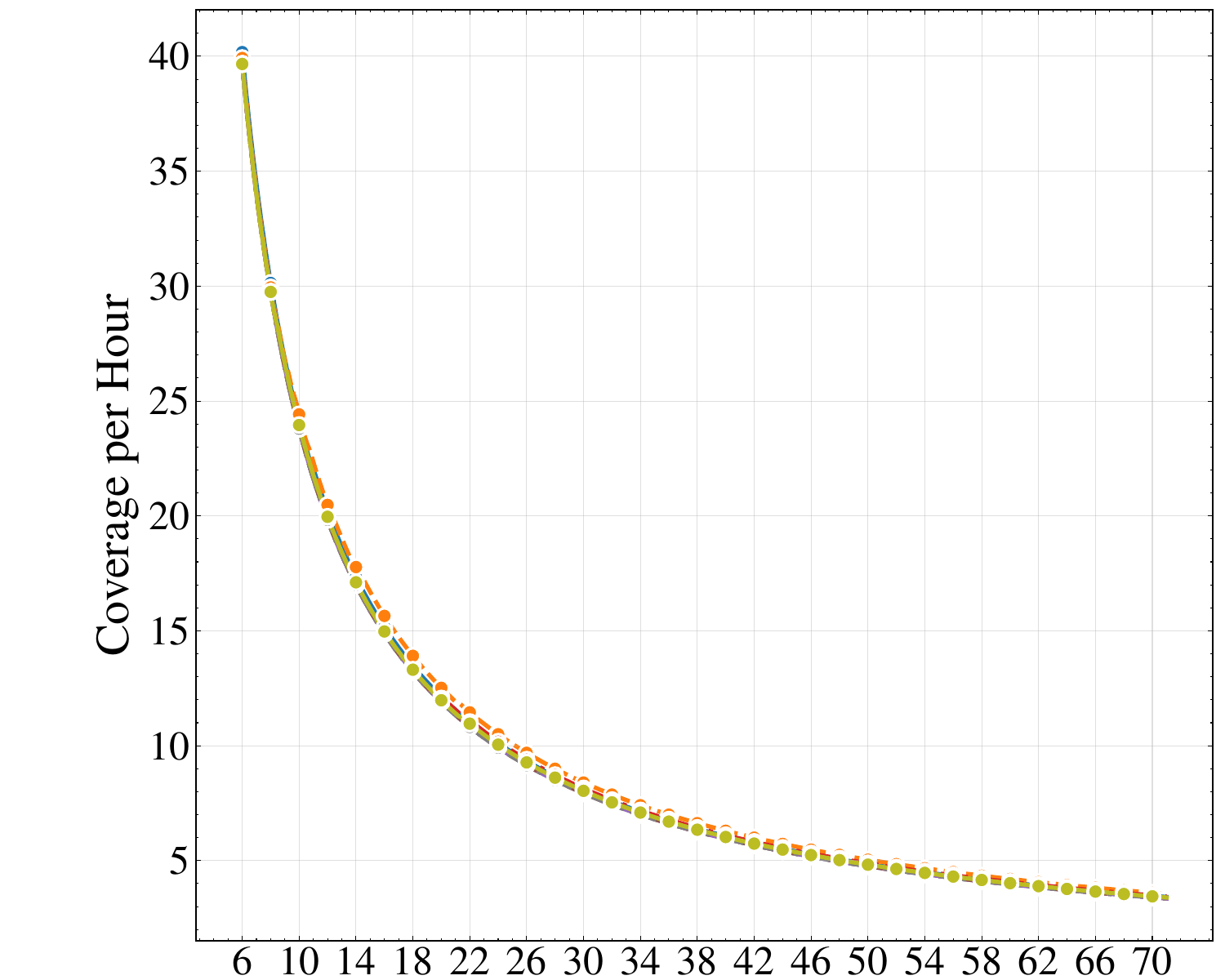} &
\includegraphics[width=0.3\textwidth]{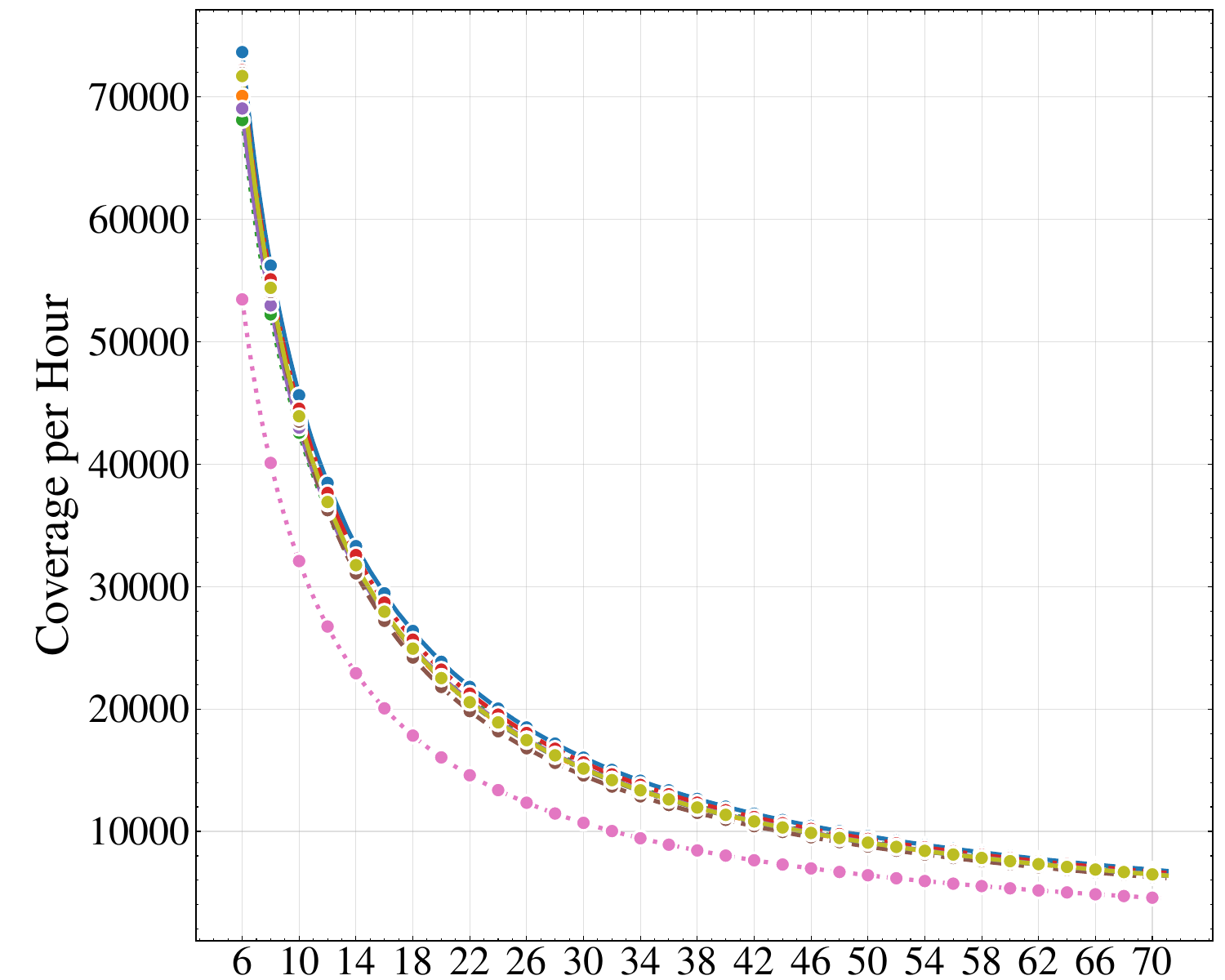} \\
(e) systemd & (f) wireshark \\

\includegraphics[width=0.3\textwidth]{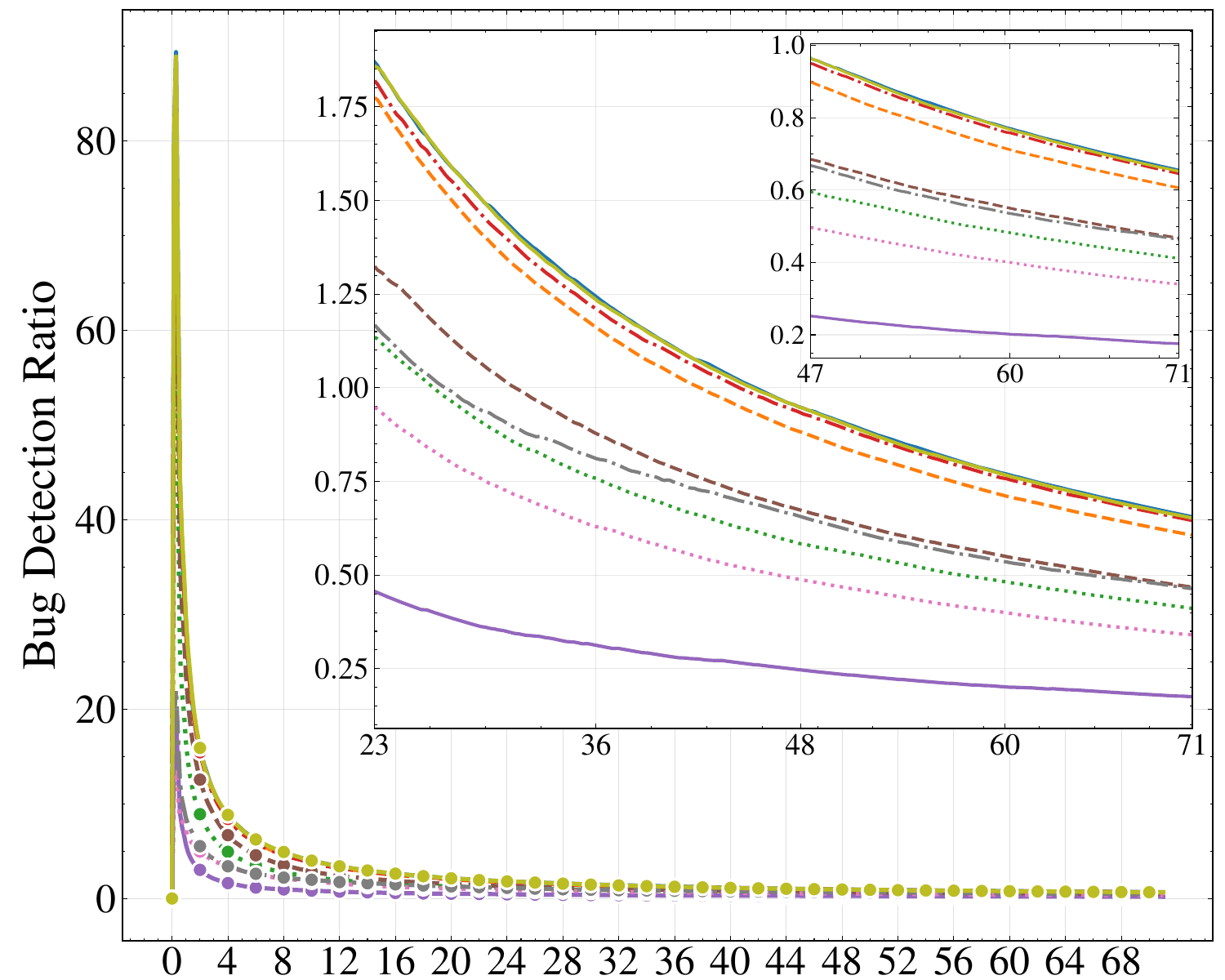} &
\includegraphics[width=0.3\textwidth]{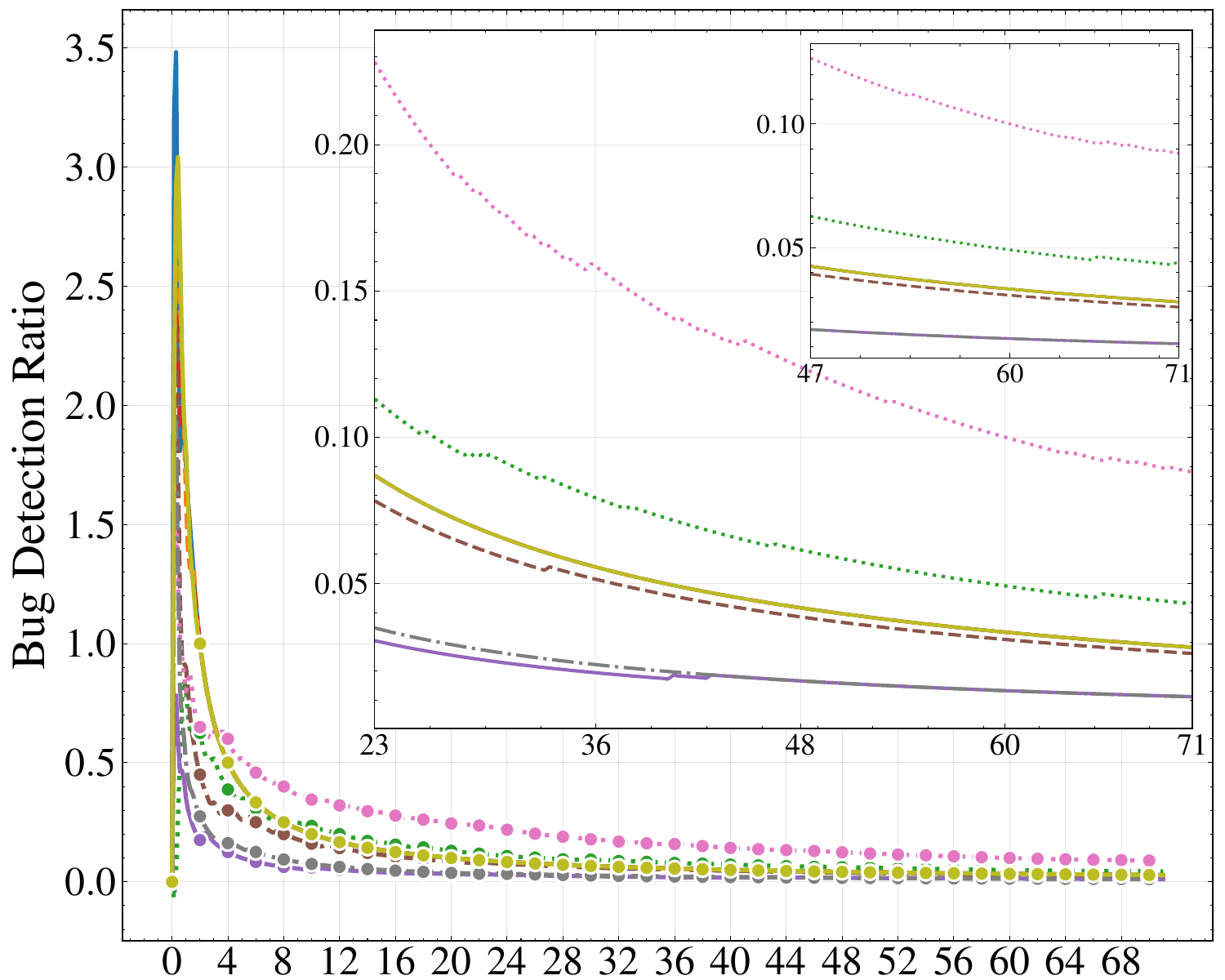} \\
(g) arrow & (h) aspell \\

\includegraphics[width=0.3\textwidth]{figs/data/parent_figure/combined_rates_ffmpeg_ffmpeg_demuxer_fuzzer_06.pdf} &
\includegraphics[width=0.3\textwidth]{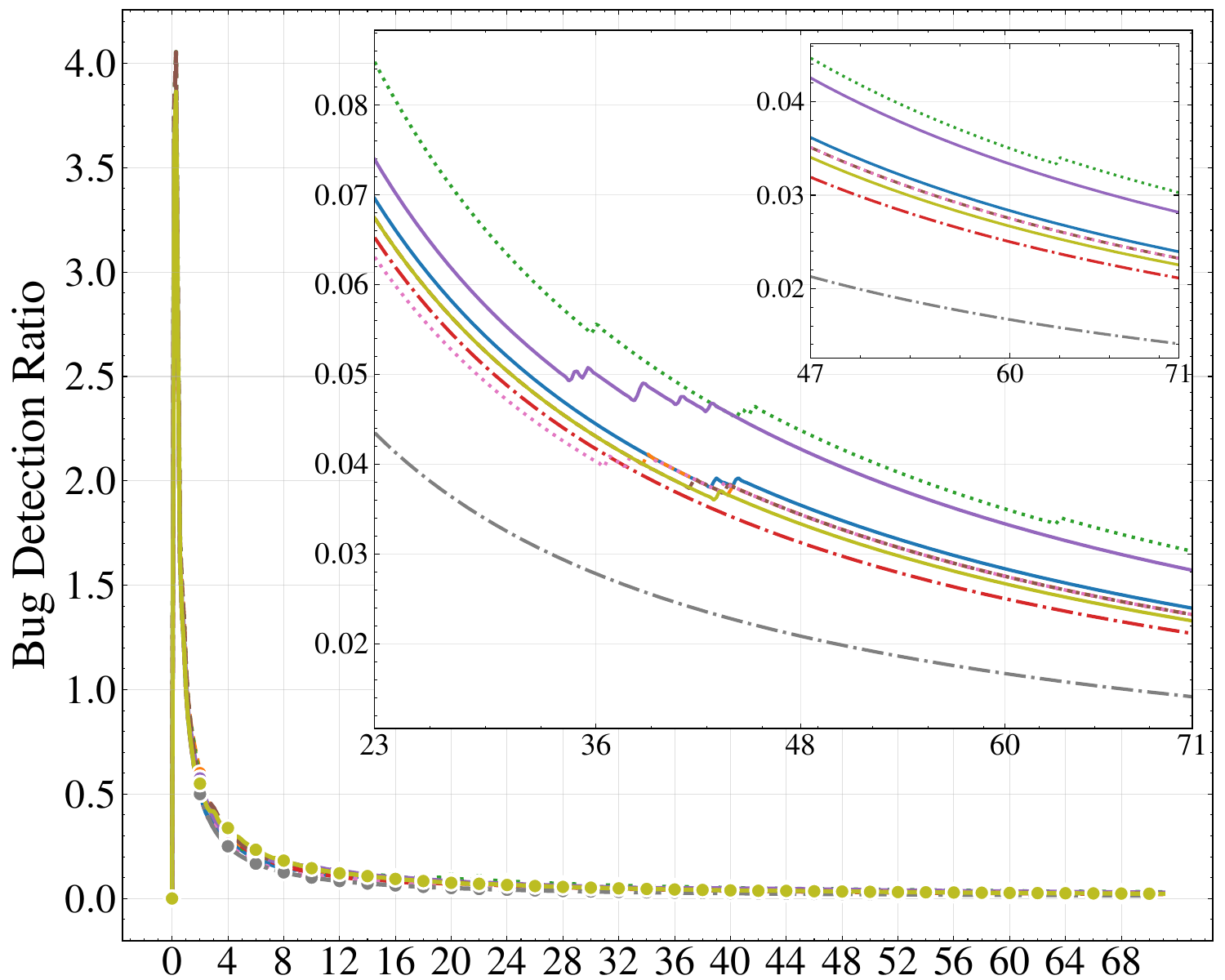} \\
(i) ffmpeg & (j) grok \\

\includegraphics[width=0.3\textwidth]{figs/data/parent_figure/combined_rates_harfbuzz_hb-subset-fuzzer_06.pdf} &
\includegraphics[width=0.3\textwidth]{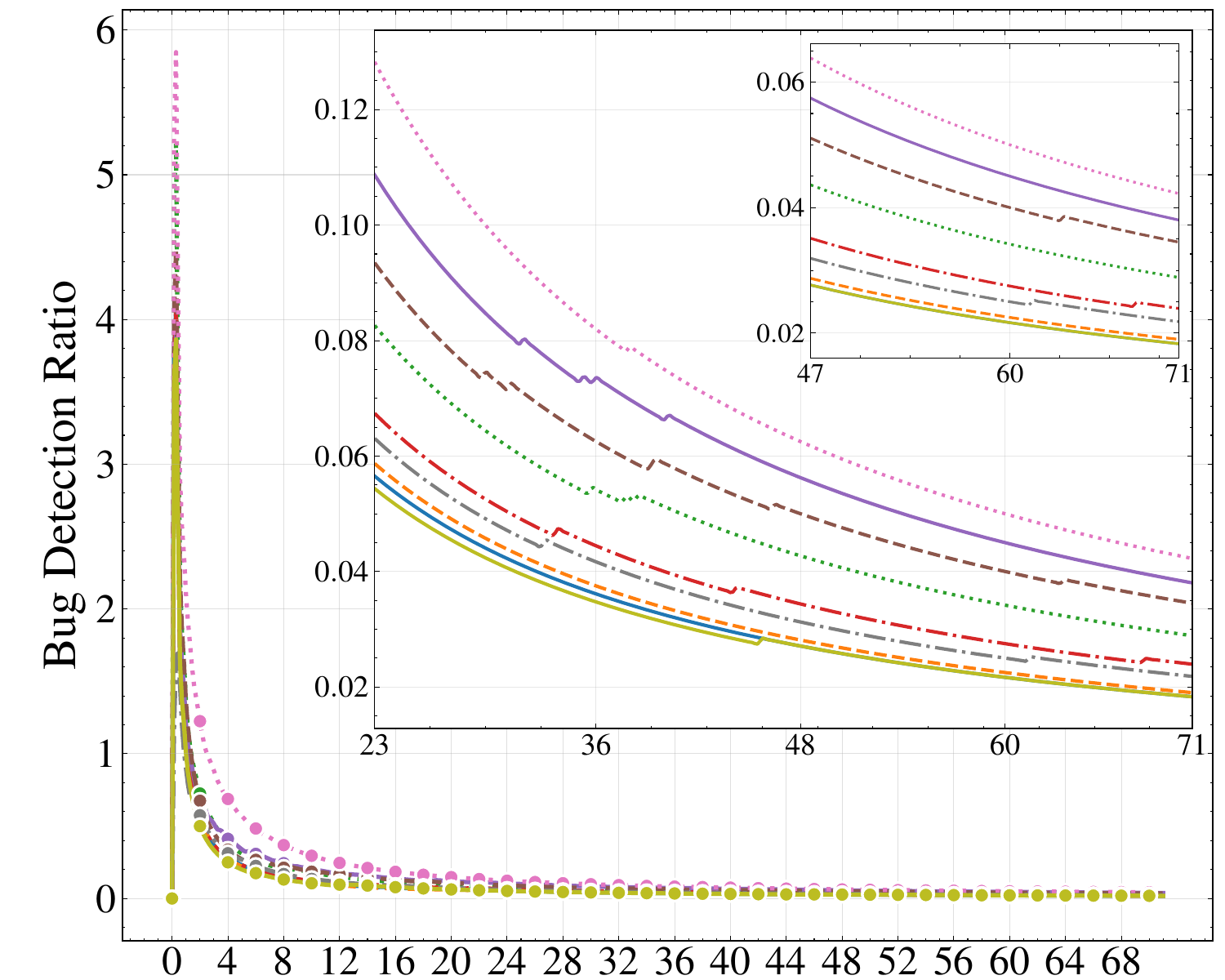} \\
(k) harfbuzz & (l) libgit2 \\
\end{tabular}

\caption{Code coverage ratio (3×2 layout) and bug detection ratio (3×2 layout) .}
\label{fig:merged_cov_rate_6x2}
\end{figure}

\begin{figure}[H]
\centering\footnotesize
\setlength{\tabcolsep}{1pt}

\begin{tabular}{cc}

\includegraphics[width=0.3\textwidth]{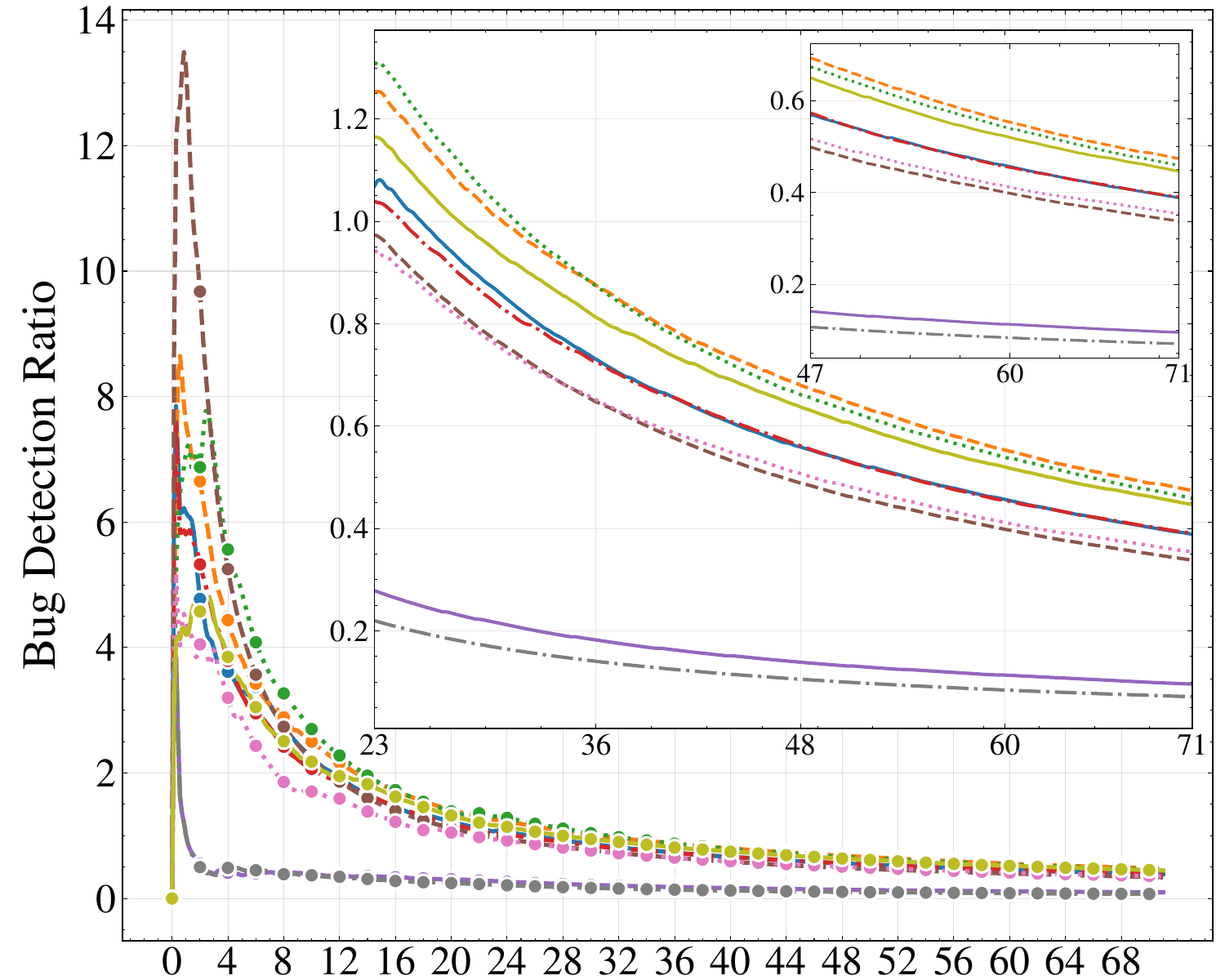} &
\includegraphics[width=0.3\textwidth]{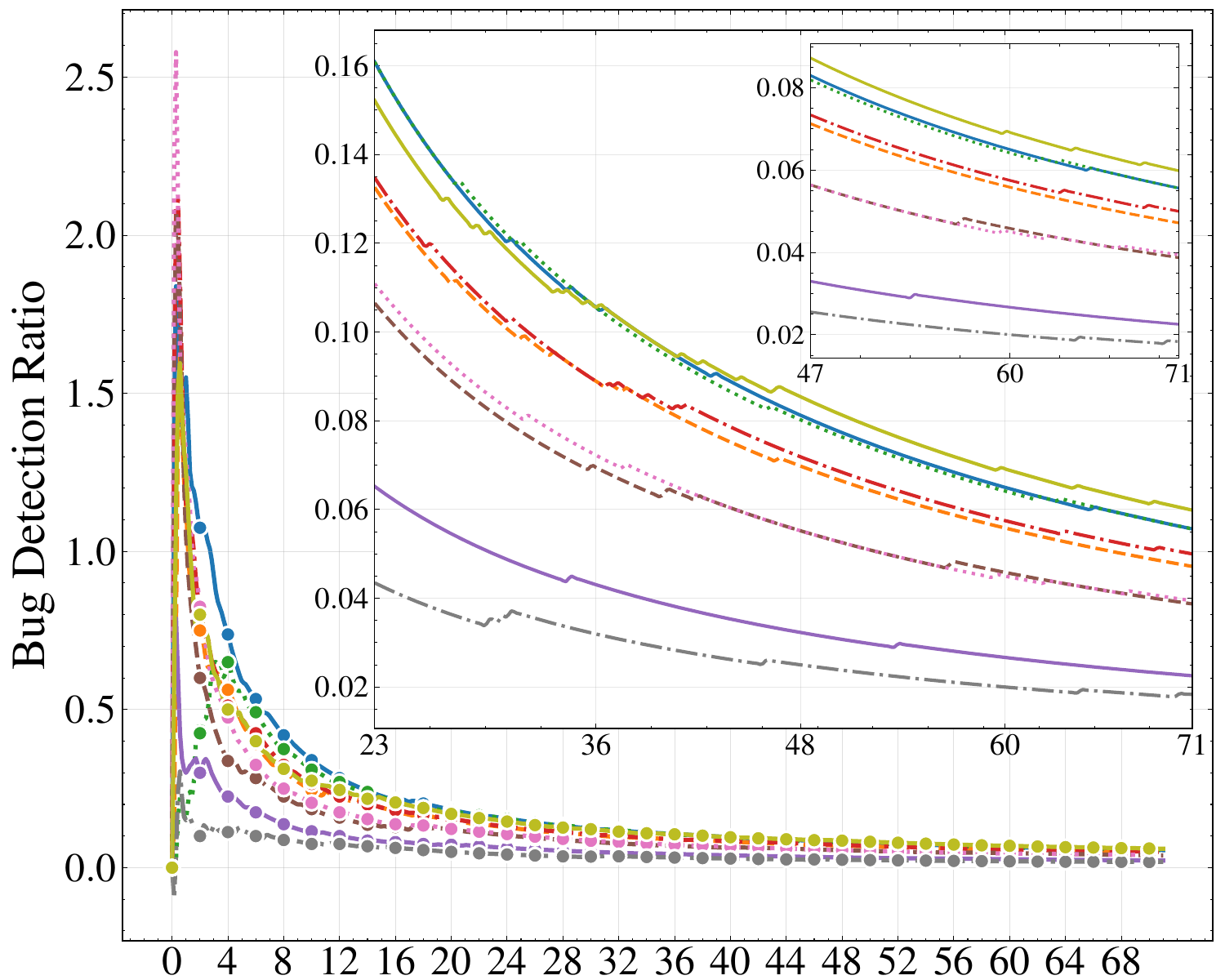} \\
(a) libhevc & (b) libhtp \\

\includegraphics[width=0.3\textwidth]{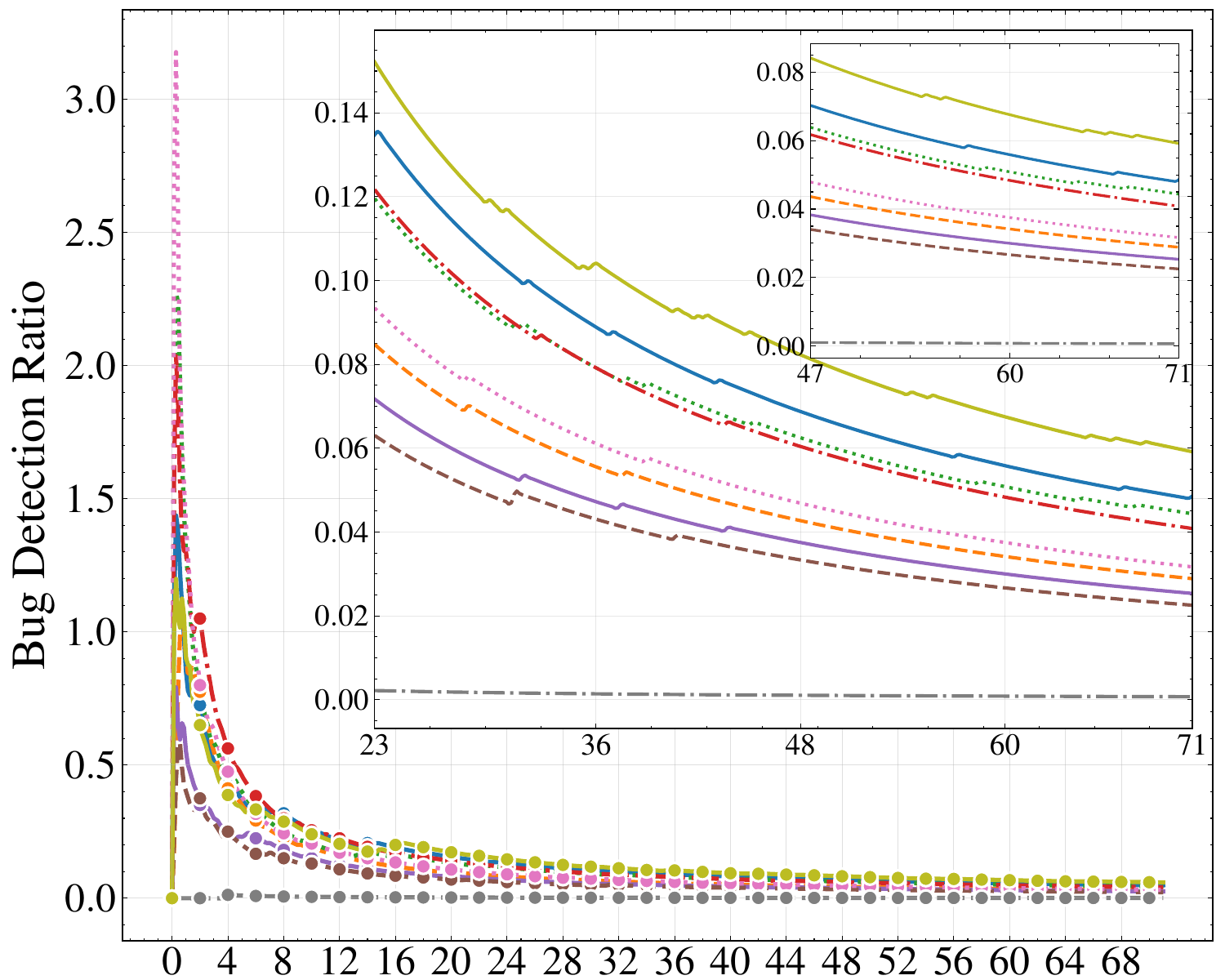} &
\includegraphics[width=0.3\textwidth]{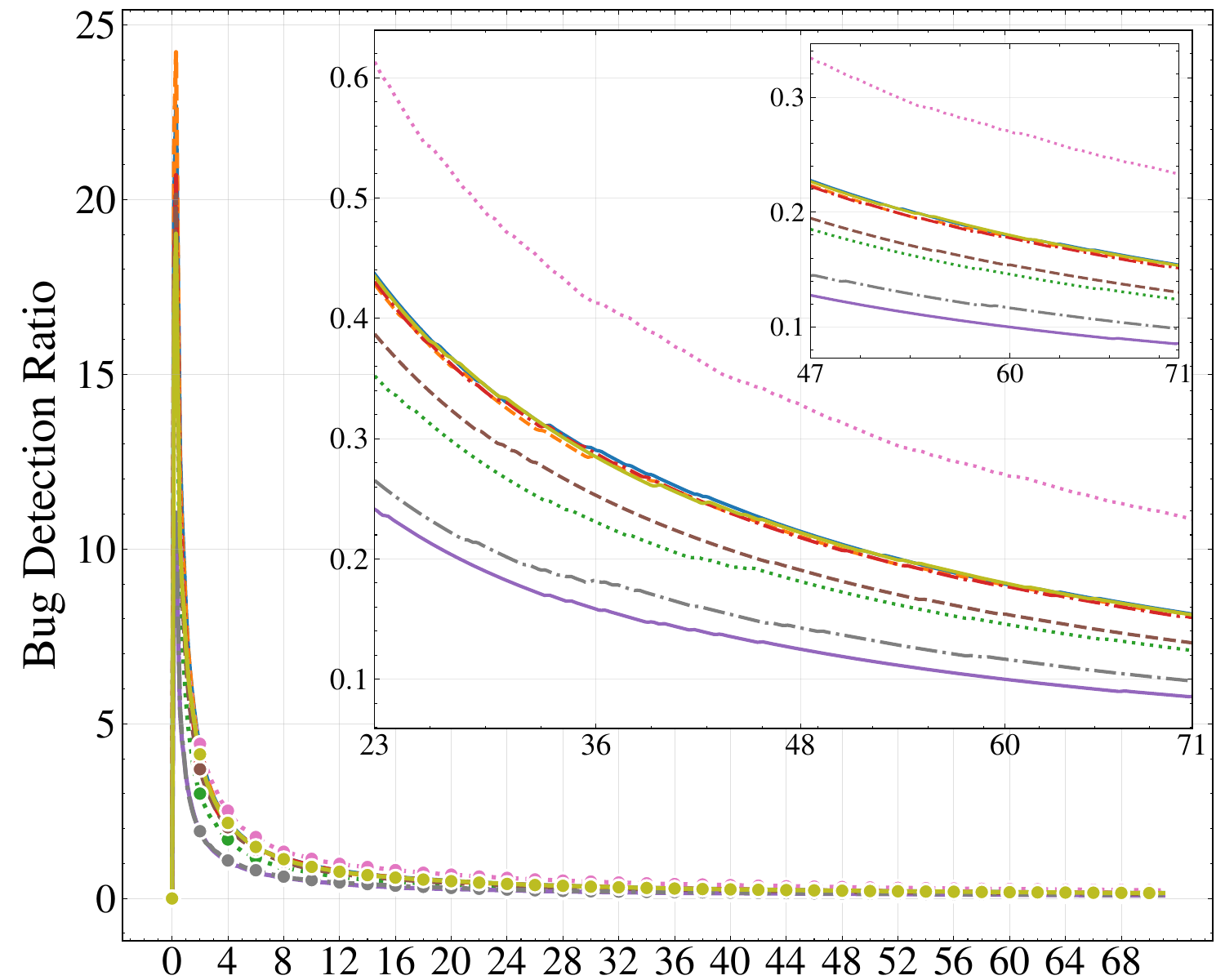} \\
(c) libxml2 & (d) matio \\

\includegraphics[width=0.3\textwidth]{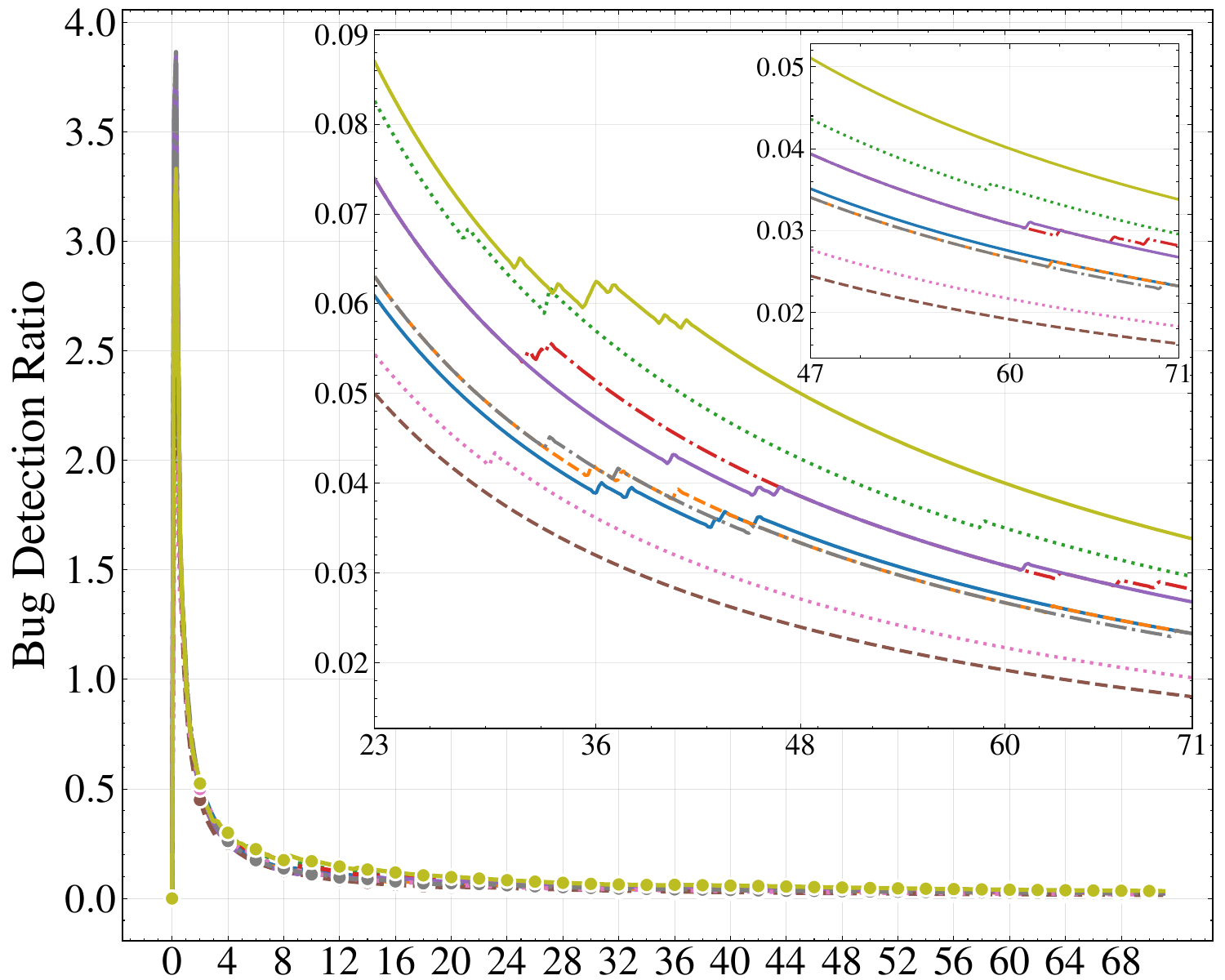} &
\includegraphics[width=0.3\textwidth]{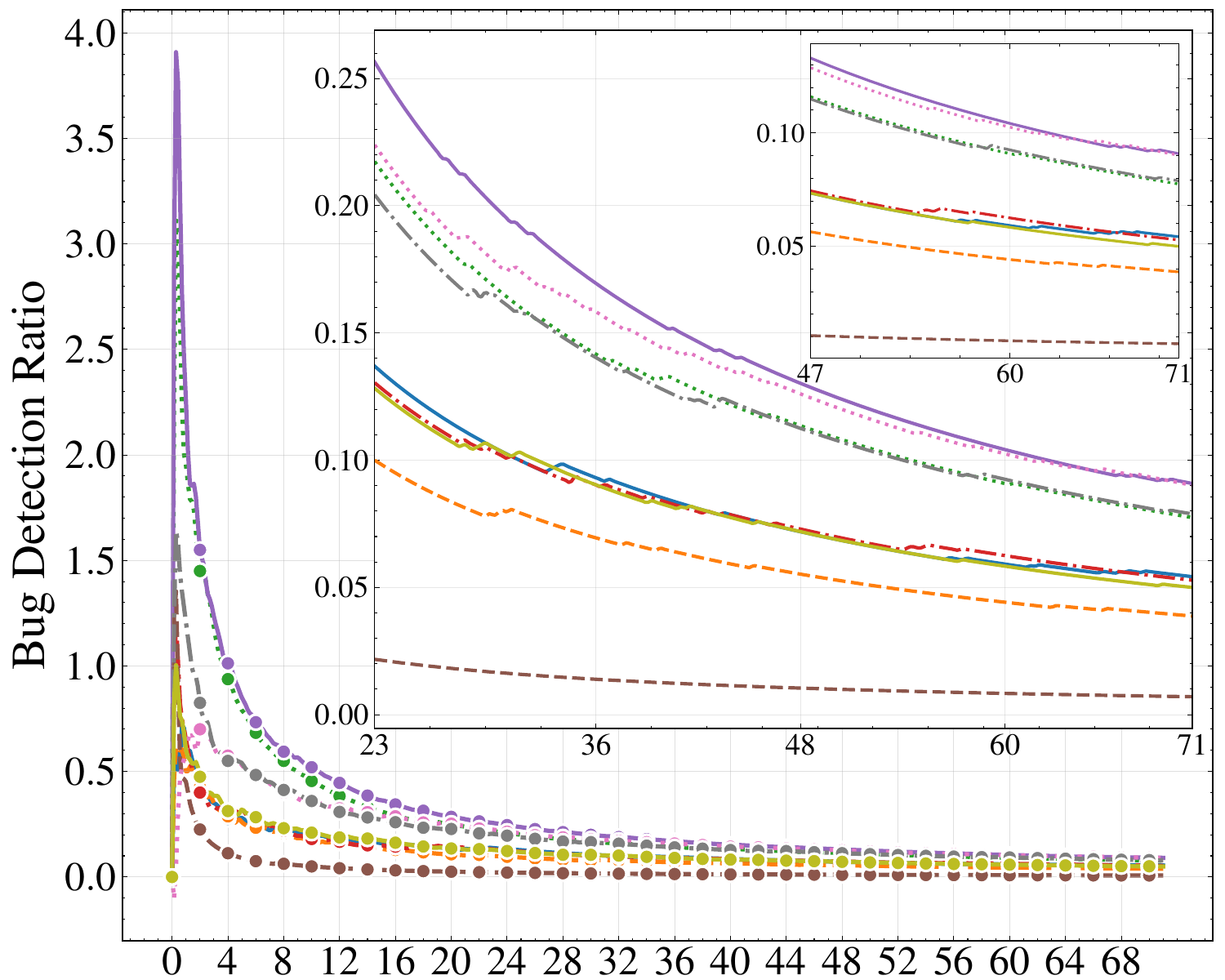} \\
(e) njs & (f) openh264 \\

\includegraphics[width=0.3\textwidth]{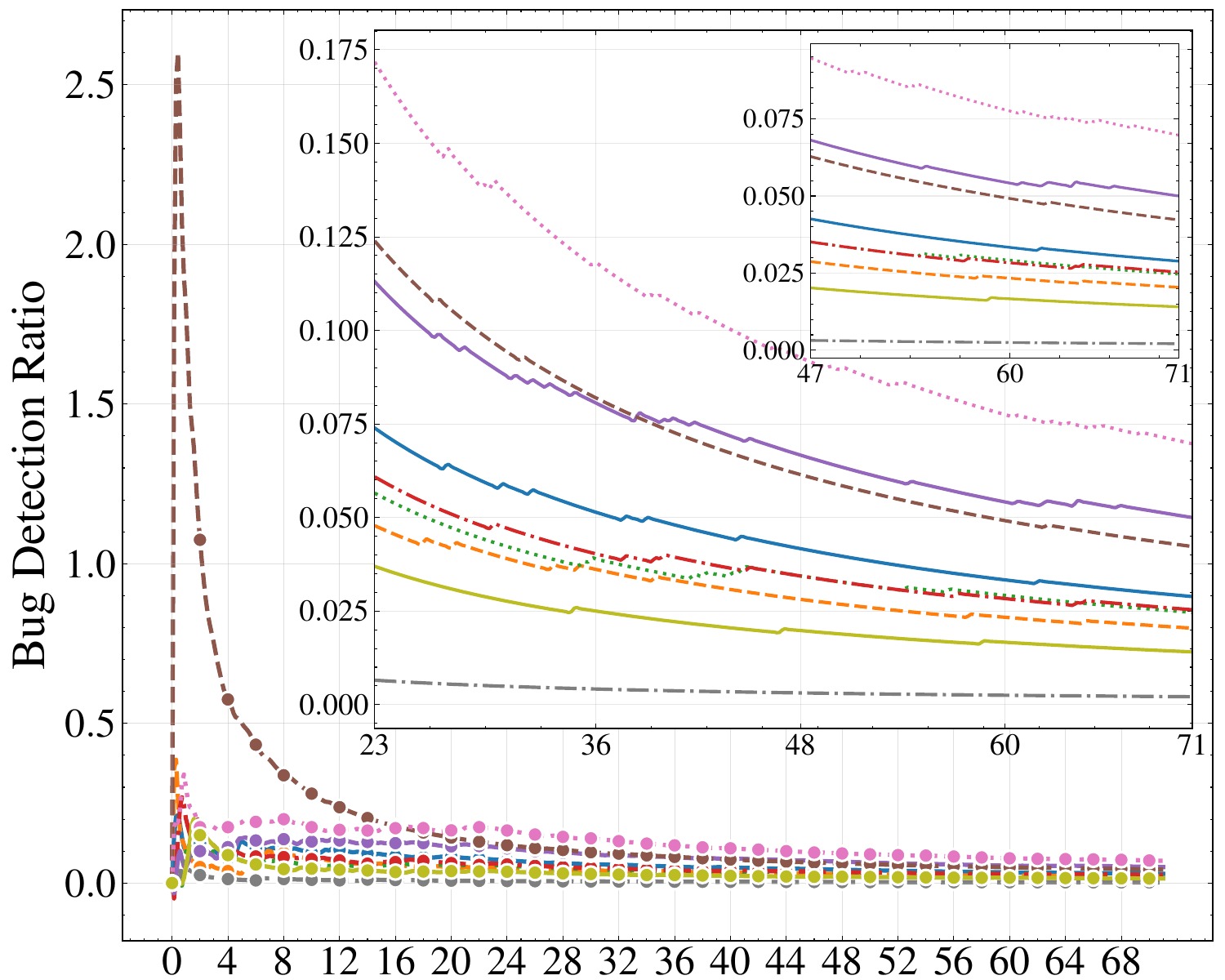} &
\includegraphics[width=0.3\textwidth]{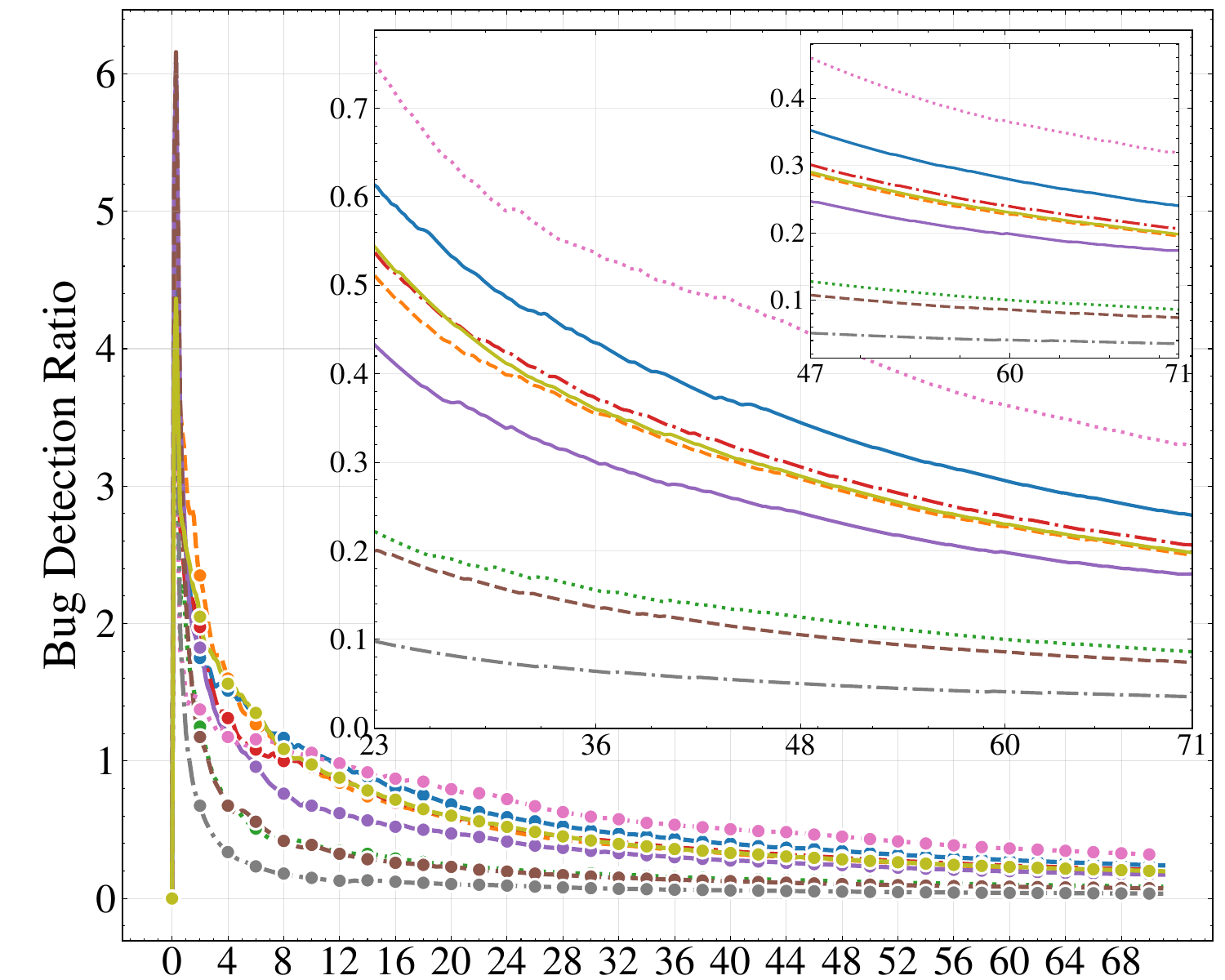} \\
(g) php & (h) poppler \\

\includegraphics[width=0.3\textwidth]{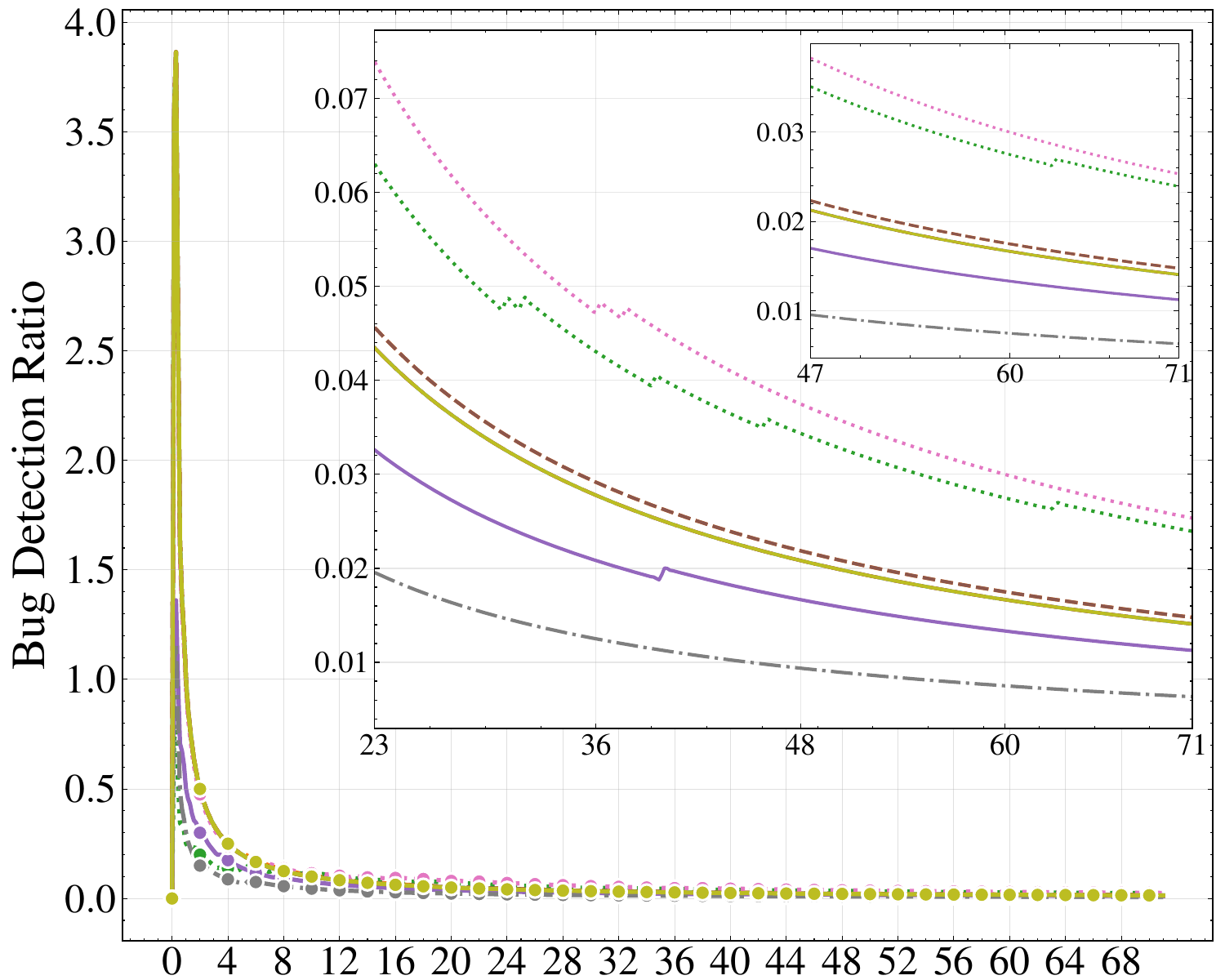} &
\includegraphics[width=0.3\textwidth]{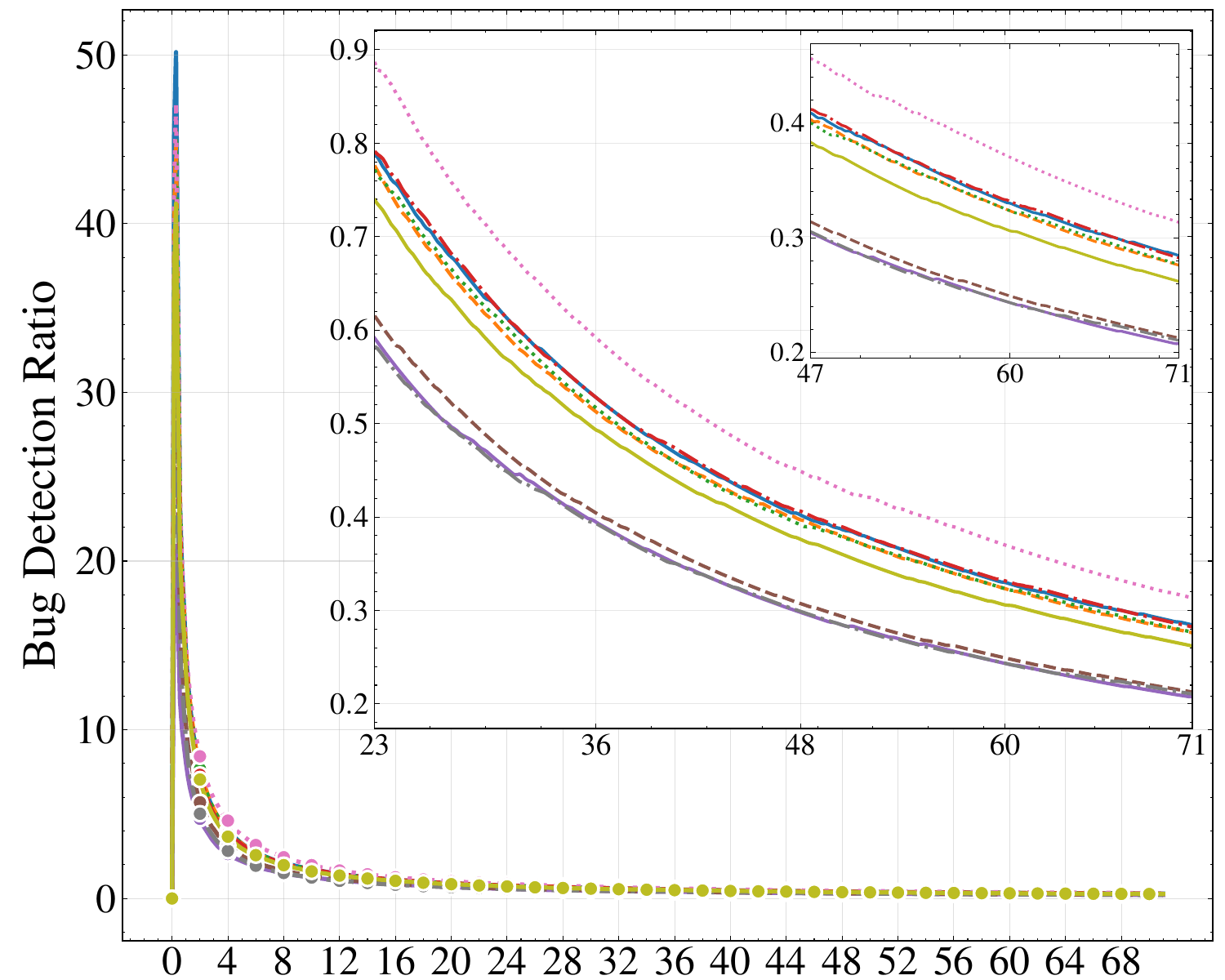} \\
(i) quickjs & (j) stb \\

\includegraphics[width=0.3\textwidth]{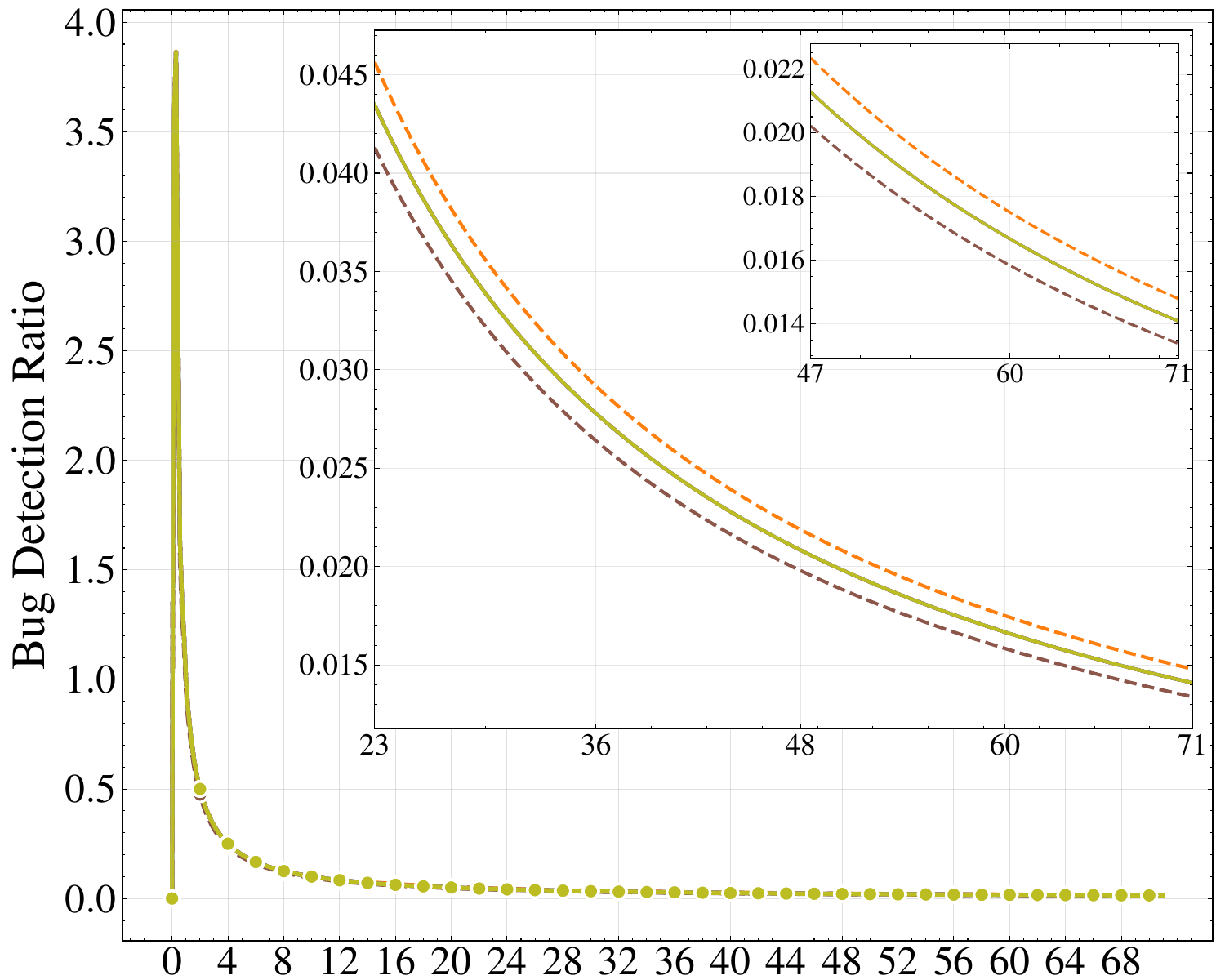} &
\includegraphics[width=0.3\textwidth]{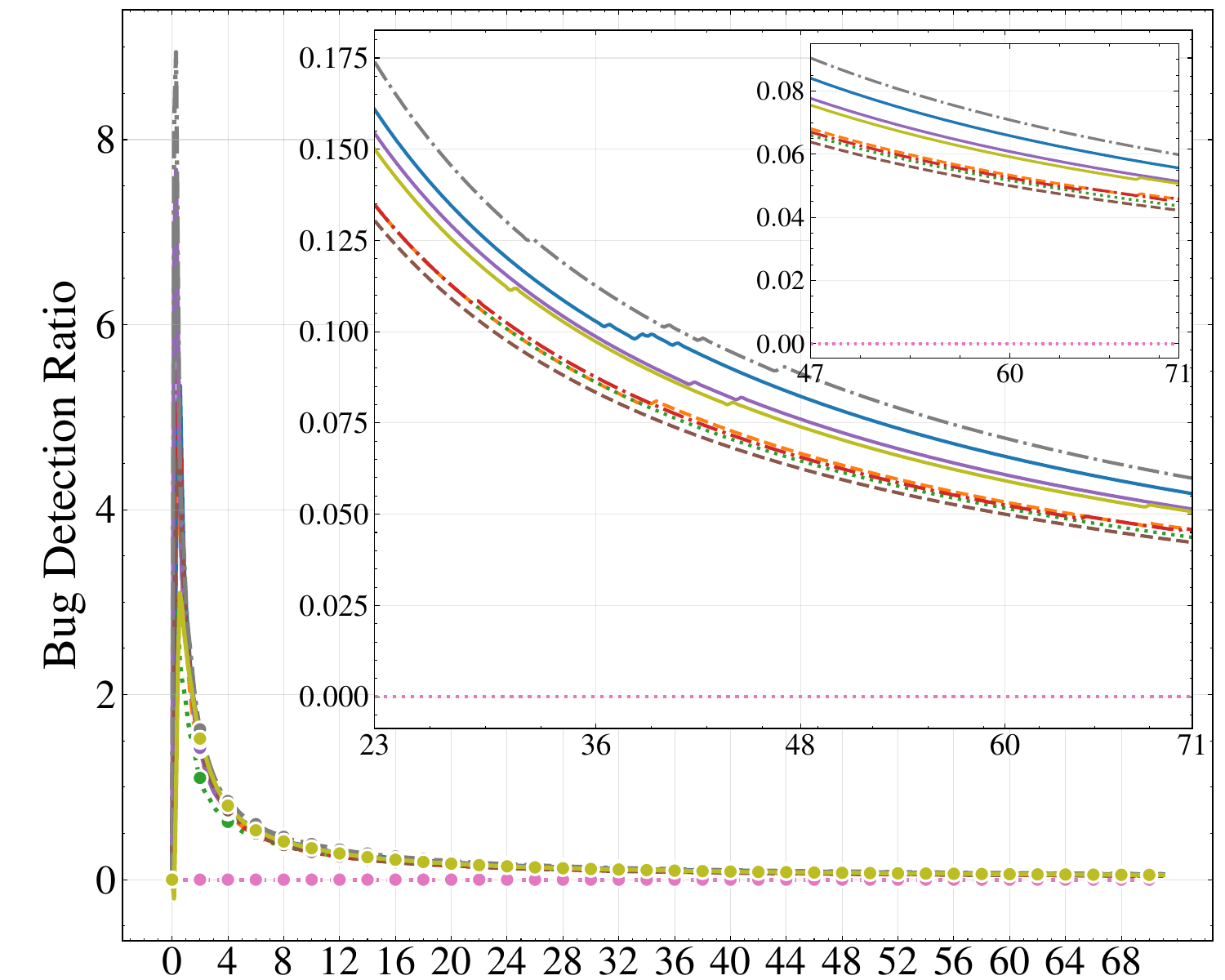} \\
(k) systemd & (l) wireshark \\
\end{tabular}

\caption{Bug detection ratio (6×2 layout).}
\label{fig:rates_combined_6x2}
\end{figure}

\newpage
\appendix
\onecolumn


\newpage
\section*{NeurIPS Paper Checklist}
\begin{enumerate}

\item {\bf Claims}
    \item[] Question: Do the main claims made in the abstract and introduction accurately reflect the paper's contributions and scope?
    \item[] Answer: \answerYes{} 
    \item[] Justification: The contributions and scope are accurately reflected in the abstract and Section 1. 
    \item[] Guidelines:
    \begin{itemize}
        \item The answer \answerNA{} means that the abstract and introduction do not include the claims made in the paper.
        \item The abstract and/or introduction should clearly state the claims made, including the contributions made in the paper and important assumptions and limitations. A \answerNo{} or \answerNA{} answer to this question will not be perceived well by the reviewers. 
        \item The claims made should match theoretical and experimental results, and reflect how much the results can be expected to generalize to other settings. 
        \item It is fine to include aspirational goals as motivation as long as it is clear that these goals are not attained by the paper. 
    \end{itemize}

\item {\bf Limitations}
    \item[] Question: Does the paper discuss the limitations of the work performed by the authors?
    \item[] Answer: \answerYes{}
    \item[] Justification: The assumptions are clearly presented and justified in Section 2 and 4. 
    \item[] Guidelines:
    \begin{itemize}
        \item The answer \answerNA{} means that the paper has no limitation while the answer \answerNo{} means that the paper has limitations, but those are not discussed in the paper. 
        \item The authors are encouraged to create a separate ``Limitations'' section in their paper.
        \item The paper should point out any strong assumptions and how robust the results are to violations of these assumptions (e.g., independence assumptions, noiseless settings, model well-specification, asymptotic approximations only holding locally). The authors should reflect on how these assumptions might be violated in practice and what the implications would be.
        \item The authors should reflect on the scope of the claims made, e.g., if the approach was only tested on a few datasets or with a few runs. In general, empirical results often depend on implicit assumptions, which should be articulated.
        \item The authors should reflect on the factors that influence the performance of the approach. For example, a facial recognition algorithm may perform poorly when image resolution is low or images are taken in low lighting. Or a speech-to-text system might not be used reliably to provide closed captions for online lectures because it fails to handle technical jargon.
        \item The authors should discuss the computational efficiency of the proposed algorithms and how they scale with dataset size.
        \item If applicable, the authors should discuss possible limitations of their approach to address problems of privacy and fairness.
        \item While the authors might fear that complete honesty about limitations might be used by reviewers as grounds for rejection, a worse outcome might be that reviewers discover limitations that aren't acknowledged in the paper. The authors should use their best judgment and recognize that individual actions in favor of transparency play an important role in developing norms that preserve the integrity of the community. Reviewers will be specifically instructed to not penalize honesty concerning limitations.
    \end{itemize}

\item {\bf Theory assumptions and proofs}
    \item[] Question: For each theoretical result, does the paper provide the full set of assumptions and a complete (and correct) proof?
    \item[] Answer: \answerYes{}
    \item[] Justification: Assumptions are presented in Section 2 and 4, and all proof are provided in Appendix. 
    \item[] Guidelines:
    \begin{itemize}
        \item The answer \answerNA{} means that the paper does not include theoretical results. 
        \item All the theorems, formulas, and proofs in the paper should be numbered and cross-referenced.
        \item All assumptions should be clearly stated or referenced in the statement of any theorems.
        \item The proofs can either appear in the main paper or the supplemental material, but if they appear in the supplemental material, the authors are encouraged to provide a short proof sketch to provide intuition. 
        \item Inversely, any informal proof provided in the core of the paper should be complemented by formal proofs provided in appendix or supplemental material.
        \item Theorems and Lemmas that the proof relies upon should be properly referenced. 
    \end{itemize}

    \item {\bf Experimental result reproducibility}
    \item[] Question: Does the paper fully disclose all the information needed to reproduce the main experimental results of the paper to the extent that it affects the main claims and/or conclusions of the paper (regardless of whether the code and data are provided or not)?
    \item[] Answer: \answerYes{}
    \item[] Justification: The experiment details are presented in Section 5 and Appendix; the code is also provided in the supplementary material. 
    \item[] Guidelines:
    \begin{itemize}
        \item The answer \answerNA{} means that the paper does not include experiments.
        \item If the paper includes experiments, a \answerNo{} answer to this question will not be perceived well by the reviewers: Making the paper reproducible is important, regardless of whether the code and data are provided or not.
        \item If the contribution is a dataset and\slash or model, the authors should describe the steps taken to make their results reproducible or verifiable. 
        \item Depending on the contribution, reproducibility can be accomplished in various ways. For example, if the contribution is a novel architecture, describing the architecture fully might suffice, or if the contribution is a specific model and empirical evaluation, it may be necessary to either make it possible for others to replicate the model with the same dataset, or provide access to the model. In general. releasing code and data is often one good way to accomplish this, but reproducibility can also be provided via detailed instructions for how to replicate the results, access to a hosted model (e.g., in the case of a large language model), releasing of a model checkpoint, or other means that are appropriate to the research performed.
        \item While NeurIPS does not require releasing code, the conference does require all submissions to provide some reasonable avenue for reproducibility, which may depend on the nature of the contribution. For example
        \begin{enumerate}
            \item If the contribution is primarily a new algorithm, the paper should make it clear how to reproduce that algorithm.
            \item If the contribution is primarily a new model architecture, the paper should describe the architecture clearly and fully.
            \item If the contribution is a new model (e.g., a large language model), then there should either be a way to access this model for reproducing the results or a way to reproduce the model (e.g., with an open-source dataset or instructions for how to construct the dataset).
            \item We recognize that reproducibility may be tricky in some cases, in which case authors are welcome to describe the particular way they provide for reproducibility. In the case of closed-source models, it may be that access to the model is limited in some way (e.g., to registered users), but it should be possible for other researchers to have some path to reproducing or verifying the results.
        \end{enumerate}
    \end{itemize}

\item {\bf Open access to data and code}
    \item[] Question: Does the paper provide open access to the data and code, with sufficient instructions to faithfully reproduce the main experimental results, as described in supplemental material?
    \item[] Answer:\answerYes{}
    \item[] Justification: The code is provided in the supplementary material and also in the anonymous open-source GitHub link.  
    \item[] Guidelines:
    \begin{itemize}
        \item The answer \answerNA{} means that paper does not include experiments requiring code.
        \item Please see the NeurIPS code and data submission guidelines (\url{https://neurips.cc/public/guides/CodeSubmissionPolicy}) for more details.
        \item While we encourage the release of code and data, we understand that this might not be possible, so \answerNo{} is an acceptable answer. Papers cannot be rejected simply for not including code, unless this is central to the contribution (e.g., for a new open-source benchmark).
        \item The instructions should contain the exact command and environment needed to run to reproduce the results. See the NeurIPS code and data submission guidelines (\url{https://neurips.cc/public/guides/CodeSubmissionPolicy}) for more details.
        \item The authors should provide instructions on data access and preparation, including how to access the raw data, preprocessed data, intermediate data, and generated data, etc.
        \item The authors should provide scripts to reproduce all experimental results for the new proposed method and baselines. If only a subset of experiments are reproducible, they should state which ones are omitted from the script and why.
        \item At submission time, to preserve anonymity, the authors should release anonymized versions (if applicable).
        \item Providing as much information as possible in supplemental material (appended to the paper) is recommended, but including URLs to data and code is permitted.
    \end{itemize}

\item {\bf Experimental setting/details}
    \item[] Question: Does the paper specify all the training and test details (e.g., data splits, hyperparameters, how they were chosen, type of optimizer) necessary to understand the results?
    \item[] Answer: \answerYes{}
    \item[] Justification: Experiment details are provided in Section 5 and Appendix. 
    \item[] Guidelines:
    \begin{itemize}
        \item The answer \answerNA{} means that the paper does not include experiments.
        \item The experimental setting should be presented in the core of the paper to a level of detail that is necessary to appreciate the results and make sense of them.
        \item The full details can be provided either with the code, in appendix, or as supplemental material.
    \end{itemize}

\item {\bf Experiment statistical significance}
    \item[] Question: Does the paper report error bars suitably and correctly defined or other appropriate information about the statistical significance of the experiments?
    \item[] Answer: \answerYes{} 
    \item[] Justification: We report error bars suitably and correctly defined in the paper.
    \item[] Guidelines:
    \begin{itemize}
        \item The answer \answerNA{} means that the paper does not include experiments.
        \item The authors should answer \answerYes{} if the results are accompanied by error bars, confidence intervals, or statistical significance tests, at least for the experiments that support the main claims of the paper.
        \item The factors of variability that the error bars are capturing should be clearly stated (for example, train/test split, initialization, random drawing of some parameter, or overall run with given experimental conditions).
        \item The method for calculating the error bars should be explained (closed form formula, call to a library function, bootstrap, etc.)
        \item The assumptions made should be given (e.g., Normally distributed errors).
        \item It should be clear whether the error bar is the standard deviation or the standard error of the mean.
        \item It is OK to report 1-sigma error bars, but one should state it. The authors should preferably report a 2-sigma error bar than state that they have a 96\% CI, if the hypothesis of Normality of errors is not verified.
        \item For asymmetric distributions, the authors should be careful not to show in tables or figures symmetric error bars that would yield results that are out of range (e.g., negative error rates).
        \item If error bars are reported in tables or plots, the authors should explain in the text how they were calculated and reference the corresponding figures or tables in the text.
    \end{itemize}

\item {\bf Experiments compute resources}
    \item[] Question: For each experiment, does the paper provide sufficient information on the computer resources (type of compute workers, memory, time of execution) needed to reproduce the experiments?
    \item[] Answer: \answerYes{}
    \item[] Justification: The details of the compute resources are provided in Appendix. 
    \item[] Guidelines:
    \begin{itemize}
        \item The answer \answerNA{} means that the paper does not include experiments.
        \item The paper should indicate the type of compute workers CPU or GPU, internal cluster, or cloud provider, including relevant memory and storage.
        \item The paper should provide the amount of compute required for each of the individual experimental runs as well as estimate the total compute. 
        \item The paper should disclose whether the full research project required more compute than the experiments reported in the paper (e.g., preliminary or failed experiments that didn't make it into the paper). 
    \end{itemize}
    
\item {\bf Code of ethics}
    \item[] Question: Does the research conducted in the paper conform, in every respect, with the NeurIPS Code of Ethics \url{https://neurips.cc/public/EthicsGuidelines}?
    \item[] Answer:  \answerYes{}
    \item[] Justification: The research fully aligns with the NeurIPS Code of Ethics. 
    \item[] Guidelines:
    \begin{itemize}
        \item The answer \answerNA{} means that the authors have not reviewed the NeurIPS Code of Ethics.
        \item If the authors answer \answerNo, they should explain the special circumstances that require a deviation from the Code of Ethics.
        \item The authors should make sure to preserve anonymity (e.g., if there is a special consideration due to laws or regulations in their jurisdiction).
    \end{itemize}

\item {\bf Broader impacts}
    \item[] Question: Does the paper discuss both potential positive societal impacts and negative societal impacts of the work performed?
    \item[] Answer: \answerYes{}
    \item[] Justification: Broader impacts are provided in Appendix. 
    \item[] Guidelines:
    \begin{itemize}
        \item The answer \answerNA{} means that there is no societal impact of the work performed.
        \item If the authors answer \answerNA{} or \answerNo, they should explain why their work has no societal impact or why the paper does not address societal impact.
        \item Examples of negative societal impacts include potential malicious or unintended uses (e.g., disinformation, generating fake profiles, surveillance), fairness considerations (e.g., deployment of technologies that could make decisions that unfairly impact specific groups), privacy considerations, and security considerations.
        \item The conference expects that many papers will be foundational research and not tied to particular applications, let alone deployments. However, if there is a direct path to any negative applications, the authors should point it out. For example, it is legitimate to point out that an improvement in the quality of generative models could be used to generate Deepfakes for disinformation. On the other hand, it is not needed to point out that a generic algorithm for optimizing neural networks could enable people to train models that generate Deepfakes faster.
        \item The authors should consider possible harms that could arise when the technology is being used as intended and functioning correctly, harms that could arise when the technology is being used as intended but gives incorrect results, and harms following from (intentional or unintentional) misuse of the technology.
        \item If there are negative societal impacts, the authors could also discuss possible mitigation strategies (e.g., gated release of models, providing defenses in addition to attacks, mechanisms for monitoring misuse, mechanisms to monitor how a system learns from feedback over time, improving the efficiency and accessibility of ML).
    \end{itemize}
    
\item {\bf Safeguards}
    \item[] Question: Does the paper describe safeguards that have been put in place for responsible release of data or models that have a high risk for misuse (e.g., pre-trained language models, image generators, or scraped datasets)?
    \item[] Answer:  \answerNA{} 
    \item[] Justification: The paper poses no such risks.
    \item[] Guidelines:
    \begin{itemize}
        \item The answer \answerNA{} means that the paper poses no such risks.
        \item Released models that have a high risk for misuse or dual-use should be released with necessary safeguards to allow for controlled use of the model, for example by requiring that users adhere to usage guidelines or restrictions to access the model or implementing safety filters. 
        \item Datasets that have been scraped from the Internet could pose safety risks. The authors should describe how they avoided releasing unsafe images.
        \item We recognize that providing effective safeguards is challenging, and many papers do not require this, but we encourage authors to take this into account and make a best faith effort.
    \end{itemize}

\item {\bf Licenses for existing assets}
    \item[] Question: Are the creators or original owners of assets (e.g., code, data, models), used in the paper, properly credited and are the license and terms of use explicitly mentioned and properly respected?
    \item[] Answer: \answerYes{}
    \item[] Justification: All third-party code, tools, and benchmark assets used in this work are publicly available, properly cited, and used in accordance with their stated licenses and terms of use.
    \item[] Guidelines:
    \begin{itemize}
        \item The answer \answerNA{} means that the paper does not use existing assets.
        \item The authors should cite the original paper that produced the code package or dataset.
        \item The authors should state which version of the asset is used and, if possible, include a URL.
        \item The name of the license (e.g., CC-BY 4.0) should be included for each asset.
        \item For scraped data from a particular source (e.g., website), the copyright and terms of service of that source should be provided.
        \item If assets are released, the license, copyright information, and terms of use in the package should be provided. For popular datasets, \url{paperswithcode.com/datasets} has curated licenses for some datasets. Their licensing guide can help determine the license of a dataset.
        \item For existing datasets that are re-packaged, both the original license and the license of the derived asset (if it has changed) should be provided.
        \item If this information is not available online, the authors are encouraged to reach out to the asset's creators.
    \end{itemize}

\item {\bf New assets}
    \item[] Question: Are new assets introduced in the paper well documented and is the documentation provided alongside the assets?
    \item[] Answer: \answerYes{}
    \item[] Justification: The code produced by the paper are well documented and provided in the supplementary material. 
    \item[] Guidelines:
    \begin{itemize}
        \item The answer \answerNA{} means that the paper does not release new assets.
        \item Researchers should communicate the details of the dataset\slash code\slash model as part of their submissions via structured templates. This includes details about training, license, limitations, etc. 
        \item The paper should discuss whether and how consent was obtained from people whose asset is used.
        \item At submission time, remember to anonymize your assets (if applicable). You can either create an anonymized URL or include an anonymized zip file.
    \end{itemize}

\item {\bf Crowdsourcing and research with human subjects}
    \item[] Question: For crowdsourcing experiments and research with human subjects, does the paper include the full text of instructions given to participants and screenshots, if applicable, as well as details about compensation (if any)? 
    \item[] Answer: \answerNA{}
    \item[] Justification: The paper does not involve crowdsourcing nor research with human subjects.
    \begin{itemize}
        \item The answer \answerNA{} means that the paper does not involve crowdsourcing nor research with human subjects.
        \item Including this information in the supplemental material is fine, but if the main contribution of the paper involves human subjects, then as much detail as possible should be included in the main paper. 
        \item According to the NeurIPS Code of Ethics, workers involved in data collection, curation, or other labor should be paid at least the minimum wage in the country of the data collector. 
    \end{itemize}

\item {\bf Institutional review board (IRB) approvals or equivalent for research with human subjects}
    \item[] Question: Does the paper describe potential risks incurred by study participants, whether such risks were disclosed to the subjects, and whether Institutional Review Board (IRB) approvals (or an equivalent approval/review based on the requirements of your country or institution) were obtained?
    \item[] Answer: \answerNA{} 
    \item[] Justification: The paper does not involve crowdsourcing nor research with human subjects.
    \item[] Guidelines:
    \begin{itemize}
        \item The answer \answerNA{} means that the paper does not involve crowdsourcing nor research with human subjects.
        \item Depending on the country in which research is conducted, IRB approval (or equivalent) may be required for any human subjects research. If you obtained IRB approval, you should clearly state this in the paper. 
        \item We recognize that the procedures for this may vary significantly between institutions and locations, and we expect authors to adhere to the NeurIPS Code of Ethics and the guidelines for their institution. 
        \item For initial submissions, do not include any information that would break anonymity (if applicable), such as the institution conducting the review.
    \end{itemize}

\item {\bf Declaration of LLM usage}
    \item[] Question: Does the paper describe the usage of LLMs if it is an important, original, or non-standard component of the core methods in this research? Note that if the LLM is used only for writing, editing, or formatting purposes and does \emph{not} impact the core methodology, scientific rigor, or originality of the research, declaration is not required.
    \item[] Answer: \answerNA{} 
    \item[] Justification: The core method development in this research does not involve LLMs as any important, original, or non-standard components.
    \item[] Guidelines:
    \begin{itemize}
        \item The answer \answerNA{} means that the core method development in this research does not involve LLMs as any important, original, or non-standard components.
        \item Please refer to our LLM policy in the NeurIPS handbook for what should or should not be described.
    \end{itemize}

\end{enumerate}

\end{document}